\documentclass[11pt,article]{article}

\usepackage{geometry}
\usepackage[square,comma,numbers,sort&compress]{natbib}
\usepackage{adjustbox}
\usepackage{makecell}
\usepackage{amsthm} %for newtheorem
\usepackage{mathtools} %for \coloneqq
\usepackage{pgfplots}
\pgfplotsset{compat=1.18} 
\usepackage{pgfplotstable}
\usepackage{silence}
\usepackage[hyphens]{url}
\usepackage[breaklinks,colorlinks]{hyperref}
\usepackage[dvipsnames]{xcolor}
\hypersetup{citecolor=blue,linkcolor=blue}
\usepackage{amsmath,amsopn,amssymb}
\usepackage{subfigure}
\usepackage{endnotes,microtype,xspace,graphicx,fancyvrb,fvextra,multirow}
\usepackage{booktabs}
\usepackage{array,underscore,relsize}
\usepackage[T1]{fontenc}
\usepackage{times}
\usepackage{fancyhdr,lastpage}
\usepackage{enumitem}
\usepackage[labelfont=bf,font=small,skip=3pt,tableposition=bottom]{caption}
\usepackage{pifont}
\usepackage{fp}
\usepackage{siunitx}

\usepackage{minted}
\usepackage{listings}
\usepackage{placeins}
\usepackage{wrapfig}

\usepackage[linesnumbered,ruled,vlined]{algorithm2e}%

\usepackage{balance}

\usepackage[acronym,nowarn]{glossaries}
\glsdisablehyper

\usepackage{colortbl}

\usepackage{pifont}
\usepackage{array}
\usepackage{adjustbox}
\usepackage{tabularx}

\usepackage[most]{tcolorbox}

\usepackage{makeidx}

\usepackage[normal]{threeparttable}

\usepackage{tikz}
\usepackage{pgf-pie}
\usepackage{pgfplots}

\newacronym{dgf}{DGF}{directed graybox fuzzer}
\newacronym{icfg}{ICFG}{interprocedural control flow graph}
\newacronym{tte}{TTE}{time to exposure}
\newacronym{uc}{UC}{unique bug crashes}

\newcommand{\DGF}{\gls{dgf}\xspace}
\newcommand{\DGFsCap}{\Glspl{dgf}\xspace}
\newcommand{\ICFG}{\gls{icfg}\xspace}
\newcommand{\UC}{\gls{uc}\xspace}

\newcommand{\TTE}{\gls{tte}\xspace}

\newcommand{\sys}{\mbox{\textsc{Atlantis}}\xspace}

\newcommand{\syscrsweb}{\mbox{\textsc{crs-webserver}}\xspace}
\newcommand{\syscpmgr}{\mbox{\textsc{cp-manager}}\xspace}

\newcommand{\syspatching}{\sys-Patching\xspace}

\newcommand{\syscontroller}{\mbox{\textsc{Controller}}\xspace}
\newcommand{\syscorpusselector}{\mbox{\textsc{CorpusSelector}}\xspace}
\newcommand{\systaskscheduler}{\mbox{\textsc{TaskScheduler}}\xspace}

\newcommand{\sysworker}{\mbox{\textsc{Worker}}\xspace}
\newcommand{\sysharnessbuilder}{\mbox{\textsc{HarnessBuilder}}\xspace}
\newcommand{\libagents}{\mbox{\textsc{LibAgents}}\xspace}
\newcommand{\deepgen}{\mbox{\textsc{DeepGenerator}}\xspace}
\newcommand{\libdeepgen}{\mbox{\textsc{LibDeepGen}}\xspace}
\newcommand{\sysllmaugmentedmutator}{\mbox{\textsc{LLMAugmentedMutator}}\xspace}
\newcommand{\sysfuzzermanager}{\mbox{\textsc{FuzzerManager}}\xspace}
\newcommand{\sysbullseye}{\mbox{\textsc{Bullseye}}\xspace}

\newcommand{\sysensembler}{\mbox{\textsc{Ensembler}}\xspace}
\newcommand{\syscrashcollector}{\mbox{\textsc{CrashCollector}}\xspace}
\newcommand{\sysseedscollector}{\mbox{\textsc{SeedsCollector}}\xspace}

\newcommand{\sysuniafl}{\mbox{\textsc{Uniafl}}\xspace}
\newcommand{\sysfuzzdb}{\mbox{\textsc{Fuzzdb}}\xspace}
\newcommand{\sysmlla}{\mbox{\textsc{Mlla}}\xspace}
\newcommand{\libfdp}{\mbox{\textsc{LibFDP}}\xspace}
\newcommand{\retriever}{\mbox{\textsc{Code Retriever}}\xspace}
\newcommand{\functracer}{\mbox{\textsc{Function Tracer}}\xspace}

\newcommand{\syscrete}{\mbox{\textsc{Crete}}\xspace}
\newcommand{\sysmartian}{\mbox{\textsc{Martian}}\xspace}
\newcommand{\sysmr}{\mbox{\textsc{MultiRetrieval}}\xspace}
\newcommand{\sysprism}{\mbox{\textsc{Prism}}\xspace}
\newcommand{\sysvincent}{\mbox{\textsc{Vincent}}\xspace}
\newcommand{\sysclaudelike}{\mbox{\textsc{ClaudeLike}}\xspace}
\newcommand{\syseraser}{\mbox{\textsc{Eraser}}\xspace}
\newcommand{\syscoderovers}{\mbox{\textsc{CodeRover-S}}\xspace}
\newcommand{\sysaider}{\mbox{\textsc{Aider}}\xspace}
\newcommand{\syssweagent}{\mbox{\textsc{SWE-Agent}}\xspace}

\makeatletter
\@ifundefined{remark*}{%
    \newtheorem*{remark*}{Remark}
}{}
\makeatother

\newcommand{\cc}[1]{\mbox{\smaller[0.5]\texttt{#1}}}
\newcommand{\ccw}[1]{{\smaller[0.5]\texttt{#1}}}
\usepackage{expl3}
\ExplSyntaxOn
\cs_new_protected:Npn \tc #1
{
  \tl_set:Nn \l_tmpa_tl {#1}
  \regex_replace_all:nnN { . } { \c{string} \0 } \l_tmpa_tl
  \tl_set:Nx \l_tmpb_tl { \l_tmpa_tl }
  \texttt{{\smaller[0.5] \l_tmpb_tl}}
}
\ExplSyntaxOff

\fvset{fontsize=\scriptsize,xleftmargin=8pt,numbers=left,numbersep=5pt}

\input{code/fmt}

\def\Snospace~{\S{}}

\newcommand{\x}{$\times$\xspace}

\if 0

\fi

\newif\ifdraft\drafttrue
\newif\ifnotes\notestrue
\ifdraft\else\notesfalse\fi

\input{glyphtounicode}
\newcolumntype{R}[1]{>{\raggedleft\let\newline\\\arraybackslash\hspace{0pt}}p{#1}}

\newcommand{\squishlist}{
\begin{itemize}[noitemsep,nolistsep,left=0pt]
  \setlength{\itemsep}{-0pt}
}
\newcommand{\squishend}{
  \end{itemize}
}
\newcommand{\squishendnum}{
  \end{enumerate}
}

\newenvironment{squishitemize}{
  \begin{itemize}[left=0pt]
}{
  \squishend
}

\newenvironment{squishenumerate}{
  \begin{enumerate}[left=0pt]
}{
  \end{enumerate}
}

\usepackage{tikz}
\usetikzlibrary{shapes, arrows.meta, positioning, fit, backgrounds, calc}
\newcommand*\WC[1]{%
\begin{tikzpicture}[baseline=(C.base)]
\node[draw,circle,inner sep=0.2pt](C) {#1};
\end{tikzpicture}}

\usepackage{xstring}
\newcommand{\PP}[1]{
  \vspace{2px}
\noindent{\bf \IfEndWith{#1}{.}{#1}{#1.}}
}

\newcommand{\PN}[1]{
 \vspace{2px}
\noindent{\bf {#1}}
}

\newcommand{\PS}[1]{
 \vspace{2px}
\noindent{\textbf{\textsc{#1}.}}
}

\newcommand{\boxbeg}{%
\vspace{2px}%
\noindent\begin{tabular}{|l|}\hline
\begin{minipage}{\columnwidth}%
\vspace{2px}%
\noindent
}

\newcommand{\boxend}{%
\vspace{2px}%
\end{minipage}\\ \hline
\end{tabular}%
\vspace{-10pt}%
}

\newcommand{\eg}{{\em e.g.,}\xspace}
\newcommand{\ie}{{\em i.e.,}\xspace}
\newcommand{\etc}{{\em etc.}\xspace}

\colorlet{bluekeywords}{blue!90!black}
\colorlet{greencomments}{green!40!black}
\colorlet{redstrings}{red!90!black}

\definecolor{promptbg}{gray}{0.97}
\definecolor{promptframe}{gray}{0.75}
\definecolor{xmltag}{RGB}{0,102,153}
\definecolor{xmlcontent}{RGB}{51,51,51}

\newenvironment{promptcontent}{%
  \fvset{xleftmargin=2pt,numbers=none,numbersep=2pt}%
}{%
}

\newtcolorbox{promptbox}[2][]{%
  colback=promptbg,
  colframe=promptframe,
  fonttitle=\bfseries\small,
  coltitle=black,
  title=#2,
  boxrule=0.5pt,
  left=2pt,
  right=2pt,
  top=2pt,
  bottom=2pt,
  width=\columnwidth,
  before={\par\medskip\noindent},
  after={\par\medskip},
  enhanced,
  arc=3pt,
  #1
}

\newtcolorbox{promptboxwrap}[2][]{%
  colback=promptbg,
  colframe=promptframe,
  fonttitle=\bfseries\small,
  title=#2,
  boxrule=0.5pt,
  left=2pt,
  right=2pt,
  top=2pt,
  bottom=2pt,
  width=\textwidth,
  before={\par\medskip\noindent},
  after={\par\medskip},
  listing only,
  listing options={
    language=XML,
    basicstyle=\ttfamily\footnotesize,
    breaklines=true,
    breakatwhitespace=true,
    columns=flexible,
    showspaces=false,
    showstringspaces=false,
    frame=none,
    numbers=none,
    xleftmargin=0pt,
    xrightmargin=0pt
  },
  #1
}

\newtcolorbox{promptboxinline}[2][]{%
  colback=promptbg,
  colframe=promptframe,
  fonttitle=\bfseries\small,
  coltitle=black,
  title=#2,
  boxrule=0.5pt,
  left=2pt,
  right=2pt,
  top=2pt,
  bottom=2pt,
  breakable,
  width=\columnwidth,
  enhanced,
  arc=3pt,
  #1
}

\newcommand{\Sound}{\textcolor{green}{\checkmark}}
\newcommand{\Fail}{\textcolor{red}{\ding{55}}}
\newcommand*{\Warning}{\textcolor{orange}{\fontencoding{U}\fontfamily{futs}\selectfont\char 66}}

\gdef\\therev{43e468b}
\gdef\\thedate{2025-09-17 21:41:43 -0400}

\begin{document}

\newcommand{\Title}{Team Atlanta}
\newcommand{\Subtitle}{\textsc{Atlantis}: AI-driven Threat Localization, Analysis, and Triage Intelligence System}

%% sync with the website
\newcommand{\Author}{Taesoo Kim,
% Atlantis-Multilang
HyungSeok Han,
Soyeon Park,
Dae R. Jeong,
Dohyeok Kim,
Dongkwan Kim,
Eunsoo Kim,
Jiho Kim,
Joshua Wang,
Kangsu Kim,
Sangwoo Ji,
Woosun Song,
% Atlantis-C
Hanqing Zhao,
Andrew Chin,
Gyejin Lee,
Kevin Stevens,
Mansour Alharthi,
Yizhuo Zhai,
% Atlantis-Java
Cen Zhang,
Joonun Jang,
Yeongjin Jang,
Ammar Askar,
Dongju Kim,
Fabian Fleischer,
Jeongin Cho,
Junsik Kim,
Kyungjoon Ko,
% Atlantis-Patch
Insu Yun,
Sangdon Park,
Dowoo Baik,
Haein Lee,
Hyeon Heo,
Minjae Gwon,
Minjae Lee,
Minwoo Baek,
Seunggi Min,
Wonyoung Kim,
Yonghwi Jin,
Younggi Park,
Yunjae Choi,
% Atlantis-SARIF
Jinho Jung,
Gwanhyun Lee,
Junyoung Jang,
Kyuheon Kim,
Yeonghyeon Cha,
and Youngjoon Kim
}
\newcommand{\Date}{\today}

\begin{titlepage}
    \raggedleft	
	\hspace{.025\textwidth}
	\parbox[b]{\textwidth}{
    \vspace{1.5cm}
		{\Huge\bfseries\Title} \\[20pt]
		{\Large\textit\Subtitle}}
		
		\vspace{4cm}
		\makebox[\textwidth][c]{%
		\colorbox[RGB]{154,175,191}{\parbox{\paperwidth}{%
		\vspace{0.5cm}
		\begin{center}
		\includegraphics[height=4cm]{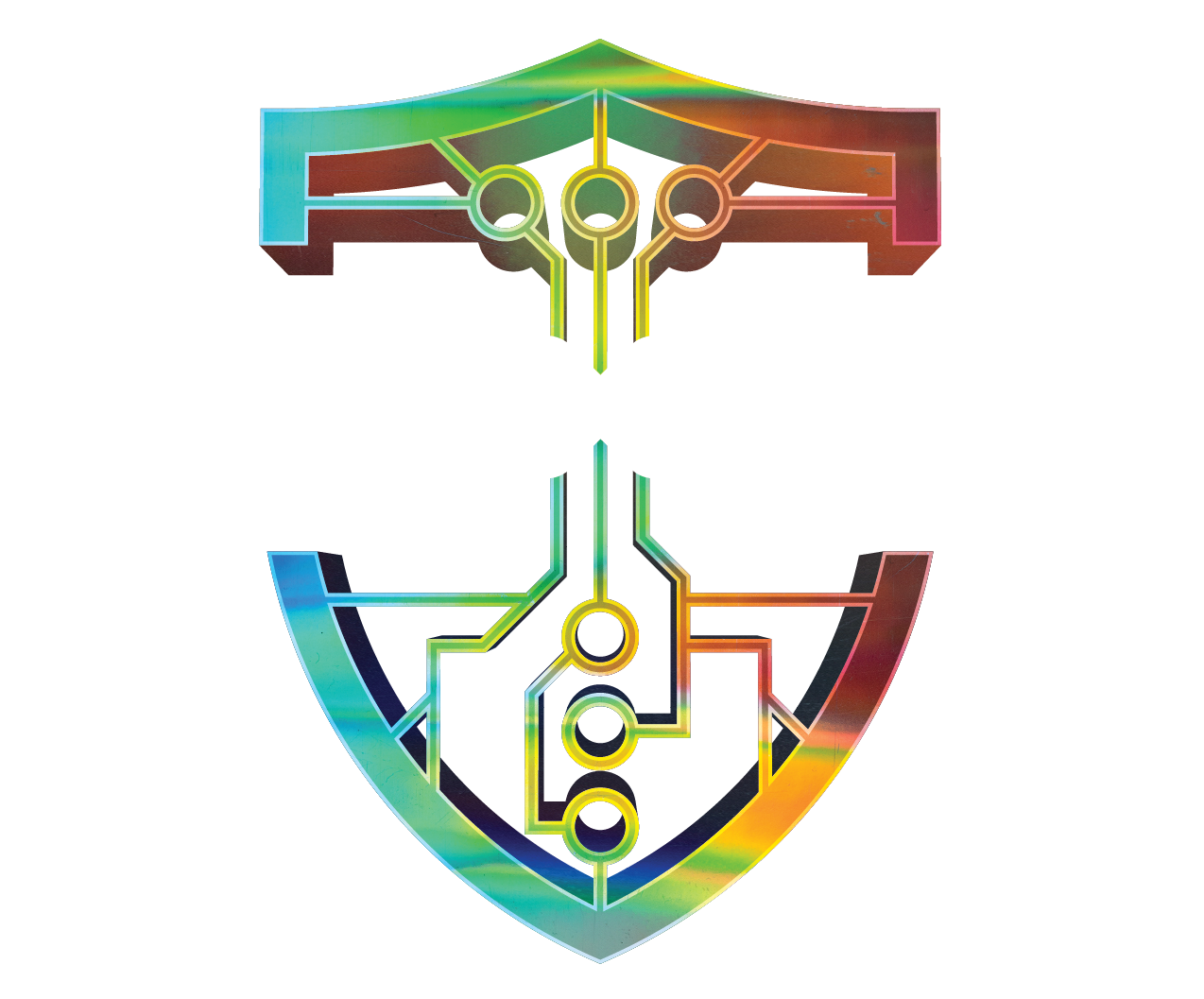}
		\hspace{1cm}
		\raisebox{1.5cm}{\Huge\bfseries +}
		\hspace{1cm}
		\includegraphics[height=4cm]{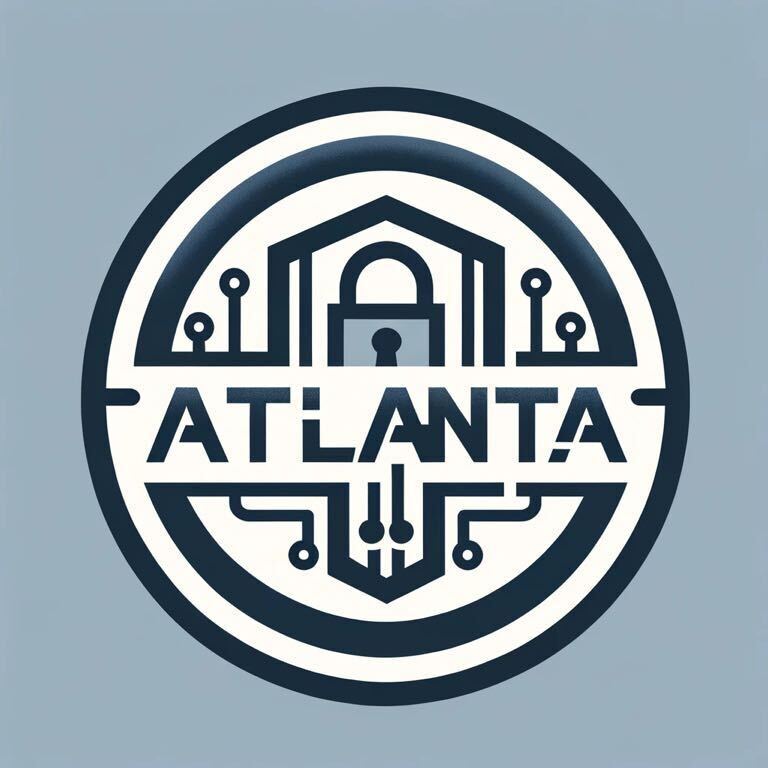}
		\end{center}
		\vspace{0.5cm}}}}
\vfill
\rule{1pt}{.30\textheight}
\hspace{.025\textwidth}
    \raisebox{10pt}{\parbox[b]{.95\textwidth}{ 
        {\small\textsc{\Author} \\[10pt]
        \large{\url{https://team-atlanta.github.io/}}} \\[25pt]
        {\Large\Date}}}

\hfill Version 1.0

%% TODO. Org logo?

\end{titlepage}

%% % Institution logos/names in a modern layout
%% \renewcommand{\arraystretch}{1.2}
%% \begin{tabular}{cc}
%% \textbf{Georgia Institute of Technology} & \textbf{Samsung Research} \\
%% \textbf{KAIST} & \textbf{POSTECH}
%% \end{tabular}

\begin{titlepage}
  \thispagestyle{empty}
  \centering
      {\large\bfseries Revision History\\[1.5em]}
      \begin{threeparttable}
        \begin{tabularx}{0.95\textwidth}{>{\centering\arraybackslash}p{0.03\textwidth}>{\centering\arraybackslash}p{0.15\textwidth}X}
          \toprule
          \textbf{Version} & \textbf{Date} & \textbf{Note} \\
          \midrule
          1.0 & 2025-09-17 & Initial public release of the \sys final report. \\
          \bottomrule
        \end{tabularx}
      \end{threeparttable}
\end{titlepage}

\tableofcontents

\sloppy
\clearpage

\section{Introduction}
\label{s:intro}

The \href{https://aicyberchallenge.com/}{Artificial Intelligence Cyber Challenge (AIxCC)}~\cite{aixcc-website},
launched by DARPA in collaboration with ARPA-H,
represents an unprecedented effort
to revolutionize cybersecurity through artificial intelligence.
This two-year competition (2023--2025),
backed by \$29.5 million in prizes
and partnerships with leading AI companies including Anthropic, Google, and OpenAI,
challenged teams to build autonomous Cyber Reasoning Systems (CRSs)
that can discover and patch vulnerabilities
at the speed and scale demanded by modern software ecosystems.
The competition addresses a critical gap:
while software complexity grows exponentially,
human-driven vulnerability discovery remains fundamentally limited,
leaving countless zero-day vulnerabilities undiscovered
in the critical infrastructure that powers our digital society.

Team Atlanta claimed victory at the AIxCC Final Competition
during DEF CON 33 in August 2025,
culminating a journey that began as one of 42 entrants
and continued as one of seven semifinalist teams.
Our collaboration unites researchers
from Georgia Institute of Technology, Samsung Research, KAIST, and POSTECH,
bringing together complementary expertise
in systems security, program analysis, and artificial intelligence.
We developed \sys{}, a system that orchestrates
state-of-the-art vulnerability discovery techniques---symbolic execution,
directed fuzzing, and static analysis---while
pioneering the deep integration of large language models
to overcome the limitations
of autonomous vulnerability discovery and patching.

This technical report documents the design philosophy,
architectural decisions, and implementation strategies
behind \sys{}.
We present our solutions to the fundamental challenges
of autonomous vulnerability discovery:
scaling across diverse codebases from C to Java,
achieving high precision while maintaining broad coverage,
and generating patches that are semantically correct.
Beyond describing our winning approach,
we share the lessons learned from pushing
the boundaries of what automated security tools can achieve
when traditional program analysis meets modern AI.
Our complete system is available as
\href{https://github.com/Team-Atlanta/aixcc-afc-atlantis}{open source},
enabling the community to build upon our work
and advance the state of automated cybersecurity.

\section{AIxCC Final Competition Overview}
\label{s:aixcc-overview}

\PP{Goals}
DARPA’s Cyber Grand Challenge (CGC, 2014–2016)
pioneered fully autonomous \emph{cyber reasoning systems}
that detected, exploited, and patched software flaws
in a controlled testbed of simplified \emph{binaries},
showcasing machine-on-machine defense using techniques like fuzzing and symbolic execution.
In contrast, the AI Cyber Challenge (AIxCC, 2023–2025) builds on that foundation
by harnessing modern AI, especially LLMs,
to secure real-world \emph{open-source software}
critical to infrastructure
such as healthcare and energy systems.
While CGC served as a proof-of-concept on a contrived computer architecture,
AIxCC emphasizes practical impact:
finalist systems are required to open-source their CRSes
and are supported for commercialization after the competition.
The shift marks a move from lab demonstrations
of autonomy to deployable, AI-powered tools designed
to strengthen national cybersecurity at scale.

\PP{Competition Structure}
The competition format differed substantially between the initial AIxCC Semifinal Competition (ASC)
and the AIxCC Final Competition (AFC) that determined the ultimate winners.
This report only describes the AFC format and our submission to it.
The final format comprised three unscored exhibition rounds
followed by a scored final round,
with each round introducing challenge projects (CPs)
derived from OSS-Fuzz~\cite{oss-fuzz} projects.
Teams were required to process CPs within strict time and budget constraints,
strategically allocating LLM API usage and Azure compute resources
to maximize their vulnerability discovery and patching capabilities.
\autoref{t:competition-rounds} summarizes key parameters
across all final competition rounds.

During competition rounds,
no internet access was available except for
AIxCC organizer-approved commercial LLM access
and necessary resources to build the CRS.
This isolated environment ensured fair competition
while providing teams with essential AI capabilities.
Note that teams are allowed to prepare their own fine-tuned custom LLMs and run these LLMs on Azure cloud during the competition.

\begin{table}[!t]
\centering
% \scriptsize
\footnotesize

\begin{threeparttable}
\begin{tabular}{@{}lcccccccccc@{}}
\toprule
    \textbf{Round} & \textbf{Date} & \textbf{Scored} & \textbf{LLM} & \textbf{Azure} & \textbf{Max Conc.}\tnote{$\uparrow$} & \textbf{Repos} & \textbf{CPs (D+F+U)\tnote{$\ast$}} & \textbf{Delta} & \textbf{Full} \\
\midrule
    Exhibition 1 & 04/01/2025 & No & \$10K & \$20K & 2 & 2 & 2~(2+0+0) & 48h & N/A \\
    Exhibition 2 & 05/06/2025 & No & \$10K & \$20K & 4 & 8 & 15~(9+6+0) & 8h & 24h \\
    Exhibition 3 & 06/05/2025 & No & \$30K & \$50K & 8 & 14 & 30~(18+9+3) & 6h & 12h \\
    \textbf{Final} & 06/26/2025 & Yes & \$50K & \$85K & 8 & 30 & 55~(33+17+5) & 6h & 12h \\
\bottomrule
\end{tabular}
\begin{tablenotes}
\item [$\uparrow$] Max Conc. refers to the maximum number of concurrent challenge projects teams could receive simultaneously.
\item [$\ast$] The number of CPs are sum of the numbers of Delta, Full, and Unharnessed CPs.
\end{tablenotes}
\end{threeparttable}

\caption{AIxCC Final Competition Round Summary}
\label{t:competition-rounds}

\end{table}

\subsection{Challenge Projects}

\PP{OSS-Fuzz Foundation}%
Challenge projects were derived from OSS-Fuzz~\cite{oss-fuzz},
Google's continuous fuzzing service that has discovered
over 10,000 vulnerabilities in critical open-source software since 2016.
OSS-Fuzz provided the foundation for AIxCC challenges through
three key components:
(1) fuzzing harnesses that exercise specific APIs and functionalities,
(2) build configurations with various sanitizers
(\eg AddressSanitizer, MemorySanitizer, and UndefinedBehaviorSanitizer for C/C++,
and Jazzer for Java),
and (3) real-world software packages ranging from libraries and parsers
to complex applications.

\PP{Challenge Types}%
The competition featured two challenge types,
known as \emph{full} and \emph{delta} modes.
In both modes,
teams began with bug discovery
by providing a bug-triggering input, known as Proof-of-Vulnerability (PoV),
using the provided fuzzing harnesses.

\begin{squishitemize}
\item \textbf{Full-scan challenges (full mode):}
  Teams were asked to discover bugs anywhere in the entire codebase
  reachable from the given harnesses.
  Unintended 0-day vulnerabilities were also scorable when demonstrated with PoVs.
\item \textbf{Delta-scan challenges (delta mode):}
  Teams were asked to identify bugs that are \emph{newly} introduced by
  provided source code changes in the form of a diff (representing a new commit or patch).
  PoVs discovered in the baseline source code were \emph{not} scorable.
\end{squishitemize}

Starting from Exhibition Round 3,
the competition also included unharnessed challenges
that lacked pre-written fuzzing harnesses,
requiring teams to generate their own test harnesses.

\PP{Challenge Project Structure}%
Each CP contained intentionally injected vulnerabilities
(``challenge project vulnerabilities'', or CPVs)
within the original codebase,
providing authentic environments for vulnerability discovery and remediation.
A typical CP consisted of:
(1) the source code repository with hidden vulnerabilities,
(2) one or more fuzzing harnesses from OSS-Fuzz (except for unharnessed challenges),
(3) build scripts and configurations,
and (4) in delta mode, a diff showing modified files.
Teams received no prior knowledge of vulnerability types or locations,
thereby simulating realistic security assessment scenarios.

\PP{Competition Tasks}%
For each CP, teams' CRSes needed to
(focusing on C/C++ and Java vulnerabilities):

\begin{squishitemize}
\item \textbf{Find PoVs:}
Generate inputs that trigger crashes to demonstrate vulnerabilities
in the challenge source code, not defects in the harnesses.
Scorable vulnerabilities included standard vulnerability categories
detectable by OSS-Fuzz and unintended 0-day vulnerabilities
that were harness-reachable.
Scorable outcomes did not need to be explicitly caused by sanitizers,
encompassing crashes, exceptions, and timeouts (if configured).
% The variant PoV system allowed different teams
% to demonstrate the same underlying vulnerability using distinct inputs,
% facilitating cross-team patch validation.
% PoVs must be reliably reproducible as they are evaluated using the libfuzzer fuzzing engine
% with x86\_64 architecture.
% buffer overflows, use-after-free errors, integer overflows,
% OS command injection, SSRF, SQL injection,
% and other memory safety and logic bugs.

\item \textbf{Generate Patches:}
Create source code modifications in unified diff format
that fix vulnerabilities while preserving intended functionality.
Patches could not modify harnesses or functional tests
(\eg \cc{test.sh} scripts or Maven test cases).
Each patch was validated independently against the original challenge
and underwent post-competition manual assessment for scoring.
% \XXX{
% Patches must pass the provided \cc{test.sh} script
% to ensure they do not break existing tests.
% Given the complexity of automated patching,
% all patches undergo post-competition manual validation
% for scoring purposes.}

\item \textbf{Assess SARIF Reports:}
Evaluate Static Analysis Results Interchange Format (SARIF) reports
broadcast during round execution.
SARIF reports describe potential vulnerabilities in active challenges,
but are not guaranteed to be accurate and contain no PoVs.
Teams submitted assessments (correct/incorrect) with justification descriptions.
Correct SARIF reports describe harness-reachable, sanitizer-triggered crashes.

\item \textbf{Bundle Submissions:}
Group related submissions to indicate relationships between discoveries.
Bundles could contain PoVs, patches, SARIF reports, and broadcast SARIF assessments.
For example, a CRS could bundle a patch with the PoV that demonstrates
the vulnerability it fixes, or combine a PoV with its SARIF assessment.
Bundles required at least two components and could be modified after submission.
\end{squishitemize}

\subsection{Scoring System}
The competition employed a comprehensive scoring system
designed to incentivize both the discovery of novel vulnerabilities
and the generation of correct, deployable patches.
Rather than attacking and analyzing other teams' binaries as in CGC,
each CRS in AIxCC operated on the organizer-provided CPs independently.
The scoring format evolved through several iterations;
% from the first draft (December 13, 2024)
% to release version 1 (March 12, 2025),
% and finally to the scoring system used in the competition (May 30, 2025).
this report describes the final ``Version 2'' methodology~\cite{aixcc-scoring}.

The final team score is calculated as:
\[
Team\ Score = \sum Challenge\ Scores
\]

The challenge score is a weighted sum of CRS performance
across vulnerability discovery, program repair, SARIF assessment,
bundling, and accuracy for all vulnerabilities in the challenge.
Each challenge score is calculated as:
\[
Challenge\ Score = AM \times (VDS + PRS + SAS + BDL)
\]
where AM is the Accuracy Multiplier,
VDS is the Vulnerability Discovery Score,
PRS is the Program Repair Score,
SAS is the SARIF Assessment Score,
and BDL is the Bundle Score.
Here, the AM emphasizes submission quality over quantity.
Any false submissions (incorrect patches, wrong SARIF assessments)
significantly degrade the overall score,
incentivizing teams to prioritize accuracy.

Points are awarded based on
uniqueness of discovered vulnerabilities,
correctness of generated patches,
accuracy of SARIF assessments,
and bonus points for bundled submissions
(PoV + patch + SARIF assessment for the same vulnerability).
Penalties apply for incorrect patches
or multiple patch submissions addressing the same root cause.
For comprehensive scoring details,
refer to the official scoring guide~\cite{aixcc-scoring}.

\subsection{Budget and Resource Constraints}
\label{ss:aixcc-budget}

\PP{Budget Allocation for Development}%
For development, teams were allocated \$100,000 in Azure compute resources
for infrastructure development and testing.
Additionally, teams received substantial support from leading AI companies:
Anthropic, Google, and OpenAI each provided \$50,000 in LLM credits per team,
totaling \$150,000 for extensive experimentation and system development.
Notably, we were allowed to exceed the given LLM credits, as long as we paid the difference ourselves.

\PP{Budget Allocation for the Competition}%
As shown in \autoref{t:competition-rounds},
each competition round provided fixed budgets for
(1) LLM API usage (\$10,000~\textasciitilde~\$50,000 per round) and
(2) Azure compute resources (\$20,000~\textasciitilde~\$85,000 per round).
Teams had to strategically allocate these resources
across multiple CPs that arrived in batches,
with overlapping deadlines
requiring careful resource management and scheduling.
For instance, Exhibition Round 3 presented a particularly challenging scenario:
with \$30K LLM budget and \$50K Azure budget,
teams received up to \emph{8 concurrent CPs}
from 14 repositories with 30 total challenges,
with tight deadlines of 6 hours for delta challenges
and 12 hours for full challenges.

\PP{Team Atlanta's Resource Strategy}%
For the final round,
we strategically allocated the \$85K Azure budget
and \$50K LLM budget
to maximize concurrent CP processing
while maintaining system reliability.
Our Azure environment utilized two key resource quotas:
57,500 vCPUs of Standard \ccw{Ddsv6} Family
for dynamically allocated CRS components,
and 300 vCPUs of Standard \ccw{NCASv3\_T4} Family with GPUs,
with 64 cores dedicated to our custom fine-tuned LLM.

We allocated 20\% of the Azure budget for infrastructure overhead,
accounting for estimated operational costs over the 10-day competition period,
including cluster management, system services, networking,
specialized hardware, monitoring, and storage.
The remaining 80\% was reserved for dynamic CP processing,
providing approximately 1,500 cores per full-mode CP (12-hour deadline)
and 3,000 cores per delta-mode CP (6-hour deadline).
Note that we dynamically allocated CPUs per CP based on the remaining budget and the number of harnesses each CP has (see \autoref{ss:cpmgr}).

\begin{table}[!t]
\centering
\footnotesize

\begin{tabular}{@{}lrrrrrr@{}}
\toprule
\textbf{Module} & \textbf{OpenAI} & \textbf{Anthropic} & \textbf{Gemini} & \textbf{Grok} & \textbf{Total (\$)} & \textbf{\%} \\
\midrule
\sys-Multilang & 8,839 & 8,839 & 884 & 0 & 18,563 & 37.1\% \\
\sys-C & 3,713 & 2,784 & 2,784 & 0 & 9,281 & 18.6\% \\
\sys-Java & 3,713 & 2,784 & 2,784 & 0 & 9,281 & 18.6\% \\
\sys-Patch & 3,713 & 7,425 & 1,238 & 0 & 12,375 & 24.8\% \\
\sys-SARIF & 75 & 100 & 250 & 75 & 500 & 1.0\% \\
\midrule
\textbf{Total} & 20,052 & 21,933 & 7,940 & 75 & 50,000 & 100\% \\
\bottomrule
\end{tabular}
\caption{Team Atlanta's LLM Budget Distribution Across Modules and Providers}
\label{t:team-atlanta-llm}
\end{table}

Our LLM allocation strategy is shown in~\autoref{t:team-atlanta-llm}.
% balanced multiple providers
% to leverage their respective strengths.
% Anthropic's Claude for complex patch generation,
% OpenAI's GPT-4 for code analysis,
% and Gemini for auxiliary tasks.
Across our five modules---vulnerability discovery (\sys-Multilang, \sys-C, \sys-Java),
program repair (\sys-Patch), and static analysis assessment (\sys-SARIF)---
the budget distribution reflected each module's computational demands
and strategic importance.
\sys-Multilang received the largest allocation (37.1\%)
due to its central role in cross-language vulnerability discovery,
while \sys-Patch received 24.8\% for generating high-quality fixes.
\sys-SARIF received a minimal allocation (1.0\%)
as static analysis assessment tasks were predictable in scope.
For detailed system architecture, please refer to \autoref{s:overview}.

\PP{LLM Rate Limit Management}%
Beyond budget allocation,
processing concurrent CPs required careful rate limit management.
While OpenAI's Tier 5 (10K RPM, 30M TPM) and
Google Gemini's Tier 3 (2K RPM, 8M TPM) limits were sufficient,
Anthropic's standard Tier 4 limits (200K ITPM, 80K OTPM) were insufficient.
AIxCC organizers negotiated enhanced Anthropic limits (over 4$\times$ increase)
as shown in \autoref{t:anthropic-rates}.

\begin{table}[!t]
\centering
\footnotesize

\begin{threeparttable}
\begin{tabular}{@{}lrrrr@{}}
\toprule
\textbf{Model} & \textbf{RPM} & \textbf{Input TPM} & \textbf{Output TPM} & \textbf{Total TPM} \\
\midrule
Claude Opus 4 & Uncapped & 875,000 & 175,000 & 1,050,000 \\
Claude Sonnet 4 & Uncapped & 1,500,000 & 275,000 & 1,775,000 \\
Claude Sonnet 3.7 & 6,250 & 125,000 & 50,000 & 175,000 \\
Claude Sonnet 3.5 & 6,250 & 500,000 & 100,000 & 600,000 \\
Claude Haiku 3.5/3 & 6,250 & 625,000 & 125,000 & 750,000 \\
Claude Opus 3 & 6,250 & 250,000 & 18,750 & 268,750 \\
\bottomrule
\end{tabular}
\begin{tablenotes}
\item RPM: Requests per minute, TPM: Tokens per minute
\end{tablenotes}
\end{threeparttable}
\caption{Enhanced Anthropic Rate Limits for AIxCC Final Competition}
\label{t:anthropic-rates}
\vspace*{-7px}
\end{table}

We implemented internal rate limits for each module
to prevent any single module from exhausting the enhanced limits.
For Sonnet 4, we set limits of
1,125K TPM for \sys-Multilang, 300K TPM for \sys-Patch,
and 375K TPM each for \sys-C and \sys-Java.
For Opus 4, we configured 750K TPM each for \sys-C and \sys-Java,
with 300K TPM each for \sys-Multilang and \sys-Patch as fallback.
This 3:1 ratio between \sys-Multilang and language-specific modules
reflected their usage patterns,
ensuring no single module could monopolize resources
and enabling consistent performance across concurrent CPs.

\subsection{Final Scores}
\label{ss:aixcc-final-score}

\begin{table}[!t]
\centering

\resizebox{\textwidth}{!}{
\begin{threeparttable}
\begin{tabular}{@{}clccccccc@{}}
\toprule
\textbf{Rank} & \textbf{Team} & \textbf{Total Score} & \textbf{Accuracy} & \textbf{VDS} & \textbf{PRS} & \textbf{SAS} & \textbf{BDL} & \textbf{Vulns Found} \\
\midrule
1 & Team Atlanta & 392.76 & 91.27\% & 79.71 & 171.1 & 5.99 & 136.38 & 43/70 (61\%) \\
2 & Trail of Bits & 219.35 & 89.33\% & 52.49 & 101.21 & 1.0 & 65.29 & 28/70 (40\%) \\
3 & Theori & 210.68 & 44.44\% & 58.12 & 110.34 & 4.97 & 53.57 & 34/70 (49\%) \\
4 & All You Need IS A Fuzzing Brain & 153.7 & 53.77\% & 54.81 & 77.6 & 6.52 & 28.28 & 28/70 (40\%) \\
5 & Shellphish & 135.89 & 94.83\% & 47.94 & 54.31 & 8.47 & 25.29 & 28/70 (40\%) \\
6 & 42-b3yond-6ug & 105.03 & 89.23\% & 70.37 & 14.22 & 9.8 & 10.97 & 41/70 (59\%) \\
7 & Lacrosse & 9.59 & 42.86\% & 1.68 & 5.43 & 0.0 & 3.62 & 1/70 (1\%) \\
\bottomrule
\end{tabular}
\begin{tablenotes}
\item VDS: Vulnerability Discovery Score,
PRS: Program Repair Score,
SAS: SARIF Assessment Score,
BDL: Bundle Score.
The Accuracy column indicates the percentage of submissions
marked as correct by the scoring system.
\end{tablenotes}
\end{threeparttable}
}
\caption{AIxCC Final Competition Results}
\label{t:final-scores}
\end{table}

The final competition round
brought together seven teams
for the ultimate test of their cyber reasoning systems.
Each team deployed their developed CRS
against 55 challenge projects from 28 repositories,
competing for the highest score
within strict time and budget constraints.
The following results demonstrate
the effectiveness
of each team's automated vulnerability discovery and patching capabilities.

\autoref{t:final-scores} presents the final competition results
for all seven participating teams in the scored final round.
Our team, Team Atlanta, achieved first place with a total score of 392.76 points,
demonstrating superior performance
across vulnerability discovery, program repair, and bundle submissions.
The results reveal significant performance variations among teams:
while several teams achieved high correctness rates for their submissions,
we distinguished ourselves through both volume and quality,
discovering 43 out of 70 total vulnerabilities (61\% coverage)
and excelling in successful patch generation
and comprehensive bundle submissions
that effectively combined PoVs, patches, and SARIF assessments.

\autoref{t:resource-utilization} reveals the detailed resource utilization
and LLM usage patterns across all teams during the final competition round.
Notably, teams had allocated budgets of \$85K for Azure compute resources and \$50K for LLM usage in the final round.
The data demonstrates clear correlations
between resource investment and performance.
Our team utilized the highest total budget (\$103.3K)
and generated the most LLM queries (696.5K),
processing over 4 billion input tokens
and generating 641.6 million output tokens.
Interestingly, second-place team Trail of Bits demonstrated
the highest input token usage (12.83B)
with relatively efficient resource allocation.
They relied on less powerful but cheaper LLMs (Claude Sonnet 4, GPT-4.1 mini, GPT-4.1)
while we used more powerful LLMs (o4-mini, GPT-4o, o3) on average.
In addition, we got more verbose outputs (\eg bug candidates and input generators) from LLMs.
This choice may have helped surface richer insights during vulnerability analysis and patch generation, ultimately leading to stronger results.

\begin{table}[!t]
\centering

\resizebox{\textwidth}{!}{
\begin{threeparttable}
\begin{tabular}{@{}rl rrr rrr@{}}
\toprule
    & & \multicolumn{3}{c}{\textbf{Budget Spending}} & \multicolumn{3}{c}{\textbf{LLM Usage}}\\\cmidrule(lr){3-5}\cmidrule(lr){6-8}
    \textbf{Rank} & \textbf{Team} & \textbf{Azure} & \textbf{LLM} & \textbf{Total} & \textbf{Queries} & \textbf{Input Tokens} & \textbf{Output Tokens} \\
\midrule
    1 & Team Atlanta & \$73.9K & \$29.4K & \$103.3K & 696.5K & 4.09B & 641.6M \\
    2 & Trail of Bits & \$18.5K & \$21.1K & \$39.6K & 613.9K & 12.83B & 402.2M \\
    3 & Theori & \$20.3K & \$11.5K & \$31.8K & 187.6K & 2.09B & 112.5M \\
    4 & All You Need IS A Fuzzing Brain & \$63.2K & \$12.2K & \$75.4K & 122.9K & 415.6M & 85.4M \\
    5 & Shellphish & \$54.9K & \$2.9K & \$57.8K & 301.0K & 4.69B & 205.1M \\
    6 & 42-b3yond-6ug & \$38.7K & \$1.1K & \$39.8K & 37.5K & 96.7M & 74.4M \\
    7 & Lacrosse & \$7.1K & \$0.7K & \$7.8K & 70.7K & 246.4M & 9.6M \\
\bottomrule
\end{tabular}
\begin{tablenotes}
\item Teams were allocated budgets of \$85K for Azure compute resources and \$50K for LLM usage in the final round.
% Token counts are in billions (B) and millions (M).
The varying resource utilization patterns reveal distinct strategic approaches
to competition challenges.
\end{tablenotes}
\end{threeparttable}
}
\caption{AIxCC Final Competition Resource Utilization. \vspace{-5pt}}
\label{t:resource-utilization}
\end{table}

\clearpage
\section{\sys}
\label{s:overview}

%- overall design figure
%- plumbing (API to the org)
%- feature comparison among CRS
%- fault tolerant
%- goals

The primary goal of the AIxCC competition is to build
a cyber reasoning system (CRS) that employs large language models (LLMs)
to automatically find proof-of-vulnerabilities (PoVs) in challenge projects (CPs),
fix the vulnerabilities,
and assess given SARIF reports.
To this end, we designed and implemented our CRS, \textbf{\sys},
while achieving the following requirements:

\PP{R1: Support multiple CPs concurrently}%
In both real-world settings and the AIxCC competition,
\sys should be able to process multiple CPs simultaneously.
When multiple CPs are provided concurrently,
\sys should scale effectively to automatically find PoVs
and fix the corresponding vulnerabilities in all CPs,
as well as assess given SARIF reports.
Notably, during the AIxCC competition,
CPs are delivered in multiple batches,
meaning that \sys must handle overlapping sets of CPs
while continuing to process newly received ones
after previous deadlines have passed.

\PP{R2: Be fail-safe as much as possible}%
While processing multiple CPs concurrently,
\sys must be designed with a fail-safe architecture
to minimize the impact of individual task failures.
For example, even if our CRS fails to handle a CP,
it must continue to process the remaining CPs reliably without interruption.

\PP{R3: Fully utilize LLM and Azure budget}%
During the AIxCC competition,
the organizers allocate fixed budgets for LLM usage and Azure cloud resources
(see \autoref{ss:aixcc-budget}).
\sys should maximize the utilization of these resources
to achieve optimal performance
while supporting multiple CPs across multiple batches,
for each round.

\PP{R4: Other requirements}%
For observability,
the AIxCC organizers required participants
to collect and submit meaningful logs
(\eg LLM requests and responses)
in the OpenTelemetry format they specified.
Furthermore, to ensure consistent deployment
and easier infrastructure management,
they mandated that our CRS be deployed on Azure with Terraform.

\begin{figure*}[!t]
  \centering
  \resizebox{\textwidth}{!}{
\begin{tikzpicture}[
  module/.style={anchor = north, rounded corners = 1mm, rectangle, draw, minimum width=4.5cm, minimum height = 1cm, thick, font=\normalsize, fill=gray!10},
  wmodule/.style={anchor = north, rectangle, draw, minimum width=4.5cm, minimum height = 0.5cm, thick, font=\small, fill=white},
  smodule/.style={anchor = north, rounded corners = 1mm, rectangle, draw, minimum width=3.5cm, minimum height = 0.5cm, thick, font=\small, fill=gray!10},
  tmodule/.style={anchor = north, rounded corners = 1mm, rectangle, draw, minimum width=3.5cm, minimum height = 0.5cm, thick, font=\footnotesize, fill=white},
  moduletxt/.style={font = \normalsize},
  apiserver/.style={cloud, draw, very thick, minimum width=3cm, minimum height=2cm, cloud puffs=10,cloud puff arc=120, aspect=2},
  arrow/.style={-{Latex[length=2mm, width=2mm]}, ultra thick},
  arrowBig/.style={-{Latex[length=3mm, width=3mm]}, line width=3pt},
  arrowBigBi/.style={{Latex[length=3mm, width=3mm]}-{Latex[length=3mm, width=3mm]}, line width=3pt},
  arrowTxtA/.style={above, font=\small, pos=0.5, align=left},
  arrowTxtR/.style={right, font=\small, pos=0.5, align=left},
  arrowTxtL/.style={left, font=\small, pos=0.5, align=left},
  wrapper/.style={rectangle, draw,dashed, minimum width=4.5cm, minimum height = 0.5cm, ultra thick},
]

\newcommand*{\ax}{0}
\newcommand*{\bx}{\ax+6.5}
\newcommand*{\cx}{\bx+6.75}
\newcommand*{\dx}{\cx+5}

\newcommand*{\ay}{0}
\newcommand*{\by}{\ay-2.5}
\newcommand*{\cy}{\by-3}

% CRS Web server
\node[module, align=center, minimum height=1.5cm] at (\bx, \ay) (web) {\textbf{\textsc{CRS Web Server}}\\\textbf{\textsc{CRS DB}}};
% Log Collector
\node[module, align=center] at (\bx, \by) (otel) {\textbf{\textsc{Log Collector}}};

% LLM Proxy
\node[wmodule, minimum height=2.75cm] at (\bx, \cy+1) (llmproxy) {};
\node[moduletxt, below] at (llmproxy.north) {\textsc{\textbf{LLM Proxy}}};
\node[smodule] at (\bx+0.2, \cy+0.2) () {};
\node[smodule] at (\bx+0.1, \cy+0.1) () {};
\node[smodule] at (\bx, \cy) (litellm) {\textbf{\textsc{LiteLLM}}};
\node[smodule] at (\bx, \cy-0.85) (customllm) {\textbf{\textsc{Custom LLM}~(\autoref{ss:custom-model})}};

% API servers
\newdimen\yWeb
\pgfextracty{\yWeb}{\pgfpointanchor{web}{west}};
\newdimen\yProxy
\pgfextracty{\yProxy}{\pgfpointanchor{llmproxy}{west}};

\node[apiserver] at (\ax, \yWeb) (aixcc) {\textbf{\textsc{AIxCC API}}};
\node[apiserver, align=center] at (\ax, \yProxy) (llm) {\textbf{\textsc{LLM}}\\\textbf{\textsc{Providers}}};

\node[wrapper, minimum width=5.4cm, minimum height = 8cm] (one) at (\bx, \by-1.15){};
\node[moduletxt, below] at (one.south) {\textbf{Nodes per CRS}};

\draw[arrowBig](aixcc)-> (web) node[arrowTxtA, pos=0.4] {Challenges};
\draw[arrowBig](otel) -| (aixcc) node[arrowTxtA, pos=0.26] {OpenTelemetry Logs};
\draw[arrow](llmproxy) -> (otel) node[arrowTxtR, pos=0.4] {Logs};
\draw[arrowBigBi](llmproxy) -> (llm);

\node[module, minimum width = 4cm, minimum height = 3cm] at (\cx, \ay) (mgr) {};
    \node[moduletxt, below] at (mgr.north) {\textbf{\textsc{CP Manager}~(\autoref{ss:cpmgr})}};
\newdimen\xMgr
\pgfextractx{\xMgr}{\pgfpointanchor{mgr}{west}};

\node[tmodule, align=center] at (\cx, \ay-0.75) {
    \textbf{\textsc{CP Builder}}\\
    \textbf{\textsc{Budget Allocator}}\\
    \textbf{\textsc{POV Verifier}}\\
    \textbf{\textsc{Task Manager}}\\
    \textbf{\textsc{Submitter}}
};

\node[module, font=\small, minimum width = 4cm, minimum height = 0.75cm] at (\dx, \ay) (sarif) {\textbf{\textsc{\sys-SARIF}~(\autoref{s:crs-sarif})}};

\newcommand*{\yPatch}{\by+0.25}
\node[module, font=\small, minimum width = 4cm, minimum height = 0.75cm] at (\dx+0.2, \yPatch+0.2) {};
\node[module, font=\small, minimum width = 4cm, minimum height = 0.75cm] at (\dx+0.1, \yPatch+0.1) {};
\node[module, font=\small, minimum width = 4cm, minimum height = 0.75cm] at (\dx, \yPatch) (patch) {\textbf{\textsc{\sys-Patch}~(\autoref{s:crs-patching})}};

\newcommand*{\yBug}{\cy+1.2}
\newcommand*{\xBug}{\cx+2+0.5}
\newcommand*{\xGap}{2}
\newcommand*{\xGapC}{2.2}
\newcommand*{\xGapM}{1.5}
\node[wmodule, minimum width=9cm, minimum height=3cm] at (\xBug, \yBug) (bugfinding) {};
\node[moduletxt, below] at (bugfinding.north) {\textsc{\textbf{Bug Finding Modules}}};
\node[smodule, minimum width=3.5cm] at (\xBug-\xGapC+0.2, \yBug-1+0.2) {};
\node[smodule, minimum width=3.5cm] at (\xBug-\xGapC+0.1, \yBug-1+0.1) {};
\node[smodule, minimum width=3.5cm] at (\xBug-\xGapC, \yBug-1) {\textbf{\textsc{\sys-C}~(\autoref{s:crs-c})}};

\node[smodule, minimum width=3.5cm] at (\xBug+\xGap+0.2, \yBug-1+0.2) {};
\node[smodule, minimum width=3.5cm] at (\xBug+\xGap+0.1, \yBug-1+0.1) {};
\node[smodule, minimum width=3.5cm] at (\xBug+\xGap, \yBug-1) {\textbf{\textsc{\sys-Java}~(\autoref{s:crs-java})}};

\node[smodule, minimum width=7.75cm] at (\xBug-0.1+0.2, \yBug-2.1+0.2) {};
\node[smodule, minimum width=7.75cm] at (\xBug-0.1+0.1, \yBug-2.1+0.1) {};
\node[smodule, minimum width=7.75cm] at (\xBug-0.1, \yBug-2.1) {\textbf{\textsc{\sys-Multilang}~(\autoref{s:crs-multilang})}};

\draw[arrowBig](web.east) -> (\xMgr, \yWeb) node[arrowTxtA, pos=0.52] {Challenge};
\draw[arrowBig](mgr.north) |- (\ax, \ay+1)  node[arrowTxtR, pos=0.35] {Submission}
-> (aixcc);

\node[wrapper, minimum width=10cm, minimum height = 8cm] (cp) at (\dx-2.4, \by-1.15){};
\node[moduletxt, below] at (cp.south) {\textbf{Nodes per CP}};

\newdimen\xCP
\pgfextractx{\xCP}{\pgfpointanchor{cp}{west}};
\newdimen\yOtel
\pgfextracty{\yOtel}{\pgfpointanchor{otel}{east}};
\draw[arrowBigBi](otel) -> (\xCP, \yOtel) node[arrowTxtA, pos=0.52] {Logs};
\draw[arrowBigBi](llmproxy) -> (\xCP, \yProxy) node[arrowTxtA, pos=0.6] {LLM\\Reqeusts};

\draw[arrow] (mgr.east) -| (sarif.200) node[arrowTxtA, pos=0.2, font=\footnotesize] {CP,\\SARIF,\\POV, Patch};
\draw[arrow] (sarif.south) |- (mgr.352) node[arrowTxtR, pos=0.25, font=\footnotesize] {SARIF\\Assessment};

\newdimen\xMgrA
\pgfextractx{\xMgrA}{\pgfpointanchor{mgr}{260}};
\draw[arrow] (mgr.260) -> (\xMgrA, \yBug) node[arrowTxtL, pos=0.5, font=\footnotesize] {CP,\\SARIF};
\draw[arrow] (\cx, \yBug) -> (mgr.south) node[arrowTxtR, pos=0.5, font=\footnotesize] {POV};

\draw[arrow](mgr.315) |- (\dx-2, \by-1.1) node[arrowTxtA, pos=0.9, font=\footnotesize]{CP, POV} -| (patch.240);
\draw[arrow](patch.south) |- (\dx-2, \by-1.3) node[arrowTxtR, pos=0.2, font=\footnotesize]{Patch} -| (mgr.310);

\node[smodule, minimum width=1cm] at (\ax-0.5, \cy-1.75) (label) {};
\node[moduletxt, right] at (label.east){\textbf{: A k8s node}};

\end{tikzpicture}
}
    \caption{The overview of \sys.}
  \label{fig:overview}
\end{figure*}

\subsection{Overview}
\label{ss:overview}

\autoref{fig:overview} presents the overall architecture of \sys.
\sys is deployed on a Kubernetes (k8s) cluster on Azure, provisioned and managed via Terraform, thereby fulfilling \textbf{R4}.
To achieve effective scalability (\textbf{R1}) and enhance fault tolerance (\textbf{R2}), \sys adopts a two-tier node architecture within the k8s cluster:
(1) CRS-level nodes, which host shared system components, and
(2) CP-level nodes, each dedicated to processing a single CP.

\sys begins by launching CRS-level nodes for three key components: \syscrsweb, \textsc{Log Collector}, and \textsc{LLM Proxy}.
The \textsc{Log Collector} is responsible for gathering and forwarding meaningful logs to the organizers (\textbf{R4}).
The \textsc{LLM Proxy}, which is based on LiteLLM, centrally manages LLM usage to stay within the allocated budget (\textbf{R3}) and logs all LLM requests and responses for observability.

\syscrsweb listens for incoming CPs or SARIF reports from the AIxCC organizers.
Upon receiving a new CP, \syscrsweb stores its metadata (e.g., repository and diff) in the database and spawns a dedicated \syscpmgr instance on a CP-level k8s node.

Each \syscpmgr instance is responsible for managing the analysis of a single CP.
It first builds the given CP and allocates both the Azure compute budget and the LLM usage budget proportionally.
Based on the allocated Azure budget and the number of fuzzing harnesses in the CP, \syscpmgr launches an appropriate number of k8s nodes for bug-finding modules.
For C-based CPs, it deploys \sys-C and \sys-Multilang; for Java-based CPs, it launches \sys-Java and \sys-Multilang.
In addition, it starts \sys-Patch and \sys-SARIF to handle patch generation and SARIF assessment, respectively.
To control LLM usage, \syscpmgr issues LiteLLM API keys for each module, with budgets calculated based on the total number of CPs and the overall LLM budget.
All modules launched under a given \syscpmgr instance use the corresponding keys to ensure budget compliance and evenly distribute RPM and TPM of LLM models for each module.
Notably, only \sys-Patch was able to utilize our custom LLM, which was fine-tuned for the patch generation.

When a bug-finding module discovers a PoV, it is sent to \syscpmgr,
which verifies whether the PoV triggers a crash and deduplicates results based on stack traces.
If the PoV is confirmed to be unique, \syscpmgr submits it to the AIxCC organizers and forwards it to \sys-Patch and \sys-SARIF for patch generation and SARIF assessment, respectively.
Once the patch and SARIF report are produced, they are returned to \syscpmgr.
It then bundles them with the PoV and submits the complete bundle package to the AIxCC organizers.
Although the rules permit submitting a patch or SARIF assessment without the corresponding PoV,
we follow this integrated process to ensure that generated patches and assessments are sufficiently precise.

\subsection{\syscpmgr}
\label{ss:cpmgr}

In this subsection, we provide more details about several subcomponents of \syscpmgr.

\PS{CP Builder}
\syscpmgr begins by compiling the given CP and making the compiled artifacts available to bug-finding modules via a shared file system within k8s.
If the target CP is written in C, it is compiled using the default configuration as well as three additional configurations (see \autoref{s:crs-multilang}): one for \sys-Multilang, one for coverage build in OSS-Fuzz, and one for symbolic execution with SymCC.
For Java-based CPs, \syscpmgr compiles the project with both the default configuration and the configuration required by \sys-Multilang.
The compiled outputs are then shared via the k8s shared file system, allowing the bug-finding modules to retrieve them and initiate the fuzzing process.
Additionally, \syscpmgr identifies all available fuzzing harnesses within the CP and finds the corresponding source code locations.
This information is essential for both accurate budget allocation and identifying the entry points for LLM-based analyses.
Note that this information is stored in the shared file system in JSON format.

\PS{Budget Allocator}
\syscpmgr is responsible for allocating both the LLM and Azure budgets for each CP.
The allocation is based on the remaining total budget and the number of CPs yet to be processed.
Once the total budget for a CP is calculated, \syscpmgr distributes it among the corresponding modules.

For the LLM budget, a fixed amount of \$500 is reserved for \sys-SARIF across all CPs,
because the number of SARIF reports is predetermined by the competition.
The remaining LLM budget is divided among the other modules as follows:
25\% to \sys-Patch,
and 37.5\% each to \sys-Multilang and \sys-C (or \sys-Java).

For the Azure budget, we first allocate a fixed number of k8s nodes per CP.
Three \ccw{Standard\_D64ds\_v6} nodes are allocated for \syscpmgr,
\sys-SARIF, and shared utilities of \sys-Multilang.
Five \ccw{Standard\_D32ds\_v6} nodes are allocated to \sys-Patch,
with each node serving one patch-generation agent.
Next, based on the deadline of the CP,
we compute how many cores can be utilized within the allocated Azure budget.
This remaining compute capacity is equally divided between \sys-Multilang and \sys-C (or \sys-Java).
For \sys-C, we allocate up to 15 nodes,
regardless of the number of fuzzing harnesses,
provided the budget allows.
For \sys-Multilang and \sys-Java,
we determine the available budget per fuzzing harness
and select appropriate node types accordingly to launch fuzzing tasks.
Finally, if any portion of the allocated budget remains unused,
it is reallocated to future CPs to maximize overall resource efficiency.

\PS{PoV Verifier}
Once a bug-finding module identifies a PoV,
it submits it to \syscpmgr.
\syscpmgr then verifies whether the PoV reliably triggers a crash
in an environment identical to that used by the AIxCC organizers.
The verifier checks for specific crash return codes:
sanitizer crashes (1 or 77),
timeouts (70, which may indicate DoS vulnerabilities
when the organizer's build configuration sets timeout limits),
and out-of-memory conditions (71).
For delta-mode CPs,
\syscpmgr additionally verifies that the PoV
does not crash on the BASE version,
ensuring the vulnerability exists only in the target changes.
Rather than relying on the AIxCC-provided PoV deduplication mechanism
based on textual similarity,
we implement a custom deduplication strategy
using stack trace analysis.
Our approach incorporates multiple heuristics tailored to specific bug types
to more accurately identify unique PoVs.
This design choice is crucial because we request patch generation only for unique PoVs,
with a rate limit of 10 patch requests per fuzzer-sanitizer combination
to manage resource allocation effectively.
Our improved PoV deduplication reduces LLM and CPU usage in \sys-Patch,
enabling more efficient budget utilization.

\PS{Task Manager}
Another key responsibility of \syscpmgr is orchestrating
complex, multi-step workflows across modules.
When \syscpmgr receives a SARIF report from \syscrsweb,
it forwards the report to \sys-SARIF to initiate SARIF assessment.
Once \sys-SARIF completes its analysis,
the result is returned to \syscpmgr,
which forwards it to the bug-finding modules to facilitate directed fuzzing.
When a PoV is determined as unique,
\syscpmgr submits it to the AIxCC organizers
and sends a patch generation request to \sys-Patch
while polling for the organizers' verification results.
If the PoV passes the organizers' verification,
\syscpmgr then initiates a PoV–SARIF mapping request to \sys-SARIF.
This design maximizes throughput
while respecting the rate limit constraints.
After receiving the generated patch from \sys-Patch,
\syscpmgr verifies the patch in the AIxCC environment
and, upon successful verification,
sends the patch to \sys-SARIF
to assist in final SARIF assessment.
All inter-module communication for task management
is implemented through web APIs exposed by each module,
with state synchronization managed through Redis
to ensure consistency across concurrent operations.

\PS{Submitter}
The primary role of \syscpmgr is to submit PoVs, patches,
and SARIF assessments to the AIxCC organizers.
However, submitting multiple patches for the same underlying bug incurs a penalty.
To avoid this, \syscpmgr employs a sophisticated bundling strategy
that dynamically manages PoV-patch-SARIF combinations.
When new patches or SARIF assessments become available,
the system evaluates existing bundles
and either updates them or recreates them
to ensure optimal point scoring.
The bundling algorithm groups related PoVs under a single representative
and manages patch selection to avoid redundant submissions.
Additionally, the competition awards more points
when PoVs, patches, and SARIF assessments are submitted as a bundled package.
To maximize points, \syscpmgr uses each PoV as a unique key
and continuously attempts to match it
with corresponding patches and SARIF assessments.
A background process monitors unfinished components
and automatically updates bundles when beneficial matches are found.
This dynamic bundling approach ensures that PoVs
are bundled with their corresponding patches and SARIF assessments whenever possible,
maximizing competition points while avoiding penalties for duplicate submissions.

\subsection{Final Competition Results by Module}
\label{ss:fcbm}

As shown in \autoref{t:module-perf},
\sys achieved competitive performance in the AIxCC final competition,
generating 1,003 PoVs with 118 passing verification,
producing 47 patches with 87.2\% success rate (41 passed),
and correctly assessing 8 out of 10 valid SARIF reports.
The system found PoVs across 62 harnesses,
with 54 harnesses producing unique (non-duplicate) results,
indicating that some vulnerable points can be reached through multiple harnesses.
The system's 89.4\% bundle submission success (42/47)
demonstrates effective end-to-end integration
across vulnerability discovery, patch generation, and assessment.

These results derive from OpenTelemetry logs provided by organizers,
with important caveats: log truncation occurred due to unstable server environments;
thus, data reflects only successfully logged operations.
Additionally, PoVs that passed the competition environment's verification
may have been identified as duplicates or failed manual assessment
during post-competition analysis.
Therefore, the results presented here differ from the final official scores
in \autoref{t:final-scores}.

\begin{table}[t]
\centering
\footnotesize

\begin{threeparttable}
\begin{tabular}{llrrrrrrrr}
\toprule
\multirow{2}{*}{Target} & \multirow{2}{*}{Module} & \multicolumn{3}{c}{PoVs} & \multicolumn{2}{c}{Patches} & \multicolumn{2}{c}{Harnesses} & \multirow{2}{*}{Contribution} \\
\cmidrule(lr){3-5} \cmidrule(lr){6-7} \cmidrule(lr){8-9}
& & Total & Passed & Dup. & Total & Passed & Affected & w/o dup & \\
\midrule
\multirow{2}{*}{C} & \sys-Multilang & 217 & 76 & 1 & 24 & 23 & 33 & 33 & 64.4\% \\
& \sys-C & 185 & 18 & 68 & 2 & 2 & 6 & 5 & 15.3\% \\
\cmidrule(lr){1-10}
\multirow{2}{*}{Java} & \sys-Java & 424 & 15 & 73 & 14 & 9 & 14 & 10 & 12.7\% \\
& \sys-Multilang & 176 & 8 & 9 & 6 & 6 & 9 & 6 & 6.8\% \\
\cmidrule(lr){1-10}
\multicolumn{2}{c}{Unknown\tnote{$\ast$}} & 1 & 1 & 0 & 1 & 1 & \multicolumn{2}{c}{---} & 0.8\% \\
\midrule
\multicolumn{2}{c}{\textbf{Total}} & \textbf{1,003} & \textbf{118} & \textbf{151} & \textbf{47} & \textbf{41} & \textbf{62} & \textbf{54} & \textbf{100.0\%} \\
\bottomrule
\end{tabular}
\begin{tablenotes}
\item [$\ast$] Due to log truncation from unstable server environments,
 the originating module for this entry could not be determined.
\end{tablenotes}
\end{threeparttable}
\caption{Performance breakdown by module in the final competition.}
\label{t:module-perf}
\end{table}

The results demonstrate how diversified AI-enhanced approaches
can effectively address complex vulnerability discovery challenges.
\sys-Multilang contributed the majority of verified PoVs through its
multi-tier LLM integration across six specialized input generation modules,
while specialized systems provided valuable complementarity:
\sys-C's multi-fuzzer ensembles with LLM-augmented seed generation
proved effective for coverage-guided scenarios,
and \sys-Java's sink-centered analysis successfully identified
security-sensitive API vulnerabilities.

The system's effectiveness emerged from complementary specialization
where each component addressed distinct vulnerability patterns
across diverse program characteristics
(detailed analysis in \autoref{s:appendix-multilang-targets}).
The results suggest that comprehensive vulnerability discovery
benefits from multiple coordinated approaches,
with AI techniques applied where they provide clear advantages.

\clearpage
\section{\sys-C}
\label{s:crs-c}

\begin{figure}[t]
  \centering
  \includegraphics[width=0.9\textwidth]{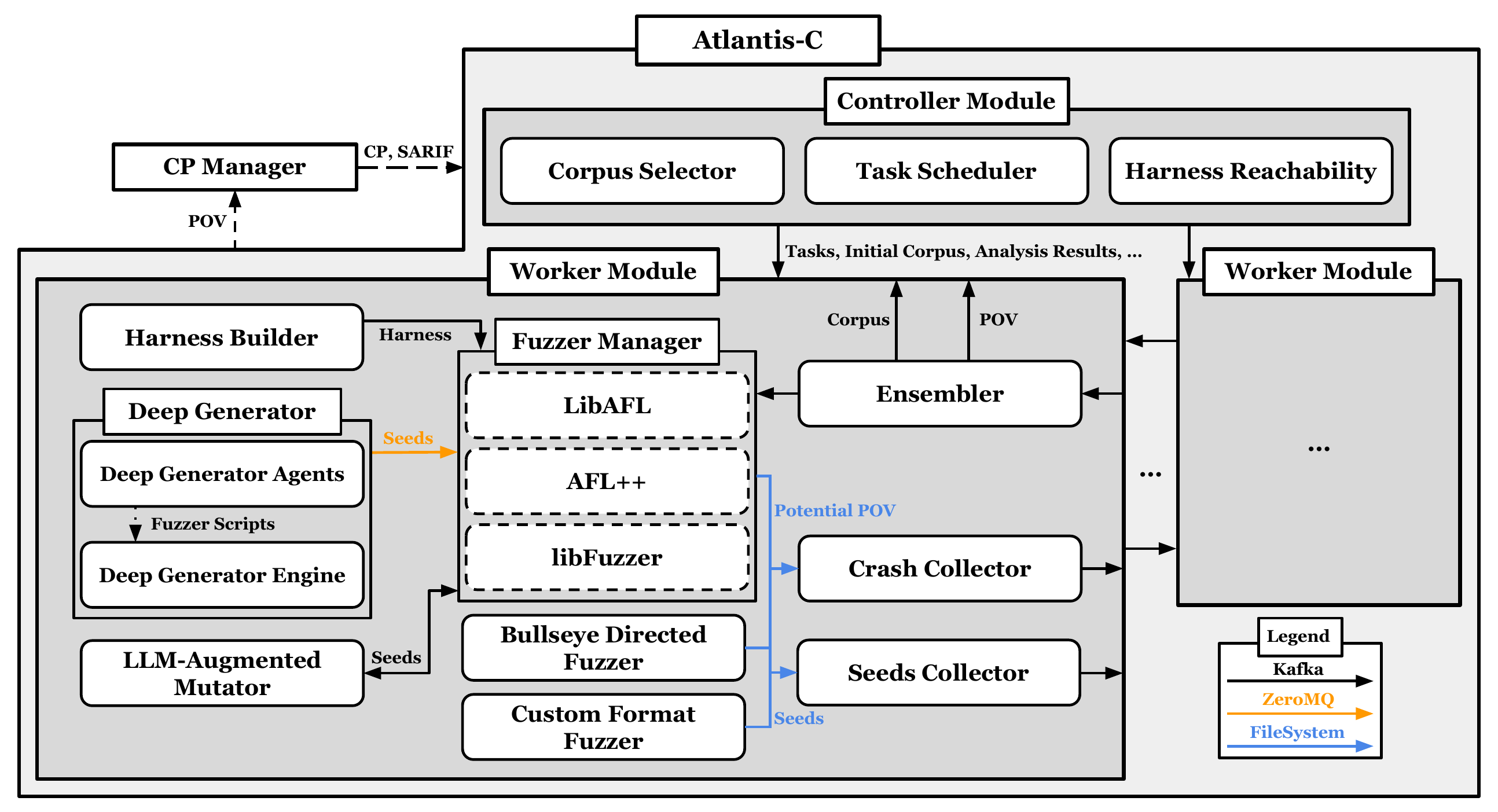}
    \caption{Overall Design of \sys-C}
  \label{fig:crs-c-overview}
\end{figure}

\sys-C targets vulnerability discovery in C and C++ challenge projects (CPs)
by orchestrating a coordinated ensemble of fuzzers.
Our experience shows that individual engines such as libAFL, AFL++, and libFuzzer respond differently
to harness structures, input formats, and coding idioms found in CPs,
leading to suboptimial performance when used alone~\cite{fu:autofz}.
To smooth out these gaps, \sys-C runs the engines side by side and wraps them
with containerized services
that can be attached to any engine to provide
shared corpus management, feedback, and automation.

\subsection{Overview}

\PP{Design.} \sys-C is an independent, multi-node, LLM-augmented fuzzing system.
For robustness and flexibility,
we design it with a Kafka-based microservice architecture.
\autoref{fig:crs-c-overview} shows the high-level architecture of \sys-C,
where each microservice is a separate containerized service.
These services can be grouped into two modules:
the \syscontroller module and the \sysworker module.
When \sys-C is deployed onto multiple Kubernetes nodes,
one of the nodes (the ``controller'' node) runs both a \syscontroller module and
a \sysworker module, and the other nodes each run a \sysworker module.

The \syscontroller module contains microservices
that are responsible for the overall task scheduling
and interaction with other systems outside of \sys-C.
It uses an epoch-based scheduling algorithm
to schedule fuzzing tasks and orchestrate other modules accordingly.
The \sysworker module contains multiple microservices
that are either part of the fuzzing pipeline
or support it.

%\autoref{fig:crs-c-overview} gives a high-level view of \sys-C.
%Its goal is to maximize bug discovery via an epoch-scheduled multi-fuzzer
%ensemble and a fast seed/crash feedback loop.
\PP{Workflow.} For each challenge, the \sysharnessbuilder
produces multiple instrumented builds per harness
(libFuzzer, LibAFL, AFL++, directed, and coverage)
and registers them as fuzzing tasks. Execution is orchestrated by the
\systaskscheduler and the \sysfuzzermanager
by assigning tasks to nodes and rebalancing priorities based on
SARIF announcements, Open Source Vulnerabilities (OSV) metadata,
reachability, crashes, and fuzzer health
signals (see \autoref{ss:fuzzer-scheduling} and \autoref{ss:harness-deprioritization}).

Seeds and crashes circulate through a central loop:
%the initial corpus is selected by the \syscorpusselector,
the \sysseedscollector and \syscrashcollector
ingest inputs from LLM generators and fuzzing engines,
and the \sysensembler deduplicates and merges seeds using
libFuzzer's merge mode, redistributes useful ones to active fuzzers,
and triages scorables (see \autoref{ss:ensembler}).
%A lightweight reachability analyzer consumes SARIF/diff results to
%deprioritize clearly unreachable harnesses and feeds decisions back to the
%scheduler (see \autoref{ss:harness-deprioritization} and
%\autoref{s:crs-sarif}).
%
%TODO: reduce redundancy
LLM components, notably \deepgen, generate high-quality seeds and
on-the-fly fuzzing scripts tailored to harness formats or patch regions
(see \autoref{ss:deepgen-c}). Communication uses two planes: Kafka provides
reliable control among microservices, while fuzzers embed ZeroMQ consumers
that use shared-memory transport for high-rate seed injection (see
\autoref{ss:zeromq}). \sys-C integrates LibAFL, AFL++, libFuzzer,
our directed fuzzer \sysbullseye (see \autoref{ss:bullseye}),
and a few project domain-specific fuzzers,
all participating in the same seed/coverage loop.

In the following sections,
we describe the features of \sys-C in detail,
along with the design choices and implementation details
for each related component.

\subsection{Multi-Fuzzer Integration}

\sys-C employs a multi-fuzzer ensemble approach that combines LibAFL, AFL++, and libFuzzer
to maximize bug discovery across diverse target programs.
Rather than selecting a single ``best'' fuzzer,
we recognize that different engines excel in different scenarios,
and their parallel execution with coordinated seed sharing
can achieve better coverage than any individual fuzzer alone.
%
%Time-based scheduling enables dynamic adaptation to runtime events
%such as SARIF reports and reachability analysis results,
%allowing the system to reprioritize computational resources
%toward the most promising harnesses.
%
%LLM-augmented components provide semantic-aware mutations
%and targeted fuzzing script generation,
%while our directed fuzzer \sysbullseye explores multiple paths
%toward known target locations to discover bug variants.

\subsubsection{Instrumentation}

\sys-C integrates a range of fuzzing engines,
including LibAFL, AFL++,
and libFuzzer~\cite{libafl, AFLplusplus-Woot20, llvm-libfuzzer},
each with distinct requirements for build-time instrumentation.
To accommodate these diverse needs,
we developed the \sysharnessbuilder module,
which orchestrates the building of fuzzing harnesses
across multiple instrumentation modes.

Instrumentation multiplexing is implemented through two primary mechanisms:

\begin{squishenumerate}
  \item Invoking the OSS-Fuzz \cc{infra/helper.py} script
  with customized environment variables
  (\eg setting \cc{CC} to a specific compiler or wrapper), and
  \item Bypassing \cc{infra/helper.py} entirely
  by executing Docker commands directly via a dedicated \cc{compile} wrapper.
\end{squishenumerate}

The second approach introduces the risk of inconsistencies or breakages,
as it deviates from the officially supported OSS-Fuzz build infrastructure.
Therefore, we apply additional validation and caution
when handling artifacts generated via the direct Docker invocation method.

In total, \sys-C supports seven instrumentation modes:

\begin{squishenumerate}
  \item LibFuzzer
  \item LibAFL
  \item Compiler Optimizations
  \item Source-Based Code Coverage
  \item AFL++
  \item Directed Fuzzer
  \item Artifact Extraction
\end{squishenumerate}

These modes differ in their degree of integration with OSS-Fuzz.
Mode 1 is essentially the standard harness build process and requires minimal deviation from it.
Modes 2--4 are implemented by
selectively overriding environment variables,
and modes 5--7 involve deeper modifications
to the build pipeline.

For instance,
the artifact extraction mode applies post-processing
to the generated harness binaries to recover the original source file paths.
It also extracts auxiliary build artifacts,
such as decompressed source directories (\eg from curl)
or consolidated source files
(\eg the monolithic amalgamation used by SQLite3).

Because OSS-Fuzz’s \cc{infra/helper.py} does not natively support
the injection of custom build steps,
the \sysharnessbuilder invokes Docker directly
while attempting to replicate the environment
that \cc{helper.py} would normally establish.
Although the artifact extraction process closely mirrors
a standard libFuzzer build under OSS-Fuzz,
we maintain it as a distinct mode
to provide redundancy and improve fault isolation.

Supporting all seven instrumentation modes
imposes a significant computational cost.
Sequentially building each variant would result in unacceptable delays
that bottleneck fuzzer startup.
To mitigate this,
\sys-C distributes build tasks across all of its available CRS nodes
by launching a separate \sysharnessbuilder instance per node.
Given the allocated computation budget,
this allows all instrumentation modes to be built in parallel,
ensuring timely fuzzer initialization without compromising throughput.

\subsubsection{Fuzzer Fallback}
\label{ss:fuzzer-fallback}

\sys-C primarily relies on a custom fuzzer built with LibAFL for high-performance fuzzing.
However, we experienced fragility issues with this fuzzer,
which could sometimes crash during instrumentation or fuzzing.
To mitigate this and prevent a complete failure of our system,
we integrated multiple fuzzing engines
with different performance and stability trade-offs.
AFL++ is more stable than our LibAFL fuzzer,
but could still fail due to our modifications
to make it run in parallel, and because the CPs
were only designed and tested with libFuzzer.
LibFuzzer is the least performant of our engines,
but is the most stable as it is the default fuzzer for OSS-Fuzz.

The interface for running all fuzzers is unified in
\sys-C's fuzzer manager.
We wrap all fuzzer engines with an interface that supports
starting the fuzzer,
stopping the fuzzer,
monitoring for crashing test cases,
monitoring for when the fuzzer successfully starts testing,
monitoring for execution errors,
and monitoring logs to extract coverage, execution, and crash statistics.
By collecting such metrics from each type of fuzzer,
we can determine when to abort the execution of a fuzzing engine
and fall back to another one.
For instance,
if LibAFL aborts more than 10 times in a single epoch,
we consider it to be too fragile,
and proceed by killing the fuzzing container
 and restarting it with AFL++.

The fuzzer not initializing is another error condition that needs to be considered.
If we had used traditional space-based partitioning,
we could assign each fuzzing engine a long initial timeout,
and if the fuzzer did not report healthy within the threshold,
we would abort and fall back.
In \sys-C's time-based harness partitioning (\autoref{ss:fuzzer-scheduling}), however,
setting a long initial timeout is not possible due to the short epoch period.
Initially we tried using hard-coded short timeouts,
but we encountered cases where harness initialization
would have succeeded but was too slow for our timeouts
and was misidentified as a failure.
Instead, we decided to simply give the fuzzer
the entire epoch to initialize,
and then handle fallback, if necessary, during the next epoch.

\subsection{Time-based Task Scheduling}
\label{ss:fuzzer-scheduling}

An important design choice for \sys-C
is the use of time-based partitioning to schedule fuzzing tasks%
---that is, instead of executing all fuzzing tasks in parallel,
\sys-C executes them based on time slices, which we refer to as ``epochs''.
Fuzzing tasks correspond to different harness binaries built by the \sysharnessbuilder
that are used for general fuzzing.
This means that for each given harness, the three instrumentation modes libFuzzer, LibAFL, and AFL++
each define a fuzzing task.

The main reason we use time-based scheduling is
for the flexibility to respond to events
that occur during the competition time, such as
SARIF report announcements, SARIF report analysis results,
diff reachability analysis results,
and fuzzer fallback-triggering events.
Specific details of these events and how they are handled
will be discussed in the following sections.

Another reason for choosing time-based scheduling
is that \sys-C's instrumentation modes
are completely separate from those of the \sys-level harness builder.
This means that, at deployment time, \sys-C will not know
the exact number of harnesses or fuzzing tasks for the challenge project it is to analyze.
Since the number of nodes is determined at deployment time,
we need a scheduling strategy able to deal with the case
where there are more fuzzing tasks than nodes allocated to \sys-C.

\subsubsection{Implementation}
\label{ss:scheduling-implementation}

The implementation of time-based scheduling is done
through the interaction between the \systaskscheduler, \sysfuzzermanager,
and other fuzzer-related modules like \sysseedscollector, \syscrashcollector, \deepgen, and \sysllmaugmentedmutator.
Every epoch is normally 20 minutes long,
but the \systaskscheduler can also end an epoch early
for a timely response to the events mentioned in \autoref{ss:fuzzer-scheduling}.

The \systaskscheduler module manages a task queue,
and an integer-value priority weight for each task.
At the start of each epoch, it pops a number of fuzzing tasks from the task queue
equal to the number of nodes allocated to \sys-C.
These fuzzing tasks are then distributed to each node
by publishing two rounds of Kafka messages.
The first round of messages stops the currently running fuzzing tasks
that are not also in the set of just-popped tasks.
These messages instruct the \sysfuzzermanager containers to exit,
and Kubernetes's restart policy will then kick in to restart
these containers in a clean state.
The second round of messages is sent to these clean \sysfuzzermanager containers
to start the new fuzzing tasks.

Once the task queue is empty, the controller will repopulate it
with weighted random sampling from the list of fuzzing tasks.
By using this repopulation strategy, the controller can prioritize certain tasks
and ensure that all fuzzing tasks with non-zero priority are eventually scheduled.
To minimize the overhead of switching epochs,
we implemented a basic optimization allowing an epoch switch to be skipped
if all of the following heuristic conditions are met:

\begin{squishitemize}
  \item \textbf{No Starvation:} there is no non-zero priority fuzzing task that has not been scheduled.
  \item \textbf{No Priority Weight Update:} no priority weight has been updated since the last epoch.
  \item \textbf{Priority Weights Already Respected:} the number of times each fuzzing task
  has been scheduled is already rankwise equal to its priority weight.
\end{squishitemize}

Whenever a \sysfuzzermanager stops and restarts with a new fuzzing task,
it will publish a Kafka message that is subscribed to by the \systaskscheduler
and other fuzzer-related modules.
The \systaskscheduler uses this information
to track the current state of \sys-C
and respond to failures in certain fuzzing tasks.
An example of such a failure would be if a harness instrumented by LibAFL
is scheduled as a new fuzzing task,
but its execution speed is extremely slow or it repeatedly crashes abnormally.
This information will be included in the message,
and allows the \systaskscheduler to trigger a fuzzer fallback event,
which will stop that specific fuzzing task,
set its priority weight to zero,
and replace it with a new one
using AFL++ instead of LibAFL.
The same flow can also apply to an AFL++ harness, resulting in a fallback to libFuzzer.
This fallback mechanism allows us to deal with
all types of instrumentation failures
and greatly enhances the robustness of \sys-C.

Other fuzzer-related modules also use this information
to modify their behavior.
For example, the \sysseedscollector uses it
to determine which path it should collect seeds from
and what harness the seeds belong to.
This is important because different fuzzing engines
have different conventions for how to store seeds in the file system.

\subsubsection{Harness Deprioritization}
\label{ss:harness-deprioritization}

\sys-C consists of multiple components that collect,
mutate, triage and distribute important seeds and results
for all active fuzzing tasks. To utilize our computational resources
efficiently, we need a smart approach for identifying which tasks%
---in particular, which harnesses---should be run with higher priority.
More specifically, for delta-mode challenges,
we want to prioritize harnesses with execution paths
that can actually reach the code affected by the delta,
and effectively ignore (``deprioritize'') all other harnesses.

While a straightforward solution would to be use precise static analysis to filter
``unreachable'' harnesses before fuzzing begins, a purely sequential approach
(static analysis first, then fuzzing) has both performance and accuracy limitations.
First, precise static analysis is computationally expensive, with runtime
varying across programs. In our experiments on (the organizer-provided version of) SQLite3,
it required several hours to complete analysis using CodeQL.
Second, static analysis tools can produce both
false positives and false negatives. While false positives would be
acceptable, false negatives are a critical issue, as misclassifying a reachable harness
as unreachable would completely prevent the opportunity to discover scorable bugs.
In fact, during our testing,
we observed such false negatives in the curl project.

To ensure correct and efficient deprioritization,
we implement a hybrid approach:
lightweight analysis is done before fuzzing starts,
and heavier analysis is done in parallel during fuzzing.
Thanks to our time-based scheduling strategy,
deprioritization based on the results of the heavier analysis can be applied dynamically
soon after those results become available.
Our approach utilizes three different sources of reachability information:

\begin{squishenumerate}
  \item Compilation-time analysis
  \item \sysbullseye's instrumentation results
  \item Analysis results from \sys-SARIF
\end{squishenumerate}

The \sysharnessbuilder features a build mode that applies aggressive compilation and linking optimization flags,
triggering dead code elimination passes in the compiler.
We use this to create a simple oracle to determine reachability.
In delta mode,
we compile all fuzzing harnesses with these aggressive flags
both before and after the provided patch is applied to the codebase.
If the two binaries (pre- and post-patch) for a given harness are identical,
it indicates that all of the code modified by the patch was eliminated as dead by the compiler,
implying that the patched code must be unreachable by that harness.
\sys-C disables all harnesses deemed irrelevant by this method,
allowing compute to be allocated to more relevant harnesses.

The other two sources of reachability information take a code location and harness as input,
and determine whether the harness can reach the location.
To select these locations,
we use an LLM agent that analyzes the diff
and selects a few locations with corresponding priority values.
%
% The intuition is that the reference diff adds or modifies
% a limited number of features to the project.
%
When passing these locations to reachability analysis,
there may be conflicting results for certain harnesses,
where some locations are reported as reachable and others as unreachable.
If too many such conflicting cases exist,
it could lead to a situation where \textit{all} harnesses become disabled and analysis cannot continue.
Prioritizing certain locations over others is key to resolving these conflicts
and avoiding that corner case.

\sys-C also leverages \sysbullseye's instrumentation to provide reachability results.
After receiving target locations from either a SARIF report or
the delta mode location-choosing agent,
\sys-C dispatches a directed fuzzing task,
which has two stages:
instrumenting and running.
If the harness instruments without error on the target location,
the directed fuzzer starts running.
If, however, a ``target not found'' error is reported by the directed fuzzer,
it means that the static analysis based on SVF determined
that the harness cannot reach the specified location.
If some harnesses report a particular location as reachable and others report it unreachable,
we disable the latter set of harnesses.
Otherwise (\ie either all harnesses or no harnesses report a location to be reachable),
no action is taken.
As results for more locations become available,
we combine them,
and stop fuzzing harnesses confirmed
to be unreachable in later epochs.

After starting the fuzzing process,
our primary source of reachability results comes from \sys-SARIF's callgraph analysis,
detailed in \autoref{s:crs-sarif}.
This hybrid approach begins with static analysis,
then monitors program execution to detect new call relationships at runtime.
When new relationships are discovered,
they are added to existing results and the entire callgraph is recalculated.
\sys-SARIF's callgraph artifacts are kept in a shared directory
that \sys-C actively monitors for real-time harness deprioritization.
\sys-C uses \cc{libSARIF},
detailed in \autoref{ss:sarif-reachability},
combined with our chosen locations to extract
an aggregate result of disabled harnesses.

This real-time, adaptive deprioritization strategy balances computational efficiency with
precision, ensuring that we optimize resource allocation without sacrificing the ability to uncover critical bugs.

\subsection{Corpus Management}
\label{ss:corpus}

Since \sys-C is essentially a distributed fuzzing system,
an important engineering task is to manage the corpora, or fuzzing seeds,
across multiple fuzzing campaigns.
Notably, \sys-C runs multiple fuzzing instances at the same time
across a number of Kubernetes nodes,
and may also fuzz the same harnesses again in future epochs.
Therefore, we need a scalable way to accumulate the progress of each campaign
across both space and time.

\subsubsection{Initial Corpus}
\label{ss:initial-corpus}

To enhance fuzzing performance,
\sys-C emphasizes the importance of a high-quality and diverse initial corpus.
A well-curated set of seed inputs significantly accelerates coverage discovery
and reduces time-to-crash by priming each fuzzer with inputs
relevant to its target.
Recognizing that generic or randomly generated inputs
are often insufficient for complex real-world targets,
\sys-C incorporates a large and semantically rich corpus
from a variety of sources.

\PP{Collection.} We manually collected seed inputs and test cases
from over 400 open-source projects,
organizing the resulting dataset into 90 semantic categories.
These categories group inputs by structure or domain,
such as SQL queries,
JavaScript programs,
and HTML documents.
This categorization enables better alignment between the fuzzing strategy
and the expected input domain of a given target.
No crashing seeds from any AFC round were included in our collection.
Seed sources include:

\begin{squishitemize}
  \item Existing corpus directories from OSS-Fuzz projects
  \item Test cases from OSS-Fuzz issue trackers
  \item Test cases from project-specific issue trackers
  \item Sample and test data found in project repositories
\end{squishitemize}

\PP{Runtime Selection.} The runtime initial corpus selection process
is implemented in the \syscorpusselector component,
which is run whenever a new challenge project is received,
via two LLM-assisted phases:

\begin{squishenumerate}
  \item \PP{Harness Analysis Phase}
  An LLM agent inspects the structure of the fuzzing harness---such as
  entry-point functions,
  data types,
  and surrounding context---to understand the expected input format
  and usage patterns.
  \item \PP{Category Matching Phase}
  Based on the extracted information,
  a simple prompt is issued to the LLM
  to suggest the most relevant seed categories from the predefined set.
  This phase does not involve complex reasoning,
  and serves as a lightweight query to guide selection.
\end{squishenumerate}

For each matched category,
\sys-C extracts the corresponding seed archive
and loads it into the fuzzers as the initial input corpus.

While supplying large corpora may incur a short-term startup overhead,
this does not affect overall system correctness,
provided the inputs are consumed within the active scheduling epoch.
After this initial corpus processing,
only interesting seeds filtered by the fuzzer will persist across
epochs for the same harness (see \autoref{ss:corpus-seed-lifecycle}).
The early performance cost is offset by
the long-term gain in fuzzer effectiveness
and faster convergence during exploration.

\subsubsection{Corpus Seed Lifecycle}
\label{ss:corpus-seed-lifecycle}
Essentially, all corpus seeds that \sys-C produces
should be consumed by the fuzzers and contribute to their coverage.
The main sources of seeds are:

\begin{squishitemize}
  \item The rich initial corpus (see \autoref{ss:corpus})
  \item Seeds generated by LLM modules (see \autoref{ss:deepgen-c}, \autoref{ss:llm-augmented mutator})
  \item Corpora from the directed fuzzer (see \autoref{ss:bullseye}) and project domain-specific fuzzers
  \item Corpora from previously or currently running fuzzing tasks
  \item Seeds shared from \sys-Multilang and \sys-SARIF
\end{squishitemize}

We must also consider that the harnesses that consume the seeds
could be instrumented with any of LibAFL, AFL++, or libFuzzer.
All of these cases were covered using three different methods.

The first method is sending the seeds via Kafka through the \sysensembler.
The \sysseedscollector is responsible for watching the seed directories
for the directed fuzzers, project domain-specific fuzzers,
and seeds shared from \sys-Multilang and \sys-SARIF.
When a new seed is added to any of these directories,
the \sysseedscollector will send it to the \sysensembler through Kafka.
These seeds will be deduplicated and sent to the fuzzers
to be consumed (see \autoref{ss:ensembler}).
This additional step prevents redundant seeds from wasting the fuzzers' computational resources.
Our LibAFL instrumentation includes a Kafka consumer,
so this method is applicable when a LibAFL harness is running.

The second method is direct filesystem dumps.
The \sysseedscollector periodically compresses the corpora of each currently running fuzzing task
and stores them in a shared directory.
The most recent version of the compressed corpora for a given harness
is essentially the deduplicated, accumulated set of seeds
from all sources for that harness up to that point in time.
Therefore, by loading this compressed corpus directly into the fuzzer's corpus directory
during fuzzing task initialization (or during runtime, in the case of libFuzzer),
we can effectively utilize all seeds we have collected so far.

The third method is using ZeroMQ to directly feed the seeds into the fuzzers.
This method is especially useful for modules that generate seeds
at a very high rate, such as \deepgen.
This method is described in more detail in \autoref{ss:zeromq}.

\subsubsection{ZeroMQ}
\label{ss:zeromq}
Although we mostly rely on Kafka for message-passing between microservices,
we use ZeroMQ for low-latency communication between fuzzers and other components
such as agents and \deepgen's dynamically generated fuzzer scripts.

\PP{ZeroMQ Fuzzer Consumer}
We need a way to send seeds to fuzzers while they are running.
To this end, we design and implement ZeroMQ consumers
as a special mutation stage in LibAFL, AFL++, and libFuzzer.
Specifically, we launch a background thread in each fuzzer
to receive external seeds and queue them in an internal buffer.
When triggering mutation,
the fuzzers first check the internal buffer for any pending external seeds,
and consume them if present.
%\compactline

\subsubsection{Ensembler}
\label{ss:ensembler}
The role of the \sysensembler is to test seeds
produced by other parts of the system,
such as fuzzers, LLM-based mutators, and \sys-Multilang.
Depending on the results,
seeds can be submitted to \syscpmgr's PoV Verifier (see \autoref{ss:cpmgr}) for scoring,
sent to fuzzers and \sys-Multilang to improve their code coverage,
or discarded.
%\compactline

libFuzzer, which is available by default for all OSS-Fuzz projects,
features a ``merge'' mode
%(\ccw{-merge=1} command-line argument),
in which,
instead of performing fuzzing,
it updates a corpus directory
by adding seeds from other provided directories.
More specifically,
it executes each seed in a set of specified folders,
measures their code coverage,
and copies any seeds that reach code not reachable by the seeds in the first folder
to that folder.
The \sysensembler uses this libFuzzer feature to continually update
its ``master corpus'',
a set of seeds intended to approximately represent all code coverage
that has been achieved by the CRS so far.
Whenever a new seed is added to the \sysensembler's master corpus directory,
it sends the seed to the fuzzers to be incorporated into their corpora,
as well as to \sys-Multilang to improve its coverage.
It also monitors libFuzzer's console output
to watch for crashes.
If any crashes are detected, basic deduplication is performed,
and the responsible seeds are submitted to \syscpmgr for scoring.

\PP{Parallel merging.}
To handle the required throughput,
the \sysensembler continually runs
multiple instances of libFuzzer in parallel,
from a pool of worker processes
that all draw seeds from a shared internal queue.
If all of the libFuzzer processes
were to update the same master corpus directory directly,
they could potentially interfere with each other,
possibly leading to loss of seeds or other issues.
%\compactline

To prevent such problems,
the \sysensembler prepares a ``snapshot'' of the master corpus directory
before each libFuzzer call.
The snapshot will not be modified while libFuzzer is running,
even if other libFuzzer instances
add new seeds to the master corpus directory
during that time.
For efficiency,
snapshot directories consist of symbolic links
to seeds in the master corpus directory
rather than full file copies.
After the libFuzzer process exits,
the \sysensembler moves any seed files that were added to the snapshot directory
to the master corpus directory.

\PP{Timeouts.}
Timeout bugs are in-scope for AIxCC.
Specifically,
inputs that take longer to execute
than a certain timeout duration
(25 seconds by default, but configurable by the competition organizers per harness)
can be submitted as PoVs for points.
Thus,
the \sysensembler must be able to run seeds
for at least that long
to determine whether a seed is scorable in this way.
However,
to maintain high throughput,
it must also avoid being slowed down too much
by any seed or batch of seeds.

To manage these conflicting goals,
the \sysensembler configures the libFuzzer merge invocations
with a short per-seed timeout of 1 second.
Any seeds that reach this timeout
are placed in a ``slow-seeds queue''
for additional processing later.
Furthermore,
to guard against large batches of slow-running seeds,
a per-\textit{batch} timeout is also used,
equal to the number of seeds in the batch times 0.5 seconds,
with a minimum of 10 seconds.
If a batch times out in this way,
its remaining seeds are discarded without being tested at all,
on the assumption that they are unlikely
to be able to contribute any useful coverage or crashes,
and are not worth spending more time on.
%\compactline

The \sysensembler executes seeds drawn from the slow-seeds queue
whenever no other internal jobs are available from its main queue.
If one of them reaches the scorable timeout duration,
it is submitted to \syscpmgr.

\subsection{LLM Components}
\label{ss:llm-components}
%TODO

\subsubsection{Agent Library: \libagents}
To facilitate agent development, we built a research-oriented library, \libagents.
This library contains a finite-state machine to split a query into sub-queries,
external plugins and LLM tools, and an answer evaluator to assess answer quality.
Users enable the desired plugins and provide a query; the library then
iterates over sub-queries and calls external plugins to obtain and evaluate answers.
Note that this library is only used by some CRS components.

\PP{Finite-State Machine}
When using the \libagents API, users first specify the plugins to enable
and the query to be solved.
\libagents selects evaluation metrics to use for the query
(\eg completeness, plurality),
then asks the LLM to answer the question.
When an answer is received, the evaluator assesses its quality based on the metrics.
If it passes, the query is solved;
otherwise, \libagents asks the LLM to reflect and generate new sub-queries.
It then iterates over the new sub-queries and adds the responses to the context.
Once either the original query is solved or the token limit is reached,
the loop is exited and the answer is returned.
Note that, to save tokens, evaluation metrics are not used for sub-queries, only for the main query.

\PP{Plugins}
To help LLMs understand external context, we provide plugins for accessing external sources.
We offer \cc{read}, \cc{ls}, \cc{grep}, and \cc{sed} for file operations,
\cc{code-browser} for code browsing,
and \cc{aider} for AI-assisted programming.
%\compactline

\PP{Answer Evaluator}
Based on the metrics selected for a query, the evaluator assesses answer quality.
In particular, we use \cc{definitive} to ensure an answer is unambiguous,
\cc{plurality} to ensure it contains sufficient information,
and \cc{completeness} to ensure coverage.

\subsubsection{Agent: \deepgen}
\label{ss:deepgen-c}
\deepgen is an LLM-based agent that can analyze a CP and generate fuzzer scripts on the fly.
It contains multiple LLM agents%
---currently, \cc{harness-analysis}, \cc{delta-scan-analysis}, and \cc{self-evolving}%
---and an engine for scheduling fuzzer-script and testcase generation.

\PP{Main Engine}
Upon receiving a new CP, the engine calls the LLM agents to generate fuzzer scripts,
which are Python scripts that produce testcases.
The engine watches for new scripts,
and runs all of them in a loop to produce testcases continuously.
All testcases are sent to the fuzzers via ZeroMQ.

\PP{Harness Analysis}
The \cc{harness-analysis} agent analyzes a fuzzing harness,
walks through the referenced source code, and infers the input format.
It then generates fuzzer scripts based on the collected information.

\PP{Delta-Scan Analysis}
For delta-scan challenges,
the \cc{delta-scan-analysis} agent analyzes the specific commit or pull request
and generates a fuzzer script focused on the relevant changes.
Specifically, it first analyzes the diff and generates a list of suspicious locations,
then generates a fuzzer script targeting them.
%\compactline

\PP{Self-Evolving Agent}
The previous agents cannot always generate high-quality fuzzer scripts.
To address this, we introduce the \cc{self-evolving} agent,
which analyzes fuzzer scripts produced by other agents
and generates new ones.
It uses predefined metrics (\eg coverage, error rate) to evaluate script quality,
then asks the LLM to propose new fuzzing strategies to improve the scripts.

\deepgen is also used in \sys-Java (see \autoref{ss:deepgen-java}).

\subsubsection{Agent: LLM-Augmented Mutator}
\label{ss:llm-augmented mutator}

Fuzzers often get stuck when exploring deep execution paths
or encountering complex program logic.
As a result, code coverage stagnates and fuzzing becomes ineffective.
This can result from several factors:
lack of semantic awareness of input structure,
dependencies on specific input data regions to pass conditional checks,
or intricate logic that standard mutation strategies fail to navigate effectively.

As a one-stop solution,
we deployed an LLM-augmented mutator as a microservice within our fuzzing system.
The primary objective is to provide fuzzers
with new mutation opportunities for stuck seeds
by leveraging program semantic understanding.
The service continuously monitors seed storage directories during fuzzing,
tracking when new seeds are added to detect potential stagnation points.
When no new seeds appear in a directory for two minutes,
we identify the fuzzer as potentially stuck,
and initiate LLM-assisted mutation on a recently added seed.

The service gathers comprehensive execution data for stuck seeds,
including complete call stack information,
source-code-level execution traces,
and variable values at function entry and exit points for the deepest executed function.
This information, along with current coverage data,
is sent to the LLM with appropriate prompts.
We find that this service is able to generate input variations that trigger new branches.

\subsection{Directed Fuzzing: \sysbullseye}
\label{ss:bullseye}
\PP{Motivation.}
\DGFsCap typically focus on reaching
a target location as quickly as possible
by pruning away paths deemed irrelevant.
While effective in reducing search space,
this strategy often leads to getting stuck in local minima,
missing alternative routes that could expose deeper or more subtle bugs.
To address this limitation, we built \sysbullseye,
an in-house \DGF that still prioritizes proximity to
the target but is explicitly designed to explore multiple
diverse paths toward it.
This increases the likelihood of triggering the
target under different program conditions.
This design is especially valuable
in the context of the AIxCC competition,
as it enhances the chance of discovering
multiple bug variants, a graded objective.

\PP{Target selection}
For full-scan challenges,
the SARIF-reported
location is used as the fuzzing target.
In delta-scan mode,
an LLM agent is used to select
one representative line per patch hunk,
focusing on additions likely to contain new logic.
These lines are
passed to \sysbullseye as fuzzing targets.
%\compactline

\begin{figure*}[t!]
  \centering
  \footnotesize
  \setlength{\tabcolsep}{3pt}
  \includegraphics[width=0.90\textwidth]{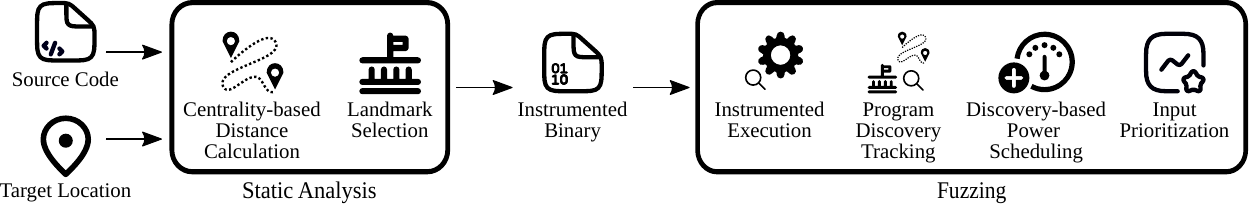}
  \caption{
    \sysbullseye architecture.
    During static analysis,
    landmark selection and distance calculation are performed,
    and the binary is instrumented with this information.
    During fuzzing,
    landmarks are used to calculate discovery,
    while the power scheduler leverages
    distance and discovery metrics to allocate
    seed energy.
    Additionally, a queue of favored seeds is maintained,
    prioritizing those with higher landmark hits and better distance scores.
  }
  \label{f:bullseye-arch}
\end{figure*}

\PP{Overview}
\sysbullseye combines static
analysis with runtime guidance
to improve target-oriented fuzzing.
As shown in~\autoref{f:bullseye-arch},
it first performs static analysis to compute
a distance score for each basic block using
closeness centrality over the \ICFG,
prioritizing blocks that are structurally
close to the target.
It also selects a fixed set of landmark
locations across the program,
which are instrumented and tracked at runtime.
The number of triggered landmarks is used
to compute a \textit{discovery} metric,
quantifying how much of the program space relevant to
the target has been explored.
During fuzzing, \sysbullseye uses these
two metrics in its power scheduler to control
how much energy is assigned to each seed,
and in its input prioritization to favor seeds
that reach more landmarks and are closer to the
target in the program structure.

\PP{Implementation}
We implemented \sysbullseye with
over 1,500 lines of code across
two main components.
The static analysis module,
written in C++ as an LLVM pass,
performs inter-procedural distance
computation and landmark selection.
The fuzzing module is built on top of
AFL++ with approximately 500 lines of
changes to integrate \sysbullseye’s
power scheduling and input prioritization.
To maximize performance,
we enabled AFL++’s persistent mode and
shared memory fuzzing,
reducing process overhead and significantly
improving execution throughput.
To our knowledge,
\sysbullseye is the first
\DGF to incorporate these
advanced optimizations,
making it well-suited for
real-world deployment.

\PP{Impact}
We evaluated \sysbullseye on
11 real-world targets provided
by the competition organizers
during preparation rounds.
Each target was fuzzed for 9 hours
across 5 independent runs,
and we report the average \TTE and \UC
in~\autoref{t:aixcc-results}.
\sysbullseye outperformed AFL++
in 7 out of 11 targets in terms of
\TTE and in 9 out of 11 in terms of \UC.
Notably, on target \#6, \sysbullseye triggered
the crash in 4 out of 5 runs,
while AFL++ failed in all runs.

\begin{table*}[!t]
  \scriptsize
  \vspace{1em}
  \centering

  \begin{tabular}{clll
    >{\centering\arraybackslash}p{0.6cm}
    >{\centering\arraybackslash}p{0.9cm}
    >{\centering\arraybackslash}p{0.6cm}
    >{\centering\arraybackslash}p{0.6cm}
    >{\centering\arraybackslash}p{0.9cm}
    >{\centering\arraybackslash}p{0.6cm}}

    \toprule

    \multirow{2}{*}{No.} &
    \multirow{2}{*}{Project} &
    \multirow{2}{*}{Harness} &
    \multirow{2}{*}{Challenge Type} &
    \multicolumn{3}{c}{\sysbullseye} &
    \multicolumn{3}{c}{AFL++} \\
    \cmidrule(lr){5-7} \cmidrule(lr){8-10}
    & & & & Runs & TTE (s) & UC & Runs & TTE (s)& UC \\
    \midrule[\heavyrulewidth]

    1 & \multirow{3}{*}{asc-nginx} & \multirow{3}{*}{pov\_harness} & \multirow{3}{*}{sarif} & 5 & \textbf{39}     & \textbf{94.2}   & 5 & 326.6  & 90.4 \\
    2 & & & & 5 & \textbf{100}    & \textbf{35.8}   & 5 & 314.2  & 13.2 \\
    3 & & & & 5 & 836.4  & \textbf{8.4}    & 5 & \textbf{398.2}  & 7.6  \\

    \cmidrule(lr){1-10}

    4 & \multirow{1}{*}{file} & \multirow{1}{*}{magic\_fuzzer\_fd} & \multirow{1}{*}{sarif}
      & 5 & 541    & \textbf{32.8}   & 5 & \textbf{533.2}  & 23.4 \\

    \cmidrule(lr){1-10}

    5 & \multirow{1}{*}{libcue} & \multirow{1}{*}{fuzz} & \multirow{1}{*}{sarif}
      & 5 & \textbf{121}    & 8.4    & 5 & 194.4  & \textbf{11}   \\

    \cmidrule(lr){1-10}

    6 & \multirow{2}{*}{libtiff} & \multirow{2}{*}{tiffcp} & \multirow{2}{*}{sarif}
      & \textbf{4} & \textbf{11665.25} & \textbf{1}     & 0 & --     & --   \\
    7 & & & & 5 & \textbf{6.8}    & 436.6 & 5 & 10     & \textbf{526}  \\

    \cmidrule(lr){1-10}

    8 & \multirow{1}{*}{libxml2} & \multirow{1}{*}{html} & \multirow{1}{*}{delta}
      & 5 & 96.8   & \textbf{66.4}   & 5 & \textbf{40.8}   & 60.8 \\

    \cmidrule(lr){1-10}

    9 & \multirow{3}{*}{libexif} & \multirow{3}{*}{\shortstack[l]{exif\_from\_data\_fuzzer\\exif\_loader\_fuzzer}} & \multirow{3}{*}{delta}
      & 5 & \textbf{62.6}   & \textbf{5.8}    & 5 & 108.8  & 4.2  \\
    10 & & & & 5 & 481    & \textbf{27}     & 5 & \textbf{386.2}  & 17   \\
    11 & & & & 3 & \textbf{4010}   & \textbf{1}      & 3 & 10804.67 & 0.8 \\

    \bottomrule
  \end{tabular}

  \caption{Fuzzing performance across AIxCC challenges using
  \sysbullseye and AFL++. Metrics shown include the number of independent
  fuzzing runs, \TTE in seconds, and \UC.}
  \label{t:aixcc-results}
\end{table*}

\clearpage
\section{\sys-Java}
\label{s:crs-java}

\sys-Java is a subsystem of \sys that focuses on Java CPV detection in the competition.
It is designed based on the observation that many Java vulnerabilities are sink-centered security issues that arise from unsafe usage of sensitive APIs.
Rather than replacing traditional coverage-based fuzzing, \sys-Java augments it with sink-aware techniques to address the unique challenges of Java vulnerability detection.
%\compactline

\begin{figure}[h]
  \centering
  \begin{promptbox}{Jenkins CPV Code Snippet I}
  \begin{promptcontent}\input{code/jenkins-doexec-commandutils.java}\end{promptcontent}
  \end{promptbox}
  \begin{promptbox}{Jenkins CPV Code Snippet II}
  \begin{promptcontent}\input{code/jenkins-createutils.java}\end{promptcontent}
  \end{promptbox}
    \caption{Example CPV from AIxCC Semifinal Jenkins CP}
  \label{fig:crs-java-motivation-example}
\end{figure}

\subsection{Motivation Example}

Our core observation is that many Java vulnerabilities stem from the unsafe usage of sink APIs, and their detection process can be modeled as a sink-centered exploration and exploitation workflow.
\autoref{fig:crs-java-motivation-example} illustrates a Jenkins CPV that demonstrates this pattern.
The vulnerability contains a backdoor that enables OS command injection when specific conditions are met.
At line 20, the \ccw{ProcessBuilder} constructor serves as a sink API, a security-sensitive operation where attacker-controllable arguments can lead to command execution.
%\compactline

From a sink-centered perspective, detecting this vulnerability involves two distinct phases:

\begin{squishitemize}
    \item 
\textbf{Sink exploration} (lines 3-8, 12-19): The fuzzer must satisfy path constraints to reach the sinkpoint, including the presence of header \ccw{x-evil-backdoor} with a value matching the SHA-256 hash of \ccw{"breakin the law"}, and non-empty command validation.
%\compactline
    \item 
\textbf{Sink exploitation} (lines 20-23): Once the sinkpoint is reached, the fuzzer must generate inputs that trigger the actual vulnerability---in this case, setting \ccw{cmds[0]} to \ccw{"jazze"} to satisfy the Jazzer detection oracle.
%\compactline
\end{squishitemize}

This two-phase pattern applies broadly to Java vulnerabilities.
Security issues typically manifest through dangerous API calls such as file operations, deserialization, command execution, network access, and template rendering.
Each vulnerability type presents unique challenges beyond simply reaching the sink.
For instance, path traversal requires constructing malicious paths that bypass sanitization; deserialization demands crafting objects that satisfy type constraints while achieving code execution; and SSRF needs URLs that bypass validation to reach external resources.
%\compactline

However, existing Java fuzzing solutions, which are mostly inherited from C/C++ fuzzers, are primarily coverage-centered and leverage only limited sink knowledge.
They treat all code paths equally, missing opportunities to prioritize security-critical operations or generate exploitation-aware inputs.
This motivates \sys-Java's design as a sink-centered framework that makes sinks first-class citizens in the fuzzing process, by combining static analysis, dynamic testing, and LLM capabilities to enhance both exploration and exploitation phases.
%\compactline

\begin{figure*}[t]
  \centering
  \includegraphics[width=\textwidth]{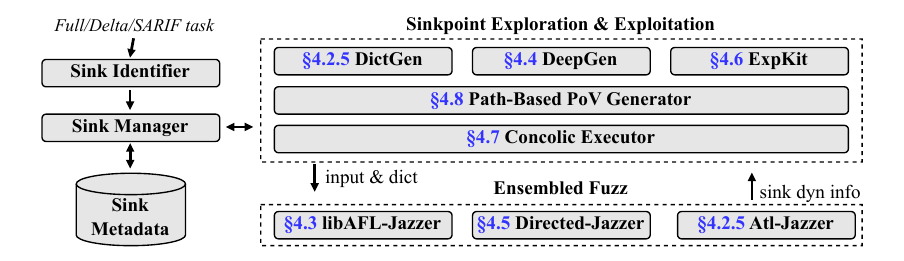}
    \caption{Overview of \sys-Java}
  \label{fig:crs-java-overview}
\end{figure*}

\subsection{System Overview}

\autoref{fig:crs-java-overview} illustrates the overall design of \sys-Java, which implements a sinkpoint-centered workflow for Java vulnerability detection.
At its foundation, \sys-Java maintains an ensemble fuzzing pipeline that serves as the base infrastructure and provides continuous input generation and execution capabilities.
Built upon this pipeline, the system performs sink analysis on target CPs to identify relevant sinkpoints through program analysis.
%\compactline

Once identified, each sinkpoint undergoes two complementary phases: \textit{sinkpoint exploration} to discover execution paths that reach the sink, and \textit{sinkpoint exploitation} to generate inputs that trigger vulnerabilities at the reached sink.
Throughout this process, sink-aware components generate specialized inputs and dictionaries that feed back into the ensemble fuzzing pipeline, while the pipeline provides dynamic execution feedback to refine sink analysis.
The system collects and maintains ``beep seeds'', inputs that successfully reach sinkpoints, which serve as valuable starting points for subsequent exploitation attempts.
%\compactline

A centralized management component tracks sinkpoint metadata, including call graphs, sink status, corpus, and crashes, and dynamically schedules system resources. Already-reached sinks skip exploration for direct exploitation, while unexploitable sinks are marked and excluded from further processing.
This bidirectional interaction between the ensemble fuzzing infrastructure and sink-centered components enables \sys-Java to effectively leverage both coverage-based and sink-aware techniques for vulnerability detection.
%\compactline

To align with the AIxCC competition format, \sys-Java transforms all three task types (Full Mode, Diff Mode, and SARIF tasks) into concrete sinkpoint lists for processing:
\ding{172} full mode tasks target all sinkpoints within the CP source tree;
\ding{173} diff mode tasks focus on sinkpoints affected by code changes;
and \ding{174} SARIF tasks prioritize sinkpoints specified in the vulnerability reports.
This transformation allows \sys-Java to use one unified sinkpoint-centered technique for handling all tasks.
%\compactline

To support various exploration and exploitation techniques within this workflow, \sys-Java provides essential infrastructure functionalities.
These components manage sinkpoint information, coordinate fuzzing efforts, and optimize resource utilization.
The following paragraphs detail each infrastructure component that enables our sinkpoint-centered approach.
%\compactline

\begin{algorithm}[t]
\footnotesize
\DontPrintSemicolon
\SetKwSty{algokeywordsty}
\SetFuncSty{algofuncsty}
\SetArgSty{algoargsty}
\SetKwFunction{fnPickSeed}{PickSeed}
\SetKwFunction{fnMutate}{Mutate}
\SetKwFunction{fnExecute}{Execute}
\SetKwFunction{fnHasNewFeedback}{HasNewFeedback}
\SetKwFunction{fnReachSinkpoint}{ReachSinkpoint}
\SetKwFunction{fnExploitationPhase}{ExploitationPhase}
\SetKwFunction{fnIsCrash}{IsCrash}
\SetKwFunction{fnSaveCrash}{SaveCrash}

\caption{Sinkpoint-Aware Fuzzing Loop}
\label{alg:sink-aware-fuzz-loop}
\KwIn{$\mathcal{P}$: Target program \newline
      $\mathcal{K}$: Set of sinkpoints}

$\mathcal{S} \gets$ \{initial seeds\}\;
$\mathcal{B} \gets \emptyset$ \tcp*{Init beep seeds set}
\While{not timeout}{
    $s \gets$ \fnPickSeed{$\mathcal{S}$}\;
    $s' \gets$ \fnMutate{$s$}\;
    $result \gets$ \fnExecute{$\mathcal{P}$, $s'$}\;
    
    \If{\fnHasNewFeedback{$result$}}{
        $\mathcal{S} \gets \mathcal{S} \cup \{s'\}$\;
    }
    \If{\fnReachSinkpoint{$result$, $\mathcal{K}$}}{
        $\mathcal{B} \gets \mathcal{B} \cup \{s'\}$ \tcp*{Add to beep seeds}
        \fnExploitationPhase{$s'$}\;
    }
    \If{\fnIsCrash{$result$}}{
        \fnSaveCrash{$s'$}\;
    }
}
\end{algorithm}

% \noindent
% \textbf{Infra. I: Sinkpoint-Aware Fuzzing Loop} \tab
\subsubsection{Infrastructure I - Sinkpoint-Aware Fuzzing Loop}
\autoref{alg:sink-aware-fuzz-loop} presents our enhanced fuzzing loop that extends traditional coverage-guided fuzzing with sinkpoint awareness.
Beyond tracking code coverage and crashes, our loop actively monitors whether test inputs reach any identified sinkpoints (line 13).
We implement this monitoring through a custom Java agent instrumentation called \ccw{CodeMarkerInstrumentation}, which we integrated into our version of Jazzer (\ccw{Atl-Jazzer}).
When the fuzzer executes an input that reaches a sinkpoint (\ie calls a sink API), the instrumentation generates a CodeMarkerHitEvent, an exception containing the complete stack trace at the sinkpoint.
This runtime information is then captured and dumped to files for other components to utilize.
%\compactline

Upon detecting sinkpoint reaching, the loop collects ``beep seeds'' (line 14), inputs that successfully reach sinkpoints, and triggers targeted exploitation attempts (line 15).
This design explicitly separates the exploitation phase from exploration, enabling specialized reasoning techniques to generate exploits using exact dynamic context.
For vulnerabilities requiring specific exploitation conditions (\eg bypassing sanitizers or satisfying complex constraints), the beep seeds provide concrete execution contexts including call stacks, reproducible input blobs, and sinkpoint vulnerability information that guide exploitation generation.
This infrastructure establishes sinkpoint reaching as a first-class feedback signal alongside coverage and crashes, forming the foundation for our sinkpoint-centered approach.
%\compactline

\subsubsection{Infrastructure II - Ensemble Fuzzing}
Ensemble fuzzing combines multiple fuzzing strategies by merging their collective efforts to improve overall bug-finding effectiveness.
In \sys-Java, the ensembler serves three key functionalities:
\ding{172} It collects and merges corpus from different fuzzer instances, performing deduplication based on coverage metrics before broadcasting the unified corpus back to all participants, enabling different strategies to benefit from each other's discoveries;
\ding{173} It incorporates inputs from non-fuzzing components such as LLM-based generators or concolic execution engines into the fuzzing workflow, effectively bridging diverse input generation techniques;
\ding{174} It facilitates synchronization of sinkpoint metadata including sink reach status and beep seed collections during corpus merging, ensuring all fuzzer instances operate with up-to-date sink information.
In general, the ensemble infrastructure serves as both a corpus synchronization layer and a metadata propagation channel across all fuzzing instances.
%\compactline

\subsubsection{Infrastructure III - Sinkpoint Identification \& Management}
Sinkpoint identification forms the foundation of our approach by statically analyzing the target CP to locate all calls to security-sensitive sink APIs.
Beyond Jazzer's built-in sink API list, we expanded our detection capabilities by collecting additional sink APIs from vulnerability benchmarks, published research papers, and static analysis tools used in the competition.
Our custom sink API list is configured through YAML files in the system's CodeQL component, with support for future extension through additional configuration files.
%\compactline

The sinkpoint management component maintains metadata for each identified sinkpoint, including call graph information, runtime status (unreached, reached, exploited, or unexploitable), associated beep seeds, and exploitation attempts.
This component dynamically schedules fuzzing resources based on sinkpoint status, such that already-reached sinks should focus more on exploitation, while sinks marked as unexploitable after thorough analysis are excluded from further processing.
By doing so, the manager minimizes redundant efforts and prioritizes high-value targets such as SARIF/diff-relevant sinkpoints.
%\compactline

All exploration and exploitation components can access this centralized sinkpoint metadata through a shared interface, enabling informed decision-making throughout the competition.
For instance, directed fuzzers only target unreached sinkpoints to guide path exploration, while exploitation agents retrieve beep seeds and stack traces for reached sinkpoints to craft targeted exploits.
This unified access to sinkpoint information ensures that every component operates with consistent, up-to-date knowledge about the target program's attack surface.
%\compactline

\subsubsection{Infrastructure IV - Distributed Design}

\begin{figure}[t]
  \centering
  \includegraphics[width=0.7\textwidth]{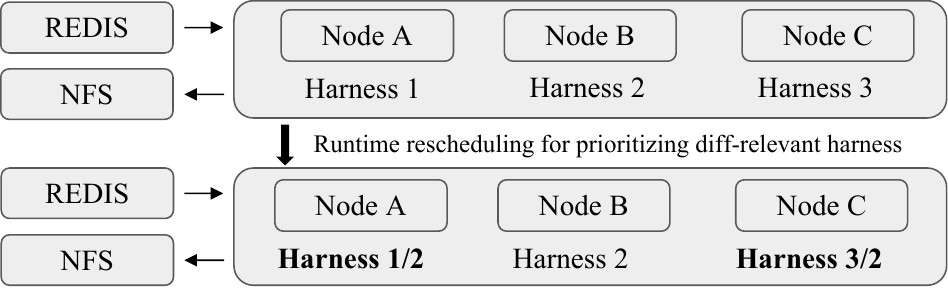}
    \caption{Runtime Rescheduling in \sys-Java for Diff Tasks}
  \label{fig:crs-java-runtime-rescheduling}
\end{figure}

Our distributed design employs a lightweight architecture that adapts to different competition task types.
In full mode, each CP harness is assigned one and only one node, with all nodes working separately until the task deadline.
During execution, NFS and Redis provide persistent caching for component outputs and system states, enabling seamless recovery from node failures or restarts.
This design ensures fault tolerance while maintaining simplicity in deployment and management.
%\compactline

For diff mode tasks, the system initially deploys identically to full mode but incorporates a dynamic rescheduling mechanism to optimize resource allocation.
As illustrated in \autoref{fig:crs-java-runtime-rescheduling}, after a predefined period (2 hours) or when static analysis results stabilize across multiple analyzers, the system performs a one-time rescheduling operation.
This rescheduling reassigns system resources by redistributing nodes to focus on diff-relevant harnesses.
For example, if Harness 2 contains critical diff-related sinkpoints, nodes originally assigned to Harnesses 1 and 3 may be reassigned to also process Harness 2, resulting in multiple nodes working on the same high-priority harness.
%\compactline

After rescheduling, when multiple nodes process the same harness, the sinkpoint manager leverages Redis to periodically synchronize sinkpoint metadata across all nodes.
This synchronization includes sink status updates, beep seed collections, and exploitation attempts, preventing duplicate efforts while maximizing resource utilization on diff-relevant vulnerabilities.
The lightweight nature of this distributed design, relying only on standard Redis and NFS infrastructure, minimizes our development and deployment efforts while providing flexibility for different competition tasks.
%\compactline

\subsubsection{Component Overview}

\begin{table}[!t]
\centering
\footnotesize
% \resizebox{0.7\columnwidth}{!}{%
\begin{tabular}{lcc}
\toprule
\textbf{Component} & \textbf{Sinkpoint Exploration} & \textbf{Sinkpoint Exploitation} \\
\midrule
\rowcolor[HTML]{EFEFEF} 
Directed Jazzer & \checkmark & \\ 
LibAFL-Based Jazzer & \checkmark & \\ 
\rowcolor[HTML]{EFEFEF} 
Atl-Jazzer & \checkmark & \\ 
Path-Based PoV Generator & \checkmark & \checkmark \\ 
\rowcolor[HTML]{EFEFEF} 
Concolic Executor & \checkmark & \checkmark \\ 
DeepGen & \checkmark & \\ 
\rowcolor[HTML]{EFEFEF} 
DictGen & \checkmark & \\ 
ExpKit & & \checkmark \\ 
\bottomrule
\end{tabular}
% }
\caption{Component Overview of \sys-Java}
\label{tab:crs-java-components}
\end{table}

Beyond the infrastructure components, \sys-Java integrates multiple specialized tools for sinkpoint exploration and exploitation as shown in \autoref{tab:crs-java-components}.
These components leverage the infrastructure to work collaboratively: exploration-focused components such as Directed Jazzer and LibAFL-based Jazzer focus on reaching sinkpoints, while exploitation-focused components like ExpKit specialize in generating vulnerability-triggering inputs for reached sinks.
Some components serve dual purposes, including the Path-Based PoV Generator and Concolic Executor, which perform both exploration and exploitation.
%\compactline

Two components deserve special mention: DictGen, adapted from \sys-Multilang, generates fuzzing dictionaries for fuzzer instances in various scenarios, including initial seed generation and beep seed exploitation phases (see \autoref{s:crs-multilang} for details).
Atl-Jazzer, our enhanced version of Jazzer, primarily adds beep seed tracking functionality alongside minor feature improvements and bug fixes.
The complete feature list is documented in our code repository.
Together, these components form a comprehensive toolkit that addresses different aspects of Java vulnerability detection within our sinkpoint-centered framework.
%\compactline

\subsection{LibAFL-Based Jazzer}

The LibAFL-based Jazzer component is designed to enhance mutation diversity beyond Jazzer's built-in mutators for generally improved sinkpoint exploration.
While Jazzer itself provides effective libFuzzer mutations, relying solely on its default strategies can limit exploration breadth.
By integrating LibAFL's advanced mutation algorithms and scheduling strategies, we aim to discover execution paths that standard Jazzer might miss, ultimately reaching more sinkpoints during the exploration phase.
%\compactline

\begin{figure}[t]
  \centering
  \begin{promptbox}{Original Jazzer Dependency Chain}
  \begin{promptcontent}\input{code/ori-jazzer-dep-chain.txt}\end{promptcontent}
  \end{promptbox}
  \begin{promptbox}{LibAFL-Based Jazzer Dependency Chain}
  \begin{promptcontent}\input{code/libafl-jazzer-dep-chain.txt}\end{promptcontent}
  \end{promptbox}
  \caption{Jazzer Dependency Chain Comparison}
\end{figure}

Our integration leverages the LibAFL-libFuzzer project \cite{libafllibfuzzer}, which provides a LibFuzzer-compatible runtime built on LibAFL's fuzzing infrastructure.
This enables a drop-in replacement architecture.
While the base LibAFL-libFuzzer project provides API compatibility, integrating it with Jazzer required addressing several technical challenges.
%\compactline

\PP{Finding Detection and Lifecycle Control}
Java fuzzing requires specialized handling of findings and fuzzer lifecycle events that differ from binary fuzzing.
We implemented custom observers (\ccw{JazzerFindingObserver} and \ccw{JazzerFuzzerStoppingObserver}) that hook into Jazzer's \ccw{__jazzer_set_death_callback} mechanism.
This integration enables proper crash artifact dumping when Java-specific vulnerabilities are detected and supports graceful fuzzer termination based on Jazzer side's state.
Without this adaptation, LibAFL would miss Java-level findings that don't manifest as traditional crashes.
%\compactline

\PP{Coverage Feedback Integration}
Jazzer's bytecode instrumentation passes program counter (PC) information to sanitizer functions for precise coverage tracking.
We extended LibAFL's coverage infrastructure with PC-aware variants (\ccw{__sanitizer_cov_trace_cmp4_with_pc}, \etc) that extract and forward this information to LibAFL's feedback mechanisms.
%\compactline

\PP{String Comparison Hooks}
Jazzer relies on specialized string comparison behaviors.
We implemented Jazzer-specific hooks including \ccw{__sanitizer_weak_hook_strstr} and \ccw{__sanitizer_weak_hook_compare_bytes} that integrate with LibAFL's comparison tracking while preserving Jazzer's requirements.
%\compactline

\PP{Mutation Strategy Control}
To fully leverage LibAFL's mutation diversity, we disabled Jazzer's custom mutations through runtime flags (\ccw{--disable_custom_mutate}), allowing LibAFL's schedulers and mutators to control the fuzzing process exclusively.
%\compactline

\PP{Discussions}
In our testing and evaluation, LibAFL-based Jazzer can show different coverage results compared to standard Jazzer in certain projects, demonstrating its complementary exploration capabilities.
We think this is because LibAFL's mutation scheduler more aggressively prioritizes unexplored code regions.
However, our current LibAFL-based Jazzer lacks support for value profile feedback, a mechanism Jazzer uses for simple exploitation scenarios (\eg gradually approaching string comparisons like ``jazze'').
Consequently, it shows weaker exploitation capabilities for vulnerabilities requiring specific value constraints.
%\compactline

\subsection{\deepgen}
\label{ss:deepgen-java}

\deepgen (previously discussed in \autoref{ss:deepgen-c}) is a high-throughput framework that leverages LLM agents to generate seed mutation/generation scripts for fuzzers, thereby producing massive amounts of diverse seeds to improve coverage exploration.
This framework represents a collaborative effort between the \sys-C and \sys-Java teams, designed to address the limitations of traditional mutation strategies through intelligent, context-aware seed generation.
The complete system consists of three core components: \libagents, \libdeepgen, and customized fuzzers with out-of-fuzzer (OOF) mutator support.
%\compactline

\begin{figure}[t]
  \centering
  \includegraphics[width=0.7\textwidth]{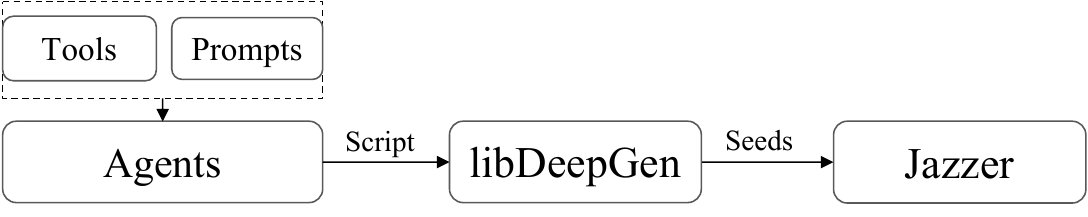}
    \caption{Overview of \deepgen}
  \label{fig:deepgen}
\end{figure}

\PP{Architecture Overview}
As illustrated in \autoref{fig:deepgen}, \deepgen employs a straightforward architecture where \libagents serves as the intelligent generation layer.
This component implements a deep research framework capable of analyzing target programs and generating context-aware Python mutation scripts.
\libagents supports various tools including common command-line utilities, and provides wrappers for advanced AI coding assistants such as Claude Code and Codex agents.
The deep research capability enables \libagents to explore the target codebase, understand API patterns, and generate scripts that produce semantically meaningful inputs.
%\compactline

The core runtime component, \libdeepgen, manages the execution and scheduling of agent-generated scripts with highly optimized performance characteristics.
Its design incorporates atomic operation-based ring buffers for script task scheduling and a combination of shared memory and ZeroMQ for seed storage and dispatch.
This architecture enables extreme throughput, supporting thousands of script switches per second and generating hundreds of thousands of seeds at peak capacity.
When \libdeepgen executes scripts produced by \libagents, the generated seeds are sent to fuzzers through shared memory coordinates transmitted via ZeroMQ messages.
%\compactline

To integrate with existing fuzzing infrastructure, \deepgen requires fuzzers to support OOFMutator (Out-Of-Fuzzer Mutator) functionality.
This involves adding a ZeroMQ client to the fuzzer's mutation engine, enabling batch retrieval of generated seeds that are then used as mutation outputs.
In \sys-Java, our Atl-Jazzer implements full OOFMutator support, allowing seamless integration with \deepgen's high-throughput seed generation pipeline.
%\compactline

\PN{\deepgen in \sys-Java}
% \noindent
% {\bf{\deepgen in \sys-Java}}
Due to competition time constraints, we focused \deepgen's deployment in \sys-Java primarily on initial corpus generation rather than its full potential for continuous mutation.
For each harness, \sys-Java consolidates relevant information including harness code and task descriptions into structured prompts for \libagents.
\libagents's deep research framework actively explores the CP codebase to gather harness-relevant context, analyzing existing test cases, API documentation, and code patterns to understand the harness generation specifics.
This contextual understanding enables the generation of Python scripts that produce semantically valid and diverse inputs, significantly outperforming random or syntax-unaware generation strategies.
\sys-Java iteratively notifies \deepgen to generate multiple Python scripts, each producing a fixed quantity of seeds.
During the execution, \deepgen will filter out invalid or ineffective scripts, ensuring that only high-quality seeds are passed to the fuzzers. 
%\compactline

\PP{Effectiveness}
As an initial corpus generator, \deepgen demonstrates remarkable effectiveness in jumpstarting fuzzing campaigns.
Our internal benchmarks reveal substantial coverage improvements when comparing 10-minute fuzzing sessions with and without \deepgen-generated initial corpora.
In the most dramatic cases, we observed coverage increases around 2,952\%, with over 20\% of projects having improvements greater than 40\%.

These results highlight \deepgen's ability to quickly explore diverse program states that would require significantly longer discovery times through traditional mutation alone.
These numbers validate the potential of LLM-guided seed generation for fuzzing.
After the competition, we will extend \deepgen's capabilities beyond initial corpus generation to support continuous, adaptive mutation strategies that evolve based on runtime feedback and discovered program behaviors.
%\compactline

\subsection{Sinkpoint-Focused Directed Fuzzing}
\label{sec:directed-fuzzing}

The sinkpoint-focused directed fuzzing component represents a combination of static analysis and runtime guidance, designed to efficiently navigate Java programs toward security-critical code locations (sinkpoints).
Although primarily developed for reaching sinkpoints, this approach proves equally valuable for exploiting sinkpoints once they are reached.
The architecture orchestrates two complementary subsystems: a comprehensive static analysis pipeline combining CodeQL and Soot frameworks, and a modified Jazzer implementation that transforms distance computations into actionable fuzzing guidance.

\begin{figure}[t]
  \centering
  \includegraphics[width=.5\columnwidth]{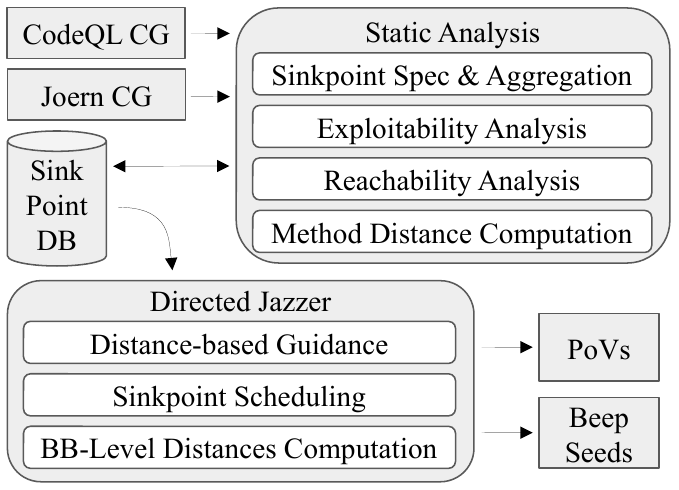}
  \caption{Overview of the Sinkpoint-focused Directed Fuzzing Architecture}
  \label{fig:directed-fuzzer-overview}
\end{figure}

An overview is available in \autoref{fig:directed-fuzzer-overview}.
The main inputs are call graphs from other CRS modules as well as data from the sinkpoint database.
To improve performance, the system performs most analyses ahead of fuzzing, in the static analysis phase.
This design choice is motivated by the constant need to compute distances from inputs to targets in directed fuzzing.
By separating static analysis from the fuzzing phase, we eliminate two major computational overheads: 
1) the cost of reconstructing the ICFG every time the fuzzer starts/restarts,
and 2) the repeated traversal of the ICFG to calculate distances during directed fuzzing.
Balancing both performance and implementation efforts, we eventually pre-compute and cache function-level distance calculations, and do basic block-level distance calculation (only requiring non-intraprocedural CFGs) during fuzzing.
In addition, the fuzzer dynamically schedules sinkpoints based on their reachability, exploitability, and competition relevance.

% TODO: Add overview graphic of directed fuzzer architecture

\subsubsection{CodeQL-Enhanced Sink Detection}

Since the vast majority of bugs in Java applications can only happen when certain Java APIs are called, we developed a framework to define those sink APIs.
While Jazzer has a list of such APIs that it sanitizes, we identified additional APIs that are likely to trigger a sanitizer.
Our CodeQL component addresses this challenge by establishing a framework for specifying security-sensitive APIs that extend beyond Jazzer's built-in sanitizers.
In particular, we added sinkpoints for calls to methods from the following classes:
\begin{squishitemize}
  \item \cc{java.math.BigDecimal}: The use of the \cc{BigDecimal} class may result in a denial of service.
  \item \cc{java.net.URL} and \cc{java.net.URI}: Additional networking APIs, which may lead to an SSRF.
  \item \cc{javax.validation.Validator}: May lead to an expression language injection, resulting in remote code execution.
  \item \cc{javax.xml.parsers.SAXParser}: Parsing an XML may result in an SSRF.
  \item \cc{org.apache.batik.transcoder.TranscoderInput}: Opening an SVG stream may lead to an SSRF.
\end{squishitemize}
%\compactline

Rather than exhaustively analyzing every library dependency, our CRS extends the list of sink APIs allowing us to only scan the challenge problem code instead of all libraries.
The strategic advantage of this approach becomes apparent when considering the resource constraints of the competition.
On our benchmarks, analyzing not only the challenge problem code but also the dependencies would take hours with Soot, which may be longer than our CRS has to analyze the challenge problem, without even starting the directed fuzzer.
However, using our custom framework, we can skip analyzing the dependencies and reduce analysis time to a matter of minutes (depending on the size of the challenge problem) on our benchmark set, leaving most of the analysis time for the directed fuzzer to run.
%\compactline

Beyond identification, the CodeQL analysis incorporates an exploitability assessment that distinguishes between theoretically reachable sinks and practically exploitable ones.
By analyzing data flow patterns, we determine whether sink API arguments may be influenced by attacker-controlled input.
When our system finds strong evidence that a sinkpoint is not exploitable, we remove it from the analysis.
We consider a sinkpoint not exploitable if, \eg the relevant argument is hard-coded or not tainted by attacker-controlled input.
This filtering transforms a large list of potential targets (hundreds to thousands) into a manageable set of high-value sinkpoints (<100 sinkpoints for 95\% of cases in our benchmark set).
%\compactline

The implementation leverages a centralized sink definition architecture built around YAML configuration files that separate CodeQL model specifications from descriptive metadata.
An example specification is given in \autoref{fig:sink-definitions} for the \cc{java.net.URL} class, which may result in an SSRF.
%\compactline

\begin{figure}[t]
  \centering
  \begin{promptbox}{Example Sink Definition}
  \begin{promptcontent}\input{code/sink_definitions.yml}\end{promptcontent}
  \end{promptbox}
    \caption{Example sink definition for \cc{java.net.URL}.}
  \label{fig:sink-definitions}
\end{figure}

\subsubsection{Static Analysis for Distance Computation}
\label{sec:directed-fuzzing:static-analysis}

The static analysis foundation tackles one of directed fuzzing's most challenging problems: accurately measuring the distance from program entry points to sinkpoints.
Our approach begins with per-harness reachability analysis, ensuring that computational resources focus exclusively on sinkpoints that are theoretically accessible from specific fuzzing entry points.
This preliminary filtering eliminates unreachable targets that would otherwise consume valuable analysis time or guide the fuzzer towards unreachable code locations.
%\compactline

Recognizing that no single static analysis tool provides complete call graph coverage, we merge call graphs from multiple other CRS components, based on analysis frameworks such as Joern and CodeQL and extended with our own analyses to resolve calls.
This multi-tool approach proves particularly valuable for handling Java's complex object-oriented features, where interface calls and reflective invocations often confound individual analysis tools.
Each framework contributes unique strengths: CodeQL excels at pattern-based detection, while Joern provides flexible code property graph analysis.
%\compactline

The merging process confronts technical challenges arising from inconsistent naming conventions across analysis tools.
Lambda expressions and nested classes receive different internal representations in various frameworks, requiring normalization and heuristics to establish consistent target identification.
Our system resolves these discrepancies by mapping all call sites to specific Jimple instructions, \ie Soot's intermediate representation, creating a unified foundation for subsequent distance calculations.
%\compactline

Distance computation itself employs a hierarchical approach that balances precision with computational efficiency.
Call graphs from static analysis provide coarse-grained method-level distances, while control flow graphs within Jazzer enable fine-grained basic block-level precision.
We only compute the call graph ahead of time and not the control flow graphs, since there are a number of limiting factors that would come with such an approach.
Statically computing distances on a basic block level would require a stable mapping of basic blocks to distances.
However, the instrumentation of Jazzer (JaCoCo) changes the bytecode and would require us to rely on heuristics to match basic blocks, which we want to avoid.
Another approach would be to map the coverage IDs to distances.
However, coverage IDs are not deterministic by default.
Coverage IDs can be made deterministic by instrumenting jar files ahead of time, but we want to avoid modifying the test artifacts as this would be a threat to stability.
The reason why we cannot directly map bytecode offsets of basic blocks from the test artifacts to the instrumented classes is because the number of bytes that the bytecode instrumentation adds is difficult to predict.
This is because every call to record a triggered coverage ID comes with a constant, which is part of the run-time constant pool.
The resulting increase in the number of constants may require certain instructions to use a different variant to support wider indices, resulting in changes even to instructions that are present in the code before the instrumentation happens.
Thus, we resort to generating the CFG at runtime of the fuzzer as the most reliable and precise approach.
%\compactline

An example result of the static analysis is illustrated in \autoref{fig:directed-analysis-result}.
It contains a regex sinkpoint in the Apache Tika library and specifies the location in the code (class name, method name, descriptor, bytecode offset) as well as metadata.
The analysis adds the reachability and exploitability results for each harness.
The sinkpoint also contains information about whether it was reached and if it is derived from a diff or SARIF report, which will be used for prioritization.

\begin{figure}[t]
  \centering
  \begin{promptbox}{Example Static Analysis Result}
  \begin{promptcontent}\input{code/directed_analysis_result.json}\end{promptcontent}
  \end{promptbox}
    \caption{Example of the static analysis result for a regex sinkpoint in Apache Tika.}
  \label{fig:directed-analysis-result}
\end{figure}

\subsubsection{Directed Jazzer}

Our version of Jazzer contains a component to guide the fuzzer towards specified code locations.
This new component transforms static analysis insights into dynamic fuzzing guidance through an adaptive scheduling system that schedules up to 15 concurrent sinkpoint targets at a time.
The scheduler continuously monitors target status, automatically removing reached sinkpoints from active consideration while promoting newly discovered targets to active status.
%\compactline

Target prioritization reflects the competitive nature of the AIxCC environment through a prioritizing round-robin algorithm that allocates double scheduling time to sinkpoints derived from diff analysis and SARIF reports.
This prioritization ensures that competition-relevant vulnerabilities receive appropriate attention while maintaining broad coverage of the overall attack surface.
The dynamic nature of this scheduling adapts to changing analysis results, automatically adjusting priorities as new data becomes available.
%\compactline

The distance computation engine confronts the practical reality that instrumentation identifiers change between execution runs, making pre-computed mappings unreliable.
Rather than modifying target JARs---which would compromise system stability and reliability---the implementation employs dominator analysis to establish runtime correlations between coverage identifiers and Soot basic blocks.
This approach proves essential because Jazzer's coverage instrumentation does not instrument every basic block, requiring intelligent inference of distances for uninstrumented code regions.
%\compactline

Our system maintains comprehensive situational awareness through continuous monitoring of distance files, automatically detecting and incorporating updates from ongoing static analysis.
This dynamic adaptation capability enables the fuzzer to respond to evolving analysis results.
Target metadata encompasses reachability status, exploitability assessments, and priority classifications, providing the scheduler with rich information for making optimal resource allocation decisions throughout extended fuzzing campaigns.
%\compactline

\subsection{ExpKit}

ExpKit is a specialized exploitation component designed to address the ``last mile'' challenge in Java vulnerability detection: cases where fuzzers successfully reach sinkpoints but fail to trigger actual exploits.
Through large-scale analysis and targeted LLM-based exploitation generation, ExpKit transforms reached-but-unexploited sinkpoints into successful vulnerability discoveries.
%\compactline

\subsubsection{Motivation: The Last Mile Challenge}

We conducted a large-scale fuzzing experiment across 42 CPVs in our benchmark, allocating 8 hours and over 100 CPU cores per harness. The results revealed a critical gap in traditional fuzzing approaches.
%\compactline

As shown in \autoref{tab:last-mile-results}, while 73.8\% of CPVs had their sinkpoints reached through fuzzing, only 35.7\% resulted in successful exploits.
This 38.1\% gap represents the ``last mile'' challenge: cases where traditional value profile feedback mechanisms prove insufficient for triggering vulnerabilities despite reaching sinkpoints.
%\compactline

\begin{table}[!t]
\centering
\footnotesize
% \resizebox{0.5\columnwidth}{!}{%
\begin{tabular}{lccc}
\toprule
\textbf{Total CPVs} & \textbf{Not Reached} & \textbf{Reached Only} & \textbf{Exploited} \\
\midrule
42 (100\%) & 11 (26.2\%) & 16 (38.1\%) & 15 (35.7\%) \\
\bottomrule
\end{tabular}%
% }
\caption{Large-scale Fuzzing Results Showing the Last Mile Challenge}
\label{tab:last-mile-results}
\end{table}

\subsubsection{Analysis of Exploitation Gaps}

The root causes of those 16 reached-but-unexploited CPVs include:
%\compactline

\begin{squishitemize}
\item
\PP{Distracted by seed explosion (4/16)}
Fuzzers generated overwhelming amounts of inputs reaching sinkpoints, but failed to focus on exploitation-relevant mutations.
%\compactline

\item
\PP{Missing sink API instrumentation (2/16)}
Some sinkpoints used APIs not covered by standard value profile instrumentation (there are significant gaps between the sink API and the Jazzer-hooked functions which can trigger value profile feedback), leaving fuzzers without exploitation guidance.
%\compactline

\item
\PP{Insufficient value profile depth (1/16)}
Complex sanitizers required longer input sequences than value profile mechanisms could effectively guide.
%\compactline

\item
\PP{Complex exploitation logic (9/16)}
This is the most common case, where exploitation required reasoning about complex conditions, multiple API interactions, or specific input formats that traditional fuzzing strategies could not handle even with the help of value profile feedback.
For instance, to trigger a XML deserialization vulnerability, the fuzzer must generate a specific XML payload that satisfies type constraints and bypasses validation checks.
%\compactline
\end{squishitemize}

These findings motivated ExpKit's design as a specialized exploitation agent that leverages beep seeds' rich execution context combined with LLM reasoning to bridge the semantic gap between reaching and exploiting sinkpoints.
%\compactline

\subsubsection{Key Design}

ExpKit operates as an autonomous exploitation agent that processes beep seeds, \ie inputs that successfully reach sinkpoints, to generate targeted exploits.
The tool is implemented with several key design principles.
%\compactline

\PP{Beep Seed Scheduling and Exploitation Loop}
ExpKit groups beep seeds by their execution context, defined as the complete stack trace at the sinkpoint.
This grouping enables fair scheduling across different execution paths.
The scheduler prioritizes contexts with fewer exploitation attempts, implementing a two-dimensional fair scheduling algorithm based on the tuple $\langle\text{attempt\_count}, \text{context\_id}\rangle$.
For each exploitation round, ExpKit selects the context with minimum attempts and randomly picks a beep seed from that context for processing.
During competition runs, ExpKit operates continuously, processing new beep seeds as they arrive from fuzzing components and immediately testing generated exploit attempts.
%\compactline

\PP{Context Collection and Prompt Generation}
For each selected beep seed, ExpKit collects context information and transforms it into structured prompts that guide LLMs through exploitation reasoning:
%\compactline
\begin{squishitemize}
\item Complete source code of all files appearing in the stack trace
%\compactline
\item Sinkpoint details including line number, sink API specification, vulnerability type, vulnerability descriptions, exploitation specifics, \etc
%\compactline
\item Hex dump of the reaching input and full stack frames
%\compactline
\item Explicit exploitation task requirements and output format specifications for direct input blob generation
%\compactline
\end{squishitemize}

The generated exploit blobs are then passed to a dedicated fuzzing phase without coverage feedback (since the sinkpoint is already reached), allowing near-successful PoCs to undergo further mutations for successful exploitation.
Subsequently, both successful exploits and failed attempts, along with their associated seeds, are shared with all fuzzer instances through the ensemble infrastructure.
%\compactline

\subsubsection{Effectiveness}

ExpKit demonstrated remarkable effectiveness in addressing the last mile challenge.
Of the 16 reached-but-unexploited CPVs, ExpKit successfully generated exploits for 13 cases (81.3\% success rate).
The success cases span diverse vulnerability types such as command injection, path traversal, and deserialization vulnerabilities, validating ExpKit's generality across different exploitation scenarios.
Particularly noteworthy is that 6 of these vulnerabilities were exploited with a single LLM query (using the OpenAI o1-preview model), demonstrating the power of combining precise execution context with language model reasoning.
%\compactline

The three remaining cases required complex multi-round iteration and specialized tools beyond ExpKit's current single-query design.
While ExpKit proves highly effective for straightforward exploitation scenarios, post-competition development will focus on evolving it into a full-featured exploitation agent with tool access (debuggers, analyzers, fuzzers), multi-round reasoning capabilities, and diverse blob generation methods including script-based approaches.
This evolution will address the remaining complex cases while improving efficiency for simpler exploitations.
%\compactline

\subsection{Concolic Execution}
\label{s:crs-java-concolic}
We utilize a Java concolic execution engine (\ie concolic executor)
to serve two primary roles in our workflow.
The executor acts as a \emph{coverage amplifier} to explore
deep and complex code paths that fuzzers cannot easily reach,
and as an \emph{exploit assistant} to help turn
a \emph{reached-but-unexploited} sink into a proof-of-vulnerability (PoV).
To this end, we developed a new concolic execution engine on
GraalVM Espresso~\cite{graal:espresso},
a JIT-compiled JVM execution environment running on GraalVM.

\subsubsection{Motivation}
Existing Java concolic engines are limited in supporting generic vulnerability discovery
because they either support only specific Java versions of the JVM (version 8 or 11) for their binary compatibility%
---\eg Java PathFinder, jCUTE, DART, JBSE%
---or often fail to run production-grade open-source software due to collisions
in their bytecode instrumentation method (\cc{asm})%
---\eg SWAT and COASTAL.
For the latter in particular, execution on critical areas such as
class initialization (\ccw{<clinit>}), use of Mockito, lambda expressions, \etc,
which are essential to explore program paths with fuzzing and testcase harnesses,
were shown to be divergent when running with the framework,
rendering them infeasible to apply to production-grade software.

Such obstacles motivated us to develop a Java concolic executor at the bytecode interpreter level
(\ie emulate bytecode instructions themselves symbolically under-the-hood) to achieve 
\textit{binary compatibility} up to the latest version of Java (25 and onward).
This way, we can \textit{run all critical executions in Java},
and \textit{scale to production-level applications} robustly.

\subsubsection{Our Approach: Interpreter-based Symbolic Emulation}
We built our concolic execution engine on \textit{GraalVM Espresso},
a high-performance (JIT-compiled) bytecode interpreter.
This approach offers several key advantages over existing frameworks:

\begin{squishitemize}
    \item \textbf{Compatibility without Instrumentation:} 
    Dynamically instrumenting bytecode allows us to hook every execution at
    the smallest possible granularity (\ie \textit{bytecode}),
    not losing any tracking of the runtime symbolic state of the target program.
    \item \textbf{State Isolation:} Espresso runs the target application in an isolated context.
    The dynamic instrumentation as well as shadow symbolic state are
    completely invisible to the target runtime,
    unlike \cc{asm}-based static instrumentation that suffers symbolic/concrete execution state contamination.
    \item \textbf{JIT-Accelerated Performance:} 
    Symbolic state emulation incurs nontrivial overhead in the target execution.
    To mitigate this overhead,
    our engine leverages Espresso's Just-In-Time (JIT) compilation for shadow symbolic execution.
    After the first visit of a code path, the code will be JIT-compiled,
    and thereby, subsequent visits to the same path are executed a near-native speed
    (for both the target execution as well as symbolic state emulation).
\end{squishitemize}

\begin{figure}[t]
  \centering
  \includegraphics[width=.7\columnwidth]{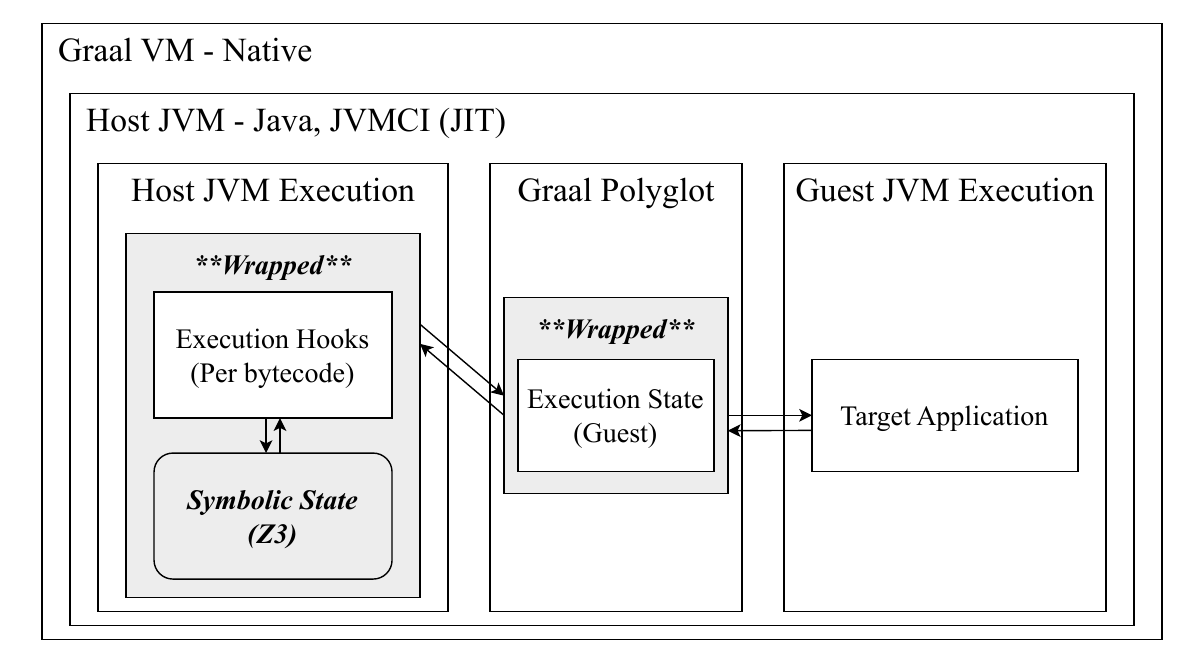}
    \caption{Concolic Executor Overview}
  \label{fig:concolic-executor}
\end{figure}

In our approach, the core concept is to ``wrap'' all Java data types---from primitives like \ccw{int} to complex \ccw{Object} instances---in custom container classes.
These containers hold both the concrete value and its corresponding symbolic representation, allowing symbolic state to be propagated naturally as the program executes.
Since this management is performed at the JVM level, no modifications to the target code are required, and the original program runs normally without knowing that symbolic execution is happening.
Z3 expressions are used to represent symbolic states, with calculations and operations performed only when necessary for efficiency.
% For instance, only the constant \ccw{5} on line 10 is converted to a Z3 expression, while the operations on lines 8 and 9 do not require symbolic representation.

\PP{Core Modifications}
To extend the framework so that both concrete values and symbolic states can be managed during the execution, three core modifications were performed.
First, data types used by the JVM (such as \ccw{Object}, \ccw{int}, \ccw{long}, \etc) are wrapped to allow simultaneous management of both concrete values and symbolic states. 
Second, data storages in the JVM, such as local variables and the operand stack, are made compatible with these wrapped data types.
Third, with data types and storage now supporting symbolic states, the bytecode handlers and other operations are extended so that operations can be performed not only on concrete values but also on symbolic values.

\PP{Fallback}
However, the structure of Espresso and the JVM is complex and extensive, and there are some challenging bytecodes, such as those for floating-point operations.
Therefore, it was not possible to wrap and extend all functionalities. When unsupported functionalities are encountered, the symbolic information is discarded and the value is downgraded to its original, unwrapped concrete form.
Although this results in the loss of symbolic state, it ensures compatibility and robustness in program execution. In particular, this strategy was highly beneficial for ensuring compatibility with internal framework code in Espresso.
During program execution, data is not only processed by bytecode handlers, but also interacts with various Espresso internal framework components, which provide optimizations and extended functionalities.
These interactions are often complex and diverse, making them difficult to fully support, yet they are largely unrelated to the symbolic states encountered during target program execution.
By downgrading in situations involving framework-internal interactions that are irrelevant to symbolic state management, it was possible to reduce implementation effort while achieving robust compatibility.

\subsubsection{Overall Workflow}
% We adopt \ccw{fuzzerTestOneInput(byte[])} as the entrypoint (calling \ccw{fuzzerInitialize()} when present). Incoming bytes are wrapped to initialize symbolic state.
% The interpreter hooks propagate symbolic information through locals, operand stacks, objects/fields, arrays, strings, and file I/O.
% Constraints are generated \emph{lazily} when control flow depends on data (branches, \ccw{switch}) and via lightweight \emph{function summaries} for common/high-impact APIs.
% A \emph{minimal related slice} of constraints around a chosen decision is solved with Z3 to produce blobs that negate the decision (exploration) or satisfy sink-oriented goals (exploitation).
% We bound growth by per-PC constraint capping and package-level filtering. Generated blobs are looped back as seeds to the fuzzer and cataloged alongside beep seeds in \sys-Java’s manager.
% % %\compactline

We adopt the fuzzer entry point, \ccw{fuzzerTestOneInput()}, as our execution root. The input is typically provided either directly as a byte array or via a \ccw{FuzzedDataProvider}.
The symbolic state is initialized by wrapping the input data. We hook bytecode execution, symbolizing the input and tracking symbolic propagation through methods, objects, fields, \etc
% During execution, the engine lazily constructs Z3 constraints for branching conditions, switch statements, and other decision points. This includes symbolic handling of primitives, objects, strings, and even file I/O.
% These collected constraints are solved using Z3 to generate negated paths or exploit-guided inputs. We apply constraint pruning, package blacklisting, and selective solving strategies for better performance.
% The resulting blobs are fed back to the fuzzer for further execution or saved as exploit candidates, increasing the chances of reaching deeper paths or triggering bugs.
To illustrate, consider the above simple Java method:

When our concolic executor runs this code, it operates as follows:
\begin{squishenumerate}
    \item \textbf{Initialization:} The value \ccw{input[0]} is loaded. Its concrete value (\eg `42') is stored, and it is assigned a symbolic variable, let's call it $S_0$. The integer `a' is now represented as a wrapped type holding (concrete: 42, symbolic: $S_0$). The integer `b' is also wrapped but initially has no symbolic state: (concrete: 5, symbolic: null).
    \item \textbf{Constraint Collection:} At the \ccw{if (a > 100)} branch, the executor records the path constraint. Since `a' has a symbolic state $S_0$, the condition $S_0 > 100$ is added to a list of constraints for the current execution path. Concretely, $42 > 100$ is false, so the `else' branch (Path 2) is taken.
    \item \textbf{Symbolic Propagation:} The operation \ccw{a - b} is performed. The executor calculates the new concrete value ($42 - 5 = 37$) and also computes the new symbolic expression by combining the symbolic states of the operands: $S_0 - 5$.
    \item \textbf{New Input Generation:} After the execution finishes, the solver can negate the collected path constraint. It asks Z3 to solve for $\neg(S_0 > 100)$, which simplifies to $S_0 \leq 100$. The solver might also be asked to find an input that satisfies the other path, $S_0 > 100$. Z3 might return a solution like $S_0 = 120$. The executor then crafts a new \ccw{input} byte array where the first byte is `120' to explore Path 1 in the next run.
\end{squishenumerate}

This process of lazily converting values to symbolic expressions only when they interact with other symbolic data avoids generating bloated and unnecessary Z3 terms, significantly improving solver performance.

\subsubsection{Symbolic State Management}

\begin{figure}[t]
  \centering
  \begin{minipage}[c]{0.45\linewidth}
    \begin{promptbox}{Concolic Execution Example}
      \begin{promptcontent}\input{code/java_concolic_example.java}\end{promptcontent}
    \end{promptbox}
  \end{minipage}\hfill
  \begin{minipage}[c]{0.53\linewidth}
    \centering
    \footnotesize
    %\resizebox{0.5\columnwidth}{!}{%
    \begin{tabular}{ll}
      \toprule
      \textbf{Type} & \textbf{Extended Class} \\
      \midrule
      \rowcolor[HTML]{EFEFEF} 
      byte    & ConcolicByte \\
      short   & ConcolicShort \\
      \rowcolor[HTML]{EFEFEF} 
      char    & ConcolicChar \\
      int     & ConcolicInt \\
      \rowcolor[HTML]{EFEFEF} 
      long    & ConcolicLong \\
      float   & ConcolicFloat \\
      \rowcolor[HTML]{EFEFEF} 
      double  & ConcolicDouble \\
      \bottomrule
    \end{tabular}
    %}
    \captionof{table}{Primitive Types and their Extended Classes}
    \label{tab:primitive-types}
  \end{minipage}
\end{figure}

\begin{figure}[t]
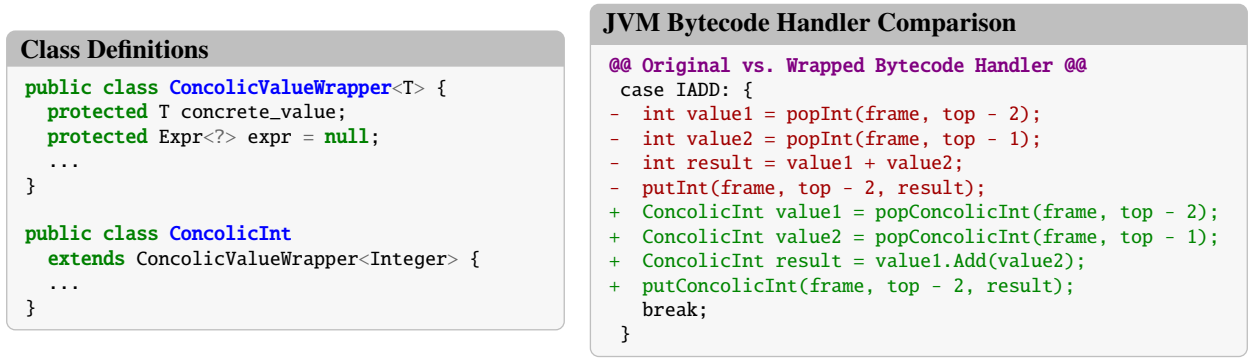

  \centering
  \begin{minipage}[c]{0.45\linewidth}
    \begin{promptbox}{Class Definitions}
      \begin{promptcontent}\input{code/concolic-class-definitions.java}\end{promptcontent}
    \end{promptbox}
  \end{minipage}\hfill
  \begin{minipage}[c]{0.53\linewidth}
    \begin{promptbox}{JVM Bytecode Handler Comparison}
      \begin{promptcontent}\input{code/iadd-handler-comparison.diff}\end{promptcontent}
    \end{promptbox}
  \end{minipage}
  \caption{JVM bytecode handler for \ccw{IADD}: Original vs. Wrapped}
  \label{fig:iadd}
\end{figure}

%\CEN{title can be more formal such as Symbolic State Management/Propagation/etc?}

To accurately track symbolic state across complex Java applications, we implement comprehensive handling for key JVM features,
combining wrapper-based value modeling for JVM values (primitives, arrays/objects, \ccw{String}) with boundary summaries (function hooks, boxed types, file I/O).

% To accurately track symbolic state across complex Java applications, we implement comprehensive handling for key JVM features:
% (A) wrapper-based value modeling for JVM values, and (B) hook-based boundary summaries at opaque or system boundaries.

% \PP{(A) Wrapper-based value modeling}

\PP{Primitives}
All primitive types (\ccw{byte}, \ccw{short}, \ccw{char}, \ccw{int}, \ccw{long}, \ccw{float}, \ccw{double}) are extended
with \ccw{Concolic*} wrappers (\autoref{tab:primitive-types}) that store a concrete value (\ccw{concrete\_value}) and a symbolic expression (\ccw{expr}).
These extended classes also expose arithmetic and, when necessary, bitwise operations that update both the concrete and symbolic parts during computation.
The bytecode handlers were updated accordingly; \autoref{fig:iadd} shows \ccw{IADD}, whose handler pops/pushes concolic values (\ccw{popConcolicInt()}, \ccw{putConcolicInt()}),
invokes the corresponding concolic operation (\eg \ccw{value1.Add()}), and updates both states.

\PP{Arrays and Objects}
Java programs are object-heavy, with frequent cross-object interactions; therefore, precise and efficient state management is essential.
We treat objects as field containers and arrays as element containers, where each slot may carry its own concrete value and symbolic state.
To keep overhead low---even for large, nested, or multidimensional structures---we lazily materialize member states on first read or write along the executed path.
Static members are tracked in a class-level symbolic map to maintain per-class semantics consistently across instances.

\PP{java.lang.String}
We treat \ccw{String} as a special object, since modeling symbolic state for strings is particularly challenging.
Two approaches are viable: (i) encode strings with dedicated string theories (\eg Z3Str/Seq) to reason directly at the symbolic-string level,
or (ii) track the backing store (\ie \ccw{String.value} as a \ccw{byte[]}) and link the underlying bytes to their symbolic states.
We adopt (ii), recording the bytes for each instance and associating them with symbolic state.
While (i) can express richer, more abstract properties, it proved difficult to implement correctly given Java's string semantics;
the backing-store approach is less expressive but substantially more robust and performant in practice.

% \PP{(B) Hook-based boundary summaries}

\PP{Function hooks and summaries}
For native or semantically dense methods, we instrument call boundaries with pre/post hooks and attach compact summaries
that capture only essential relations (e.g., prefixes, lengths, bounds, aliasing) rather than expanding full method bodies.
In practice, we provide manual summaries for frequently invoked native methods---\ccw{System.arraycopy()}, \ccw{Object.clone()},
and \ccw{Unsafe} methods---that preserve copy length, bounds checks, and element-wise aliasing.
To ensure full compatibility with Jazzer, we also model the \ccw{FuzzedDataProvider} APIs,
propagating symbolic provenance through calls such as \ccw{consumeBytes}, \ccw{consumeRemainingAsBytes}, and \ccw{consumeString}.

\PP{Boxed Types (Autoboxing/Unboxing)}
Java provides wrapper classes for primitives, each of which encapsulates its corresponding primitive value.
At autoboxing/unboxing boundaries (\eg \ccw{Integer.valueOf} / \ccw{intValue}), we preserve the link to the underlying primitive's concolic state.
Because boxed values may be cached, we re-inject the symbolic state after \ccw{valueOf} to avoid stale or missing expressions.

\PP{File I/O}
To maintain symbolic integrity across file operations, we extended \ccw{FileInputStream}, \ccw{FileOutputStream}, and \ccw{RandomAccessFile}, together with the required dispatchers.
These wrappers preserve the symbolic state of file contents, file descriptors, and memory-mapped regions, and they include symbolic handling for file-pointer adjustments.
This allows the executor to track whether symbolic input data is written to or read from files, ensuring that symbolic reasoning remains intact throughout I/O interactions.

\subsubsection{Constraint Solving}
During execution, whenever the program reaches an exploration target, the executor collects the associated path constraints.
After each execution, it solves these collected constraints to synthesize new inputs that either steer execution into previously unexplored branches or bias execution toward exploitation-oriented behaviors at vulnerability-related sinks.

\PP{(A) Constraint Solving for Exploration}
When a new branch predicate is observed, the executor negates it and solves a \emph{minimal related slice} of constraints---\ie only those constraints that influence the predicate---to produce an input taking the opposite outcome.
This raises coverage and input diversity while limiting incidental changes to unrelated bytes.

To check the contribution of concolic execution to exploration, we ran it on top of a corpus collected after 1 hour of fuzzing \ccw{org.apache.commons.imaging.Imaging.getBufferedImage()} in \textit{Apache Commons Imaging}.
This run showed an increase of about 9.5\% branch coverage over the fuzzing-only baseline. The following excerpt illustrates part of this improvement.

In the excerpt above, guard~(1) is a range check on \ccw{destinationIdentifier}, and guard~(2) checks whether \ccw{marker == JpegConstants.DHT\_MARKER}.
With fuzzing alone, the corpus repeatedly triggers the exception at~(1), leaving the following block unexecuted; similarly, (2) never holds, so the \ccw{DhtSegment} path is uncovered.
This reflects a common trajectory in mature fuzzing campaigns: once inputs are locally ``good enough'' to reach a region, the chance of stumbling into \emph{new} branches inside that region drops, especially when deeper progress requires coordinated changes across multiple bytes and checks.

\begin{figure}[t]
\centering
\begin{promptbox}{Concolic Solving Example}
\begin{promptcontent}\input{code/concolic-solving-example.java}\end{promptcontent}
\end{promptbox}
\end{figure}

With concolic execution, condition~(1) was negated to satisfy the range check by solving only the bytes influencing \ccw{destinationIdentifier}.  
This ensured $0 \le \ccw{destinationIdentifier} < \ccw{quantizationTables.length}$, and allowed us to visit a block that had remained unexecuted under fuzzing alone.  
Likewise, condition~(2) was flipped by solving the marker bytes so that \ccw{marker == DHT\_MARKER}, which enabled the \ccw{DhtSegment} branch to be explored for the first time.

\PP{(B) Constraint Solving for Exploitation}
We further target \emph{sink-aware} generation: when symbolic values are present at a sink,
the executor solves constraints to bias inputs toward exploit-relevant behaviors---for example, allocating arrays large enough to induce an \ccw{OutOfMemoryError}
or constructing values that trigger Jazzer's detectors. This is motivated by cases where fuzzers reached a vulnerability-related sink but failed to trigger actual exploitation.

Jazzer can also guide mutations toward such sinks via compare/substring feedback (\eg \ccw{traceMemcmp}, \ccw{traceStrstr}).
This works when the compared value is derived almost directly from raw input, but it weakens after non-linear transforms (hashes/CRCs/bit-twiddling), multi-step encodings, or table lookups;
and because the signal is local, it struggles to satisfy multiple, distant preconditions.

To address these limits, our executor adds minimal payload constraints and solves them after reachability is established.
We primarily leverage Jazzer's sentinels and model constraints for hooked methods such as \ccw{java.lang.ProcessBuilder.start()} and \ccw{java.net.Socket.connect()}.
we also target methods not directly hooked by Jazzer with early, non-sanitizer constraints, increasing the overall potential for exploitation.
There remains ample opportunity to extend these ideas to other forms of exploit-oriented generation.

\smallskip
\PP{Out-Of-Memory via Oversized Array Allocation}
We identify array allocations through the \ccw{NEWARRAY} and \ccw{ANEWARRAY} bytecode instructions,
and use Z3's \ccw{Optimize} to maximize the allocation length under the feasible path conditions.
This steers execution toward inputs most likely to provoke an \ccw{OutOfMemoryError}.

Assume the program computes \ccw{len} from four input bytes at offset $p$ and allocates \ccw{new byte[len]}.
Here, the four bytes $b_p,\dots,b_{p+3}$ are combined into a 32-bit \ccw{int} in big-endian order.
We then solve:
\[
\begin{aligned}
& b_i \in [0,255] \quad (i \in \{p,\dots,p{+}3\}),\\
& \ccw{len} = (b_p \ll 24)\;|\;(b_{p+1} \ll 16)\;|\;(b_{p+2} \ll 8)\;|\;b_{p+3},\\
& \textbf{path conditions: }\; C_1(b_\cdot) \land \cdots \land C_m(b_\cdot),\\
& \textbf{objective: }\; \max\, \ccw{len}.
\end{aligned}
\]
where each $C_i$ denotes the instantiated outcome of the $i$-th branch predicate on the chosen path, such as checksum/range checks that gate the allocation.
The dependency slice includes only the bytes influencing \ccw{len} and the relevant guards, keeping solving fast and localized.

\begin{figure}[t]
\centering
\begin{promptbox}{Out-Of-Memory in LZWInputStream.initializeTables()}
\begin{promptcontent}\input{code/concolic-solving-oom.java}\end{promptcontent}
\end{promptbox}
\end{figure}

This case is drawn from \ccw{Apache Commons Compress} in competition round~3.
As in the exploration section, the array allocation is guarded by conditions that the executor should satisfy.
Beyond satisfying these guards, the executor synthesizes inputs that maximize the subsequent allocation size where \ccw{maxTableSize = 1 <}\ccw{< maxCodeSize}, thereby provoking an \ccw{OutOfMemoryError}.
Although the path is relatively deep, we found that the executor collected the constraints along the way and synthesized an input that reached the allocation site and triggered the issue.

\smallskip
\PP{OS Command Injection}
For command injection,
Jazzer instruments \ccw{java.lang.ProcessBuilder.start()} and reports an \ccw{OsCommandInjection} finding when the executable filename is recognized as the sentinel \ccw{"jazze"}.
In practice, multiple preconditions must be satisfied before this sink is reached,
and in our experiments a coverage-guided fuzzer alone often struggled to trigger the detector---even on paths that met the validations---and in some cases failed to discover it at all.

\begin{promptbox}{OS Command Injection in GzipCompressorInputStream.init()}
\begin{promptcontent}\begin{Verbatim}[commandchars=\\\{\},codes={\catcode`\$=3\catcode`\^=7\catcode`\_=8\relax}]
\PY{k+kd}{final}\PY{+w}{ }\PY{n}{DataInput}\PY{+w}{ }\PY{n}{inData}\PY{+w}{ }\PY{o}{=}\PY{+w}{ }\PY{k}{new}\PY{+w}{ }\PY{n}{DataInputStream}\PY{p}{(}\PY{n}{in}\PY{p}{)}\PY{p}{;}\PY{+w}{             }\PY{c+c1}{// Input}
\PY{k+kd}{final}\PY{+w}{ }\PY{k+kt}{int}\PY{+w}{ }\PY{n}{method}\PY{+w}{ }\PY{o}{=}\PY{+w}{ }\PY{n}{inData}\PY{p}{.}\PY{n+na}{readUnsignedByte}\PY{p}{(}\PY{p}{)}\PY{p}{;}
\PY{k}{if}\PY{+w}{ }\PY{p}{(}\PY{n}{method}\PY{+w}{ }\PY{o}{!}\PY{o}{=}\PY{+w}{ }\PY{n}{Deflater}\PY{p}{.}\PY{n+na}{DEFLATED}\PY{p}{)}\PY{+w}{ }\PY{p}{\PYZob{}}\PY{+w}{                            }\PY{c+c1}{// Validation 1}
\PY{+w}{    }\PY{k}{throw}\PY{+w}{ }\PY{k}{new}\PY{+w}{ }\PY{n}{IOException}\PY{p}{(}\PY{l+s}{\PYZdq{}}\PY{l+s}{Unsupported compression method }\PY{l+s}{\PYZdq{}}\PY{+w}{ }\PY{o}{+}\PY{+w}{ }\PY{n}{method}\PY{+w}{ }\PY{o}{+}\PY{+w}{ }\PY{l+s}{\PYZdq{}}\PY{l+s}{ in the .gz header}\PY{l+s}{\PYZdq{}}\PY{p}{)}\PY{p}{;}
\PY{p}{\PYZcb{}}
\PY{k+kd}{final}\PY{+w}{ }\PY{k+kt}{int}\PY{+w}{ }\PY{n}{flg}\PY{+w}{ }\PY{o}{=}\PY{+w}{ }\PY{n}{inData}\PY{p}{.}\PY{n+na}{readUnsignedByte}\PY{p}{(}\PY{p}{)}\PY{p}{;}
\PY{k}{if}\PY{+w}{ }\PY{p}{(}\PY{p}{(}\PY{n}{flg}\PY{+w}{ }\PY{o}{\PYZam{}}\PY{+w}{ }\PY{n}{GzipUtils}\PY{p}{.}\PY{n+na}{FRESERVED}\PY{p}{)}\PY{+w}{ }\PY{o}{!}\PY{o}{=}\PY{+w}{ }\PY{l+m+mi}{0}\PY{p}{)}\PY{+w}{ }\PY{p}{\PYZob{}}\PY{+w}{                       }\PY{c+c1}{// Validation 2}
\PY{+w}{    }\PY{k}{throw}\PY{+w}{ }\PY{k}{new}\PY{+w}{ }\PY{n}{IOException}\PY{p}{(}\PY{l+s}{\PYZdq{}}\PY{l+s}{Reserved flags are set in the .gz header.}\PY{l+s}{\PYZdq{}}\PY{p}{)}\PY{p}{;}
\PY{p}{\PYZcb{}}
\PY{k+kt}{long}\PY{+w}{ }\PY{n}{modTime}\PY{+w}{ }\PY{o}{=}\PY{+w}{ }\PY{n}{ByteUtils}\PY{p}{.}\PY{n+na}{fromLittleEndian}\PY{p}{(}\PY{n}{inData}\PY{p}{,}\PY{+w}{ }\PY{l+m+mi}{4}\PY{p}{)}\PY{p}{;}
\PY{p}{.}\PY{p}{.}\PY{p}{.}
\PY{n}{String}\PY{+w}{ }\PY{n}{fname}\PY{+w}{ }\PY{o}{=}\PY{+w}{ }\PY{k+kc}{null}\PY{p}{;}
\PY{c+c1}{// Original file name}
\PY{k}{if}\PY{+w}{ }\PY{p}{(}\PY{p}{(}\PY{n}{flg}\PY{+w}{ }\PY{o}{\PYZam{}}\PY{+w}{ }\PY{n}{GzipUtils}\PY{p}{.}\PY{n+na}{FNAME}\PY{p}{)}\PY{+w}{ }\PY{o}{!}\PY{o}{=}\PY{+w}{ }\PY{l+m+mi}{0}\PY{p}{)}\PY{+w}{ }\PY{p}{\PYZob{}}\PY{+w}{                           }\PY{c+c1}{// Validation 3}
\PY{+w}{    }\PY{n}{fname}\PY{+w}{ }\PY{o}{=}\PY{+w}{ }\PY{k}{new}\PY{+w}{ }\PY{n}{String}\PY{p}{(}\PY{n}{readToNull}\PY{p}{(}\PY{n}{inData}\PY{p}{)}\PY{p}{,}\PY{+w}{ }\PY{n}{parameters}\PY{p}{.}\PY{n+na}{getFileNameCharset}\PY{p}{(}\PY{p}{)}\PY{p}{)}\PY{p}{;}
\PY{+w}{    }\PY{n}{parameters}\PY{p}{.}\PY{n+na}{setFileName}\PY{p}{(}\PY{n}{fname}\PY{p}{)}\PY{p}{;}
\PY{p}{\PYZcb{}}
\PY{k}{if}\PY{+w}{ }\PY{p}{(}\PY{n}{modTime}\PY{+w}{ }\PY{o}{=}\PY{o}{=}\PY{+w}{ }\PY{l+m+mi}{1731695077L}\PY{+w}{ }\PY{o}{\PYZam{}}\PY{o}{\PYZam{}}\PY{+w}{ }\PY{n}{fname}\PY{+w}{ }\PY{o}{!}\PY{o}{=}\PY{+w}{ }\PY{k+kc}{null}\PY{p}{)}\PY{+w}{ }\PY{p}{\PYZob{}}\PY{+w}{                }\PY{c+c1}{// Validation 4}
\PY{+w}{    }\PY{k}{new}\PY{+w}{ }\PY{n}{ProcessBuilder}\PY{p}{(}\PY{n}{fname}\PY{p}{)}\PY{p}{.}\PY{n+na}{start}\PY{p}{(}\PY{p}{)}\PY{p}{;}\PY{+w}{                        }\PY{c+c1}{// Sink}
\PY{p}{\PYZcb{}}
\end{Verbatim}
\end{promptcontent}
\end{promptbox}

In another case from \ccw{Apache Commons Compress} in competition round~3, once validations 1-4 are satisfied
and \ccw{new ProcessBuilder(fname).start()} is invoked with a filename recognized as the sentinel, Jazzer reports an OS command-injection finding.
On the executor side, we use a simple sufficient precondition for solving: \ccw{fname} equals \ccw{"jazze"} or ends with \ccw{"/jazze"}.
This accommodates path-shaped inputs (\eg \ccw{"/tmp/jazze"}) while leaving filename resolution to the runtime and preserving the detector's semantics.

In addition to OS command injection, we apply a similar approach to other Jazzer-hooked sinks---such as Path Traversal or Reflection---tailoring the constraints to each vulnerability signature.

% Let $\{b_k\}$ denote the bytes that feed \ccw{s}. 
% We impose:
% \[
% \begin{aligned}
% & \ccw{s} = \ccw{prefix} \;\|\; \ccw{cmd} \\
% & \ccw{cmd} \text{ endsWith } \ccw{"jazze"} \\
% & \text{guards on parsing/validation along the path to } \ccw{ProcessBuilder.start()}
% \end{aligned}
% \]
% Byte-wise, the suffix constraint is encoded as
% \[
% (b_{n-5},\dots,b_{n-1}) = (\text{'j'},\text{'a'},\text{'z'},\text{'z'},\text{'e'})
% \]
% with $n$ the end index of \ccw{cmd}. 
% We leave \ccw{prefix} largely unconstrained (or lightly restricted to valid filename/argument characters) to preserve solvability, and we add only the minimal guards necessary for the code to reach \ccw{start()}.
% No fixed pattern beyond the suffix is required---the solver fills in the remaining bytes to satisfy the path conditions.

% \CEN{can u give an example here how the constraints for exploiting specific sink are modeled and solvable by z3? This is a fundamental aspect about how this type of generation works but it is unclear from current description.}

\subsubsection{Optimization}

%\CEN{title use noun phrase, \eg Constraint Solving/etc?}

%\CEN{You can add a leading paragraph to give an overview of the constraint solving, \eg u are using z3 to solve the constraints, and due to the complexity of the constraints, u proposed several strategies to boost the solving process, and the followings list the main strategies u used, etc.}

%\CEN{If possible, you should mention how much those strategies improve the solving process quantatively, some rough metrics or numbers is also helpful for readers to understand how necessary those strategies are.}

In concolic execution, solving all collected constraints can be highly time-consuming, and in practice it is rarely feasible to resolve every constraint within a limited budget.
As execution progresses, constraints not only become more complex but also grow in number, which can lead to severe inefficiency or solver failures.  
To mitigate these challenges and make constraint solving more effective, we introduce several optimization strategies.
The following describes the primary approaches employed in our system.

\PP{Constraint Limiting}
The concolic execution handles symbolic variables that may appear anywhere in the code, including as comparison conditions inside loops executed hundreds of thousands of times.
Each comparison generates constraints, resulting in an enormous number of constraints that complicate later solving and may even cause memory exhaustion, halting execution.

To address this, we limit the number of constraints that can be generated from the same program counter (PC) to a small number (\eg 5).
This was highly effective because it prevents the constraints from being created as many times as the number of iterations; in one test case, it reduced the number of SAT constraints from over 5,000, preventing an \ccw{OutOfMemoryError}.
The trade-off is a potential loss of information, as the same PC might be reachable via different contexts, but this was crucial for stability.

\PP{Package Blacklisting}
Since symbolic states can propagate through most JVM operations, constraints can also be generated by common Java packages and libraries.
However, this sometimes results in the creation of excessive constraints, making constraint solving difficult due to overconstraint.
To address this, we implemented filtering for certain packages to prevent the generation of unnecessary constraints.

\PP{Coverage-Guided Prioritization}
The number of constraints could be substantial, and considering the limited computational time and resources available, solving all constraints may be infeasible.
We observed cases where the solver attempted to solve constraints that did not directly increase block coverage because their successor blocks had already been visited.
While such efforts may indirectly contribute to coverage improvement, it is more efficient to prioritize constraints that directly lead to new block coverage given resource constraints.
Therefore, before constraint solving, we employ a heuristic-based prioritization strategy to maximize new block coverage:

\begin{squishenumerate}
    \item Prioritize branches that have never been attempted before, as their successful solving has the potential to yield new coverage gains since their feasibility and coverage impact remain unexplored.
    \item Next, prioritize branches whose successor blocks have not been visited, as solving these constraints can directly contribute to block coverage expansion by enabling access to previously unreachable code blocks.
    \item Finally, prioritize deeper branches, as they are more likely to have been recently discovered, thus representing untapped opportunities for coverage enhancement.
\end{squishenumerate}

This reordering does not guarantee optimality. For example, branches that have been attempted previously may contribute differently to coverage depending on the execution context.
Our strategy does not account for this contextual variation, potentially under-prioritizing such constraints.
Nevertheless, it effectively directs solving efforts toward the most promising opportunities for direct coverage improvement.

\autoref{fig:sample-branch-blocks} shows the executed (visited) blocks and the order in which the constraints are solved.
The branches were executed in the order \ccw{B1}, \ccw{B2}, and \ccw{B3}; however, with our strategy, the constraints are solved in a different order: \ccw{B3}, \ccw{B1}, and \ccw{B2}.
% \CEN{will this reordering and partial solve has drawbacks? \eg generating inputs that are not feasible? If so, please mention it here.}
% \CEN{U need to tell readers the challenge u can considering (aka context), and then reason in high level how to solve the challenge, and finally list ur solution. Currently, it is just directly enumerating the solutions, which can be refined.}

\autoref{fig:scheduler} illustrates the scheduler functionality designed to prioritize constraints. The scheduler maintains up-to-date block coverage information by updating the coverage map based on the given inputs.
With this up-to-date coverage data, it assigns priorities to constraints in order to guide the concolic executor more effectively.

\usetikzlibrary{positioning,matrix}
\begin{figure}[t]
\centering
\begin{minipage}[t]{0.48\linewidth}
\centering
\begin{tikzpicture}[
    scale=0.95,
    every node/.style={scale=0.95, outer sep=0pt},
    font=\small,
    node distance=0.30cm and 0.52cm,
    block/.style={rectangle, draw, minimum width=1.28cm, minimum height=0.62cm, inner sep=1.1pt},
    visited/.style={fill=gray!30},
    unvisited/.style={fill=none},
    arrow/.style={-Stealth, line width=0.5pt}
]

\node[block, visited] (B1) {B1};
\node[block, unvisited, above right=of B1] (B1L) {};
\node[block, visited,   below right=of B1] (B2) {B2};

\node[block, visited,   above right=of B2] (B2L) {};
\node[block, visited,   below right=of B2] (B3) {B3};

\node[block, unvisited, above right=of B3] (B3L) {};
\node[block, visited,   below right=of B3] (B3R) {};

\draw[arrow] (B1) -- (B1L);
\draw[arrow] (B1) -- (B2);
\draw[arrow] (B2) -- (B2L);
\draw[arrow] (B2) -- (B3);
\draw[arrow] (B3) -- (B3L);
\draw[arrow] (B3) -- (B3R);

\begin{scope}[overlay]
  \matrix[%
      matrix of nodes,
      nodes in empty cells,
      anchor=south west,
      inner sep=0pt,
      row sep=1mm,
      column sep=1.2mm
  ] at ($(current bounding box.south west)+(2.2mm,2.0mm)$) {
      \node[block, unvisited, minimum width=0.80cm, minimum height=0.40cm, inner sep=0pt] {}; &
      \node[anchor=west, font=\scriptsize] {unvisited}; \\
      \node[block, visited,   minimum width=0.80cm, minimum height=0.40cm, inner sep=0pt] {}; &
      \node[anchor=west, font=\scriptsize] {visited}; \\
  };
\end{scope}
\end{tikzpicture}
\vspace{1mm}
\caption{Sample branch blocks}
\label{fig:sample-branch-blocks}
\end{minipage}%
\hfill%
\begin{minipage}[t]{0.48\linewidth}
\centering
\begin{tikzpicture}[
  scale=0.95,
  every node/.style={scale=0.95, outer sep=0pt},
  font=\footnotesize\rmfamily,
  >=Latex,
  every path/.style={line width=0.6pt},
  node distance=6mm and 9mm,
  box/.style={ % Exec
    draw, rectangle, minimum width=28mm, minimum height=9mm,
    inner sep=2pt, align=center
  },
  boxio/.style={ % Input/Output
    draw, rectangle, minimum width=16mm, minimum height=7mm,
    inner sep=2pt, align=center,
  },
  sbox/.style={ % Scheduler
    draw, rectangle, minimum width=28mm, minimum height=8mm,
    inner sep=1.5pt, align=center, font=\scriptsize\rmfamily
  },
  cbox/.style={ % Cov.map
    draw, rectangle, minimum width=16mm, minimum height=5.5mm,
    inner sep=1.2pt, align=center, font=\scriptsize\rmfamily
  },
  lab/.style={inner sep=1pt}
]

\node[box] (exec) {Concolic Executor};

\node[sbox, above=6mm of exec] (sched) {Scheduler};
\node[cbox, above=4.5mm of sched] (cov) {Cov.\ map};

\path let \p1=(exec.west), \p2=(sched.west) in
      coordinate (midwes) at ($(\p1)!0.5!(\p2)$);
\node[boxio, anchor=east] (input) at ($(midwes)+(-9mm,0)$) {Input};

\coordinate (outbase) at ($(exec.east)+(8mm,0)$);
\node[draw, fill=white, anchor=west,
      minimum width=16mm, minimum height=8mm] (outback2)
      at ($(outbase)+( 2.0mm,-1.6mm)$) {};
\node[draw, fill=white, anchor=west,
      minimum width=16mm, minimum height=8mm] (outback1)
      at ($(outbase)+( 1.0mm,-0.8mm)$) {};
\node[boxio, fill=white, anchor=west,
      minimum width=16mm, minimum height=8mm] (outtop) at (outbase) {Output};

\draw[->] (exec) -- (outtop.west);
\draw[->] (input.east) -- (exec.west);
\draw[->] (input.east) -- (sched.west);
\draw[->] (sched) -- (exec);
\path (sched) -- (exec) coordinate[pos=0.52] (midSE);
\node[lab, right=2mm of midSE, anchor=west] {Reordered constraints};
\draw[->] (sched) -- (cov);

\end{tikzpicture}
\vspace{1mm}
\caption{Scheduler}
\label{fig:scheduler}
\end{minipage}%
\end{figure}

\subsection{Path-Based PoV Generator}
% \CEN{change tool name}
% \CEN{After reading this section, I think it needs major revision if not rewrite. Now I just list about the key points I can think of:
% }

% \CEN{
% 1st/leading paragraph:
% - (what is this tool, what's its main functionality) this is a pov generation tool, this tool aims to genreate PoVs for the given harness..., its intermediate results will also be used as seeds for fuzzing to maximize the ...
% }
The Path-Based PoV Generator is an AI agent that produces inputs capable of triggering vulnerabilities in Java programs through a given harness.
It operates by identifying sinkpoints within the provided code, finding a path from the harness to each of them, and then generating a PoV for each path.
%\compactline

% \CEN{
% running example:
% - example figure
% - the example is showing ur path-based PoV generation process (finding the call path form harnes to sink, and then generatin PoV based on the path)
% }

\begin{figure}[!t]
  \centering
  \begin{promptbox}{Running Example for Path-Based PoV Generator}
  \begin{promptcontent}\input{code/java_pg_running_example.java}\end{promptcontent}
  \end{promptbox}
  \caption{Running example for Path-Based PoV Generator}
  \label{fig:pg-ex}
\end{figure}

\autoref{fig:pg-ex} is an example code snippet to explain how this tool works.
This code is a simplified version of a Jenkins example, originally presented for the ASC, where command injection can be triggered at lines 33 and 36.
To address this vulnerability, the tool first identifies all function calls in the code that are related to command injection, which is line 36 in this example.
Then it statically explores the execution path from the parameters of the harness entry point, \ccw{fuzzerTestOneInput}, to these calls.
In the example, this call path is $Line\ 1 \to Line\ 5 \to Line\ 19 \to Line\ 29 \to Line\ 36$.
Once a path is identified, the tool collects all relevant code along it by concatenating related method bodies, and provides it to the LLM.
The LLM then generates an input value capable of triggering command injection.
While traditional fuzzing would struggle to generate inputs that can satisfy complex constraints such as the conditional statement involving a hash function on line 24, an LLM can readily generate a PoV.
This is achieved by providing the LLM with only the necessary code snippets, allowing it to infer the required input.
Of course, a correct value may not be generated on the first try due to inherent limitations of LLMs, such as hallucination.
For this reason, the tool uses an iterative process that attempts to generate the value several times with some feedback.
Because most of these intermediate data blobs can aid in exploring new code regions and navigating areas closer to the suspected vulnerable locations, all intermediate outputs are used as fuzzing seeds for the harness.
%\compactline

This tool utilizes code property graphs (CPGs) and Joern queries for code analysis.
CPGs are powerful data structures that represent source code in a comprehensive, graph-based format.
Each node in a CPG represents a specific element of the source code, such as classes, methods, instructions, and variables.
Joern queries operate on a CPG to return nodes that satisfy specific conditions.
For example, the query \ccw{cpg.call.name("A")} will return all CPG nodes associated with statements that call function \ccw{A}.
The agent operates by first generating a CPG for the given code.
Whenever code exploration is needed, it dynamically creates Joern queries and runs them on the CPG.
%\compactline

% \CEN{
% overall workflow:
% - workflow figure (largely improved version of figure 19 is an option)
% - carefully consider how many components this tool logically has (add section labels for each component)
% }

% \CEN{
%           5
%           |
% 1 -> 2 -> 3 -> 4
% }

\begin{figure}[!t]
  \centering
  \tikzstyle{process} = [rectangle, minimum width=3.2cm, text centered, draw=black, fill=white]
  \tikzstyle{inner-process} = [rectangle, rounded corners, minimum width=2.8cm, text centered, draw=black, fill=white]
  \tikzstyle{arrow} = [thick,->,>=stealth]
  \tikzstyle{line} = [thick]

  \begin{tikzpicture}[node distance=0.1cm, every node/.style={font=\small}]

  \node (sf-label) {\textbf{Sink Finder \autoref{sss:pg-sf}}};
  \node (scanner) [inner-process, below=of sf-label] {Sink Scanning};
  \node (loader) [inner-process, below=of scanner, node distance=0.5cm] {Sink Synchronization};
  \node (sink-finder) [draw, fit=(sf-label) (scanner) (loader), inner ysep=0.1cm] {};

  \node (sm-label) [right=of sf-label, xshift=1cm] {\textbf{Sink Manager \autoref{sss:pg-sm}}};
  \node (sched) [inner-process, below=of sm-label] {Sink Scheduling};
  \node (mgmt) [inner-process, below=of sched, node distance=0.5cm] {Sink Management};
  \node (sink-manager) [draw, fit=(sm-label) (sched) (mgmt), inner ysep=0.1cm] {};

  \node (pf-label) [right=of sm-label, xshift=1cm] {\textbf{Path Finder \autoref{sss:pg-pf}}};
  \node (sta) [inner-process, below=of pf-label] {Static Taint Analysis};
  \node (cga) [inner-process, below=of sta, node distance=0.5cm] {Call Graph Analysis};
  \node (path-finder) [draw, fit=(pf-label) (sta) (cga), inner ysep=0.1cm] {};

  \node (pov-label) [right=of pf-label, xshift=1cm] {\textbf{PoV Generator \autoref{sss:pg-pg}}};
  \node (gen) [inner-process, below=of pov-label] {Generator};
  \node (eval) [inner-process, below=of gen, yshift=-0.2cm] {Evaluator (JDB)};
  \node (exp) [inner-process, below=of eval, yshift=-0.2cm] {Code Expander};
  \node (pov-generator) [draw, fit=(pov-label) (gen) (eval) (exp), inner ysep=0.1cm] {};

  \node (call-graph-manager) [process, below=of path-finder, yshift=-0.32cm] {\textbf{Call Graph Manager \autoref{sss:pg-mgr}}};

  \coordinate (h-center) at ($(sink-finder.center)!0.5!(sink-manager.center)$);
  \node (misc-label) [left=of call-graph-manager, xshift=-6cm] {\textbf{Misc \autoref{sss:pg-misc}}};
  \node (diff) [inner-process, right=of misc-label, inner ysep=0.07cm] {Diff Analyzer};
  \node (cache) [inner-process, right=of diff, inner ysep=0.1cm] {Cache};
  \node (misc) [draw, fit=(misc-label) (diff) (cache), inner ysep=0.01cm] {};

  \draw [arrow] (sink-finder.east) -- (sink-manager.west |- sink-finder.east);
  \draw [arrow] (sink-manager.east) -- (path-finder.west);
  \draw [arrow] (path-finder.east) -- (pov-generator.west |- path-finder.east);
  \draw [line] (call-graph-manager) -- (path-finder.south);

  \draw [arrow] (gen) -- (eval);
  \draw [arrow] (eval) -- (exp);
  \draw [arrow] (exp.west) to[bend left=45] (gen.west);

  \end{tikzpicture}
  \caption{Overview of Path-Based PoV Generator}
  \label{fig:pg-overview}
\end{figure}
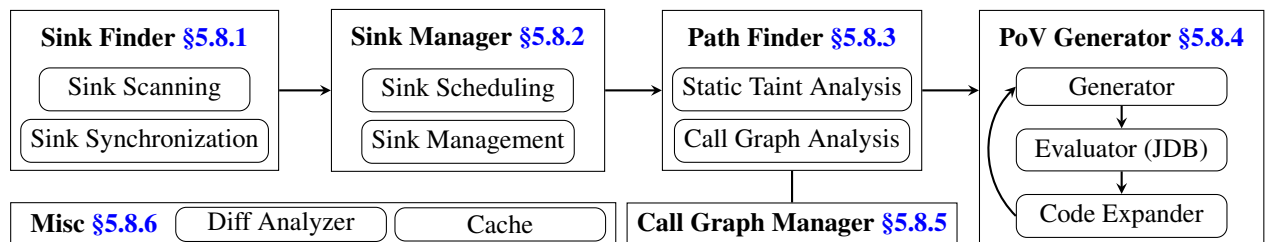

\autoref{fig:pg-overview} illustrates our tool's workflow, which is described in detail in the following subsections.
%\compactline

\subsubsection{Sink Finder} \label{sss:pg-sf}
% \CEN{
% component 1 (sink scanner)
% - the sink list (jazzer sanitizer, also from other sources like crs, customized sink API list)
% - and the query in Joern (what APIs in Joern, example)
% }
The Sink Finder is responsible for finding sinkpoints within the code and forwarding them to the Sink Manager.
It uses two methods to identify sinkpoints.

The first method is to run a Joern query that contains a customized list of potential sinkpoint APIs, which was created by combining methods hooked by the Jazzer sanitizer with other library APIs that could potentially call these methods.
% \CEN{how do u pick other lib APIs, any criteria behind?}
Other library APIs were added by referencing static analysis tools like CodeQL and various vulnerability benchmarks.
\autoref{fig:cmdi} is an example Joern query to find instructions calling functions that may relate to command injection.
This agent has a collection of similar queries for all vulnerability types targeted by the Jazzer sanitizer, such as deserialization, SSRF, and path traversal.
These queries are executed sequentially to identify sinkpoints in the given codebase.
%\compactline

The second method involves periodically referencing a sinkpoint list provided by \sys-Java.
The sinkpoint list contains function names that are used to dynamically create a Joern query, like \ccw{cpg.call.name("Function Name")}.
This query is executed on the CPG to identify all CPG nodes that call these functions.
%\compactline

\begin{figure}[!t]
  \centering
  \begin{promptbox}{Sink Scanning Query Example for Command Injection}
  \begin{promptcontent}\input{code/java_pg_cpg_query.scala}\end{promptcontent}
  \end{promptbox}
  \caption{A CPG query in Scala to find potential command injection sinks.}
  \label{fig:cmdi}
\end{figure}

\subsubsection{Sink Manager} \label{sss:pg-sm}
% \CEN{
% component 2 (sink manager)
% - sink schedule strategy (why need schedule, prioritize the sinks)
% - merge sink from sarif/diff/crs
% }
While sink-based bug finding is viable for Java programs, as they have far fewer potential sinks than C programs, a significant challenge remains.
The sheer number of sinks, combined with the fact that most are not vulnerable, makes exhaustive analysis inefficient.
The process of generating a PoV for each sink is time-intensive, and creates a bottleneck because it requires multiple LLM interactions.
To address this, the Sink Manager is tasked with the primary role of sink scheduling.
This scheduling is vital for prioritizing promising sinks to accelerate the discovery of actual vulnerabilities.
First, the Sink Manager begins by filtering out sinks whose call arguments cannot be manipulated by attackers.
To avoid the possibility of falsely removing exploitable sinks, we only do conservative filtering by excluding sinks whose arguments are literal values or final variables.
Although simple, this filtering can reduce a large number of sinks to be processed, and is particularly useful for injection-related sinks such as command or regex injection because developers often use hard-coded strings for parameters.
%\compactline

Next, for delta-mode challenges, the Sink Manager utilizes the Diff Analyzer (see \autoref{sss:pg-misc}) to elevate the priority of diff-related sinks.
Specifically, sinks that were newly added by the diff, or those associated with methods added or modified by the diff, are scheduled to run first.
Finally, sinks related to SARIF reports are also prioritized since we deem SARIF reports strong indicators of potential vulnerabilities.
%\compactline

In addition to scheduling, the Sink Manager is responsible for state management.
It tracks sinks for which a PoV has already been found or where PoV generation has failed, preventing the redundant processing of the same sink.
It also manages sinks for which a path from the entry point could not be found, allowing for re-analysis if the call graph is updated in the future.
%\compactline

\subsubsection{Path Finder} \label{sss:pg-pf}
% \CEN{
% component 3 (path finder)
% - clarify what is the path here (the sequence of functions from harness to sink, give an example), and describe how you get the path technically via Joern (mention the key search query, etc)
% - main challenge: path selection strategy (there can have many paths for one given <source, sink>, data-flow taint analysis paths + shortest path is the strategy, due to prompt size consideration, might be the easiest way for the generation task, generally, also discuss its limitations with example)
% - discuss about the benefit of merged CG thing (from all static analyzers \& fuzzer), there are more paths and solved more CPVs)
% }
A ``path'' represents the sequence of code execution from a harness to a sinkpoint.
Note that a path does not encompass all executed code; rather, it consists of a relevant subset of code identified by the specific path-finding method employed.
The Path Finder identifies paths using two methods: static taint analysis and call graph analysis.

\PP{Static Taint Analysis}
Static Taint Analysis is an algorithm that operates by recursively tracing data flow backward from a sinkpoint, identifying variables that can influence its value.
If this recursive trace determines that an argument of the harness can affect the sinkpoint, a valid path is considered to exist.
The sequence of code instructions that form the basis for this determination constitutes the path.
While static taint analysis provides high-fidelity results due to its data-flow-centric approach, it suffers from scalability issues, as its computational complexity makes it impractical for large codebases.
This tool leverages Joern's Static Taint Analysis module for this analysis.
%\compactline

\begin{figure}[!t]
  \centering
  \begin{promptbox}{Static Taint Analysis Query Example}
  \begin{promptcontent}\input{code/java_pg_sta.scala}\end{promptcontent}
  \end{promptbox}
  \caption{Example of static taint analysis using Joern}
  \label{fig:pg-sta}
\end{figure}

\autoref{fig:pg-sta} illustrates the methodology by which this tool employs Joern's Static Taint Analysis.
In this code, \ccw{entry} indicates the entry method of the harness, and \ccw{src} refers to the parameters of that method, which are the designated targets of the taint analysis.
\ccw{sink} represents the sinkpoint, which involves a simple conversion procedure where the CPG nodes for the sinkpoints are secured using their corresponding CPG IDs received from the Sink Manager.
Finally, the static taint analysis is executed seamlessly via the API provided by Joern.
This API returns an empty list when no path is detected, or a list of paths if one or more can be identified.
In the latter case, each sublist contains the specific instructions through which the taint has propagated for each path.
However, static taint analysis suffers from a scalability problem: Joern's default call depth limit is four, which is not practical.
Attempts to increase this depth resulted in excessive resource consumption, leading to timeouts or out-of-memory errors in most cases.
Consequently, static taint analysis is employed only for relatively accessible sinkpoints within a call depth of four.
For deeper sinkpoints, call graph analysis is used as an alternative.

\PP{Call Graph Analysis}
Call Graph Analysis is a process that identifies a path from a source to a sink within a call graph, which is constructed based on caller-callee relationships.
In this context, the harness method serves as the source, and the method containing the sinkpoint acts as the sink.
If a path from source to sink is found, the tool collects the list of methods visited along the route,
and appends the sinkpoint-containing method itself to the end of the list to form the final path.
This method is more scalable than static taint analysis, but suffers from a great number of false positive results
because the existence of a caller-callee relationship in the code does not guarantee that a call is actually feasible at runtime.
We adopted this approach based on the belief that LLMs could more practically determine the feasibility of such calls compared to conventional methods.
Therefore, the Path Finder was designed to identify as many potential paths as possible, even at the risk of including false positives.

\begin{figure}[!t]
\centering
\begin{tikzpicture}[
    node distance=0.5cm and 0.5cm,
    codeblock/.style={
        rectangle,
        draw,
        thick,
        align=left,
        font=\scriptsize\ttfamily
    }
]

\node[codeblock] (node1) {
    fuzzerTestOneInput(...) \{ \\
    \quad if(...) \{ \\
    \quad \quad set\_condition(); \\
    \quad \} \\
    \quad vulnerable(); \\
    \}
};

\node[codeblock, below right=of node1] (node2) {
    set\_condition(...) \{ \\
    \quad trigger\_flag = true; \\
    \}
};

\node[codeblock, below=of node1, yshift=-1.5cm] (node3) {
    vulnerable(...) \{ \\
    \quad if(trigger\_flag == true) \{\\
    \quad \quad trigger(...); \\
    \quad \} \\
    \quad ... \\
    \}
};

\draw[-{Stealth[length=2mm]}, thick, blue] (node1.east) -- (node2.north);
\draw[-{Stealth[length=2mm]}, thick, blue] (node2.south) -- (node3.east);
\draw[-{Stealth[length=2mm]}, thick, red] (node1.south) -- (node3.north);

\begin{scope}[every node/.style={font=\scriptsize}]
    \node (legend) at ([xshift=0.9cm, yshift=-0.5cm]node3.east) [anchor=west, rectangle, align=left, minimum width=2.2cm] {
        \begin{tikzpicture}[x=1em]
            \draw[-{Stealth[length=2mm]}, thick, blue] (0,0) -- (2,0);
            \node[right, xshift=1mm] at (1.3,0) {Vulnerable Path};
            \draw[-{Stealth[length=2mm]}, thick, red] (0,-0.5) -- (2,-0.5);
            \node[right, xshift=1mm] at (0.8,-0.5) {Shortest Path};
        \end{tikzpicture}
    };
\end{scope}

\end{tikzpicture}

\caption{Limitation of the shortest path}
\label{fig:pg-issue}
\end{figure}
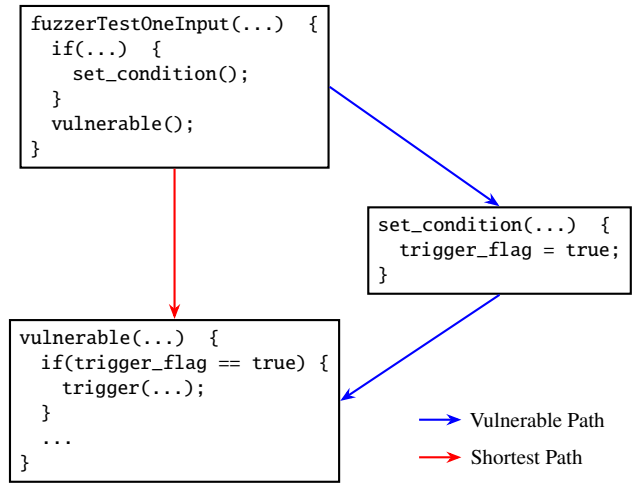

Although the Path Finder identifies a large number of potential paths, it ultimately utilizes only the single shortest path, for two reasons.
First, there may be too many paths to process within a reasonable time limit.
Second, since these paths are used by the PoV Generator to create the code snippets provided to the LLM, an excessive amount of code cannot be included due to constraints on prompt size.
However, as illustrated in \autoref{fig:pg-issue}, it is possible for the shortest path to fail to trigger the vulnerability.
This limitation necessitates further research into alternative methods to overcome this issue.

\subsubsection{PoV Generator} \label{sss:pg-pg}
% \CEN{
% component 4 (pov generator)
% - PoV generation, beepseed generation (prompt template, generation models used in competition)
% - generation approach (direct ask LLM generate blob/script, discuss their pros \& cons)
% - debugger (example?)
% }
% \begin{figure}[h!]
%     \centering
%     \includegraphics[width=\columnwidth]{fig/crs-java/lpg-prompt.pdf}
%     \caption{\textbf{LLMPOCGEN Prompt Concept} The prompt for LLMPOCGEN is composed of Instruct, Sink, Blob, Script, Feedback, and Code. Sink, Blob, Script, and Feedback refer to elements generated from previous attempts. Code is extracted in the format of FilePath and Method Body for each method constituting the path, specifically prefixed with LineNumber.\CEN{overall idea is good but can be refined, too fuzzy and small, the font color and style can be adjusted for highlighting the content of each part of prompt}}
%     \label{fig:lpg-prompt}
% \end{figure}
\begin{algorithm}[!t]
\footnotesize
\DontPrintSemicolon
\SetKwSty{algokeywordsty}
\SetFuncSty{algofuncsty}
\SetArgSty{algoargsty}
\SetKwFunction{fnGetCode}{GETCODE}
\SetKwFunction{fnGenerate}{GENERATE}
\SetKwFunction{fnEvaluate}{EVALUATE}
\SetKwFunction{fnFeedback}{FEEDBACK}
\SetKwFunction{fnExpand}{EXPAND}
\caption{PoV Generation}
\label{alg:pg-pov}
\KwIn{$\mathcal{P}$: Path}
$models \gets$ \{claude-4.0-opus, claude-4.0-sonnet, o3, gemini-2.5-pro\}\;
$prevScore \gets$ \text{-0.1}\;
$score \gets$ \text{0.0}\;
$feedback \gets$ \text{""}\;
$script \gets$ \text{""}\;
$blob \gets$ \text{""}\;
$code \gets$ \fnGetCode{$\mathcal{P}$}\;
\While{$score$ > $prevScore$}{
    $prevScore \gets$ $score$\;
    \For{$model \gets models$} {
        $script', blob' \gets$ \fnGenerate{$model, \mathcal{P}, feedback, script, blob$}\;        
        $score' \gets$ \fnEvaluate{$\mathcal{P}, blob$}\;
        \If{$score' == \text{1.0}$} {
          \Return $blob$\;
        }
        \If{$score' > score$} {
          $score \gets score'$\;
          $script \gets script'$\;
          $blob \gets blob'$\;
        }
    }
    $feedback \gets$ \fnFeedback{$\mathcal{P}, blob$}\;
    $code \gets$ \fnExpand{$\mathcal{P}$}\;
}

\end{algorithm}
The PoV Generator interacts with an LLM to generate a PoV for each path.
\autoref{alg:pg-pov} shows how the PoV Generator works.
The process begins with getting a code snippet based on the given path.
This essentially includes the entire harness code, and the code for all other methods along the path.
To enable the LLM to reference specific locations within the code accurately, all extracted code snippets are annotated with their corresponding file paths and line numbers, as shown in \autoref{fig:pg-llm-prompt}.
%\compactline

\begin{figure}[!t]
  \centering
  \begin{promptbox}{PoV Generation Prompt}
  \input{code/pba-prompt.txt}
  \end{promptbox}
  \caption{Example prompt for PoV Generation}
  \label{fig:pg-llm-prompt}
\end{figure}

\PP{Generate}
In this step, a prompt is constructed and sent to the LLM to produce a data blob.
\autoref{fig:pg-llm-prompt} shows an example prompt for this step.
The core instruction of this prompt is a straightforward request: to create an input that explores the path from the harness to the sink, thereby triggering the vulnerability.
To guide the LLM effectively, some key information is also provided: the sink, the sentinel, the previous script and blob, feedback, and the code.
The sink specifies the precise location of the sinkpoint within the code.
Its purpose is to direct the LLM to generate a value that will trigger the vulnerability at that specific program location.
The sentinel is a predefined canary value, engineered to be detected by the Jazzer sanitizer.
It is included so that the LLM can construct a PoV that explicitly triggers the sanitizer, confirming the vulnerability's discovery.
As the process is iterative, the script and blob from the previous attempt are included in the subsequent prompt.
This allows the LLM to refine and build upon its previous generations.
Finally, feedback from the execution of the previous generated input is provided, which helps to guide the LLM's next generation cycle toward a successful PoV.
%\compactline

\PP{Evaluate}
This step assesses the data blob generated by the LLM, with the primary objective of measuring path coverage.
This is accomplished using the Java Debugger.
The tool sets breakpoints at each code location corresponding to the given path and then executes the program with the LLM-generated blob.
By checking which breakpoints were hit, the tool can identify the last visited code location within the path.
A score is calculated based on this last visited location.
The scoring logic is specifically designed to differentiate between an input that reaches the sinkpoint without triggering the vulnerability and one that successfully causes exploitation.
For this reason, the score is normalized against the total number of path locations plus one.
For example, on a path with four code locations, if the generated input only reaches the second location, it receives a score of 0.4.
If the input successfully traverses the entire path to the sink but fails to trigger the vulnerability, it is awarded a score of 0.8.
A perfect score of 1.0 is reserved for inputs that not only reach the sink but also successfully trigger the vulnerability, indicating that the final objective has been fully achieved.
%\compactline

\PP{Feedback}
This step is responsible for generating informational feedback to be used in the subsequent Generate step.
In a process identical to that of the Evaluation step, the tool first identifies the last code location on the path that was visited during the execution.
Based on this finding, a feedback prompt is constructed.
This prompt explicitly states that a problem was encountered in reaching the next intended code location in the path from the last successfully visited one, thereby guiding the LLM's next attempt to overcome the specific point of failure.
%\compactline

\PP{Expand}
This step attempts to augment the code context provided to the LLM.
As illustrated in \autoref{fig:pg-issue}, constructing the initial context from only the shortest path may omit functions that, while not on the direct path, are essential for generating a valid PoV.
Furthermore, because our initial code extraction is performed at the method level, critical context can be lost.
For instance, if an analyzed method references member variables, their initialization values or other methods that modify their state can be indispensable for correct analysis.
Without this information, the LLM is unlikely to infer the correct input values, even across multiple attempts.
To address this issue, the Expand step operates by systematically providing the LLM with the full contents of each file related to the path, excluding the harness file.
The LLM is then instructed to prune these files, retaining only the code segments it deems necessary for the task.
This remaining code, identified by the LLM as relevant, is then used to update the existing code context for the next Generate prompt.
While this method is often effective, it has notable limitations.
First, it cannot incorporate necessary code from files that are not already associated with the path, limiting its scope.
Second, and more critically, providing entire files to the LLM results in significant token waste and risks exceeding the model's context size limit, which can lead to operational failure.
A more advantageous approach might be to query the LLM for the specific names of methods or variables it requires.
Following this, only the relevant code snippets could be selectively retrieved and used to construct a more targeted and token-efficient prompt.
%\compactline

\subsubsection{Call Graph Manager} \label{sss:pg-mgr}
% \CEN{
% component 5 (call graph manager)
% - Use Joern to build the CG, but the CG is maintained by us instead of Joern (so easily to update/merge new CG info from other static analyzers, fuzzer, etc)
% - recalculate the unreachable sink (failed to find path -> may not reachable)
% - Have specifically solved reflection issue
% }
Following construction of the CPG, the Call Graph Manager becomes active.
It traverses the call graph using depth-first search, starting from the entry method of each harness, to generate a call tree, which is then saved as a separate file for use by the Path Finder.
A separate call tree file is used, rather than the CPG directly, to facilitate easier updates.
\sys-Java utilizes a unified call graph, which is an aggregation of the call graphs from all modules that generate and use them.
This unified approach allows for the addition of edges that may not be discoverable through static analysis alone due to implementation issues in the CPG or inherent limitations of static analysis, thereby necessitating updates.
Specifically, call graphs generated through dynamic function call tracing contain edge information that is often invisible to static analysis, such as calls made via reflection.
By incorporating this dynamic information, the unified call graph enables the discovery of previously untraceable paths, significantly enhancing the path finding capabilities of the system.
%\compactline

\begin{figure}[!t]
  \centering
  \begin{promptbox}{Reflection Solver Prompt}
  \input{code/java_reflection_solver.scala}
  \end{promptbox}
  \caption{Example of reflection solver in Call Graph Manager}
  \label{fig:pg-ex-reflect}
\end{figure}

Additionally, for simple cases of reflection, the Call Graph Manager is equipped with a feature to connect call edges based on candidate functions identified by an LLM.
For instance, when reflection is performed as shown on line 3 of \autoref{fig:pg-ex-reflect}, the tool queries the LLM to determine which method, belonging to which class and with what specific parameters, is likely to be invoked.
Upon receiving the LLM's response, the tool uses this information to formulate a Joern query.
This query searches for all matching methods, and the tool then establishes call edges to each of them.
In the provided example, the LLM might determine that a method within the \ccw{UserRemoteConfig} class that accepts a single \ccw{String} parameter can be called.
A Joern query is subsequently executed to extract all methods satisfying this signature, and related call edges are updated.
%\compactline

\subsubsection{Misc} \label{sss:pg-misc}
% \CEN{
% Misc:
% Local and NFS cache
% diff analyzer
% - harness reachability
% }
This section introduces additional functionalities of this tool that extend beyond its basic workflow.
These capabilities leverage existing components and are specifically designed for use within \sys-Java.
%\compactline

\PP{Diff Analyzer}
Diff Analyzer operates on delta-mode challenges by taking a diff file as input.
The diff file is first provided to an LLM to identify code segments that could potentially introduce vulnerabilities.
The identified code is then converted into a node on the CPG and passed to the Sink Manager.
By default, this tool does not designate code related to out-of-memory or timeout errors as sinkpoints, because the vast number of code segments that could potentially cause such issues leads to an excessive number of paths.
However, by specifically using the LLM to find these instances within the confines of the diff, the tool overcomes the limitation of being unable to handle certain types of vulnerabilities.
Similarly, if the Sink Manager contains sinks that are relevant to the diff, their priority is elevated to ensure they are processed first.
Finally, the tool identifies methods that have been added or modified by the diff.
It then utilizes the Path Finder to identify any harnesses capable of reaching these methods.
This allows \sys-Java to allocate more resources to and prioritize the fuzzing of these identified harnesses.
%\compactline

\PP{Cache}
In certain scenarios, \sys-Java may re-execute a node for a specific purpose.
When this occurs, the entire process must restart from the beginning, which can be time-consuming due to the data generation required in the initial stages.
CPG creation and the call tree database are notable examples.
To address this issue, the tool incorporates a feature that periodically updates and saves this data to an NFS store provided by \sys-Java.
Upon re-execution, the tool first checks the NFS for existing data.
If found, it downloads the data, bypassing the initial setup phases.
However, the processes of identifying paths and generating PoVs for each one may still need to be re-executed, which can lead to significant consumption of LLM tokens and time.
To mitigate this, the tool also caches all interactions with the LLM as files that are also stored on the NFS.
This feature ensures that upon restarting, all previous LLM interactions can be rapidly re-processed without additional token consumption, effectively resolving the problem of redundant LLM computation.

\clearpage
\section{\sys-Multilang}
\label{s:crs-multilang}

\begin{figure*}[htbp]
  \centering
  \resizebox{\textwidth}{!}{
\begin{tikzpicture}[
  module/.style={anchor=north west, rectangle, draw, minimum width=4.5cm, minimum height = 0.5cm, very thick, font=\small, fill=gray!10},
  bmodule/.style={anchor=north west, rectangle, draw, minimum width=9cm, minimum height = 0.5cm, very thick, font=\small, fill=gray!10},
  moduletxt/.style={font = \small},
  smodule/.style={anchor=north west, rectangle, draw, minimum width=3.75cm, minimum height = 0.5cm, very thick, font=\footnotesize, fill=white},
  smoduletxt/.style={font = \footnotesize},
  ssmodule/.style={rectangle, draw, rounded corners = 1mm, minimum width=3.25cm, , minimum height=0.5cm, thick, font=\scriptsize, align=center, fill=gray!10},
  ssmoduleLLM/.style={ssmodule, fill=GreenYellow},
  arrow/.style={-{Latex[length=2mm, width=2mm]}, ultra thick},
  arrowBig/.style={-{Latex[length=3mm, width=3mm]}, line width=3pt},
  arrowBigBi/.style={{Latex[length=3mm, width=3mm]}-{Latex[length=3mm, width=3mm]}, line width=3pt},
  arrowTxtR/.style={right, font=\scriptsize, pos=0.5, align=left},
  arrowTxtB/.style={right, font=\footnotesize, pos=0.5, align=left},
  db/.style={cylinder, draw, minimum width = 4.5cm, minimum height = 5cm, anchor=north west, very thick, shape border rotate=90, shape aspect=.90, font=\small, fill=gray!10}
]

\newcommand*{\yInitGap}{0.8}
\newcommand*{\yGap}{1}
\newcommand*{\ax}{-7}
\newcommand*{\bx}{0}
\newcommand*{\cx}{10}

\newcommand*{\txtX}{0.05}

\newcommand*{\intY}{1.1}
\newcommand*{\ay}{0}

% Corpus Manager
\newcommand*{\yMgr}{\ay}
\node[module, minimum height = 3cm] at (\ax, \yMgr) (mgr) {};
\node[moduletxt, below] at (mgr.north) {\textsc{\textbf{Corpus Manager}}};
\newdimen\xMgr
\pgfextractx{\xMgr}{\pgfpointanchor{mgr}{north}};

\node[smodule, align=center, anchor=north] at (\xMgr, \yMgr-\yInitGap) {
  \textbf{Seed Loader/ Sharer}
};

\node[smodule, align=center, anchor=north] at (\xMgr, \yMgr-\yInitGap-0.9) {
    \textbf{Seed Scheduler}\\
  (Directed Fuzzing)
};

\newcommand*{\yExecutor}{\ay-4}
\node[module, minimum height = 4.6cm] at (\ax, \yExecutor) (executor) {};
\node[moduletxt, below] at (executor.north) {\textsc{\textbf{Executor}}};
\newdimen\xExecutor
\pgfextractx{\xExecutor}{\pgfpointanchor{executor}{north}};

\node[smodule, anchor=north, align=center] at (\xExecutor, \yExecutor-\yInitGap)(scriptExec) {\textbf{Script Executor}};
\node[smodule, anchor=north, align=center] at (\xExecutor, \yExecutor-\yInitGap-1.15)(inputExec) {\textbf{Input Executor}\\(libFuzzer/Jazzer-based) };
\node[smodule, anchor=north, align=center] at (\xExecutor, \yExecutor-\yInitGap-2.65)(symbolizer) {\textbf{Coverage Symbolizer}};

\draw[arrow](scriptExec) -> (inputExec) node[arrowTxtR] {Gen./Mut. Inputs};
\draw[arrow](inputExec) -> (symbolizer) node[arrowTxtR] {Edge Coverage};
\draw[arrow](symbolizer) -> (executor.south) node[arrowTxtR] {Line Coverage};

% Input Generators
\newcommand*{\iya}{\ay}
\newcommand*{\iyb}{\iya - 0.75}
\newcommand*{\iyc}{\iyb - 1.35}
\newcommand*{\iyd}{\iyc - 2.9}

\newcommand*{\ixa}{\bx+0.5}
\newcommand*{\ixb}{\ixa + 4.25}

\node[bmodule, minimum height=8.6cm] at (\bx, \iya) (generators) {};
\node[moduletxt, below] at (generators.north) {\textsc{\textbf{Input Generators}}};

% Given Fuzzer
\node[smodule, align=center] at (\ixa, \iyb)(given) {
  \textbf{Given Fuzzer}\\
  (libFuzzer, Jazzer)
};

% Dictionary-based
\newcommand*{\dictY}{\iyc}
\node[smodule, align=center, minimum height=2.5cm] at (\ixa, \dictY)(dict) {};
\node[smoduletxt, below] at (dict.north) {
  \textbf{Dictionary-based}~(\autoref{ss:dict})
};
\newdimen\xDict
\pgfextractx{\xDict}{\pgfpointanchor{dict}{north}};
\node[ssmoduleLLM] at (\xDict, \dictY-\yInitGap)(dictgen) {\textbf{Dict. Generator}};
\node[ssmodule] at (\xDict, \dictY-\yInitGap-\yGap)(dictmut) {\textbf{Mutator}};
\draw[arrow] (dictgen.191) -> (dictmut.169) node[arrowTxtR]{Function-level Dictionary};
\draw[arrow] (dictmut.191) -> (dictmut.191  |- dict.south) node[arrowTxtR]{New Inputs};

% Testlang-based
\newcommand*{\testlangY}{\iyd}
\node[smodule, align=center, minimum height=2.5cm] at (\ixa, \testlangY)(testlang) {};
\node[smoduletxt, below] at (testlang.north) {
  \textbf{Testlang-based}~(\autoref{ss:testlang})
};
\newdimen\xTestlang
\pgfextractx{\xTestlang}{\pgfpointanchor{testlang}{north}};
\node[ssmoduleLLM] at (\xDict, \testlangY-\yInitGap)(reverser) {\textbf{Harness Reverser}};
\node[ssmodule] at (\xDict, \testlangY-\yInitGap-\yGap)(testlangGen) {\textbf{Mutator / Generator}};
\draw[arrow] (reverser.191) -> (testlangGen.169) node[arrowTxtR]{Testlang};
\draw[arrow] (testlangGen.191) -> (testlangGen.191  |- testlang.south) node[arrowTxtR]{New Inputs};

% Concolic
\node[smodule, align=center] at (\ixb, \iyb)(concolic) {
  \textbf{Hybrid Fuzzer}~(\autoref{ss:concolic})\\
  (SymCC, SymQEMU)
};

% MLLA Standalone
\newcommand*{\mllaSY}{\iyc}
\node[smodule, align=center, minimum height=1.8cm] at (\ixb, \mllaSY)(mllaS) {};
\node[smoduletxt, below] at (mllaS.north) {
  \textbf{MLLA Standalone}~(\autoref{ss:mlla-standalone})
};
\newdimen\xMllaS
\pgfextractx{\xMllaS}{\pgfpointanchor{mllaS}{north}};
\node[ssmoduleLLM] at (\xMllaS, \mllaSY-\yInitGap)(bgaS) {\textbf{BGA}};
\draw[arrow] (bgaS.191) -> (bgaS.191  |- mllaS.south) node[arrowTxtR]{
  New Inputs\\
  Gen./Mut. Scripts
};

\newcommand*{\mllaY}{\mllaSY-2.25}
\node[smodule, align=center, minimum height=3.8cm] at (\ixb, \mllaY)(mlla) {};
\node[smoduletxt, below] at (mlla.north) {
  \textbf{MLLA}~(\autoref{ss:mlla})
};
\node[ssmoduleLLM] at (\xMllaS, \mllaY-\yInitGap)(mcga) {\textbf{MCGA}};
\node[ssmoduleLLM] at (\xMllaS, \mllaY-\yInitGap-\yGap)(bcda) {\textbf{BCDA}};
\node[ssmoduleLLM] at (\xMllaS, \mllaY-\yInitGap-\yGap-\yGap)(bga) {\textbf{BGA}};
\draw[arrow] (mcga.191) -> (bcda.169) node[arrowTxtR]{Tainted Call Graph};
\draw[arrow] (bcda.191) -> (bga.169) node[arrowTxtR]{Bug Candidates};
\draw[arrow] (bga.191) -> (bga.191  |- mlla.south) node[arrowTxtR]{
  New Inputs\\
  Gen./Mut. Scripts
};

% Shared Utils
\newcommand*{\yUtil}{\ay}
\node[module, minimum height = 3cm] at (\cx, \yUtil) (utils) {};
\node[moduletxt, below] at (utils.north) {\textsc{\textbf{Shared Utils} (\autoref{ss:shared-utils})}};
\newdimen\xUtils
\pgfextractx{\xUtils}{\pgfpointanchor{utils}{north}};
\node[smodule, align=center, anchor=north] at (\xUtils, \yUtil-\yInitGap)(tracer) {
  \textbf{Function Tracer}
};
\node[smodule, align=center, anchor=north] at (\xUtils, \yUtil-\yInitGap-0.9)(tracer) {
  \textbf{Code Retriever}\\
  (Joern, LSP, Code Indexer)
};

% FuzzDB
\newcommand*{\yDB}{\ay-4.6}
\newcommand*{\dbInitGap}{0.75}
\newcommand*{\dbGap}{0.75}
\node[db, minimum height = 4.75cm] at (\cx, \yDB) (db) {};
\newdimen\xDB
\pgfextractx{\xDB}{\pgfpointanchor{db}{north}};
\node[moduletxt, below] at (\xDB, \yDB-0.1) {\textsc{\textbf{FuzzDB}}};
\node[smodule, align=center, anchor=north] at (\xDB, \yDB-\dbInitGap) {
  \textbf{Courpus}
};
\node[smodule, align=center, anchor=north] at (\xDB, \yDB-\dbInitGap-\dbGap) {
  \textbf{POVs}
};
\node[smodule, align=center, anchor=north] at (\xDB, \yDB-\dbInitGap-\dbGap * 2) {
  \textbf{Edge Coverage}
};
\node[smodule, align=center, anchor=north] at (\xDB, \yDB-\dbInitGap-\dbGap * 3) {
  \textbf{Line Coverage}
};
\node[smodule, align=center, anchor=north] at (\xDB, \yDB-\dbInitGap-\dbGap * 4) {
  \textbf{Bug Candidates}
};

\draw[arrowBig](mgr) -> (mgr -| generators.west) node[arrowTxtB, above] {Scheduled Seed};
\draw[arrowBig](executor) -> (mgr);
\newdimen\xWestGen
\pgfextractx{\xWestGen}{\pgfpointanchor{generators}{west}};
\draw[arrowBig](\xExecutor, \yExecutor+0.3) -> (\xWestGen, \yExecutor+0.3) node[arrowTxtB, above, pos=0.3] {Execution Result};
\draw[arrowBig](inputExec.east -| generators.west) -> (inputExec.east) node[arrowTxtB, above, pos=0.43]{New Inputs};
\draw[arrowBig](scriptExec.east -| generators.west) -> (scriptExec.east)node[arrowTxtB, above, pos=0.43] {Gen./Mut. Scripts};
\draw[arrowBigBi] (utils) -> (utils -| generators.east);

\newcommand*{\gap}{0.5}
\draw[ultra thick, rounded corners=5pt]
     (\ax-\gap,\ay+\gap) -- (\cx+4.5+\gap,\ay+\gap) -- (\cx+4.5+\gap, \ay-3.5)
     -- (\cx-0.5,\ay-3.5) -- (\cx-0.5, \yExecutor-4.7-\gap) -- (\ax-\gap, \yExecutor-4.7-\gap)
     -- cycle;

\newdimen\xGen
\pgfextractx{\xGen}{\pgfpointanchor{generators}{north}};
\node[smodule, align=center, anchor=center, font=\normalsize] at (\xGen, \ay+\gap) {\textsc{\textbf{UniAFL}~(\autoref{ss:uniafl})}};
\draw[arrowBigBi] (db.north) -> (\xDB,\ay-3.5);

\end{tikzpicture}
}
    \caption{The overview of \sys-Multilang. The green boxes indicate LLM-powered modules.}
    \vspace*{-2.5px}
  \label{fig:multilang}
\end{figure*}
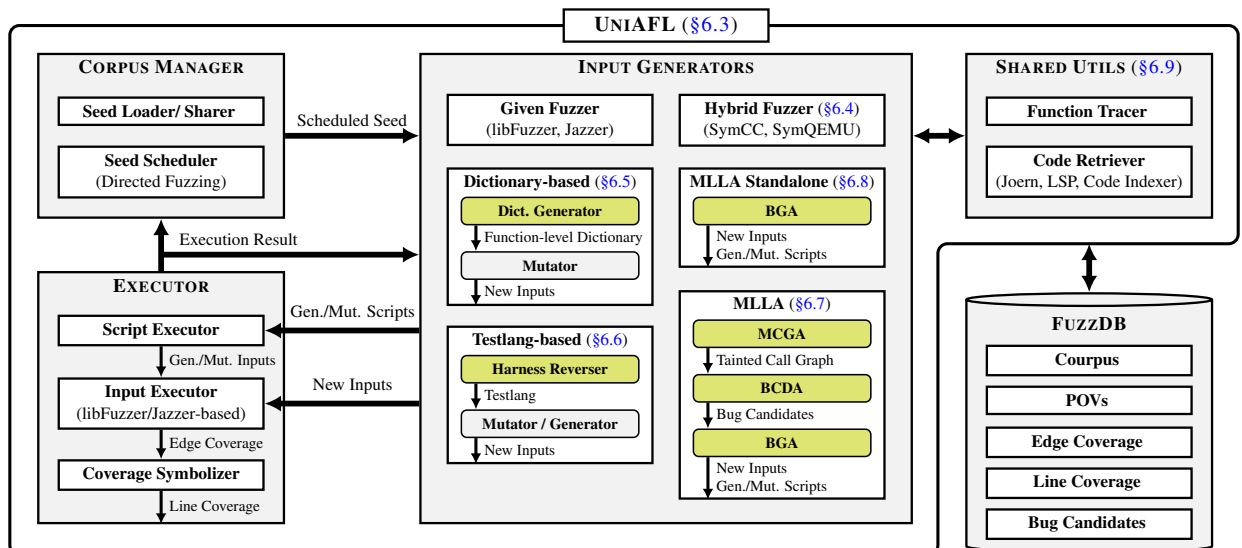

The primary goal of \sys-Multilang is to automatically find vulnerabilities in the target CPs while achieving the following objectives:

\PP{G1: Support fuzzing CPs written in any language}
The target CPs are provided in the OSS-Fuzz project format, where projects can be implemented in various languages (e.g., C and Java).
Therefore, our objective is to enable fuzzing for CPs regardless of their implementation language.
It is worth noting that while the AIxCC competition provided only CPs written in C or Java,
our approach is designed to support any language compatible with the OSS-Fuzz project format (see \autoref{ss:uniafl}).

\PP{G2: Improve fuzzing performance by using LLMs}
There has been much research on enhancing fuzzing performance, particularly in generating or mutating inputs more effectively.
However, most existing tools are limited to specific programming languages or even to programs compiled with particular compiler versions.
Furthermore, their performance improvements are often insufficient to automatically find vulnerabilities in complex programs, especially those requiring highly structured inputs.
To address these challenges, we aimed to leverage LLMs to enhance fuzzing performance, with a focus on more effective input generation and mutation strategies.
Notably, due to constraints on LLM usage during the competition, we designed \sys-Multilang with modular components utilizing various levels of LLM usage
(see \autoref{ss:concolic}-\autoref{ss:mlla-standalone}).

\PP{G3: Optimize fuzzing pipelines and development}
\sys-Multilang is designed to run on multi-core machines.
Therefore, it is important to optimize the fuzzing pipelines to minimize overhead when synchronizing fuzzing states across multiple fuzzing processes.
In addition, it is crucial to streamline the development process because \sys-Multilang must support the target CPs written in various languages.
To achieve this, we modularized \sys-Multilang as much as possible, ensuring flexibility, maintainability, and easier integration of language-specific components (see \autoref{ss:uniafl}).

\subsection{Overview}
\label{ss:multilang-overview}

\autoref{fig:multilang} presents the overall architecture of \sys-Multilang.
The system is primarily composed of two components: \sysuniafl and \sysfuzzdb.
\sysuniafl is responsible for fuzzing while addressing the three aforementioned goals, and \sysfuzzdb stores intermediate outputs (corpus, POVs, edge coverage, line coverage, and bug candidates) produced by \sysuniafl.

At a high level, \sysuniafl follows the same workflow as traditional fuzzers.
\textsc{corpus manager} selects a seed, and \textsc{input generators} produce new inputs by either mutating the seed or generating inputs from scratch.
Notably, some \textsc{input generators} produce Python scripts that themselves generate or mutate inputs.
\textsc{Script Executor} then runs these Python scripts to produce new inputs.
After that, \textsc{Input Executor} runs the new inputs under the target harness and outputs execution results along with edge coverage information.
Finally, \textsc{Coverage Symbolizer} translates the edge coverage into line coverage, which LLM-powered modules require.
\sysuniafl repeats this process until the timeout.

\PS{Input Generators}
The core of \sysuniafl lies in not only \sysuniafl infrastructure (\autoref{ss:uniafl}) but also its six input generation modules.
These modules are intentionally designed with varying levels of LLM dependence to ensure resilience when LLM usage is constrained, either due to budget limitations or infrastructure issues.
First, there are two modules that do not use LLMs at all.
\textsc{Given Fuzzer} simply runs the target harness, while \textsc{Hybrid Fuzzer} (\autoref{ss:concolic}) performs hybrid fuzzing based on concolic execution.
Note that we used LLMs to symbolically model functions while developing \textsc{Hybrid Fuzzer}.
Second, there are three modules that incorporate limited LLM usage.
\textsc{Dictionary-Based} (\autoref{ss:dict}) employs an LLM to infer the relevant dictionary for a given function, then performs function-level dictionary-based mutation.
\textsc{Testlang-Based} (\autoref{ss:testlang}) uses an LLM to figure out the input format of the target harness and express it in \textsc{testlang}, then performs \textsc{testlang}-based generation and mutation.
\textsc{Mlla-Standalone} (\autoref{ss:mlla-standalone}) leverages an LLM to generate Python scripts that produce new inputs.
Lastly, \sysmlla (\autoref{ss:mlla}) is the most LLM-intensive module.
It utilizes LLMs to derive tainted call graphs, identify bug candidates, and then generate inputs and Python scripts that create or mutate inputs targeting those bug candidates.

\begin{table}[t]
\centering
\footnotesize

\begin{threeparttable}
\begin{tabular}{clrrrrrrrr}
\toprule
\multirow{2}{*}{LLM Usage} & \multirow{2}{*}{Module} & \multicolumn{3}{c}{PoVs} & \multicolumn{2}{c}{Patches} & \multicolumn{2}{c}{Harnesses} & \multirow{2}{*}{Contribution\tnote{$\dagger$}} \\
\cmidrule(lr){3-5} \cmidrule(lr){6-7} \cmidrule(lr){8-9}
& & Total & Passed & Dup. & Total & Passed & Affected & w/o dup & \\
\midrule
None & Given Corpus & 9 & 9 & 0 & 2 & 2 & 7 & 7 & 7.6\% \\
None & Given Fuzzer & 280 & 58 & 0 & 21 & 20 & 30 & 30 & 49.2\% \\
None & Hybrid Fuzzer\tnote{$\ddagger$} & 4 & 2 & 0 & 1 & 1 & 2 & 2 & 1.7\% \\
Low & Dictionary-based & 6 & 0 & 0 & 0 & 0 & 0 & 0 & 0.0\% \\
Mid & Testlang-based & 32 & 8 & 0 & 2 & 2 & 5 & 5 & 6.8\% \\
High & \sysmlla & 39 & 7 & 0 & 4 & 4 & 5 & 5 & 5.9\% \\
Other & Shared Corpus & 23 & 0 & 10 & 0 & 0 & 5 & 0 & 0.0\% \\
\midrule
\multicolumn{2}{c}{\textbf{Total Multilang}} & \textbf{393} & \textbf{84} & \textbf{10} & \textbf{30} & \textbf{29} & \textbf{54} & \textbf{49} & \textbf{71.2\%} \\
\bottomrule
\end{tabular}
\begin{tablenotes}
\item [$\dagger$] Percentage of total system PoVs (118 total across all \sys modules).
\item [$\ddagger$] Hybrid Fuzzer used LLMs during development for symbolic modeling but not during runtime execution.
\end{tablenotes}
\end{threeparttable}
\caption{Performance breakdown of \sys-Multilang modules in the final competition.}
\label{t:multilang-finders}
    \vspace*{5px}
\end{table}

\subsection{Final Competition Results by \sys-Multilang}
\label{ss:multilang-performance}

\sys-Multilang is the primary vulnerability discovery engine in \sys, contributing 71.2\% of all verified PoVs in the AIxCC final competition.
As the cornerstone of our multi-language fuzzing capability, it combines traditional methods with LLM-powered techniques to discover vulnerabilities in C and Java codebases.
This subsection analyzes how our multi-tier LLM integration balanced reliability and innovation under budget constraints.
\autoref{t:multilang-finders} summarizes performance by LLM usage level and affected harnesses.

The results reveal a clear performance hierarchy across LLM integration levels.
However, these metrics show only final PoV attribution
and do not fully capture the incremental contributions
of all submodules during the fuzzing process.
Throughout the competition, including Exhibition rounds 1-3
and internal evaluations, we observed that each submodule
contributed to progressively expanding coverage,
with earlier discoveries enabling subsequent vulnerability findings
by other modules, including the given fuzzer.

Traditional approaches formed the foundation: Given Corpus (7.6\%), representing
the initial seed corpus provided with each challenge, and Given Fuzzer (49.2\%)
combined for 56.8\%, proving that quality seeds and robust fuzzing infrastructure
remain the most reliable vulnerability discovery methods under competitive pressure.
Hybrid Fuzzer contributed an additional 1.7\% through offline LLM-derived symbolic models,
bridging traditional and LLM-powered approaches without runtime LLM overhead.

LLM integration showed domain-specific effectiveness.
Dictionary-based mutations achieved 0.0\%, revealing that simple function-level
enhancements require deeper contextual understanding to succeed.
In contrast, Testlang-based generation delivered 6.8\% by leveraging LLM
pattern recognition for structured input formats where LLM excels.
\sysmlla contributed 5.9\% through comprehensive LLM integration,
discovering complex vulnerabilities beyond traditional fuzzing capabilities
despite significant computational overhead.

The Shared Corpus row shows 23 PoVs that originated from \sys-C or \sys-Java,
which incorrectly placed their discovered PoVs into the shared directory
alongside seeds.
According to our design, the shared directory should only contain seeds
for cross-module coordination, not PoVs.
This implementation error by other modules resulted in \sys-Multilang
inadvertently reporting these external PoVs,
with 10 being duplicates of vulnerabilities already found.

This multi-tier architecture validates a strategic approach: establish robust
traditional foundations (56.8\%), selectively deploy LLM where domain advantages
justify costs (6.8\%), and maintain research components for frontier exploration (5.9\%).
The key insight is that comprehensive vulnerability discovery requires
solid engineering fundamentals before adding LLM.
Individual harness performance details are in \autoref{s:appendix-multilang-targets}.

% \vspace*{-4px}
\subsection{\sysuniafl Infrastructure \& Directed Fuzzing}
\label{ss:uniafl}
In this subsection, we describe how \sysuniafl supports fuzzing CPs written in any language (\textbf{G1}) and how we optimize its fuzzing pipeline and development (\textbf{G3}).

\PP{Microservice-based Fuzzing}
To optimize the fuzzing pipeline for full multi-core utilization and accelerate the development process (\textbf{G3}), we designed \sysuniafl as a microservice-based fuzzer, inspired by $\mu$Fuzz~\cite{chen:mufuzz}.

Running each fuzzing instance on a separate core requires frequent synchronization of fuzzing states across instances.
This also requires maintaining the fuzzing state within each process.
These introduce significant overhead in terms of CPU and memory usage.
Furthermore, this setup is not fail-safe: if one input generator crashes, its process terminates and its core becomes idle.

To address these issues, we employed a microservice-based architecture.
Each component of \sysuniafl runs as a standalone process, with multiple threads internally leveraging multi-core resources.
This design allows each component to maintain its state in memory without duplicated memory usage and costly synchronization between multiple processes.
It also improves fault tolerance: if one input generator crashes, the remaining generators continue to produce inputs without disrupting the overall fuzzing workflow.
Note that each process communicates through shared memory and controls the execution of threads using semaphores.

\PS{Input Executor}
The core component for supporting CPs written in any language is \textsc{Input Executor}.
We designed a protocol that delivers inputs for execution via shared memory and retrieves execution results (e.g., code coverage and crash logs) via shared memory.
We chose shared memory instead of the file system to significantly reduce overhead.
To enable this, we modified existing fuzzers, libFuzzer and Jazzer, which serve as the foundation for the target harnesses.
Because this protocol is target-agnostic, \sysuniafl can seamlessly support any CP written in any language.

\PS{Script Executor}
\sysuniafl repeatedly executes Python scripts for mutating the seed or generating inputs from scratch, where the scripts are created by \textsc{Input Generators}.
By leveraging our shared-memory-based protocol, \textsc{Script Executor} writes new inputs directly into shared memory to deliver to \textsc{Executor},
rather than passing them to the target harness through the file system.
This allows us to avoid the significant overhead associated with using the file system.

\PS{Coverage Symbolizer}
Line coverage is essential for communication with LLMs because LLMs cannot interpret edge coverage, which consists of basic block addresses.
However, most fuzzers produce only edge coverage, which is enough to filter out interesting seeds, for efficiency.
To bridge this gap, \textsc{Coverage Symbolizer} gathers line coverage for all seeds and all POVs
which is JSON including which functions, lines, and files are executed.
While the line coverage format is language-agnostic, \textsc{Coverage Symbolizer} relies on language-specific instrumentation to gather line coverage.
The following describes how we implemented line coverage support for both C and Java:

%\autoref{fig:coverageformat} illustrates how the LLM module \sysmlla consumes coverage data.
%Below, we outline how \sysuniafl generates coverage for C/C++ and Java.

\PS{Coverage Symbolizer for C}
We primarily use the \texttt{coverage} build feature in OSS-Fuzz, which relies on coverage sanitizer in LLVM.
However, some CPs cannot be compiled with the \texttt{coverage} build option.
If so, we recover line coverage from edge coverage, which libFuzzer already produced, by using \texttt{llvm-cov},
which maps addresses to their corresponding source code lines.
Note that line coverage from \texttt{coverage} build is more precise than \texttt{llvm-cov}-based approach.

Obtaining line coverage for normal inputs is straightforward using the \texttt{coverage} build feature.
However, this approach cannot capture coverage for POVs, as they often crash the target harness before line coverage is produced.
To ensure execution halts immediately upon detecting memory corruption, we attach sanitizers to the harnesses built with \texttt{coverage} feature.
Additionally, we patched the LLVM source to hook an internal function, ensuring that coverage data is flushed to disk even in the event of a crash or abort.
As a result, we can utilize coverage of POVs to de-prioritize already triggered bug candidates in our directed fuzzing.

If the target CP cannot be compiled with the \texttt{coverage} build option, we instead convert edge coverage from libFuzzer into line coverage.
First, we compute the CFG of the target program before fuzzing begins.
Using the collected edge coverage, we determine which basic blocks were executed and then translate this basic block coverage into line coverage by invoking \texttt{llvm-cov}.
To improve efficiency, we further optimize this process by applying heuristics, implementing caching, and tailoring \texttt{llvm-cov} for our own purpose.

\PS{Coverage Symbolizer for Java}
Jazzer, the foundational fuzzer for Java in OSS-Fuzz, employs the widely used Java code coverage tool JaCoCo to get edge coverage.
Thanks to this JaCoCo, we can easily translate the edge coverage into line coverage by utilizing the translation features in the JaCoCo library.

\PP{Directed Fuzzing}
To better focus on generating POVs, \sysuniafl employs directed fuzzing for the bug candidates identified by \sysmlla.
The set of targets is dynamically updated as new bug candidates discovered by \sysmlla, and previously triggered bug candidates are removed from the target set.
However, existing directed fuzzers, which typically rely on custom instrumentation, do not support such dynamic targeting.
Some of them also cannot handle multiple targets simultaneously while we often have multiple bug candidates.
Lastely, custom instrumentation may not work with OSS-Fuzz CPs.

To address this limitation, we assign scores to seeds based on their line coverage.
Seeds that execute lines associated with bug candidates receive higher scores.
This approach is effective because each bug candidate from \sysmlla includes not only the vulnerable lines but also the key lines that must be executed to reach the vulnerable code and trigger the bug.
Based on this approach, we were able to fully exploit the chain-of-thought reasoning embedded in \sysmlla, enabling step-by-step directed fuzzing that incrementally guides execution toward the vulnerable code.
Finally, we schedule seeds for mutation using a weighted-random strategy, where the weight corresponds to a score of each seed.
As a result, seeds that cover more lines related to a bug candidate are selected more frequently, ultimately increasing the likelihood of triggering the vulnerability.

Unfortunately, directed fuzzing is not always ideal.
For example, \sysmlla may return incorrect bug candidates, or it may be necessary to mutate seeds that have not yet touched any key lines.
In addition, this method risks becoming overfitted to bug candidates that contain a large number of key lines.
To mitigate these issues, we adopt a mixed seed-selection strategy:
\begin{squishitemize}
    \vspace*{-5pt}
    \setlength{\itemsep}{-1pt}
    \setlength{\leftskip}{10pt}
    \item 25\%: Select a seed completely at random.
    \item 25\%: Randomly select one among seeds that already touched at least a key line.
    \item 50\%: Select a seed by score-based weighted random.
\end{squishitemize}
\clearpage
\subsection{Hybrid Fuzzing}
\label{ss:concolic}

\newcommand{\ntotalprojects}{551}
\newcommand{\ninvalidprojects}{130}
\newcommand{\ninvalidprojectsresolved}{128}
\newcommand{\nvalidprojects}{411}
\newcommand{\successfulprojectpcnt}{60.8\%}
\newcommand{\nsuccessfulprojects}{250}
\newcommand{\nfailedprojects}{161}
\newcommand{\nsuccessfulprojectsafter}{409}
\newcommand{\symstate}{\mbox{\textsc{SymState}}\xspace}
\newcommand{\crossover}{\emph{fusing mutation}}
\begin{figure}[t]
  \centering
  \input{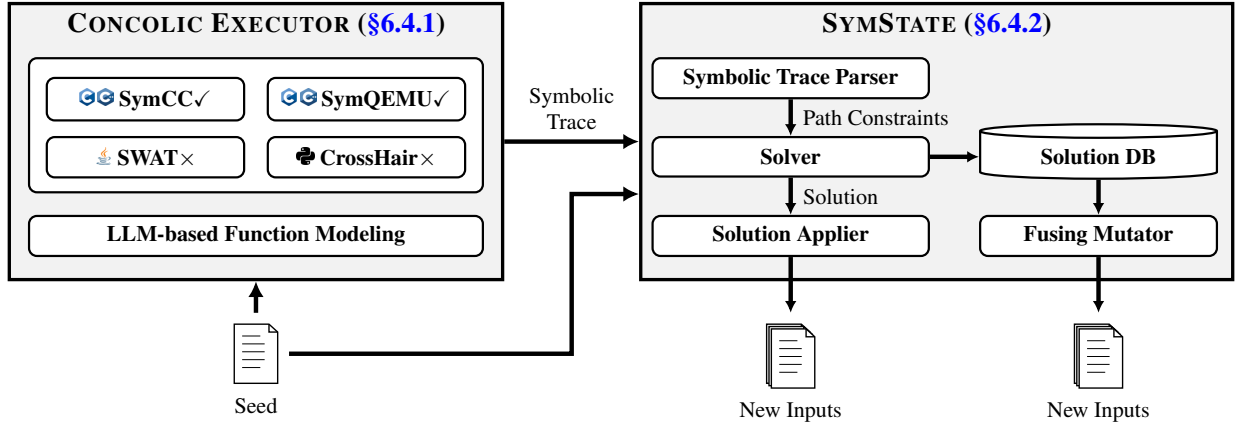}
    \vspace{-4px}
    \caption{The overview of our \textsc{Hybrid Fuzzer}. Check-marked boxes represent executors that were integrated into \sys-multilang for the final submission.} 
  \label{fig:concolic}
\end{figure}

We integrated a hybrid fuzzer based on concolic execution to 
facilitate the exploration of tight and complex branch conditions.
\autoref{fig:concolic} depicts the architecture of our hybrid fuzzer, 
which primarily comprises two modules: the \textsc{Concolic Executor} 
and \symstate. The concolic executor executes the instrumented 
target program and produces a symbolic trace. \symstate 
processes symbolic traces generated by the executor and produces new inputs.

To support multiple languages (\textbf{G1}), 
we built upon existing hybrid fuzzing frameworks 
\cite{symcc,symqemu,swat,crosshair},
but faced two major challenges:
\ding{172} they had limited support for 
functions residing in external libraries; \ding{173} they could 
not be integrated into a single framework due to
differences in their representation of path constraints.

\PP{C1. Lack of Modeling for External Functions}
Existing hybrid fuzzers lack comprehensive support for external function modeling,
leading to \emph{out-of-instrumentation} (OOI) functions, which are external functions that bypass
the instrumentation. 
For instance, SymCC models only 30 libc functions, 
omitting many functions responsible for 
string manipulation (e.g., \cc{strcmp}) and I/O operations (e.g., \cc{recv}). 
To address this gap, we developed an offline LLM-powered module that
systematically identifies and models OOI functions.

\PP{C2. Multiple Language Support} 
Integrating multiple hybrid fuzzers into a single framework 
was challenging due to variations in their path constraint representations.
For instance, SymCC maintained path constraints in memory
as \cc{Z3\_ast} structures, whereas SWAT produced an array of SMT-LIB2 strings.
To integrate multiple language targets into
a single framework, we decoupled the hybrid fuzzer into 
the executor and the solver (\symstate). By segregating 
the language-dependent executor from the language-independent \symstate,
we significantly reduced engineering costs as all language targets 
share the same \symstate module and benefit from the same optimizations. 
%One example of such optimization is \emph{fusing mutation}, 
%a novel technique that combines concolic solutions from multiple seeds.

%Both the executor and \symstate module comprises multiple submodules,
%which are duplicated in each core with the exception of the \emph{Solution DB}, which is shared across all cores.
%The hybrid fuzzer processes seeds received from the scheduler and generates new inputs. 
%If the seed's unique ID (\textsc{InputID}) is not present in the Solution DB,
%indicating that it was not observed yet by any core,
%the executor executes the instrumented harness with the seed and generates a symbolic trace.
%Then, \symstate processes the symbolic trace to produce new inputs.
%If it the seed was observed by any other core, \symstate bypasses the executor 
%and performs a lightweight \crossover using the solutions stored in the Solution DB.

\subsubsection{Concolic Executor}
\label{sss:concolic}
In this section, we discuss the design of the \textsc{Concolic Executor} and how it addresses
the challenge of incomplete modeling for external library functions (\textbf{C1}) via 
\emph{offline LLM-based function modeling}. We also discuss how we 
made the concolic executor robust to compilation errors by modifying the
SymCC toolchain.
%While building upon frameworks including SymCC, 
%SymQEMU, SWAT, and CrossHair. However, using them without modification 
%exhibited suboptimal performance due to two key limitations: 
%first, the frameworks provided limited support for external library functions; 
%second, they frequently failed to compile the target due to internal bugs.
%We addressed the first issue through an offline LLM-based function modeling technique. 
%We tackled the second issue by testing our compilation workflow against all OSS-Fuzz projects 
%to ensure robustness across diverse codebases. 

\PP{Offline LLM-based Function Modeling.}
We developed an offline LLM-based module that systematically 
identifies and models OOI functions. This is done in 
three phases: \ding{172} we conduct a three-way differential analysis to pinpoint
OOI functions and the values they over-concretize, i.e., values that 
should be symbolic but were rendered concrete; 
\ding{173} we prompt the LLM to model OOI function(s) using python code;
\ding{174} we verify the generated model by repeating the differential analysis 
with the model applied.

\begin{algorithm}[t]
\footnotesize
\DontPrintSemicolon
\SetKwProg{Fn}{function}{}{}
\SetKwSty{algokeywordsty}
\SetKw{KwBreak}{break}
\SetKw{KwContinue}{continue}
\SetFuncSty{algofuncsty}
\SetArgSty{algoargsty}
\SetKwFunction{fnConcolicExecute}{ConcolicExecute}
\SetKwFunction{fnGetIthExprAndTaken}{GetIthExprAndTaken}
\SetKwFunction{fnGetCalledFunctions}{GetCalledFunctions}
\SetKwFunction{fnGetDifferentValues}{GetDifferentValues}

\caption{Out-Of-Instrumentation (OOI) Function Detection}
\label{alg:differential}
\KwIn{Target program $P$, Two different inputs $I_A$ and $I_B$}
\KwOut{Set of OOI functions $\mathcal{F}$ \newline
  Over-concretized values $\mathcal{V}$
}
\BlankLine
\Fn{\textsc{DetectOOIFunctions}($P$, $I_A$, $I_B$)}{
    $\mathcal{F} \leftarrow \emptyset$, $\mathcal{V} \leftarrow \emptyset$\;
    $\Phi^A \leftarrow$ \fnConcolicExecute{$P$, $I_A$}\;
    $\Phi^{A'} \leftarrow$ \fnConcolicExecute{$P$, $I_A$}\;
    $\Phi^{B} \leftarrow$ \fnConcolicExecute{$P$, $I_B$}\;
    \For{$i = 0$ \KwTo $|\Phi^A| - 1$}{\label{ln:differential-l6}
        $(e_i^{A}, t_i^{A}) \leftarrow$ \fnGetIthExprAndTaken{$\Phi^A$, $i$}\;
        $(e_i^{B}, t_i^{B}) \leftarrow$ \fnGetIthExprAndTaken{$\Phi^{B}$, $i$}\;
        $(e_i^{A'}, t_i^{A'}) \leftarrow$ \fnGetIthExprAndTaken{$\Phi^{A'}$, $i$}\;
        \label{ln:differential-l9}
        $\mathcal{F} \leftarrow \mathcal{F} \cup$ \fnGetCalledFunctions{$e_i^{A}$}\;
        \If{$t_i^{A} \neq t_i^{B}$}{\label{ln:differential-l10}
            \KwBreak
        }
        \If{$e_i^{A} \neq e_i^{A'}$}{\label{ln:differential-l12}
            \KwContinue
        }
        \If{$e_i^{A} \neq e_i^{B}$}{\label{ln:differential-l16}
            $\mathcal{V} \leftarrow \mathcal{V} \cup$ \fnGetDifferentValues{$e_i^{A}$, $e_i^{B}$}\;
            \KwBreak
        }
    }
    \Return{$\mathcal{F}$, $\mathcal{V}$}\;
}
\end{algorithm}

\begin{figure}[t]
    \centering
\newcommand*{\codeX}{0cm}
\newcommand*{\codeY}{0cm}
\newcommand*{\codeWidth}{4.5cm}
\newcommand*{\pcWidth}{3.75cm}
\newcommand*{\xgap}{0.3cm}

\begin{tikzpicture}[
  code/.style={anchor=north west, rectangle, draw, minimum width=\codeWidth, minimum height = 4.8cm, fill=white},
  input/.style={anchor=north west, rectangle, minimum width=\pcWidth, minimum height = 0.75cm, font=\fontsize{8pt}{8pt}, fill=white},
  inputtight/.style={anchor=north west, rectangle, minimum width=\pcWidth - 0.45cm, minimum height = 0.75cm, font=\fontsize{8pt}{8pt}, fill=white},
  pc/.style={anchor=west, rectangle, draw, minimum width=\pcWidth, minimum height = 0.75cm, font=\fontsize{8pt}{8pt}, inner xsep=0pt, fill=white},
  pctight/.style={anchor=west, rectangle, draw, minimum width=\pcWidth - 0.45cm, minimum height = 0.75cm, font=\fontsize{8pt}{8pt}, inner xsep=0pt, fill=white},
>={Stealth[length=2pt,width=2pt]},
]
\node[code] at (\codeX, \codeY) (code) {

\begin{minipage}{\codeWidth}%
\begin{Verbatim}[commandchars=\\\{\},codes={\catcode`\$=3\catcode`\^=7\catcode`\_=8\relax}, fontsize=\fontsize{9pt}{9pt}\selectfont]
\PY{k+kt}{int}\PY{+w}{ }\PY{n+nf}{check}\PY{p}{(}\PY{p}{)}\PY{+w}{ }\PY{p}{\PYZob{}}
\PY{+w}{  }\PY{k+kt}{int}\PY{+w}{ }\PY{n}{fd}\PY{p}{;}
\PY{+w}{  }\PY{k+kt}{char}\PY{+w}{ }\PY{o}{*}\PY{n}{buf}\PY{p}{;}
\PY{+w}{  }\PY{n}{fd}\PY{+w}{ }\PY{o}{=}\PY{+w}{ }\PY{n}{socket}\PY{p}{(}\PY{c+cm}{/*..*/}\PY{p}{)}\PY{p}{;}
\PY{+w}{  }\PY{k}{if}\PY{+w}{ }\PY{p}{(}\PY{n}{fd}\PY{+w}{ }\PY{o}{\PYZlt{}}\PY{+w}{ }\PY{l+m+mi}{0}\PY{p}{)}\PY{+w}{ }\PY{p}{\PYZob{}}
\PY{+w}{    }\PY{k}{return}\PY{+w}{ }\PY{l+m+mi}{\PYZhy{}1}\PY{p}{;}
\PY{+w}{  }\PY{p}{\PYZcb{}}

\PY{+w}{  }\PY{n}{buf}\PY{+w}{ }\PY{o}{=}\PY{+w}{ }\PY{n}{malloc}\PY{p}{(}\PY{l+m+mi}{1024}\PY{p}{)}\PY{p}{;}
\PY{+w}{  }\PY{k}{if}\PY{+w}{ }\PY{p}{(}\PY{n}{buf}\PY{+w}{ }\PY{o}{=}\PY{o}{=}\PY{+w}{ }\PY{n+nb}{NULL}\PY{p}{)}\PY{+w}{ }\PY{p}{\PYZob{}}
\PY{+w}{    }\PY{k}{return}\PY{+w}{ }\PY{l+m+mi}{\PYZhy{}1}\PY{p}{;}
\PY{+w}{  }\PY{p}{\PYZcb{}}

\PY{+w}{  }\PY{n}{recv}\PY{p}{(}\PY{n}{fd}\PY{p}{,}\PY{+w}{ }\PY{n}{buf}\PY{p}{,}\PY{+w}{ }\PY{l+m+mi}{1024}\PY{p}{,}\PY{+w}{ }\PY{l+m+mi}{0}\PY{p}{)}\PY{p}{;}
\PY{+w}{  }\PY{k}{if}\PY{+w}{ }\PY{p}{(}\PY{n}{buf}\PY{p}{[}\PY{l+m+mi}{0}\PY{p}{]}\PY{+w}{ }\PY{o}{=}\PY{o}{=}\PY{+w}{ }\PY{l+s+sc}{\PYZsq{}}\PY{l+s+sc}{X}\PY{l+s+sc}{\PYZsq{}}\PY{p}{)}\PY{+w}{ }\PY{p}{\PYZob{}}
\PY{+w}{    }\PY{k}{return}\PY{+w}{ }\PY{l+m+mi}{\PYZhy{}1}\PY{p}{;}
\PY{+w}{  }\PY{p}{\PYZcb{}}
\PY{+w}{  }\PY{k}{return}\PY{+w}{ }\PY{l+m+mi}{0}\PY{p}{;}
\PY{p}{\PYZcb{}}
\end{Verbatim} 
\end{minipage}%

};

\newdimen\codeBottomY
\pgfextracty{\codeBottomY}{\pgfpointanchor{code}{south}};
\newdimen\codeRealWidth
\pgfextractx{\codeRealWidth}{\pgfpointanchor{code}{east}};

\newcommand*{\inputAX}{\codeX + \codeRealWidth + \xgap}
\newcommand*{\inputAY}{0cm}
\newcommand*{\inputBX}{\inputAX + \pcWidth + \xgap}

\newcommand*{\brAX}{\codeRealWidth - 0.2cm}
\newcommand*{\brBX}{\codeRealWidth - 0.2cm}
\newcommand*{\brCX}{\codeRealWidth - 0.2cm}
\newcommand*{\brAY}{\codeBottomY * 5 / 20}
\newcommand*{\brBY}{\codeBottomY * 10 / 20}
\newcommand*{\brCY}{\codeBottomY * 15 / 20}

\node[input] at (\inputAX, \inputAY) (inputA) {
    $I_A$\texttt{: "AAAA"}%
};
\node[inputtight] at (\inputBX, \inputAY) (inputB) {
  $I_B$\texttt{: "BBBB"}%
};

\node[pc] at (\inputAX, \brAY) (pcAA) {
\begin{minipage}[c]{\pcWidth}%
    \centering
    $e_0^{A}$\texttt{: (lt \#x3 \#x0)}\\%
    $e_0^{A'}$\texttt{: (lt \#x3 \#x0)}%
    \end{minipage}
};

\node[pctight] at (\inputBX, \brAY) (pcBA) {$e_0^{B} = \texttt{(lt \#x3 \#x0)}$};

\node[pc] at (\inputAX, \brBY) (pcAB) {
    \begin{minipage}[c]{\pcWidth}%
    \centering
    $e_1^{A}$\texttt{: (= \textcolor{red}{\#x672f2a0} \#x0)}\\%
    $e_1^{A'}$\texttt{: (= \textcolor{red}{\#xebf12a0} \#x0)}%
    \end{minipage}
};

\node[pc] at (\inputAX, \brCY) (pcAC) {
 \begin{minipage}[c]{\pcWidth}%
    \centering
    $e_2^{A}$\texttt{: (lt \textcolor{red}{\#x41} \#x0)}\\%
    $e_2^{A'}$\texttt{: (lt \#x41 \#x0)}%
    \end{minipage}
};
\node[pctight] at (\inputBX, \brCY) (pcBC) {$e_2^{B} = \texttt{(lt \textcolor{red}{\#x42} \#x0)}$};

% Arrows connecting execution flow vertically
\newdimen\pcABottomY
\pgfextracty{\pcABottomY}{\pgfpointanchor{pcAA}{south}};
\newdimen\pcBABottomY
\pgfextracty{\pcBABottomY}{\pgfpointanchor{pcBA}{south}};
\newdimen\pcBTopY
\pgfextracty{\pcBTopY}{\pgfpointanchor{pcAB}{north}};
\newdimen\pcBBottomY
\pgfextracty{\pcBBottomY}{\pgfpointanchor{pcAB}{south}};

% use north in case lengths of A and B differ
\newdimen\pcAX
\pgfextractx{\pcAX}{\pgfpointanchor{pcAA}{north}};
\newdimen\pcAXE
\pgfextractx{\pcAXE}{\pgfpointanchor{pcAA}{east}};
\newdimen\pcABMidY
\pgfextracty{\pcABMidY}{\pgfpointanchor{pcAB}{east}};
\newdimen\pcBX
\pgfextractx{\pcBX}{\pgfpointanchor{pcBA}{north}};
% doing -\xdist will cause an error
\newcommand*{\xdist}{-\pcBX + \pcAX}

% zig-zag arrows
%\draw[->, thick] (pcAA.east) -- (pcBA.west);
%\draw[->, thick] (pcBA.south) -- ++(0cm, \pcBTopY * 0.5 + \pcABottomY * 0.5 - \pcBABottomY) -- ++(\xdist,0cm) -- (pcAB.north);
%\draw[->, thick] (pcAB.south) -- (pcAC.north);
%\draw[->, thick] (pcAC.east) -- (pcBC.west);

\draw[->, thick] (pcAA.south) -- (pcAB.north);
\draw[->, thick] (pcAB.south) -- (pcAC.north);
\draw[->, thick] (pcBA.south) -- (pcBC.north);

% Black dots to indicate branch locations 
\fill[black] (\brAX, \brAY) circle (0.8pt);
\fill[black] (\brBX, \brBY) circle (0.8pt);
\fill[black] (\brCX, \brCY) circle (0.8pt);

% Dotted lines from pc squares to closest dots
\newcommand*{\distXDotted}{\pcBX - \pcAXE}
\draw[dotted] (pcAA.west) -- (\brAX, \brAY);
\draw[dotted] (pcAB.west) -- (\brBX, \brBY);
\draw[dotted] (pcAC.west) -- (\brCX, \brCY);
\draw[dotted] (pcAA.east) -- (pcBA.west);
\draw[dotted] (pcAB.east) -- ++(\distXDotted, 0cm);
\fill[black] (\pcBX, \pcABMidY) circle (0.8pt);
\draw[dotted] (pcAC.east) -- (pcBC.west);

% Dotted lines from inputs to first path constraints
\draw[dotted] (inputA) -- (pcAA);
\draw[dotted] (inputB) -- (pcBA);

\end{tikzpicture}

\caption{Three cases handled by our differential analysis algorithm within a single function.}

\label{f:differential-cases}
\end{figure}
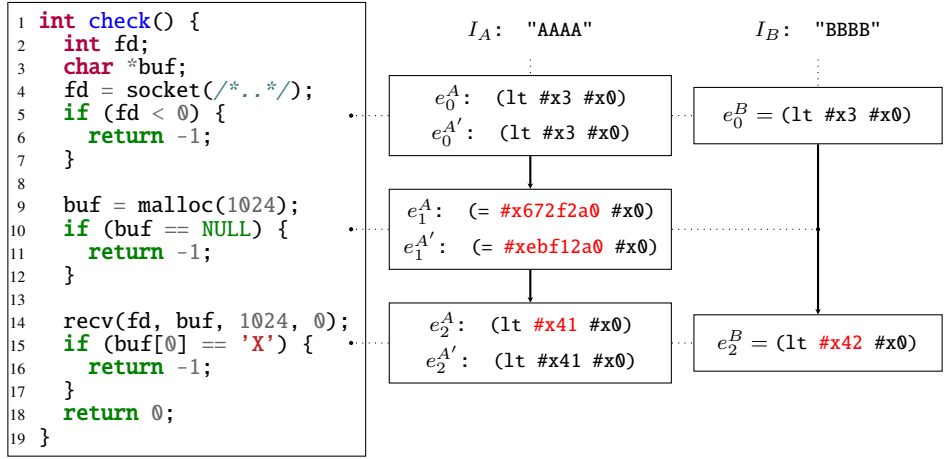

\PP{P1. Detect OOI Functions via Differential Analysis}
Our key insight is that a difference in symbolic expressions
($e_i^{A} \neq e_i^{B})$ indicates an OOI function. 
The bottom-most branch in \autoref{f:differential-cases} exemplifies this:
\cc{recv} is OOI, causing symbolic expressions
to be different for inputs \ccw{"AAAA"} and \ccw{"BBBB"} 
(\ccw{\#x41} vs. \ccw{\#x42}).
However, a difference in symbolic expressions may stem from non-determinstic behavior, 
rather than an OOI function.
To filter out such cases, we skip if the symbolic expressions 
for the same input are different ($e_i^{A} \neq e_i^{A'}$). 
The middle case in \autoref{f:differential-cases} illustrates this:
symbolic expressions are different due to the non-deterministic behavior of \cc{malloc}, 
allowing us to skip the branch.

%Our three-way differential analysis (\autoref{alg:differential}) executes the target program 
%with two different inputs, $I_A$ and $I_B$, and then executes it again with $I_A$. The 
%procedure produces a list of OOI functions and
%any values that were over-concretized. This information is later used for
%prompting the LLM to model those functions. 

To formalize this idea, we introduce the notion of a \emph{symbmlic trace},
$\Phi = (\phi_0, \phi_1, \cdots \phi_n)$, a sequence of path constraints generated during a single execution. 
Each \emph{path constraint} is a tuple $\phi_i = (e_i, t_i)$ where 
$e_i$ is a boolean symbolic expression and $t_i$ is its concrete evaluation,
indicating whether the branch was taken.
The leaf nodes of $e_i$ are either symbolic variables (e.g., \cc{data[0]}) 
or concrete values (e.g., \cc{\#x41414141}). 
%For instance, the symbolic expression 
%for \cc{data[0:3] == 0x41414141} is a binary comparison 
%where the right-hand side is a concrete value \cc{\#x41414141} and
%the left-hand side is a concatenation expression composed of
%symbolic variables \cc{data[0], .. data[1]}.

Based on these observations and definitions, we constructed a differential analysis algorithm (\autoref{alg:differential}).
The algorithm iterates over the path constraints of three executions:
$\Phi_A$ and $\Phi_{A'}$ for input $A$, and $\Phi_B$ for input $B$. 
In each iteration, it records all functions 
that were executed up to the current branch (\autoref{ln:differential-l9}). 
It loops until either \ding{172} branches are fully exhausted (\autoref{ln:differential-l6}),
\ding{173} control flow diverges between the two inputs (\autoref{ln:differential-l10}),
or \ding{174} an OOI function is detected (\autoref{ln:differential-l16}).
To account for non-deterministic behavior, we compare the symbolic expressions for the same input (\autoref{ln:differential-l12}).
We consider the program to be correctly instrumented if there are no over-concretized values ($\mathcal{V} = \emptyset$). 

\PP{P2. Model OOI functions by Prompting the LLM}
We present the information gathered from the differential analysis to the LLM and request it to 
model the OOI function(s) using Python code. The prompt includes the list 
of OOI functions and their call sites, 
the source location of the over-concretized values and their concrete values,
and the problematic branch site. \autoref{fig:concolic-prompt-example} shows an example of the prompt 
based on the differential analysis performed in \autoref{f:differential-cases}, and 
\autoref{fig:concolic-response-example} shows the LLM-generated response, which 
correctly models the memory writing behavior of the \cc{recv} function.

\begin{figure}[t]
\centering
\subfigure[Prompt requesting the LLM to generate a model for \cc{recv}.]{
\label{fig:concolic-prompt-example}
\begin{promptbox}{LLM-Based Function Modeling Prompt}
\begin{promptcontent}\input{code/llm-prompt-concolic-plain.txt}\end{promptcontent}
\end{promptbox}
}

\subfigure[LLM-generated Python model for the \cc{recv} function.]{
\label{fig:concolic-response-example}
\begin{promptbox}{LLM-Based Function Modeling Response}
\begin{promptcontent}\input{code/llm-response-concolic.py}\end{promptcontent}
\end{promptbox}
}

\caption{Example of LLM-based function modeling for OOI functions.}
\label{fig:concolic-llm-example}
\vspace*{-10px}
\end{figure}

\PP{P3. Verify the Model}
We verify the generated model by repeating the differential analysis 
with the model applied. We consider the model to be incorrect if the 
previous over-concretized values are still present ($\mathcal{V}_{prev} \cap \mathcal{V}_{after} \neq  \emptyset$).
When this occurs, we terminate the automated modeling process and resort to manual function analysis. 
Conversely, if the model successfully eliminates all over-concretized values for that branch,
we permanently add it to the target program's instrumentation. We repeat \textbf{P2} and \textbf{P3} until 
the differential analysis terminates with an empty set of over-concretized values ($\mathcal{V} = \emptyset$).

\PP{Compilation Robustness}
We designed the workflow to be robust against compilation failures by testing
against all OSS-Fuzz projects. The existing frameworks were susceptible to a 
significant number of compilation failures.
For instance, when we compiled the \nvalidprojects{}  C/C++ projects in OSS-Fuzz 
(Out of a total of \ntotalprojects{} C/C++ projects,
\ninvalidprojects{} of them were excluded as they failed to compile the given fuzzer)
using SymCC,
\nfailedprojects{} failed to compile 
due to internal bugs. After fixing the issues, we 
were able to compile \nsuccessfulprojectsafter{} of the \nvalidprojects{} projects, a significant improvement from 
the original \nsuccessfulprojects. 

%\PP{SymQEMU}
We also implemented a SymQEMU based fallback mechanism for projects that failed to compile with SymCC.
This was done by executing the libfuzzer harnesses, which are guaranteed to compile, with SymQEMU. 
However, runing the libfuzzer harnesses with SymQEMU resulted in suboptimal performance 
due to overhead from TCG instrumentation. To reduce the overhead, 
we ran the harness in fuzzing mode (without any seed directory argument)
and hooked into the \cc{LLVMFuzzerTestOneInput} function to inject the input bytes directly into memory.
This approach allowed us to re-use the TCG instrumentation for multiple inputs. 
Additionally, we patched SymQEMU to skip the instrumentation of 
SanitizerCoverage and AddressSanitizer runtime functions. 

\subsubsection{SymState}
\label{sss:symstate}

In this section, we describe the architecture of our solving module \symstate.
%The separation of the language-dependent executor from the language-independent \symstate module
%allowed us to implement a multi-lingual hybrid fuzzer (\textbf{C2}) with reduced engineering costs.
\symstate comprises four submodules, three of which are duplicated in each core: 
the \textsc{Symbolic Trace Parser}, 
the \textsc{Solver}, 
and the \textsc{Solution Applier}. 
The \textsc{Solution DB}, which is shared across all cores,
allows \symstate to switch between two input generation regimes, depending 
on whether the seed has been observed by any other core.

For unobserved seeds, \symstate follows a three-stage pipeline:
the \textsc{Symbolic Trace Parser} parses language-specific symbolic traces 
(C/C++: Z3 AST; Java: SMT-LIB2 strings) into a unified path constraint format.
Then, \textsc{Solver} generates satisfying assignments for these constraints,
and \textsc{Solution Applier} transforms these assignments into concrete input bytes.
Conversely, seeds previously processed by other cores bypass this pipeline entirely,
leveraging our novel \emph{fusing mutator} for efficient cross-core solution reuse.

\PP{Fusing Mutator}
One major limitation of existing concolic executors is that they rely solely on the symbolic trace of the current input when performing mutations, without leveraging traces from other seeds.
In theory, this design choice would be ideal if precise symbolic traces were always available.
In practice, however, obtaining precise traces is often infeasible due to concretization issues.

To address this, we introduce the \textsc{Fusing Mutator}, a novel hybrid fuzzing technique which utilizes solutions from multiple seeds.
The \textsc{Fusing Mutator} first selects random solutions from the \textsc{Solution DB}.
After that, it sequentially applies cached solutions to an input,
generating a set of new inputs where each application builds upon the previous result.
This technique has two advantages that were not present in existing hybrid fuzzers:
\ding{172} The fuzzer can create new inputs that satisfy path constraint properties of multiple seeds, allowing it to explore novel paths.
\ding{173} It improves performance by reusing solutions from other cores, eliminating redundant execution and solving overhead. 

\PP{Auxiliary Symbols}
Our \textsc{Solution Applier} supports path constraints expressed in terms of 
higher-level symbols beyond individual bytes (e.g., \cc{data[0]}).
These auxiliary symbols significantly improve \symstate's performance by eliminating the need 
to represent all path constraints at the byte level.

For example, we introduced the $\cc{ScanfExtract}_{f,s,e,i}$ 
symbol, which represents the value of the $i$-th variadic argument in a \cc{sscanf} function call 
where the format string is \cc{f} and the input string is \cc{data[s:e]}.
This symbol allows the solver to reason in terms of \cc{sscanf} arguments directly
rather than individual input bytes, avoiding the need to encode complex 
string parsing rules into SMT expressions. After the solver solves in 
terms of \cc{ScanfExtract}, and the \cc{Solution Applier} uses 
the inverse operation \cc{sprintf(output\_str, fmt, arg1, arg2, ...)} to 
produce byte-level replacements.

\subsubsection{Discussion}

\PP{LLM-Based Function Modeling}
We ran our LLM-based function modeling 
pre-competition, successfully identifying and modeling 13 functions
which were permanently integrated to our hybrid fuzzer. 
However, We decided not to run our LLM-based function modeling 
during the competition for the strategic reasons.  
Our decision was based on two considerations.
\ding{172} The LLM-generated models frequently produced incorrect results,
prompting manual analysis and correction.
Future work includes utilizing
prompt engineering techniques such as few-shot prompting and chain-of-thought for accuracy improvement,
and using a feedback loop between the verification and prompting to enable incremental revision.
\ding{173} Our differential analysis was resource intensive in terms of both time and memory. 
This happened because we disabled the \emph{concreteness checks} feature of SymCC,
an optimization that skips path constraints with fully concrete operands,
leading to a significant increase (10x-100x) in the number of path constraints.
We plan to mitigate this issue by making our concolic executor more efficient 
using techniques from existing literature \cite{symsan,leansym}.

\PP{Additional Language Support}
Due to time constraints, we could not integrate SWAT into our hybrid fuzzer. 
Additionally, we did not integrate concolic executors for other languages due to the narrowed competition scope.
Future work includes integrating them
to support all languages included in the OSS-Fuzz benchmarks.
% END

\subsection{Function-level Dictionary-based Input Generation}
\label{ss:dict}

Dictionary-based fuzzing improves effectiveness by mutating inputs with tokens or keywords frequently used by the target.
The overall effectiveness of this approach depends heavily on the quality of the dictionary and the strategies used to apply its entries during mutation.
In practice, dictionary-based fuzzing achieves better coverage when combined with other techniques.
For example, while dictionary-based fuzzing emphasizes exploration of the value space, grammar-based fuzzing (e.g., \textsc{testlang-based} fuzzing in \autoref{ss:testlang}) focuses on structural aspects of the input, making the two approaches complementary.

%The dictionary-based mutation module enhances fuzzing effectiveness by
%applying intelligent, structure-aware mutations to inputs.  It serves
%as a complementary component to structured input generators, such as
%the \textsc{Testlang-Based} module (\autoref{ss:testlang}) and
%\sysmlla (\autoref{ss:mlla}).  While those modules focus on generating
%syntactically and semantically valid inputs that conform to a target's
%complex format, the dictionary-based module is responsible for
%mutating these inputs in a way that explores the value space more
%thoroughly without compromising their structural integrity.

\PP{Function-level Dictionary-based Fuzzing}
Effective use of dictionaries is critical for the success of dictionary-based fuzzing.
Traditional approaches typically construct a single dictionary for the entire program and use it during mutation.
However, this often leads to low efficiency, since many dictionary entries may be unrelated to a given input.
To overcome this limitation, we introduce a function-level dictionary approach.
By generating dictionaries for individual functions and applying only those associated with the functions actually touched by an input, we significantly improve the efficiency of the fuzzing process.

\PP{On-the-fly Context-aware Dictionary Generation}
The crux of this module is its use of LLMs to generate context-aware dictionaries on-the-fly for individual functions.
Unlike traditional approaches that rely on a single static dictionary of magic values, our method dynamically creates a dictionary for each function.
This is powerful because LLMs can analyze the source code to infer the function's purpose and the types of data it processes.
For example, for a function that parses network protocols, an LLM-generated dictionary might include common protocol identifiers, boundary values, and even well-formed but adversarial data snippets.
The mutator can then use this dictionary to replace parts of the input, while preserving the overall structure.

\begin{figure}[t]
  \centering
  \resizebox{0.9\textwidth}{!}{
\begin{tikzpicture}[
  module/.style={rectangle, draw, minimum width=4cm, minimum height = 0.75cm, thick, font=\small, fill=gray!10},
  db/.style={cylinder, draw, minimum width = 3.5cm, minimum height = 2cm, thick, shape border rotate=90, shape aspect=.60, font=\small, fill=white},
  moduletxt/.style={font = \small},
  smodule/.style={rectangle, draw, minimum width=3.5cm, minimum height = 0.5cm, thick, font=\small, fill=white},
  line/.style={-, thick},
  arrow/.style={-{Latex[length=2mm, width=2mm]}, thick},
  arrowBig/.style={-{Latex[length=3mm, width=3mm]}, line width=2pt},
  arrowBi/.style={{Latex[length=2mm, width=2mm]}-{Latex[length=2mm, width=2mm]}, thick},
  arrowTxtR/.style={right, font=\scriptsize, pos=0.5, align=left},
  arrowTxtL/.style={left, font=\scriptsize, pos=0.5, align=right},
  arrowTxtB/.style={right, font=\footnotesize, pos=0.5, align=left}
]

\newcommand*{\ax}{0}
\newcommand*{\bx}{\ax+5.5}
\newcommand*{\cx}{\bx+5.5}

\newcommand*{\ay}{0}
\newcommand*{\by}{\ay-1.4}
\newcommand*{\cy}{\by-3.75}

\node[module, align=center, minimum width=5cm, minimum height = 4cm] at (\ax, \ay) (gen) {};
\node[moduletxt, align=center,below] at (gen.north) {
    \textbf{Dictionary Generator}\\
    \textbf{(\autoref{sss:dict-gen})}
};
\node[smodule] at (\ax, \ay+0.5) (funcsel) {
    \textbf{Function Selector}
};

\node[smodule, align=center, fill=GreenYellow] at (\ax, \ay+0.5-1.5) (dictgen) {
    \textbf{Function-level}\\
    \textbf{Dictionary Generator}
};

\node[db, align=center, minimum height=2.75cm] at (\bx, \by+0.6) (dict) {
};
\node[moduletxt, align=center,below, font=\footnotesize] at ($(dict.north) + (0, -0.1)$) {
    \textbf{Function-level}\\
    \textbf{Dictionary}
};

\newcommand*{\yGenInit}{-0.75}
\newcommand*{\yGenGap}{0.5}
\node[smodule, font=\scriptsize, minimum width=3cm] at (\bx, \by-\yGenInit) {
    \textbf{Comparison Operand}
};
\node[smodule, font=\scriptsize, minimum width=3cm] at (\bx, \by-\yGenInit-\yGenGap) {
    \textbf{Bug Triggering Token}
};
\node[smodule, font=\scriptsize, minimum width=3cm] at (\bx, \by-\yGenInit-\yGenGap*2) {
    \textbf{Formatted String}
};

\node[module, align=center, minimum width=5cm, minimum height = 4cm] at (\cx, \ay) (mut) {};
\node[moduletxt, align=center, below] at (mut.north) {
    \textbf{Dictionary-based Mutator}\\
    \textbf{(\autoref{sss:dict-mut})}
};

\newcommand*{\yMutInit}{0.5}
\newcommand*{\yMutGap}{0.9}
\node[smodule] at (\bx, \ay+1.5) (mutsel) {
    \textbf{Dictionary Selector}
};

\node[smodule, minimum width=4cm] at (\cx, \ay+\yMutInit) {
    \textbf{Token Insertion}
};
\node[smodule, minimum width=4cm] at (\cx, \ay+\yMutInit-\yMutGap) {
    \textbf{Token Replacement}
};
\node[smodule, minimum width=4cm] at (\cx, \ay+\yMutInit-2*\yMutGap) {
    \textbf{Byte Replacement}
};

\draw[arrowBig] ($(gen.95) + (0, 1)$) -> (gen.95);
\node[moduletxt, anchor=west] at ($(gen.95) + (0.15, 0.8)$) (input) {\textbf{Input}};
\node[moduletxt, anchor=west] at ($(gen.95) + (0.15, 0.3)$) (cov) {\textbf{Func. Coverage}};

\draw[arrowBig] ($(gen.137) + (0, 1)$) -> (gen.137);
\node[moduletxt, align=left, anchor=west] at ($(gen.137) + (0.1, 0.5)$) {\textbf{Challenge}\\\textbf{Project}};

\draw[arrowBig] (funcsel) -> (dictgen);
\node[moduletxt, anchor=west, font=\footnotesize] at ($(funcsel.south) + (0.2, -0.4)$) {\textbf{Function Body}};

\draw[arrowBig] (gen.east |- dict.west) -> (dict.west);
\draw[arrowBig] (dict) -> (mutsel);
\draw[arrowBig] (mutsel.east) -> (mutsel.east -| mut.west);
\draw[arrowBig] (cov.east) -| (mutsel.north);
\draw[arrowBig] (input.east) -| (mut.110);
\draw[arrowBig] (mut.north) ->($(mut.north)+(0, 1)$);
\node[moduletxt, anchor=west] at ($(mut.north) + (0.15, 0.4)$) {\textbf{New Inputs}};
\end{tikzpicture}
}

  \caption{An overview of the function-level dictionary-based input generation.}
  \label{fig:dict-gen}
  % \vspace{-6px}
\end{figure}
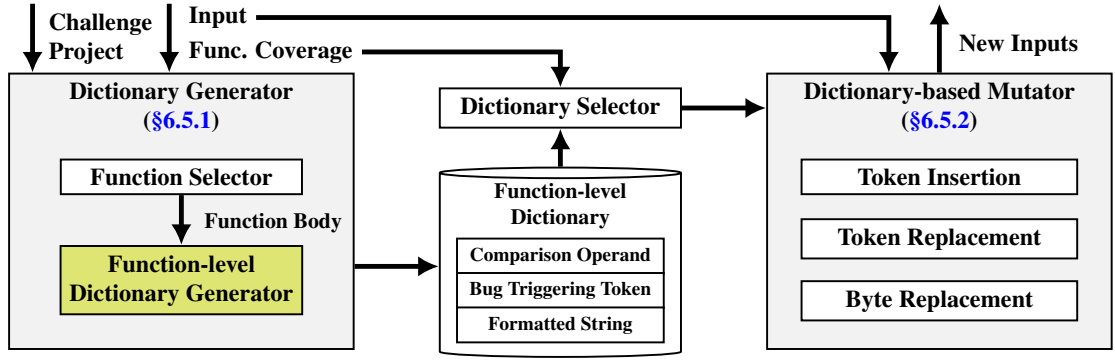

\PP{Workflow}
\autoref{fig:dict-gen} illustrates the overall architecture and workflow of our function-level dictionary-based fuzzing.
First, the \textsc{Dictionary Generator} takes as input the source code of the challenge project along with an input and its function coverage information provided by the \textsc{Corpus Manager} of UniAFL.
It then selects which functions within the coverage to target for dictionary generation, guided by the suspicious functions identified by MLLA in \autoref{ss:mlla}.
This filtering step is particularly useful, as the suspicious functions significantly reduce the search space of dictionary generators. 
For each selected function, the generator uses LLMs to produce interesting dictionary entries (see~\autoref{sss:dict-gen}).
Next, the \textsc{Dictionary Selector} leverages the function coverage of the given input to determine which of these entries will be used for mutation. 
Finally, the \textsc{Dictionary-based Mutator} applies the selected entries to mutate the input through several mutation strategies, such as token insertion, token replacement, and byte replacement.

%\begin{figure}[t]
  %\centering
  %\subfigure[Structured input generated by other components] {
    %\begin{promptbox}[width=0.45\textwidth]{Original Input}
%GET /index.html HTTP/1.1\\
%Host: example.com\\
%User-Agent: a-fuzzer
    %\end{promptbox}
  %}
  %\subfigure[Mutated input with preserved structure] {
    %\begin{promptbox}[width=0.45\textwidth]{Mutated Input}
%POST /api/submit HTTP/1.1\\
%Host: example.com\\
%User-Agent: a-smarter-fuzzer
    %\end{promptbox}
    %}
 %\end{figure}

\subsubsection{Function-level Dictionary Generator}
\label{sss:dict-gen}
This module generates a dictionary for a function provided by \textsc{Function Selector}, which selects it based on input coverage and bug candidate information.
It first determines which token types would be most beneficial.
For example, it prioritizes bug-triggering tokens (\eg long string) in functions that invoke \texttt{strcpy}.
It then passes the function body with a prompt describing the relevant token type to the LLM to generate tokens.
To maximize throughput and performance of the LLM, as shown in \autoref{fig:dictgen-prompts}, we employed GPT-4o with comprehensive prompt engineering such as \textit{Chain-of-Thought}, \textit{role-playing}, and \textit{few-shot example}.
The following lists the token types supported in our system along with their descriptions.

\PP{Comparison Operand}
The generator identifies comparison operations within the function's source code and uses LLMs to infer the operands.
To enhance reliability and filter out irrelevant tokens, this process is repeated multiple times.
The final set of operands is the intersection of the tokens collected from each trial,
ensuring that only the most consistent and relevant values are included in the dictionary.
For example, if three trials for a function handling HTTP headers yield \{\cc{"Host"}, \cc{"User-Agent"}\}, \{\cc{"Host"}, \cc{"User-Agent"}\}, and \{\cc{"Host"}, \cc{"User-Agent"}, \cc{"Referer"}\},
the final dictionary would contain \{\cc{"Host"}, \cc{"User-Agent"}\}.
In this example, the token \cc{"Referer"} is spurious, generated as an artifact of the LLM's stochastic nature.

\PP{Bug Triggering Token}
This module prompts the LLM to generate inputs that are likely to trigger common vulnerabilities.
For example, if the target function handles SQL queries,
LLMs will produce SQL injection payloads (\eg `' OR '1'='1'`) as tokens.
This allows the fuzzer to proactively test for known bug classes.

\PP{Formatted String}
For functions that perform parsing, this module leverages the LLM's
understanding of common data formats to generate strings that are
syntactically valid and likely to be successfully parsed (\eg for a
function that parses User-Agent strings, it might generate
\ccw{"Mozilla/5.0 (X11; Linux x86_64)"}). This helps the fuzzer
bypass initial validation checks and explore deeper logic within the
parsing function.

\begin{figure}[!t]
  \centering
      \begin{minipage}{0.85\textwidth}

    % \scalebox{0.85}{
    \begin{promptbox}{Comparison Operand Extraction Prompt}
    \begin{promptcontent}\input{code/dictgen-constant-focused.txt}\end{promptcontent}
    \end{promptbox}
    % }
    % \scalebox{0.85}{
    \begin{promptbox}{Bug Triggering Token Extraction Prompt}
    \begin{promptcontent}\input{code/dictgen-trigger-focused.txt}\end{promptcontent}
    \end{promptbox}
    % }
    % \scalebox{0.85}{
    \begin{promptbox}{Formatted String Extraction Prompt}
    \begin{promptcontent}\input{code/dictgen-parsable-focused.txt}\end{promptcontent}
    \end{promptbox}
    % }
  \end{minipage}

  \caption{Summarized dictionary generation prompts}
  \label{fig:dictgen-prompts}
\end{figure}

\smallskip

\subsubsection{Function-level Dictionary-based Mutator}
\label{sss:dict-mut}
%
%The mutation engine orchestrates the entire dictionary-based mutation
%process, which can be initiated in one of two ways. For targeted delta
%analysis, the engine can be provided with a diff file. In this mode,
%it begins by identifying all functions modified within the diff and
%immediately invokes the \ccw{Function-level Dictionary Generator}
%to pre-populate the dictionary cache for them. This front-loading
%allows the fuzzer to focus its efforts on recent code changes from the
%very beginning.

%During the main fuzzing loop (or if no diff file is provided), the
%process is driven by coverage. For each function covered by a given
%seed input, the mutator consults the global cache. If a dictionary for
%a function already exists (a cache hit), it is used directly. On a
%cache miss, the mutator invokes the generator. To prevent the costly,
%LLM-based generation process from blocking the entire fuzzing
%campaign, only one mutation process is allowed to invoke the generator
%at any given time; other processes continue to mutate inputs using
%existing dictionaries or traditional methods. Once the new dictionary
%is generated, it is stored in the cache for all processes to use.

Based on the function coverage of the input, \textsc{Dictionary Selector} filters out the dictionary used for mutating the input.
To maximize the chances of discovering new coverage and vulnerabilities,
the mutator employs a hybrid approach that combines dictionary-based strategies with traditional random mutations.
With a high probability, it applies one of the dictionary-based strategies. 
With a small probability, however, it falls back to a traditional fuzzing strategy, such as performing a random bit or byte flip.
The primary mutation strategies are as follows:

\PP{Token Insertion}
The mutator randomly selects a token from the dictionary and inserts it at a random location in the input.
This strategy is effective at testing how the program handles inputs including unexpected data or additional tokens.

\PP{Token Replacement}
%A random chunk of the input is replaced with a randomly selected token from the dictionary.
%This helps to test the program's response to semantically relevant but different data.
A random chunk of the input is replaced with a randomly selected token from the dictionary.
This produces structurally valid but semantically different data, which might be helpful for testing whether the program handles unexpected tokens correctly.

\PP{Byte Replacement}
This is a more granular version of token replacement, where a small, random number of bytes in the input are replaced by a token from the dictionary.
This can create subtle changes that may trigger edge-case behaviors.

\clearpage

\subsection{Testlang-based Input Generation}
\label{ss:testlang}
One of the key challenges in fuzzing is how to randomly generate good inputs that are highly structured, so that the target program is more likely to accept them.
Existing tools and research mainly focus on reverse-engineering the target program to recover its input format.
In this subsection, we introduce our Testlang-based input generator and explain how it addresses this challenge.
Our approach goes one big step further by not only considering input format inference but also the following:

\PP{Optimizing structural specification complexity}
Some input formats are extremely complex, making it impractical to fully specify them.
By leveraging well-known format specifications and LLM-generated Python scripts, we can balance specification completeness with manageability.
Also, partial specifications can kick in to make the LLM focus on the most interesting parts of the input format.

\PP{Handling complex codebases}
Static analysis often struggles with complex or large codebases.
By a simple touch, utilizing LLMs to reverse-engineer code behavior, we can reduce the computational cost of interpreting complex logic and reasoning about input structure identification.

\PP{Guiding fuzzing with LLM}
LLMs can be used not only for reversing the input format, but also for evaluating fuzzing status using fuzzer feedback.
This allows us to systematically prioritize fuzzing directions (e.g., Testlang variants) that align with the analysis strategy of the Harness Reverser. We call this \textit{LLM-opinionated}.

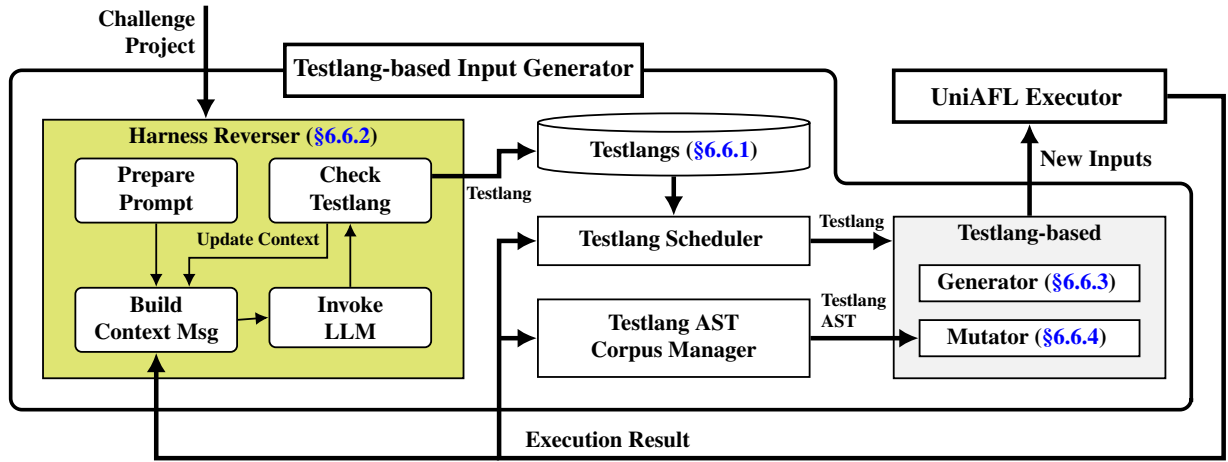
\begin{figure}[t]
  \centering
  \resizebox{0.99\textwidth}{!}{
\begin{tikzpicture}[
  module/.style={rectangle, draw, minimum width=4.2cm, minimum height = 0.75cm, thick, font=\small, fill=gray!10},
  db/.style={cylinder, draw, minimum width = 4.2cm, minimum height = 0.75cm, thick, shape border rotate=90, shape aspect=.10, font=\small, fill=white},
  moduletxt/.style={font = \small},
  smodule/.style={rectangle, draw, minimum width=3.4cm, minimum height = 0.5cm, thick, font=\small, fill=white},
  ssmodule/.style={rectangle, draw, rounded corners = 1mm, minimum width=2.5cm, minimum height = 0.4cm, thick, font=\small, fill=white},
  line/.style={-, thick},
  arrow/.style={-{Latex[length=2mm, width=2mm]}, thick},
  arrowBig/.style={-{Latex[length=3mm, width=3mm]}, line width=2pt},
  arrowBi/.style={{Latex[length=2mm, width=2mm]}-{Latex[length=2mm, width=2mm]}, thick},
  arrowTxtR/.style={right, font=\scriptsize, pos=0.5, align=left},
  arrowTxtL/.style={left, font=\scriptsize, pos=0.5, align=right},
  arrowTxtB/.style={right, font=\footnotesize, pos=0.5, align=left}
]

\newcommand*{\ax}{0}
\newcommand*{\bx}{\ax+6.5}
\newcommand*{\cx}{\bx+5.5}

\newcommand*{\ay}{0}
\newcommand*{\by}{\ay-1}
\newcommand*{\cy}{\by-3.75}

\node[module, align=center, fill=GreenYellow, minimum width=6.5cm, minimum height = 4cm, anchor=north] at (\ax, \ay+0.5) (reverser) {};
\node[moduletxt, below] at (reverser.north) {
    \textbf{Harness Reverser~(\autoref{sss:reverser})}
};

% Reverser submodules
\newcommand*{\rax}{\ax-1.5}
\newcommand*{\rbx}{\ax+1.5}
\newcommand*{\ry}{\ay-0.1}
\newcommand*{\dy}{2}
\node[ssmodule, align=center, fill=White, anchor=north] at (\rax, \ry) (prompt) {
  \textbf{Prepare}\\\textbf{Prompt}
};

% \node[db, minimum width=1.5cm, align=center, anchor=north, font=\small] at (\rax, \ry-1.5) (codedb) {
%   \textbf{Code DB}
% };

\node[ssmodule, align=center, fill=White, anchor=north] at (\rax, \ry-\dy*1) (build) {
  \textbf{Build}\\\textbf{Context Msg}
};

\node[ssmodule, align=center, fill=White, anchor=north] at (\rbx, \ry-\dy*1) (invoke) {
  \textbf{Invoke}\\\textbf{LLM}
};

\node[ssmodule, align=center, fill=White, anchor=north] at (\rbx, \ry) (check) {
  \textbf{Check}\\\textbf{Testlang}
};

\draw[arrow] (prompt.south) -- (build.north);
\draw[arrow] (build.east) -- (invoke.west);
\draw[arrow] (invoke.north) -- (check.south);
\draw[arrow] (check.235) -- ($(check.235 |- 0, \ry-1.55 )$) -| (build.45)  node[above, font=\scriptsize, pos=0.25, align=center] {\textbf{Update Context}} ;
% \node[ssmodule, align=center, fill=White, anchor=north] at (\rbx, \ry-\dy*3) (update) {
%   \textbf{Update Context}
% };
% Reverser End

\node[module, align=center, anchor=north, minimum height = 2.5cm] at (\cx, \by) (inputgen) {};
\node[moduletxt, below] at (inputgen.north) {\textbf{Testlang-based}};

\node[smodule, align=center, anchor=north] at (\cx, \by-0.75) (gen) {
    \textbf{Generator~(\autoref{sss:testlanggen})}
};

\node[smodule, align=center, anchor=north] at (\cx, \by-1.6) (mut) {
    \textbf{Mutator~(\autoref{sss:testlangmut})}
};

\newdimen\yInputGen
\pgfextracty{\yInputGen}{\pgfpointanchor{inputgen}{north}};

\newdimen\yMut
\pgfextracty{\yMut}{\pgfpointanchor{mut}{center}};

\node[module, align=center, fill=White, ultra thick, font=\normalsize, minimum height=0.8cm] at (\cx, \ay+0.85) (executor) {
  \textbf{UniAFL Executor}
};

\node[module, align=center, fill=White, anchor=north] at (\bx, \yInputGen) (scheduler) {
  \textbf{Testlang Scheduler}
};

\node[module, align=center, fill=White, minimum height=1.2cm] at (\bx, \yMut) (cmgr) {
  \textbf{Testlang AST}\\
  \textbf{Corpus Manager}
};
\newdimen\yCmgr
\pgfextracty{\yCmgr}{\pgfpointanchor{cmgr}{west}};

\newdimen\yScheduler
\pgfextracty{\yScheduler}{\pgfpointanchor{scheduler}{west}};
\newdimen\xRev
\pgfextractx{\xRev}{\pgfpointanchor{reverser}{east}};

\node[db] at (\bx, \ay) (testlang) {
    \textbf{Testlangs~(\autoref{sss:testlang})}
};

\draw[arrowBig] (check.10) -- (check.10 -| \bx-2.7, 0)  node[arrowTxtR, below, pos=1] {\textbf{Testlang}} |- (testlang.west);
\draw[arrowBig] (testlang) -> (scheduler);
\draw[arrowBig] (scheduler.east) -> (scheduler.east -| inputgen.west) node[arrowTxtR, above] {\textbf{Testlang}};
\draw[arrowBig] (cmgr.east) -> (cmgr.east -| mut.west) node[arrowTxtR, above,pos=0.4] {\textbf{Testlang}\\\textbf{AST}};
\draw[arrowBig] (inputgen) -> (executor) node[arrowTxtR, pos=0.6,font=\small] {\textbf{New Inputs}};
\draw[arrowBig] (executor.east) -| (\cx+3,\cy) -> node[arrowTxtL, above, pos=0.85, font=\small]{\textbf{Execution Result}} (\bx -2.7, \cy) |- (scheduler.west);
\draw[arrowBig] (\bx - 2.7, \yCmgr ) -> (cmgr.west);
\draw[arrowBig] (\bx - 2.7, \cy ) -| (build.south);
%\draw[arrowBig] (\bx - 2.7, \yScheduler ) -> (\xRev, \yScheduler);
\draw[arrowBig] ($(reverser.110) + (0,1.75)$) -> (reverser.110) node[arrowTxtL, pos=0.25, font=\small]{\textbf{Challenge}\\\textbf{Project}};

\newcommand*{\xgap}{2.5}
\newcommand*{\ygapU}{1.25}
\newcommand*{\ygapB}{3}
\draw[ultra thick, rounded corners=5pt]
  (\ax-\xgap-1.25, \ay+\ygapU) -- (\bx+\xgap, \ay+\ygapU)
  |- (\cx+\xgap, \by+0.5) |- (\cx+\xgap, \by-\ygapB)
  -- (\ax-\xgap-1.25, \by-\ygapB)
  -- cycle;
\node[rectangle, draw, fill=White, ultra thick, font=\normalsize, minimum width=4.5cm, minimum height=0.8cm] at (\ax * 0.5 + \bx * 0.5, \ay+\ygapU) {\textbf{Testlang-based Input Generator}};

\end{tikzpicture}
}
  \caption{Overview of the Testlang-based input generator.}
  \label{fig:testlang-overview}
\end{figure}

As shown in \autoref{fig:testlang-overview}, the Testlang-based input generator contains four major components: the Harness Reverser that extracts input formats, Testlang and Python files which are outputs from the Harness Reverser, and generation and mutation engines that operate on these Testlang-based structured representations.
When the challenge project (CP) is given, the Harness Reverser analyzes the target harness code and the broader CP codebase to understand how the harness processes input data, producing a comprehensive Testlang as output.
In addition, the Harness Reverser may also write Python scripts that contain tailored generators and encoders to handle complex formats.
This will be explained further in \autoref{sss:reverser}.
Testlangs, the outputs of the Harness Reverser, capture the hierarchical structure, field types, constraints, and relationships discovered during this analysis process.
Testlangs are managed with extra metadata such as CP source coverage target lines, deprioritization, and usage metrics, which are used in the Testlang Scheduler. These will be explained further in \autoref{sss:testlang}.
The Testlang and Python files are then fed into the Testlang-based generation and mutation engines to produce new test inputs that are semantically meaningful and structurally valid.
The generated inputs are then executed on the target harness by the UniAFL Executor, and the resulting code coverage and program behaviors are monitored and fed back to evaluate and improve the effectiveness of the module.
Detailed processes will be explained in \autoref{sss:testlanggen} and \autoref{sss:testlangmut}.

\subsubsection{Testlang}
\label{sss:testlang}

Testlang is a domain-specific language designed to formally describe arbitrary data structures and input formats for fuzzing purposes.
Grammar-based generation and mutation itself is not a new idea, and there are several existing tools and research that utilize grammar-based approaches for structured fuzzing~\cite{holler2012fuzzing, aschermann2019nautilus}.
However, Testlang introduces several novel features and capabilities that distinguish it from prior works.

\PP{LLM-friendly design for on-the-fly generation}%
Testlang is designed to be easily generated and modified by LLMs.
Its syntax is based on JSON schema, which is both human-readable and machine-parsable, with strong support for hierarchical structures, field types, constraints, and relationships.
By using popular formats like JSON, we make it straightforward for LLMs to produce valid Testlang.
This JSON-based design also provides additional benefits.
It enables seamless integration with Testlang-based generation and mutation because these structured representations can be directly interpreted.
It also reduces development overhead when modifying features of Testlang as the changes can be easily reflected in the schema.
As a result, Testlang achieves both flexibility and maintainability while remaining accessible to both humans and machines.

\PP{Semantics of data fields}%
Testlang can describe not only basic data types but also semantic relationships between fields, thereby ensuring that generated inputs remain structurally valid.
Specifically, it supports data types (\eg integers, strings, arrays), size constraints (fixed, variable), value constraints (ranges, enumerations), inter-field relationships (\eg dependency, record reuse), and hierarchical structures (nested objects, arrays of objects).
For example, in the simple Testlang shown in \autoref{fig:testlang-simple-example},
the record named "\ccw{Lookup}" contains a field named "\ccw{table}" which refers to the preceding field named "\ccw{table_size}" as its size.

\begin{figure}[t]
  \centering
  \subfigure[Simple Testlang Example]{
    \begin{minipage}[t]{0.48\textwidth}
      \scalebox{0.9}{
      \begin{promptbox}{}
      \begin{promptcontent}\input{code/testlang-example.json}\end{promptcontent}
      \end{promptbox}
      }
    \end{minipage}
    \label{fig:testlang-simple-example}
  }
  \hfill
  \subfigure[Grammar-based Custom Generator]{
    \begin{minipage}[t]{0.48\textwidth}
      \scalebox{0.9}{
      \begin{promptbox}{}
      \begin{promptcontent}\input{code/testlang-custom-grammar-example.json}\end{promptcontent}
      \end{promptbox}
      }
    \end{minipage}
    \label{fig:testlang-custom-grammar-example}
  }
  \caption{Examples of Testlang output formats.}
  \label{fig:testlang-examples-basic}
\end{figure}

\PP{Generation extensions using external tools}%
Fields can be associated with custom Python generators that produce semantically valid content, and encoders that transform data during serialization.
This enables support for complex formats like compressed data, encrypted content, or domain-specific encodings.
Some well-known text-based formats (e.g., XML, HTTP, HTML) may also be handled with a bundled grammar-based generator called Grammarinator~\cite{hodovan2018grammarinator}, inducing the LLM to focus more on finding interesting parts of the input than on the complex input format itself.
For example, in \autoref{fig:testlang-custom-grammar-example}, the record named "\ccw{CSVCustom}" contains a field named "\ccw{csv\_data}" which uses a custom generator that leverages Grammarinator to produce valid CSV content.
Another example is shown in \autoref{fig:testlang-custom-python-example}, where the record named "\ccw{CSVData}" contains a field named "\ccw{csv\_data}" which uses a Python-based generator named "\ccw{CSVGenerator}" in the corresponding Python code.
Details about generating Python-based custom generators will be explained in \autoref{sss:reverser}.

\PP{Oracle provision for output quality evaluation}%
Testlang holds metadata that indicate which lines in the CP source code are expected to be covered when inputs generated from this Testlang are executed.
This allows the system to evaluate the quality of a Testlang based on how well it achieves the intended coverage targets.
This feedback can be used to refine and improve Testlang over time.

\begin{figure}[t]
  \centering

  \subfigure[Python-based Custom Generator]{
    \resizebox{0.48\textwidth}{!}{
    \begin{minipage}{0.6\textwidth}
      \begin{promptbox}[width=\linewidth]{}
        \begin{promptcontent}
          \input{code/testlang-custom-python-example.json}
        \end{promptcontent}
      \end{promptbox}

      \begin{promptbox}[width=\linewidth]{}
        \begin{promptcontent}
          \input{code/testlang-custom-python-example.py}
        \end{promptcontent}
      \end{promptbox}
    \end{minipage}%
    }
    \label{fig:testlang-custom-python-example}
  }
  ~
  \subfigure[Partial Update Example]{
    \resizebox{0.445\textwidth}{!}{
    \begin{minipage}{0.48\textwidth}
      \begin{promptbox}[width=\linewidth]{}
        \begin{promptcontent}
          \input{code/testlang-partial-example.json}
        \end{promptcontent}
      \end{promptbox}
      \label{fig:testlang-partial-example}
    \end{minipage}
    }
  }

  \caption{Advanced Testlang features.}
  \label{fig:testlang-examples-advanced}
\end{figure}

\PP{Partial update support}%
Testlang supports partial specifications that can be incrementally refined.
This allows LLMs to start with a basic structure and progressively add details, constraints, and relationships as they gain a deeper understanding of the input format.
Also, this enables the LLM to focus on the most interesting parts of the input format by reducing analysis context rather than attempting to fully specify every aspect.
For example, in \autoref{fig:testlang-partial-example}, the Testlang is marked as a partial specification with the field "\ccw{is\_partial}" set to true.
If this partial Testlang is used with \autoref{fig:testlang-simple-example}, the final output Testlang will be a combination of both Testlangs, where the "\ccw{INPUT}" record is taken from \autoref{fig:testlang-simple-example}, and the others are taken from \autoref{fig:testlang-partial-example}.

\PP{Focus alignment of LLM analysis and fuzzing}%
Testlang storages have metadata that guides both the LLM’s analysis strategy and the prioritization of fuzzing tasks:
(1) deprioritization tags for cases where a Testlang does not align with the current analysis strategy and should therefore be downweighted in future fuzzing cycles;
(2) usage metrics such as the frequency of use and the coverage achieved.
Leveraging this information, the Testlang Selector can make informed choices about which Testlangs to generate or mutate, ensuring that the fuzzing process remains aligned with the analysis objectives defined by the Harness Reverser.

\subsubsection{Harness Reverser}
\label{sss:reverser}
% \TODO{
%     Key contribution?: On-the-fly generation of grammar specifications
%     How harness reverser figures out three key contribution of testlang.
%     And how to leverage execution results (coverage) and target lines to refine the testlang or generate bettter testlang.
% }

% \begin{figure}[!htbp]
%   \centering
%   \input{fig/reverser-overview}
%     \caption{The overview of harness reverser.}
%   \label{fig:reverser-overview}
% \end{figure}

% Introduction to Harness Reverser
The Harness Reverser is an LLM-powered system that automatically analyzes target programs to extract input format specifications and generate corresponding Testlang descriptions. This component addresses the fundamental challenge of manual format reverse engineering, which is both time-consuming and error-prone. As LLMs have demonstrated strong capabilities in code understanding and generation, the Harness Reverser leverages these strengths to systematically analyze input processing logic within target programs. Also, as the correctness and level of detail of the generated Testlang are crucial for effective testing, the Harness Reverser employs a rigorous analysis process to ensure high-quality output.

By using LLMs, this system systematically analyzes how the harness processes input data, producing a comprehensive Testlang as output.
It takes the target harness code and the CP codebase as input.
Starting from the harness entry point, the reverser traces data flow through parsing routines, validation logic, and data structure transformations to understand the expected input format.
The Harness Reverser also proposes auxiliary Python generators to handle complex field types and encodings, ensuring that the generated Testlang captures both structural and semantic aspects of the input format.
This process is iterative, with the LLM actively selecting and refining the code scope to analyze (functions, files, and specific regions) as the investigation progresses.
In addition, by leveraging runtime feedback (coverage and crashes), 
the Harness Reverser refines the generated Testlang, enhancing accuracy and effectiveness in vulnerability discovery.

% Key Features of Harness Reverser
The Harness Reverser incorporates several key features that improve both the accuracy and effectiveness of Testlang generation:
\begin{squishitemize}
\item \textbf{Automated Input Format Extraction:} It automatically identifies and extracts input format specifications from the target harness code and CP code base, achieving advanced structural fuzzing capabilities without requiring manual intervention.
\item \textbf{LLM‑Based Custom Generator Support:} Leveraging LLMs' capability, it recognizes well‑known formats and attaches existing custom fuzzers instead of re‑specifying them in Testlang. Additionally, for formats that are hard or impractical to model in Testlang alone (e.g., stateful protocols, nested compression, checksums), the LLM synthesizes Python generators that encapsulate the semantics and expose them as Testlang nodes, boosting expressiveness, validity, and coverage.
\item \textbf{LLM-opinionated Analysis with Runtime Feedback:} It employs an \textit{LLM-opinionated} analysis strategy where the LLM actively directs its own investigation, targeting code sections most likely to yield security-relevant insights. Also, the system integrates runtime feedback from fuzzing (code coverage and crash logs) to evaluate and refine the Testlang iteratively.
\end{squishitemize}

\PP{Overview}
The green box in \autoref{fig:testlang-overview} illustrates the architecture of the Harness Reverser, which operates through a sophisticated multi-stage analysis pipeline that progressively evolves the Testlang through continuous LLM interaction and context refinement.
The Harness Reverser consists of five primary stages: Prepare Prompt, Build Context Messages, Invoke LLM, Check Testlang, and Update Context. Each stage plays a critical role in establishing the analysis context, guiding LLM behavior, validating outputs, and refining the understanding of the target program's input format.
The following subsections provide a detailed explanation of each stage and its function within the overall analysis workflow.

\PP{Prepare Prompt}
In the Prepare Prompt stage, the system constructs a base prompt that defines the LLM's role as a reverse-engineering expert and establishes the foundational context for analysis. This prompt includes instructions on how to request code snippets, analyze input processing logic, and express findings in Testlang format. The prompt also incorporates representative examples of analysis patterns and expected output formats to guide the LLM's reasoning process.

This step begins by loading and preprocessing the harness code from the specified file path, creating a structured \ccw{Code} object that includes full location metadata, such as file path and line number ranges. For \ccw{Code} objects, the system applies code filtering based on coverage information to identify and remove unused code sections. The filtering stage extracts compiled line information from the target binary to determine which lines are actually compiled and executed, and systematically eliminates preprocessor conditional blocks that can never be reached given the build configuration. This preprocessing ensures the LLM focuses its analysis only on relevant, executable code paths.

In delta mode, the system processes \textit{diff} information to extract additional code blocks that represent changes in the broader codebase, providing context about modifications that may affect input processing logics. The diff processing identifies modified functions, added data structures, and changed parsing logic that could impact input format requirements. This integration allows the reverser to understand how recent changes to the codebase might have introduced new input format requirements or modified existing parsing logics.
Consequently, this stage constructs a comprehensive message consisting of four primary components:

\begin{squishitemize}
%\vspace*{-5pt}
\item \textbf{System Message:} Defines the LLM's role and reverse engineering capabilities.
\item \textbf{Grammar Specification Message:} Contains the complete Testlang JSON schema dynamically populated with available custom generator definitions from the \ccw{customgen} module.
\item \textbf{Examples Message:} Provides representative analysis patterns and expected output formats.
\item \textbf{Target Message:} Contains the preprocessed and filtered harness code with any relevant diff information.
\end{squishitemize}

To optimize token usage and response latency across multiple analysis iterations, the system applies strategic message caching to the examples and target messages. The caching strategy leverages Anthropic's extended cache-TTL beta features to reduce API costs and improve response times for repeated analyses of similar code patterns. This optimization is particularly important for iterative analysis workflows where similar context is repeatedly processed.
Additionally, when the Harness Reverser is invoked multiple times during a single analysis session, it loads previously generated Testlang from prior iterations to provide continuity and context.

\PP{Build Context Messages}
This stage builds the context messages that enable the LLM to make the next analysis and generation decisions.
The context contains
(1) the current Testlang with its associated Python generator code;
(2) LLM‑requested code blocks with file/line metadata;
(3) runtime feedback (code coverage and crash logs); 
and (4) warnings and error messages to guide the LLM.
The output is a set of structured messages that provide situational awareness for the LLM's next invocation.
This adaptive context building process operates as the critical bridge between the static code analysis phase and the dynamic fuzzing execution phase, ensuring LLMs receive complete situational awareness for informed decision-making.

The context construction process begins with Testlang state management. For initial analysis sessions, the system presents an empty Testlang context with explicit guidance to create a default \textit{INPUT} record structure with appropriate mode and endianness settings.
For ongoing sessions, the method formats the current validated Testlang with its versioned identifier, enabling LLMs to understand the evolution of their analysis.
Associated Python generator codes are included as structured code blocks with names and implementations, allowing LLMs to reference and modify existing generators during iterative refinement.

Next, the system processes LLM code search requests to retrieve specific code blocks for analysis. It interprets structured queries from the LLM that specify function names, file paths, and line numbers, and extracts the corresponding code snippets from the preprocessed harness and project code. Each retrieved code block is annotated with full location metadata to provide context for the LLM's analysis.

Furthermore, Testlang is refined using runtime feedback such as code coverage and crash logs.
As the LLM selects target lines to cover in the Testlang metadata, the system correlates coverage data with these targets to determine which lines were successfully exercised by generated inputs.
This provides precise feedback on Testlang’s effectiveness in meeting coverage goals.
In addition, crash logs are parsed to extract details about discovered vulnerabilities, allowing the system to guide the LLM away from retargeting known issues and toward unexplored code paths.

Lastly, the system synthesizes various warnings and error messages to guide the LLM's analysis strategy. These messages include validation warnings from the Testlang checking stage, execution errors of Python generators, and strategic advisories based on runtime feedback. These messages are structured to provide clear, actionable insights that help the LLM refine its analysis and generation approach. Notably, the LLM may ignore certain warnings if it deems them irrelevant to its current analysis strategy, allowing for flexible decision-making. If the LLM chooses to disregard specific warnings, the system respects this decision and does not reintroduce those warnings in subsequent iterations, preventing redundant feedback loops.

% As a result, this stage produces four distinct categories of messages:
% \begin{squishitemize}
% \item \textbf{Specification state context:} Current Testlang specifications and Python generators with version tracking and evolution history
% \item \textbf{Code discovery context:} Dynamically discovered code blocks with metadata, location information, and analysis provenance
% \item \textbf{Security vulnerability context:} Parsed crash logs with deduplication and intelligent trimming, cached for performance optimization
% \item \textbf{Execution feedback context:} Line-by-line coverage data correlated with Testlang iterations, enabling precise coverage attribution
% \end{squishitemize}

\PP{Invoke LLM}
In this stage, the system invokes the LLM with the prepared base prompt and dynamically constructed context messages. The invocation process implements a robust architecture that handles transient errors and maintains operational continuity during extended analysis sessions. Since the system operates in automated fuzzing environments without human supervision, robustness is essential due to the unpredictability of LLM responses, which may include incomplete outputs, inconsistent formatting, or unexpected analytical directions that could disrupt the iterative analysis pipeline.

The invocation architecture uses a context-based interaction model that reconstructs complete context for each LLM invocation rather than maintaining conversation history. Each call provides base prompts defining the analysis objectives along with comprehensive context messages containing current analysis state. This approach prevents information bloat by filtering out irrelevant historical exchanges while ensuring the LLM receives the information needed for informed decision-making. The conversation-history-free design enables deterministic analysis outcomes, reduces token consumption through strategic information curation, and maintains control over analytical direction by presenting only refined, contextually relevant information.

The invocation process employs error recovery and resource management mechanisms for LLM interaction robustness during extended analysis sessions:
\begin{squishitemize}
\item \textbf{Context window overflow:} Content reduction that replaces detailed code implementations with standardized placeholders while preserving essential metadata such as function names, file paths, and locations.
\item \textbf{Token limit violations:} Reallocation of output tokens based on available context capacity, with automatic content reduction when limits are exceeded.
\item \textbf{Model switching failures:} Automatic switching between different LLM models when encountering failures or timeouts to maintain continuous operation.
\item \textbf{LLM infrastructure issues:} Handling of LLM errors, API compatibility changes, and caching optimizations to ensure continuous operation despite external service constraints.
\end{squishitemize}

\PP{Check Testlang}
In this stage, the system validates LLM-generated Testlang to ensure syntactic correctness, semantic validity, and effectiveness in producing test inputs for vulnerability discovery.
This validation also checks the quality of both the Testlang structure and its associated Python generator code, guaranteeing compatibility with Testlang-based input generation and mutation.

First, the generated Testlang undergoes basic JSON schema validation to ensure syntactic correctness and semantic validity.
The system checks for required fields, data types, and hierarchical relationships defined in the Testlang schema.
Any deviations from the schema result in immediate rejection of the Testlang, with detailed error messages provided to guide LLM corrections.
Furthermore, during the validation process, the system generates various warnings to guide LLMs in refining the Testlang.
These warnings may include suggestions for improving clarity, enhancing coverage, or addressing potential ambiguities in the generated Testlang.
For example, if a field lacks sufficient constraints, the system may issue a warning recommending additional analyses to define appropriate value ranges or dependencies.
In addition, depending on the field types and constraints specified in the Testlang, the system may suggest additional code analysis to identify relevant parsing logic or validation routines that can inform more precise field definitions.

% Testlang specification processing begins with JSON extraction from the LLM output and incremental updating, where new specifications are merged with existing validated versions using the \ccw{testlang.update()} method to handle record additions, modifications, and deletions. The system enforces strict structural constraints on Python generator codes, requiring each generator to target exactly one Testlang field to prevent conflicts and ensure clear field-to-generator mappings that maintain specification integrity and avoid duplicate field generation.

Second, Python generator code associated with Testlang fields also undergoes rigorous validation to ensure correctness and effectiveness.
The system checks for syntax errors, runtime exceptions, and logical consistency within the generator implementations.
Each generator is executed in a controlled environment with randomized input data to verify that it produces valid outputs without errors.
This execution-based validation helps identify issues that static analysis alone may miss, such as unhandled edge cases or incorrect assumptions about input data.
These validation results are used to generate warnings that inform the LLM about potential issues in the generator code, guiding improvements and refinements of Python generators.

Since Python generators play a crucial role in producing actual bug-triggering inputs, the system has additional guidance and validation mechanisms specifically for them.
The LLM assigns security severity scores to the vulnerability classes that each generator targets, so that the system can prioritize generators that focus on high-severity vulnerabilities.
It also assigns trigger probabilities indicating how likely the generator is to produce inputs that can trigger the targeted vulnerabilities.
These scores are used to prioritize generators during fuzzing and to guide LLM analysis toward the most promising directions.

\PP{Update Context}
Lastly, the system updates its internal context state to reflect the outcomes of the current analysis iteration and prepare for subsequent cycles. This stage involves persisting validated Testlang, managing warning states, and implementing error recovery mechanisms to ensure continuous progress through the iterative analysis pipeline.
For critical errors that prevent further analysis (e.g., LLM failures, validation errors, code search failures), the system preserves the previous LLM output and error context while clearing warnings. This prevents cascading failures from terminating analysis prematurely and provides LLMs with sufficient context to understand and resolve errors prior to the next iteration.

\subsubsection{Testlang-based Generation}
\label{sss:testlanggen}
Based on the Testlang generated by the Harness Reverser, this component generates new inputs
focusing on comprehensive exploration of the specified input format.% while targeting security-critical execution paths.

\PP{Semantics-aware}%
Rather than generating random byte sequences, the system produces inputs that are semantically meaningful within the context of the target application.
This includes respecting harnesses that use FuzzedDataProvider (FDP) by incorporating \libfdp (see \autoref{ss:libfdp}) so that generated inputs are consumed by the target harness properly.

\PP{Balancing coverage and error findings}%
The generation process employs two complementary strategies.
Coverage-targeted generation produces valid inputs that adhere to Testlang constraints and explore the input space to maximize code coverage.
Crash-targeted generation intentionally applies out-of-range values to trigger exceptions and expose vulnerabilities.
Together, these strategies balance breadth and depth in testing.

\PP{Integrating Custom Generators}%
The system seamlessly integrates custom grammar-based and Python-based generators that can generate inputs with complex formats.
These generators leverage existing parsing libraries to ensure format validity while introducing controlled variations to explore edge cases.
Furthermore, they can incorporate domain knowledge (\eg known file formats) to create realistic test cases.

\PP{Contributing to the Main Seed Pool}%
As described in \autoref{sss:testlangmut}, we maintain separate seed pools for Testlang-based inputs to enhance mutation effectiveness.
However, rather than using those inputs for Testlang-based mutations only, the generated inputs are also added to the UniAFL corpus if they are interesting (e.g., discover new coverage) to further enhance the overall fuzzing process.

\PP{Focus alignment of LLM analysis and fuzzing}%
Before generating new inputs based on Testlang, the Testlang Selector prioritizes Testlangs using metadata and achieved coverage.
The Harness Reverser may deprioritize Testlangs that do not align with the current analysis strategy or fail to deliver sufficient coverage,
while recently created Testlangs are favored as they better reflect the current analysis focus and help avoid rediscovering already found crashes.
Usage metrics such as how frequently a Testlang has been used are also considered to promote diversity in exploration.
By leveraging this information, the Testlang Selector improves fuzzing performance and ensures that the process stays aligned with the overall analysis goals.

\subsubsection{Testlang-based Mutation}
\label{sss:testlangmut}
Based on the Testlang from the Harness Reverser, this component mutates inputs while preserving their structural validity and effectively exploring the input space.

\PP{Seed Pool Separation}%
To enhance mutation effectiveness, the system maintains separate seed pools for Testlang-based inputs and byte-level inputs from other input generators.
It is very challenging to restore the structure of an input by only using the input definition described in Testlang.
Thus, we separated seed pools to ensure that Testlang-based mutations are applied to inputs that are already structured and semantically meaningful according to the Testlang.
Mutated inputs are also added to the main UniAFL seed pool if they are interesting.
We still allow Testlang-based mutations to be applied to inputs from other input generators, which is called \textit{AST-Free Mutations}, to seek more opportunities in exploring the codebase.

\PP{Structure-aware Mutations}%
Unlike byte-level mutations that can easily break input format constraints, which is not a mainly desired behavior for this module, Testlang-based mutation operates at the semantic level.
Mutations respect field types, size constraints, and inter-field relationships, ensuring that mutated inputs remain structurally valid while exploring different value combinations.
The system implements multiple mutation strategies tailored to different field types:
\begin{squishitemize}
    \vspace*{-5pt}
    \setlength{\itemsep}{0pt}
    \item \textbf{Boundary Value Mutation:} Replace numeric fields with values near boundaries (\eg INT\_MAX).
    \item \textbf{Type-Aware Mutation:} Apply domain-specific mutations (e.g., path traversal strings for filename fields, SQL injection patterns for text fields).
    \item \textbf{Constraint Violation Mutation:} Deliberately violate specified constraints to test error handling paths.
    \item \textbf{Cross-Field Mutation:} Modify relationships between fields to test validation logic.
    \item \textbf{AST-Free Mutation:} Mutate inputs from non-Testlang based input generators by inserting or replacing part of the inputs with Testlang-based structures.
\end{squishitemize}

The Testlang-based approach demonstrates significant advantages over traditional fuzzing methods, particularly for programs requiring structured inputs and domain-specific formats.
By operating at the semantic level rather than the byte level, the system achieves higher code coverage, discovers vulnerabilities more efficiently, and reduces the time spent on invalid inputs that are immediately rejected by input validation routines.
LLM-powered analysis is especially crucial for understanding complex domain-specific schemas such as HTML/XML, SQL queries, configuration file formats, and protocol specifications, where traditional fuzzing approaches struggle to generate valid inputs that can penetrate deep into application logic.

\clearpage

\begin{figure}[t]
    \centering
    \input{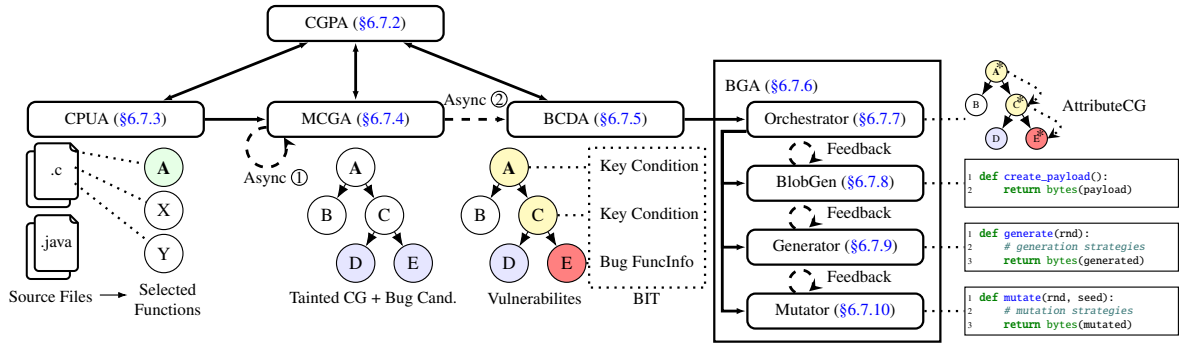}
      \caption{The overall architecture of \sysmlla.}
    \label{fig:mlla}
  \end{figure}

\subsection{Multilang-LLM-Agent (\sysmlla)}
\label{ss:mlla}

This section describes the Multi-Language LLM Agent (\sysmlla),
which is the most LLM-intensive input generation module
within the \sys-Multilang system (\autoref{s:crs-multilang}).

Traditional fuzzers rely on syntax-unaware mutations
that fail to generate the semantically-valid inputs
required for modern vulnerabilities which are often hidden behind
complex data structures, format validation,
and are incapable of handling stateful execution sequences.
\sysmlla addresses this semantic gap
by leveraging LLMs' code comprehension capabilities,
employing a coordinated multi-agent architecture
that transforms vulnerability discovery
from blind mutation to intelligent, state-aware, and targeted exploitation.
While LLMs offer semantic understanding, 
applying them to vulnerability discovery introduces fundamental challenges
that \sysmlla addresses through specialized agent design:

\PP{C1: Scale and Context Limitations}
Analyzing large, multi-language codebases
strains LLM context windows and computational resources.
Complex call chains and deep function hierarchies
exceed token limits, while function resolution ambiguity
creates uncertainty when multiple candidates match queries.

\PP{C2: Precision and Validation Challenges}
Non-determinism and hallucinations of LLMs make reliable vulnerability detection difficult.
Distinguishing genuine vulnerabilities from hallucinations
requires validation mechanisms,
while target-reaching accuracy depends on
understanding complex triggering conditions.

\PP{C3: Integration and Coordination Complexity}
Coordinating many tools such as code retrievers, static analyzers, script execution environment
and input executor introduces systemic complexity.
Furthermore, it requires robust design to orchestrate agents with different strategies
while maintaining fault tolerance and resource management.

\subsubsection{Overview}
\sysmlla operates in two primary modes:
a lightweight standalone mode for rapid seed generation,
and a full pipeline mode for systematic vulnerability exploitation.
The standalone mode leverages LLMs to generate diverse fuzzing seeds
directly from harness analysis (detailed in~\autoref{ss:mlla-standalone}).
The full pipeline mode orchestrates specialized agents
through the workflow illustrated in~\autoref{fig:mlla},
where each agent addresses traditional fuzzing limitations
while overcoming LLM-specific challenges (\textbf{C1}-\textbf{C3}).

\PP{Call Graph Parser Agent (CGPA)} 
This agent resolves ambiguous or incomplete
function information into precise, structured representations (\ccw{FuncInfo}),
leveraging Joern, LSP, code indexer, and AST-based search. This mitigates
inaccuracies in LLM-based function resolution while addressing \textbf{C1} by ensuring
consistent function resolution across large, multi-language codebases. Therefore,
preventing propagation of hallucinated or incorrect function definitions into downstream agents
(see \autoref{ss:cgpa}).

\PP{CP Understanding Agent (CPUA)}
This agent analyzes harness files 
to identify functions that need to be analyzed with taint flow information.
This enables semantic code scoping beyond traditional fuzzing capabilities
while managing \textbf{C1} by transforming overwhelming codebase complexity
into context-window-compatible analysis targets
(see \autoref{ss:cpua}).

\PP{Make Call Graph Agent (MCGA)}
This agent constructs interprocedural call graphs
while detecting vulnerable sinks within each function.
This overcomes cross-language call relationship mapping limitations
while addressing \textbf{C1} through recursive decomposition,
caching, and tool integration 
that validates LLM responses and prevents hallucinated function calls
(see \autoref{ss:mcga}).

\PP{Bug Candidate Detection Agent (BCDA)}
This agent validates vulnerabilities
by analyzing execution paths to distinguish genuine issues from false positives,
extracting concrete triggering conditions.
This provides semantic vulnerability understanding beyond traditional capabilities
while solving \textbf{C2} through multi-stage analysis
that combines LLM reasoning with execution verification
(see \autoref{ss:bcda}).

\PP{Blob Generation Agent (BGA)} 
This agent generates targeted payloads
through script-based creation of format-compliant exploits.
This transcends syntax-unaware mutations with semantically-valid inputs
while addressing \textbf{C2} and \textbf{C3}.
Basically, BGA consists of four sub-agents where
the \emph{Orchestrator Agent} coordinates three sub-agents:
\emph{BlobGen} creates single payloads through iterative refinement with coverage feedback,
\emph{Generator} produces multiple payloads to increase target-reaching probability, 
and \emph{Mutator} applies focused mutations for individual function transitions when context is complex
(see \autoref{ss:bga}--\autoref{ss:mutator}).

These agents operate with execution validation, where some operate asynchronously,
mitigating non-determinism of LLM by
feedback loops, tool verification, and adaptive refinement.

\PP{Tool Integration}
The architecture incorporates sophisticated tool integration
enabling capabilities beyond pure LLM reasoning.
MCGA and BCDA utilizes the CGPA
in a self-loop configuration for precise function analysis,
while all agents benefit from shared tooling infrastructure
including static analysis tools, language servers, and state management systems
(see \autoref{ss:cgpa} and \autoref{ss:shared-utils}).

\PP{Diff-aware Analysis}
When provided with diff files,
\sysmlla leverages LSP-based function-level diff analysis
to intelligently prioritize modified functions.
The system maps diff hunks to specific function boundaries using Language Server Protocol (LSP),
enabling targeted vulnerability analysis of code changes
that are most likely to contain newly introduced vulnerabilities.
For instance, MCGA integrates diff context directly into function analysis,
treating modified functions as high-priority targets with comprehensive taint
tracking.

\vspace*{-5px}
\subsubsection{Call Graph Parser Agent (CGPA)}
\label{ss:cgpa}
\vspace*{-5px}

CGPA addresses the challenge of resolving 
ambiguous or incomplete function information in large, multi-language codebases.
Because CPUA and MCGA often operate on partial or uncertain function metadata 
(\eg a function name without full signature, or a callsite without definition),
CGPA provides a reliable mechanism for mapping such partial information 
to precise, structured function representations.
This capability is essential to prevent hallucinated callees 
and ensure downstream agents operate on accurate code contexts.

\PP{Functionality and Workflow}%
Given partial input such as a function name, callsite, or file path, 
CGPA queries multiple backends described in \autoref{ss:shared-utils} to collect candidate definitions:
Joern for graph-based static queries, 
LSP servers for symbol resolution, 
a lightweight code indexer for efficient search, 
and AST-grep for syntax-level matching.
CGPA aggregates these results, removes duplicates, 
and applies heuristic or LLM-based selection to choose the most relevant definition. 
The output is a structured \texttt{FuncInfo} object 
that contains the file path, function signature, code body, 
and other metadata required for accurate call graph construction.
This standardized output format enables consistency across agents, 
allowing MCGA and BCDA to consume function information seamlessly 
without having to implement redundant resolution logic.
When multiple definitions remain ambiguous, 
CGPA invokes LLM-based reasoning to rank candidates, 
leveraging context such as caller function body or surrounding imports.
Results are cached in Redis to minimize repeated computation 
and accelerate future queries, especially in iterative or recursive analyses.

\PP{Integration and Benefits}%
CGPA serves as a shared utility within \sysmlla, 
supporting both CPUA and MCGA by providing precise function resolution. 
By combining multiple backends with LLM validation, 
CGPA mitigates limitations of purely static or purely LLM-based reasoning, 
ensuring accurate resolution in the face of overloads, or reflection.
This prevents error propagation in call graph expansion 
and strengthens downstream vulnerability detection. 
Ultimately, CGPA transforms incomplete and noisy code references 
into consistent, analyzable function objects, 
allowing the overall system to maintain robustness 
across diverse languages and complex codebases.

\subsubsection{Challenge Project Understanding Agent (CPUA)}
\label{ss:cpua}

CPUA addresses the fundamental
scoping problem in automated vulnerability analysis and \textbf{C1}
by determining which functions in a large codebase deserve detailed investigation.
Instead of attempting to analyze all functions indiscriminately,
CPUA focuses the analysis pipeline by interpreting harness files and
extracting high-value entry points that are most likely to process fuzzed input or expose vulnerabilities.
This targeted approach enables \sysmlla to allocate resources efficiently and reduce noise in downstream analyses.

\PP{Functionality and Workflow}%
Given the contents of a harness file, CPUA applies LLM-based reasoning to identify interesting candidate functions.
These include functions that (1) directly receive fuzzer-provided input, or (2)
propagate data to downstream libraries.
For each candidate, CPUA generates rich metadata including priority level, call sites, and tainted arguments.
The output is a prioritized list of target functions annotated with metadata, which downstream agents such as MCGA consume to construct detailed call graphs.

\PP{Integration and Benefits}%
A major strength of CPUA lies in its language-agnostic design.
By relying on the LLM’s semantic understanding rather than on language-specific ASTs or hardcoded parsing rules, CPUA can operate seamlessly across C, C++, Java, and other supported languages without modification.
This abstraction makes CPUA effective even in projects that rely on reflection or dynamic dispatch, where traditional static analyzers struggle.
However, its effectiveness depends on both the quality of the harness and the reliability of the LLM’s reasoning.
When harnesses expose limited or indirect entry points, CPUA may yield
incomplete coverage.
For example, in projects like \texttt{nginx}, communication often occurs through
sockets rather than direct function calls.
In such cases, the fuzzing harness can interact with the program externally
without ever invoking its internal functions, leaving CPUA with little semantic
signal to analyze.

\subsubsection{Make Call Graph Agent (MCGA)}
\label{ss:mcga}
MCGA addresses the fundamental challenge of mapping function call relationships in large, multi-language codebases.
Based on interesting candidate functions identified by CPUA,
MCGA constructs precise call graphs that 
capture all direct and indirect callees while simultaneously detecting vulnerable sinks.
To enhance accuracy beyond static analysis alone,
MCGA optionally integrates \functracer for dynamic callsite discovery.
While static analysis provides comprehensive coverage of potential call relationships,
\functracer supplies runtime execution traces that validate actual function calls (see~\autoref{ss:shared-utils}),
enabling MCGA to distinguish between theoretically possible and practically reachable paths.
This hybrid approach significantly improves call graph precision
by incorporating real execution behavior into the analysis.
This dual responsibility forms the backbone of the static analysis layer,
enabling downstream components to reason about interprocedural control flow,
identify vulnerable sinks, and guide payload generation strategies toward security-relevant code paths.

\PP{Functionality and Workflow}%
Given a function and its metadata (\eg name, location, tainted arguments), 
MCGA performs a recursive analysis to build a call graph
where each node represents a function and edges denote callsite relationships. 
\autoref{alg:mcga} describes the recursive process of the MCGA,
which builds a call graph rooted at a given target function. 

For every visited function, MCGA performs two tasks:
(1) determine whether the function contains vulnerable sinks, 
and (2) identify its callees. 
MCGA begins by invoking CGPA to resolve the exact definition of each function
(\autoref{line:mcga-resolve-func-info}), using a code retriever described in \autoref{ss:retriever}.
Once resolved, 
MCGA checks each function for vulnerability patterns such as
unsafe deserialization, unchecked system calls, or improper input validation.
Sink detection is performed via LLM prompts augmented with static context
(\autoref{line:mcga-detect-vuln-sink}).
Simultaneously, 
the function body is parsed to extract all call sites
(\autoref{line:mcga-extract-callees}).
For each callee, MCGA repeats the resolution process and expands the graph
recursively (\autoref{line:mcga-dispatch-async}).
It incorporates cycle detection and depth limits to ensure termination.
For functions with recent modifications,
MCGA integrates diff context directly into the function body analysis,
treating all function arguments as tainted when diff information is present
to ensure comprehensive vulnerability detection in recently changed code.
If a function is marked as potentially vulnerable, 
it is asynchronously forwarded to BCDA for further analysis.

\PP{Asynchronous and Cached Execution}
MCGA invokes both itself (\WC{1} in \autoref{fig:mlla}) and BCDA (\WC{2} in \autoref{fig:mlla})
asynchronously to improve analysis throughput and reduce latency in
vulnerability detection. 
When MCGA encounters a function during call graph expansion, 
it may reinvoke itself recursively as a non-blocking task, 
enabling parallel exploration of multiple call paths. 
Additionally, if a function is found to contain a potential vulnerability, 
it is immediately dispatched to BCDA 
without waiting for the full call graph to complete. 
This asynchronous design allows \sysmlla to prioritize 
and analyze high risk regions earlier, 
overlap LLM I/O and compute-intensive tasks, 
and avoid bottlenecks that would arise from sequential traversal, 
especially in large, deeply nested codebases.

\begin{algorithm}[!t]
\footnotesize
\DontPrintSemicolon
\SetKwSty{algokeywordsty}
\SetFuncSty{algofuncsty}
\SetArgSty{algoargsty}
\SetKwFunction{fnResolveFuncInfo}{ResolveFuncInfo}
\SetKwFunction{fnDetectVulnSink}{DetectVulnSink}
\SetKwFunction{fnDispatchAsync}{DispatchAsync}
\SetKwFunction{fnExtractCallees}{ExtractCallees}
\SetKwFunction{fnCache}{Cache}

\caption{MCGA: Make Call Graph Agent}
\label{alg:mcga}
\KwIn{Target function $f$ with metadata (name, file, code, tainted args)}
\KwOut{Call graph rooted at $f$ with sink annotations}

$node \leftarrow$ \fnResolveFuncInfo{$f$} via CGPA\;\label{line:mcga-resolve-func-info}
\If{$node$ is already visited}{
    \Return{cached result}\;
}

$sinks \leftarrow$ \fnDetectVulnSink{$node$}\;\label{line:mcga-detect-vuln-sink}
\If{$sinks \neq \emptyset$}{
    \fnDispatchAsync{BCDA, $node$}\;
}

$callees \leftarrow$ \fnExtractCallees{$node.code$}\;\label{line:mcga-extract-callees}
\ForEach{$callee \in callees$}{
    $calleeInfo \leftarrow$ \fnResolveFuncInfo{$callee$}\;
    \fnDispatchAsync{MCGA, $calleeInfo$}\;\label{line:mcga-dispatch-async}
    Add $calleeInfo$ as child of $node$\;
}

\fnCache{$node$}\;
\Return{$node$}\;
\end{algorithm}

\subsubsection{Bug Candidate Detection Agent (BCDA)}
\label{ss:bcda}

BCDA solves the critical challenge of distinguishing true vulnerabilities from false positives in the call graphs from MCGA.
When MCGA identifies functions with potential vulnerable sinks, the key question is whether these sinks represent exploitable vulnerabilities and, if so, under what conditions they can be triggered.

BCDA addresses this challenge through an LLM-powered vulnerability analysis system
that processes extracted execution paths into structured Bug-Inducing Things (BITs).
By taking as input a function previously identified in MCGA as potentially containing a vulnerable sink,
BCDA performs a multi-stage analysis 
to determine whether a vulnerability exists within a given code path.
Beyond simple detection, BCDA identifies the specific conditions (\eg if-else
branches or exception handling) necessary to trigger the discovered vulnerabilities,
providing concrete guidance for directed fuzzing and payload generation stages.
%
% To achieve this, the system uses LLM-driven analysis 
% to examine both data flow and control flow within and across functions, 
% enabling it to identify critical decision points 
% and conditions along vulnerability-inducing paths.

% The resulting Bug Inducing Things (BITs) are prioritized vulnerability candidates
% representing specific code paths or function transitions
% likely to trigger security issues.
% BITs provide structured information including the vulnerability location (\ie file path and line numbers), trigger conditions, 
% detailed analysis messages, and potential sanitizer types.
% This information supports subsequent LLM-based payload generation 
% and fuzzing stages, helping to effectively target specific vulnerabilities.

% The following paragraphs provide a detailed explanation of each step within the BCDA process.

\PP{Functionality and Workflow}%
BCDA operates in four main phases.
First, it receives candidate sinks and corresponding call graphs from MCGA and
forms source-to-sink execution paths.
It then performs path expansion and pruning to enrich the call path with
relevant functions while discarding irrelevant ones.
Next, BCDA applies vulnerability classification, using sanitizer-specific
prompts and domain knowledge to decide whether the path indeed contains a
vulnerability.
If a vulnerability is detected, BCDA performs key condition extraction to
identify the critical branching conditions that must be satisfied to reach the
sink.
Finally, BCDA compiles the results into structured Bug Inducing Things (BITs), which encapsulate vulnerability type, location, trigger conditions, and priority.
These BITs serve as actionable vulnerability candidates for downstream fuzzing agents.

\PP{Path Expansion and Pruning}
Initially, after MCGA detects a potential sink, 
BCDA receives the complete call graph. 
Before BCDA's main procedure begins, 
it filters out paths that have already been analyzed 
or are currently under analysis, retaining only the unexplored ones.
Then, BCDA starts its analysis by forming a call path 
from the source function (typically the harness) to the sink function.

BCDA expands execution paths to address the issue 
of incomplete information in call stack–like paths, 
which often lack details about out-of-stack functions.
To enrich the execution path with the required functions, 
we provide all available functions to the LLM and retain only the necessary ones.
Using Tree-sitter–based code parsing, 
BCDA extracts all function calls within each node of the execution path.
This expansion process ensures that the analysis has access 
to the full context of data transformations and validation logic 
that occur along the vulnerability path.
 
To avoid analyzing irrelevant code, 
BCDA uses LLM-powered pruning to selectively retain 
only the functions likely to be relevant for vulnerability detection.
To improve the effectiveness of this pruning, 
each additionally attached function is tagged with its own ID, 
and the LLM is prompted to output the tags 
corresponding to the necessary functions.
As a result, the expanded path includes 
more context needed to determine the vulnerability, 
ensuring that subsequent analysis remains focused on vulnerability-relevant code.
 
\PP{Vulnerability Classification}
The subsequent step for BCDA involves classifying the expanded path to determine the presence of the vulnerability.
The classification is guided by the potential sanitizer type (i.e., bug type) identified through the sink classification in MCGA.
BCDA then formulates a sanitizer-specific prompt 
that provides detailed explanations of relevant vulnerability information, 
common patterns, and effective detection strategies
(leveraging the domain knowledge described in~\autoref{ss:domain}).
This prompt provides the LLM with concrete guidance 
to analyze the expanded code path 
and accurately determine whether it contains the specified vulnerability or not.

\PP{Key Condition and Trigger Path Extraction}
The key condition extraction step is performed on the expanded path 
if it is determined to contain a vulnerability.
This step identifies important conditional statements along the expanded path 
that must be satisfied to reach the vulnerable location.
For example, an if statement may need to evaluate to true to reach the vulnerability, 
while a try-catch block might need to fail 
in order to enter the error-handling code.

BCDA analyzes each function-level transition individually, rather than extracting conditions from the entire path.
This allows the LLM to focus on a single transition between functions, making it easier to accurately identify critical decision points.
For each transition, BCDA constructs detailed prompts 
that include call flow information, relevant additionally expanded functions, 
and context of the source and target locations. 
This enables the LLM to understand 
the specific conditions required to trigger a vulnerability.

\PP{BIT Generation}
The final output of the BCDA analysis is a Bug Inducing Thing (BIT) data structure, 
which includes all relevant information about identified vulnerabilities 
in a format optimized for subsequent LLM-based analysis and fuzzing stages.
Each BIT contains the vulnerability type, location of vulnerability, 
priority level, all identified key conditions, 
and analysis messages from each step.
Priority levels are assigned based on factors such as the presence of interesting code changes (\eg recent modifications that may have introduced vulnerabilities).
Each BIT is then forwarded to the next agent and to the fuzzers.

\subsubsection{Blob Generation Agent (BGA) Framework}
\label{ss:bga}

BGA framework comprises the payload generation stage of \sysmlla,
transforming LLM-based vulnerability exploitation into a coordinated multi-agent system.
The fundamental insight driving our design is that effective vulnerability exploitation
requires not just generating payloads,
but creating an intelligent feedback loop between LLM reasoning, code execution, and domain knowledge
(see \autoref{ss:domain}).

\PP{Key Innovation: Script-Based Payload Generation}
Traditional fuzzing approaches lack semantic understanding 
of code paths and data structures required to trigger vulnerabilities,
while single-shot LLM payload generation often fails 
due to non-determinism and complexity.
The BGA framework addresses this by having LLMs generate Python scripts 
that serve as ``executable exploit recipes'',
which programmatically construct precise payloads 
while documenting the reasoning process.
This script-based approach enables complex payload construction,
format-aware generation,
and systematic exploration of vulnerability-triggering conditions.

\PP{Multi-Agent Architecture}
The framework employs four specialized agents that work synergistically to overcome distinct challenges in automated exploitation:
The \emph{Orchestrator Agent} serves as the coordination hub,
receiving Bug Inducing Things (BITs) and call graphs from BCDA,
applying intelligent filtering to eliminate redundant work,
and dispatching contexts to specialized agents 
through concurrent \ccw{asyncio} execution.
The \emph{BlobGen Agent} creates single targeted payloads 
through iterative refinement,
generating \ccw{create_payload() -> bytes} functions 
that evolve based on coverage feedback
to systematically overcome obstacles preventing successful exploitation.
The \emph{Generator Agent} employs probabilistic target-reaching,
producing \ccw{generate(rnd: random.Random) -> bytes} functions 
that create multiple payloads
to dramatically increase the likelihood of reaching target vulnerabilities.
The \emph{Mutator Agent} achieves surgical precision in complex call paths,
generating \ccw{mutate(rnd: random.Random, seed: bytes) -> bytes} functions
that focus on individual function transitions 
when full context would exceed LLM limitations.

\PP{Synergistic Integration}
This multi-agent design creates complementary exploitation strategies:
BlobGen provides depth through iterative single-payload refinement,
Generator offers breadth through probabilistic variation,
and Mutator enables precision through focused transition analysis.
The Orchestrator unifies these approaches, ensuring that each vulnerability
receives the most appropriate exploitation strategy based on its characteristics.

The framework seamlessly integrates 
with the \sysuniafl infrastructure (described in~\autoref{ss:uniafl}).
BlobGen's binary blobs feed directly to the \textsc{Input Executor},
while Generator and Mutator functions execute via the \textsc{Script Executor}
to produce continuous streams of payloads.
Successful payloads that explore new paths become seeds for broader fuzzing campaigns,
enabling the BGA framework to contribute 
alongside other input generators in the \sys-Multilang system.

All agents utilize Claude Sonnet 4 (\ccw{claude-sonnet-4-20250514}) 
with temperature \ccw{0.4} for consistent code generation,
operating through Docker-based execution environments 
with comprehensive sanitizer integration
including Jazzer for JVM targets and AddressSanitizer for native code.
We selected Claude Sonnet 4 after evaluating performance-cost trade-offs:
during our internal evaluation,
we found that while Claude Opus 4 demonstrated slightly better performance,
its 5$\times$ higher cost rendered Claude Sonnet 4 
the optimal choice for our multi-agent vulnerability analysis pipeline.
We also compared Claude models with other LLMs;
for detailed evaluation results across different vulnerability types and languages,
see \autoref{tab:domain-knowledge-evaluation}.

\subsubsection{BGA: Orchestrator Agent}
\label{ss:oa}

The Orchestrator Agent serves as the coordination hub for the BGA framework,
receiving \emph{Bug Inducing Things (BITs)} and call graphs from BCDA
and intelligently dispatching them 
to the three specialized payload generation agents.
Its primary role is to maximize exploitation effectiveness 
through strategic work distribution and resource management.

\PP{Intelligent Filtering and Prioritization}
The Orchestrator applies structured filtering logic before dispatching work to downstream agents.
It eliminates redundant efforts by filtering out:
(i) transitions already covered by previous fuzzing inputs,
ensuring agents focus on unexplored vulnerability paths;
(ii) duplicated transitions across call graphs,
preventing multiple agents from targeting identical code paths;
and (iii) transitions lacking conditional branches,
as these offer limited mutation opportunities.
Additionally, the Orchestrator duplicates high-priority BITs
to increase the probability of successful exploitation for critical vulnerabilities.
Historical crash data is reintegrated to avoid redundant exploration
of previously triggered vulnerabilities.

\PP{Concurrent Execution Architecture}
Implemented using \ccw{asyncio}, the Orchestrator enables concurrent execution
of all three payload generation agents while maintaining system stability.
It employs semaphore-based concurrency control to prevent resource exhaustion,
logs detailed performance metrics for analysis,
and implements fault isolation to ensure that failures in one agent
do not cascade to others.

\PP{Context Transformation}
The Orchestrator transforms raw BITs and call graphs 
into structured execution contexts tailored for each agent's specialization.
For BlobGen, it provides complete vulnerability context for iterative refinement;
for Generator, it supplies source and destination function information 
for probabilistic exploration;
and for Mutator, it identifies specific function transitions for surgical precision.
This context-aware dispatch ensures that each agent receives 
the information format best suited to its exploitation strategy.

\subsubsection{BGA: BlobGen Agent}
\label{ss:blobgen}

\PP{Key Idea}
The BlobGen Agent makes a fundamental shift in automated payload generation
by adopting a script-based approach.
Its core innovation is \emph{generating Python scripts to create payloads},
rather than generating the raw payloads directly.
The agent prompts an LLM to create a Python function that programmatically constructs the exploit payload.
This method offers several advantages:
it allows for precise control over complex data structures and formats,
enables incorporation of dynamic values,
provides a reproducible recipe for exploit generation,
and facilitates intelligent feedback loops 
through readable code that documents the exploitation logic.
The generated scripts serve as ``executable exploit recipes'' that capture
both the payload construction logic and the reasoning behind it,
enabling iterative refinement based on execution feedback.% and coverage analysis.

\PP{Workflow}
The BlobGen Agent follows a systematic 5-step workflow for payload generation:
(1) \emph{Sanitizer Selection} (optional): chooses appropriate sanitizers
if not pre-configured based on target language and vulnerability context
by leveraging the vulnerability categorization described in~\autoref{ss:domain},
(2) \emph{Payload Script Generation}: creates a Python function \ccw{create_payload()}
using LLM-powered analysis of vulnerability context,
incorporating domain knowledge about exploit guides and data structures 
(\autoref{ss:domain}) to define the logic for building the exploit payload,
(3) \emph{Coverage Collection}: executes the generated script in a Docker sandbox environment
to produce binary payloads and collects execution data,
(4) \emph{Failure Analysis}: analyzes coverage data and crash information
to assess effectiveness and understand why payloads failed to trigger vulnerabilities, and
(5) \emph{Iterative Refinement}: uses execution coverage feedback 
to generate improved scripts,
progressively refining the exploitation strategy through multiple iterations.

\begin{figure}[!t]
  \centering
    \scalebox{0.9}{
      \begin{promptbox}[width=1.05\columnwidth]{BlobGen System Prompt}
        \begin{promptcontent}\input{code/blobgen-system-prompt.xml}\end{promptcontent}
      \end{promptbox}
    }

    \scalebox{0.9}{
      \begin{promptbox}[width=1.05\columnwidth]{BlobGen Source Code Prompt}
        \begin{promptcontent}\input{code/blobgen-source-code-prompt.xml}\end{promptcontent}
      \end{promptbox}
    }
  \caption{An example of system prompt and source code prompt used in BlobGen Agent.}
  \label{fig:blobgen-prompt}
  \vspace*{-20px}
\end{figure}

\PP{System Prompt Design}
The BlobGen system prompt follows key design principles
that maximize LLM effectiveness for vulnerability exploitation.
Our approach employs hierarchical structure,
context-before-complexity,
annotation clarity,
and XML organization~\cite{openai-prompt-engineering,anthropic-prompt-engineering,anthropic-xml}.
\autoref{fig:blobgen-prompt} demonstrates these principles in practice.
The \cc{<role>} section establishes the LLM as a security researcher,
priming it for technical exploitation challenges.
The \cc{<final_objective>} enforces critical constraints,
mandating a \ccw{create_payload} function
that returns only single bytes objects using built-in Python libraries,
ensuring seamless integration with fuzzing infrastructure.
The \cc{<context>} section provides upfront target binding
with project name and sanitizer type before introducing code complexity,
preventing focus dilution across irrelevant possibilities.
The \cc{<code_annotations>} section explains our annotation system:
\cc{@BUG_HERE} and \cc{@KEY_CONDITION} markers derived from BCDA's BITs,
with \cc{@} prefix chosen to avoid confusion with developer comments.

\PP{Source Code Prompt Design}
\autoref{fig:blobgen-prompt} shows how the BlobGen delivers source code
as a prompt.
Within \cc{<SOURCE_CODE_INFO>}, each line uses \cc{[n]:} formatting
inspired by RustAssistant~\cite{deligiannis2024rustassistant},
where brackets and colon distinguish line numbers from code literals.
The XML structure with \cc{<FUNCTION>} tags provides clear organization
for both automated parsing and human readability.

Beyond the visible elements, the full prompt includes methodology guidance
for systematic vulnerability analysis and output format specifications
using additional XML tags that enable reliable response parsing.

\PP{Coverage Feedback Mechanism}
The BlobGen Agent employs systematic coverage feedback
to drive iterative payload refinement.
When a generated payload is executed,
the agent collects runtime coverage data
and compares executed lines against BIT-identified key conditions and vulnerability locations,
dynamically adding \cc{@VISITED} markers to indicate which conditions were reached.

The agent analyzes coverage results to locate gaps between executed lines and the conditions required to trigger vulnerabilities.
Through selective source code inclusion, feedback prompts present only vulnerability-relevant annotated lines, marked with \cc{@VISITED}.
This strategy filters out irrelevant code paths and directs the LLM’s attention to critical decision points where payload modifications are most impactful.
Coverage gaps are then systematically reported to guide targeted script refinements in subsequent iterations.

\begin{figure}[t]
  \centering
    \scalebox{0.9}{
  \begin{promptbox}{BlobGen Coverage Feedback Prompt}
  \begin{promptcontent}\input{code/blobgen-coverage-prompt.xml}\end{promptcontent}
  \end{promptbox}
    }
  \caption{An example coverage integration showing XML-structured feedback with source code annotations}
  \label{fig:coverage-integration}
\end{figure}

\autoref{fig:coverage-integration} demonstrates the structured feedback format
showing partial success in reaching vulnerability conditions.
The \cc{<COVERAGE_INFO_FOR_KEY_CONDITIONS>} section reveals that the payload
successfully reached the key condition at line 21 (marked \cc{@VISITED})
but failed to trigger the vulnerability at line 24.
The \cc{<XXD_OUTPUT_FOR_PAYLOAD_BLOB>} section enables payload debugging
through hexadecimal analysis of the generated data structure.
The \cc{<STDERR_FOR_PAYLOAD_BLOB>} section captures the resulting exception,
confirming payload execution reached the target code path but with incorrect data format.

This coverage feedback mechanism integrates with the iterative refinement workflow,
systematically guiding the LLM through evaluation of coverage gaps
and vulnerability-specific constraints.
The structured feedback enables targeted script modifications in subsequent iterations,
with the agent repeating up to four refinement cycles
until successful vulnerability triggering or maximum iterations are reached.

\begin{figure}[t]
  \centering
    \scalebox{0.9}{
  \begin{promptbox}{BlobGen Agent Response}
  \begin{promptcontent}\input{code/blobgen_example1.py}\end{promptcontent}
  \end{promptbox}
    }
  \caption{An example of a single payload generation script from the BlobGen Agent}
  \label{fig:blobgen}
\end{figure}

\PP{Validation \& Integration}
The BlobGen Agent operates within a Docker-based execution environment
that provides security isolation and resource management.
All generated scripts undergo comprehensive validation before execution,
including syntax checking, import verification for built-in Python libraries only,
and function signature validation to ensure \ccw{create_payload() -> bytes} compliance.
Failed validation triggers automatic script revision until validation passes.

The execution environment enforces strict resource constraints:
a 1GB memory limit prevents runaway allocations,
a 1MB blob size limit ensures payloads remain manageable,
and execution timeouts prevent infinite loops or resource exhaustion.
The system systematically categorizes failures into syntax errors,
runtime exceptions, and target-specific validation failures,
enabling precise feedback for script refinement.

Integration with the \sysuniafl infrastructure occurs through the \textsc{Input
Executor}~(\autoref{ss:uniafl}),
which executes generated binary blobs and promotes successful payloads as seeds
if they trigger new code paths or explore previously unreached program states,
enabling BlobGen outputs to contribute to the broader fuzzing campaign.

\PP{Example}
\autoref{fig:blobgen} illustrates a generated script 
that constructs a GZIP archive to exploit
an OS command injection vulnerability 
in \ccw{apache-commons-compress}'s \ccw{CompressorGzipFuzzer}
in the competition round 2.
The generated script exhibits sophisticated understanding of the GZIP format structure,
including magic bytes (\ccw{0x1f, 0x8b}), compression methods, 
and the specific conditions required to trigger the vulnerability: 
an exact modification time (\ccw{1731695077}) and filename (\ccw{"jazze"}).
The agent is guided to generate extensive comments that serve as documented reasoning,
providing context summaries and step-by-step explanations
consistent with the prompt design principles described earlier,
facilitating the feedback loop and enabling iterative refinement of the payload generation strategy.
% aixcc/jvm/r2-apache-commons-compress
% CompressorGzipFuzzer

\subsubsection{BGA: Generator Agent}
\label{ss:generator}

\PP{Key Idea}
The Generator Agent introduces a \emph{probabilistic target-reaching} approach to LLM-powered vulnerability discovery,
addressing the common failure where single LLM-generated payloads do not reach their targets because of non-deterministic generation.
Rather than issuing a single payloads, the agent produces Python generator functions of the form \ccw{generate(rnd:random.Random) -> bytes} that emit multiple, structurally valid payloads under controlled randomness.
Although any single payload may fail, generating many correlated variations dramatically increases the probability that at least one will traverse the correct execution path and trigger the vulnerability.
The agent runs in two modes to maximize adaptability:
the \emph{guided mode} leverages 
not only specified source and sink functions for targeted exploitation,
but also AttributeCGs, call graphs augmented with vulnerability annotations, taint information, and security-relevant metadata;
the \emph{standalone mode} (see \autoref{ss:mlla-standalone}) generates inputs directly from source code for broad vulnerability discovery.
This design lets the agent support both targeted exploitation and exploratory fuzzing campaigns.

\PP{Workflow}
The Generator Agent operates through a systematic 6-step process
within an iterative refinement loop:
(1) \emph{Sanitizer Selection} (optional): chooses appropriate sanitizers
for standalone mode operation,
(2) \emph{Generator Planning}: analyzes code paths and vulnerability requirements
to create a comprehensive generation strategy
by leveraging vulnerability patterns and exploit guides from~\autoref{ss:domain},
(3) \emph{Generator Creation}: creates Python \ccw{generate} functions
using LLM analysis to produce multiple structurally valid payloads,
(4) \emph{Coverage Collection}: executes multiple generated payloads (typically 20)
and collects merged coverage data across all payloads,
(5) \emph{Function Context Update}: identifies and updates functions showing promising coverage patterns
for subsequent iterations, and
(6) \emph{Coverage Analysis and Refinement}: analyzes effectiveness and provides feedback
for iterative improvement,
repeating up to four rounds to refine the generation strategy.

\begin{figure}[t]
  \centering
    \scalebox{0.9}{
  \begin{promptbox}{Generator Coverage Feedback Prompt}
  \begin{promptcontent}\input{code/generator-coverage-summary.xml}\end{promptcontent}
  \end{promptbox}
    }
  \caption{An example coverage feedback for Generator Agent showing structured summary and analysis prompts}
  \label{fig:generator-coverage}
\end{figure}

\PP{System Prompt Design}
The Generator Agent adopts the prompt engineering principles described in \autoref{ss:blobgen}
with adaptations for probabilistic multi-variation generation.
The prompt specifies the \ccw{generate(rnd: random.Random) -> bytes} function signature,
requiring controlled randomness for reproducible variation generation.
A two-phase approach structures the generation strategy:
Phase 1 navigates validation checks to reach the destination function,
while Phase 2 targets vulnerability exploitation through strategic mutations.
The prompt integrates a four-step refinement workflow
(planning, creation, analysis, and improvement),
guiding systematic exploration while maintaining format validity.

\begin{figure}[t]
  \centering
    \scalebox{0.9}{
  \begin{promptbox}{Generator Agent Response}
  \begin{promptcontent}\input{code/generator_example1.py}\end{promptcontent}
  \end{promptbox}
    }
  \caption{An example of multiple payload generation script from the Generator Agent}
  \label{fig:generator}
\end{figure}

\PP{Multi-Variation Coverage Analysis}
The Generator Agent employs merged coverage analysis
across multiple payloads (typically 20 executions)
to evaluate collective progress toward vulnerability exploitation.
\autoref{fig:generator-coverage} demonstrates the dual-level reporting structure:
\emph{primary coverage} tracks target vulnerability path functions,
while \emph{entire coverage} encompasses all explored code.
The XML-structured feedback combines quantitative summaries with detailed function-level changes,
enabling systematic evaluation of variation effectiveness.
The analysis identifies successful variation patterns
and performs intelligent function prioritization,
selecting the top 3 most promising unexplored functions
based on proximity to vulnerability paths and security-sensitive operations.
This systematic approach enables iterative refinement
through collective insights from multiple payloads.

\PP{Validation \& Integration}
Similar to the BlobGen Agent described in \autoref{ss:blobgen},
all generated functions undergo comprehensive validation to ensure
\ccw{generate(rnd: random.Random) -> bytes} function signature compliance and 
correct variation generation across multiple executions.
The validation process executes sample payloads (typically 20 iterations) 
to verify the correctness of function and the diversity of payload.
Failed validation triggers iterative refinement with categorized feedback for syntax
errors, execution issues, and target-specific problems.

Integration with the \sysuniafl infrastructure occurs through the \textsc{Script
Executor}~(\autoref{ss:uniafl}), which continuously executes validated generator
functions throughout the fuzzing campaign. The system promotes successful
payloads as seeds when they trigger new code paths or explore
previously unreached program states, enabling Generator outputs to contribute to
the broader fuzzing campaign.

\PP{Example}
\autoref{fig:generator} presents an example 
exploiting XML External Entity (XXE) vulnerabilities
using diverse attack strategies encapsulated in a ZIP archive.
The generator targets a vulnerability 
that can be triggered from \ccw{apache-tika}'s \ccw{ThreeDXMLParserFuzzer} 
in competition round 3.
The LLM implements multiple attack vectors:
basic XXE with external entity declarations,
XInclude-based inclusion attacks,
and external DTD references.
The generator dynamically selects among these attack vectors 
using controlled randomness,
demonstrating sophisticated understanding 
of both ZIP file structure and XML parsing vulnerabilities.
It creates valid archives containing malicious XML files
that target external entity resolution 
to trigger network requests to \ccw{jazzer.com}.
This example illustrates the agent's ability 
to synthesize complex, multi-layered payloads
that combine format compliance with vulnerability exploitation logic.

% \PP{Mutator Agent}
\subsubsection{BGA: Mutator Agent}
\label{ss:mutator}

\PP{Key Idea}
The Mutator Agent addresses a critical challenge 
in LLM-based vulnerability exploitation:
context limitations that arise when analyzing complex call paths.
Its core innovation is \emph{focused transition analysis}
that concentrates on individual function transitions 
to enable precise mutations
when the overall vulnerability context would exceed LLM token limits.
Rather than attempting to comprehend entire call graphs,
the agent operates on single transitions $(f_{src}, f_{dst})$,
allowing the LLM to deeply understand the specific data flow
and transformation requirements between two functions.
This surgical precision approach is particularly valuable 
for complex vulnerabilities that require navigating 
through specific function sequences
where comprehensive context analysis would be computationally prohibitive.
The agent generates \ccw{mutate(rnd: random.Random, seed: bytes) -> bytes} functions
that surgically modify existing seeds to explore targeted transitions,
ensuring that mutations are precisely tailored 
to the specific vulnerability path
rather than applying generic transformation strategies.

\PP{Workflow}
The Mutator Agent operates through a systematic 3-step process
within an iterative refinement loop:
(i) \emph{Mutation Planning}: analyzes the AttributeCG 
to understand the specific transition context between source and destination functions,
devising a targeted strategy for data transformation and condition satisfaction
(utilizing vulnerability descriptions and exploit guides from~\autoref{ss:domain}),
(ii) \emph{Mutator Creation}: generates the actual 
\ccw{mutate(rnd: random.Random, seed: bytes) -> bytes} function
based on the planned strategy,
synthesizing transformation logic that surgically modifies existing seeds
to explore the targeted transition, and
(iii) \emph{Mutator Analysis and Refinement}: analyzes whether mutations 
successfully traverse the targeted transition and provides feedback for improvement,
with ineffective strategies discarded 
and successful patterns enhanced in subsequent iterations.
This process continues until a satisfactory mutator is created
or the maximum number of iterations is reached.

\PP{Validation \& Integration}
Similar to the BlobGen Agent (\autoref{ss:blobgen}), all generated mutator functions undergo comprehensive
validation to ensure \ccw{mutate(rnd: random.Random, seed: bytes) -> bytes} function
signature compliance and correct mutation behavior across targeted transitions.
The validation process verifies that mutations maintain seed structure integrity
while introducing targeted modifications for transition exploration.
Failed validation triggers iterative refinement
with categorized feedback for syntax errors, execution issues, and transition-specific problems.

Integration with the \sysuniafl infrastructure occurs through the \textsc{Script Executor}~(\autoref{ss:uniafl}),
which executes validated mutator functions throughout the fuzzing campaign.
The system systematically explores targeted transitions with surgical precision,
enabling Mutator outputs to contribute focused vulnerability path navigation
to the broader fuzzing campaign.

\begin{figure}[t]
  \centering
  \scalebox{0.95}{
  \begin{promptbox}{Mutator Agent Response}
  \begin{promptcontent}\input{code/mutator_example1.py}\end{promptcontent}
  \end{promptbox}
  }
  \caption{An example of seed blob mutation script from the Mutator Agent}
  \label{fig:mutator}
\end{figure}

\PP{Example}
\autoref{fig:mutator} demonstrates a case 
where the agent targets EXIF metadata,
crafting mutations that manipulate headers and offset fields
to provoke memory corruption reached from \ccw{libexif}'s \ccw{exif_from_data_fuzzer}
in competition round 3.
The LLM demonstrates sophisticated understanding of EXIF file structure,
including Makernote section detection, directory entry parsing, 
and endianness considerations.
The mutation strategies target specific vulnerability patterns:
malformed directory counts, oversized data fields, and invalid offsets
that commonly lead to buffer overflows and memory corruption 
in image processing libraries.
This example illustrates the agent's ability to synthesize
highly targeted mutations that combine deep format understanding
with precise vulnerability exploitation logic,
demonstrating how focused transition analysis enables
surgical precision in complex vulnerability scenarios.
% aixcc/c/r3-libexif-delta-01
% exif_from_data_fuzzer

\subsubsection{Domain Knowledge Integration}
\label{ss:domain}

The BGA framework incorporates systematic domain knowledge 
across two critical dimensions that enable effective LLM-based vulnerability exploitation: 
comprehensive sanitizer-based vulnerability detection 
and sophisticated handling of challenging data structures 
that traditionally impede LLM fuzzing effectiveness.

\PP{Vulnerability-Aware Payload Generation}
The system provides multi-sanitizer awareness spanning
\ccw{Jazzer} for Java targets and \ccw{AddressSanitizer}, \ccw{MemorySanitizer}, and \ccw{UndefinedBehaviorSanitizer} for native C code,
enabling agents to understand and target vulnerability patterns
across different execution contexts.

\begin{figure}[t]
  \centering
    \scalebox{0.9}{
  \begin{promptbox}{Exploit Guide Prompt (OS Command Injection)}
  \begin{promptcontent}\input{code/os_cmd_injection.xml}\end{promptcontent}
  \end{promptbox}
    }
  \caption{An example of vulnerability description and exploit guide for command injection vulnerabilities}
  \label{fig:oscmdinjection}
\end{figure}

The framework categorizes vulnerabilities into distinct classes:
memory corruption vulnerabilities (\eg buffer-overflow, use-after-free, double-free),
injection attacks (\eg SQL injection, OS command injection, XPath injection),
remote code execution vulnerabilities (\eg deserialization, expression language injection),
information disclosure issues (\eg uninitialized memory access), and
denial of service conditions (\eg timeout, out-of-memory).
The framework includes targeted detection patterns for AIxCC-introduced timeout bugs,
identifying unusual timeout behaviors that require specialized exploitation techniques.

Each vulnerability category is accompanied by structured exploit guidance
that provides agents with concrete patterns and triggering mechanisms.
For example, OS command injection exploitation incorporates knowledge
of \ccw{Runtime.exec()} and \ccw{ProcessBuilder} usage patterns,
enabling agents to craft payloads that execute the target command \ccw{"jazze"}
for vulnerability confirmation.
\autoref{fig:oscmdinjection} illustrates this structured guidance approach.
This domain knowledge is then integrated into LLM prompts using XML formatting,
showing the complete structure from vulnerability categorization
to concrete exploitation guidance.

\begin{figure}[t]
  \centering
    \scalebox{1}{
  \begin{promptbox}{Structure Guide Prompt (LibFDP)}
  \begin{promptcontent}\input{code/data-structure-guide-example.xml}\end{promptcontent}
  \end{promptbox}
    }
  \caption{An example data structure guidance integration showing selective FDP method mapping}
  \label{fig:data-structure-guide}
\end{figure}

\PP{Structured Data Format Handling}
The BGA framework addresses three categories of challenging data structures
that require specialized approaches for effective LLM-based payload generation.
These structures present unique format requirements
that traditional byte-level fuzzing approaches cannot adequately handle.

\textbf{(1)} \emph{FuzzedDataProvider structures} pose significant challenges
due to their complex data consumption behaviors:
consuming primitive types from the end of data buffers
while consuming structured data from the beginning,
with specialized methods like \ccw{consumeInt(min, max)} for bounded value generation.
The framework overcomes these challenges through \libfdp described in \autoref{ss:libfdp}.
\libfdp abstracts away implementation details by giving LLMs simple functions to call
rather than asking them to understand underlying binary format specifications.
This enables proper payload encoding using selective function mapping
that focuses only on methods relevant to the target source code.

\autoref{fig:data-structure-guide} shows how data structure handling guidance
is integrated into LLM prompts,
demonstrating the selective method mapping approach
that focuses only on FuzzedDataProvider methods detected in the target source code.
The guidance provides language-specific encoder usage:
\ccw{libFDP.JazzerFdpEncoder()} for Java targets and \ccw{libFDP.LlvmFdpEncoder()} for C/C++ targets,
with method mappings filtered to include only the consumption patterns
actually present in the analyzed code to avoid unnecessary complexity.

\textbf{(2)} \emph{Java ByteBuffer formats} require precise endianness handling
for multi-byte integer consumption patterns.
The framework guides agents in understanding ByteBuffer's endianness, which is big-endian,
enabling proper payload construction for methods like \ccw{getInt()} and \ccw{getLong()}.
Agents must generate payloads with correct big-endian byte ordering
to match ByteBuffer's default byte order specifications,
such as transforming \tc{b'\r\x00\x00\x00\x01\x00\x00\x00'} to \tc{b'\x00\x00\x00\r\x00\x00\x00\x01'}
for proper ByteBuffer consumption sequences.

\textbf{(3)} \emph{Application-specific data structures} encompass domain-specific formats
like \ccw{ServletFileUpload} (in \ccw{Jenkins}) for multipart-based file upload processing.
\autoref{fig:servletfileupload} demonstrates the structured guidance approach
for understanding multipart/form-data parsing and FileItem processing patterns.

\begin{figure}[t]
  \centering
    \scalebox{0.9}{
  \begin{promptbox}{Structure Guide Prompt (ServletFileUpload)}
  \begin{promptcontent}\input{code/servlet_file_upload.xml}\end{promptcontent}
  \end{promptbox}
    }
  \caption{An example data structure guide for ServletFileUpload multipart processing}
  \label{fig:servletfileupload}
\end{figure}

\PP{Adaptive Knowledge Integration}
The BGA framework uses adaptive knowledge integration to balance domain expertise with computational efficiency through context-aware prompt generation.
Instead of overloading LLMs with exhaustive knowledge, it selectively incorporates vulnerability patterns and data structure insights derived from target-specific analysis and detected behaviors.

The integration strategy follows two principles.
\emph{Contextual relevance} ensures that domain knowledge aligns with the vulnerability context and target characteristics,
while \emph{selective application} prevents overload by focusing only on detected patterns instead of entire knowledge bases.
Guided by BCDA’s vulnerability categorization, the system selects appropriate exploit patterns,
and detected data structures activate relevant strategies for BlobGen agent, BGA Generator agent, and BGA Mutator agent.
Targeted prompts then embed only the most pertinent patterns and structural constraints, giving LLMs focused guidance without exceeding context limits.
This adaptive strategy represents a key advance in LLM-based security analysis.
Rather than relying on static knowledge application, it delivers dynamic, context-sensitive guidance that improves exploitation effectiveness.
At the same time, it maintains scalability across diverse vulnerability types, programming languages, and challenge projects.

\PP{Domain Knowledge Evolution}%
Our domain knowledge integration approach 
evolved systematically through competition rounds:

\begin{squishitemize}
\item \emph{By round 1}: Basic vulnerability categorization 
using sanitizer outputs,
systematic sentinel handling (\ccw{"jazze"}),
and initial LLM-based vulnerability detection frameworks.

\item \emph{After round 2 (R2.5)}: Enhanced with direct vulnerability descriptions 
and concrete examples,
separate exploit guides with structured triggering mechanisms,
and specialized handling for AIxCC-introduced timeout vulnerabilities.

\item \emph{By round 4 (Final)}: Mature vulnerability categorization 
based on human security expertise rather than sanitizer output alone,
with concise descriptions and LLM-optimized exploit guides.
\end{squishitemize}

\PP{Evaluation of Domain Knowledge Integration}%
To demonstrate the effectiveness 
of our domain knowledge integration techniques,
we conducted systematic evaluation on JenkinsThree,
one of the Java projects in our benchmark.
JenkinsThree is fundamentally a Jenkins repository 
tailored for 11 vulnerability types that Jazzer can detect,
with an independent harness for each vulnerability.
For the evaluation, we executed 10 runs for each harness 
(\ie each vulnerability type) across multiple LLM models,
comparing performance before (R2.5 baseline) and after (Final)
applying domain knowledge improvements.

\begin{table}[!t]
\centering
\footnotesize

\begin{threeparttable}
\begin{tabular}{lrrrrrr}
\toprule
{Vulnerability Type} & {Claude-4} & {Claude-4} & {Claude-3.7} & {Claude-3.7} & {Gemini-2.5} & {O4-Mini} \\
{Knowledge Context Version} & {(R2.5)} & $\rightarrow$ {(Final)} & {(R2.5)} & $\rightarrow$ {(Final)} & {(Final)} & {(Final)} \\
\midrule
XPath Injection & 10/10 & 10/10 & 4/10 & \textbf{5/10} & 10/10 & 10/10 \\
OS Command Injection & 0/10 & \textbf{10/10} & 0/10 & \textbf{10/10} & 10/10 & 10/10 \\
Server Side Request Forgery & 6/10 & \textbf{8/10} & 10/10 & 6/10 & 10/10 & 10/10 \\
Regex Injection & 3/10 & \textbf{10/10} & 7/10 & \textbf{10/10} & 10/10 & 8/10 \\
Remote JNDI Lookup & 0/10 & \textbf{10/10} & 0/10 & \textbf{10/10} & 1/10 & 8/10 \\
Reflective Call & 0/10 & \textbf{10/10} & 0/10 & \textbf{10/10} & 9/10 & 5/10 \\
SQL Injection & 0/10 & \textbf{10/10} & 0/10 & \textbf{3/10} & 10/10 & 9/10 \\
Script Engine Injection & 10/10 & 10/10 & 10/10 & 10/10 & 10/10 & 10/10 \\
LDAP Injection & 3/10 & \textbf{4/10} & 7/10 & \textbf{10/10} & 6/10 & 6/10 \\
Remote Code Execution & 0/10 & \textbf{10/10} & 0/10 & \textbf{10/10} & 10/10 & 9/10 \\
File Path Traversal & 4/10 & \textbf{3/10}\tnote{$\dagger$} & 8/10 & 8/10 & 8/10 & 9/10 \\
\midrule[\heavyrulewidth]
{Successful Exploits} & 36/110 & \textbf{95/110} & 49/110 & \textbf{89/110} & 92/110 & 93/110 \\
{Success Rate} & 32.7\% & \textbf{86.4\%} & 44.5\% & \textbf{80.9\%} & 83.6\% & 84.5\% \\
{Execution Time (s)} & 1066.54 & \textbf{467.55} & 1525.45 & \textbf{490.59} & 2231.88 & 1227.99 \\
{Input Tokens} & 2.37M & \textbf{1.18M} & 2.27M & \textbf{1.24M} & 1.22M & 1.39M \\
{Output Tokens} & 337K & \textbf{165K} & 347K & \textbf{189K} & 1.31M & 924K \\
{Total Tokens} & 2.71M & \textbf{1.34M} & 2.61M & \textbf{1.43M} & 2.53M & 2.31M \\
{Total Cost (\$)} & 8.05 & \textbf{3.99} & 8.09 & \textbf{4.36} & 14.23 & 4.68 \\
\bottomrule
\end{tabular}
\begin{tablenotes}
\item [$\dagger$] Including the description of \ccw{ServletFileUpload} improved the results from 3/10 to 9/10,
demonstrating the effectiveness of application-specific data structure integration.
\end{tablenotes}
\end{threeparttable}

\caption{Domain Knowledge Integration: Evaluation Results and Performance Metrics}
\label{tab:domain-knowledge-evaluation}

\end{table}

The results validate the effectiveness 
of our domain knowledge integration approach.
Most significantly, we observed breakthrough improvements (0/10 → 10/10)
for six vulnerability types across both models,
including OS Command Injection, Remote JNDI Lookup, 
and Reflective Call vulnerabilities
that were previously impossible to exploit.
These breakthroughs demonstrate that structured exploit guidance
combined with vulnerability-specific context templates
enables LLMs to reliably generate working exploits
for complex security vulnerabilities.
Additionally, the effectiveness of application-specific data structure integration
is demonstrated by a separate experiment with ServletFileUpload context,
which improved File Path Traversal success from 3/10 to 9/10,
validating our approach of providing targeted data structure summaries
for domain-specific format handling.

Among the evaluated models, Claude-4 demonstrated 
the most favorable cost-performance trade-off,
achieving the highest success rate 
(86.4\% and 91.8\%, without and with \ccw{ServletFileUpload}, respectively)
while maintaining reasonable computational costs (\$3.99).
This superior performance, combined with substantial reductions
in execution time and token consumption compared to other models,
led us to select Claude-4 as our primary model 
for the final competition rounds.

\subsection{\sysmlla Standalone}
\label{ss:mlla-standalone}

\sysmlla Standalone tackles a core limitation of traditional fuzzing,
the difficulty of generating semantically meaningful inputs to trigger complex vulnerabilities.
Its goal is to accelerate the early fuzzing process 
by producing diverse, semantically informed seeds
without relying on heavy static analysis.
Unlike the full \sysmlla, which coordinates multiple agents for targeted exploitation,
standalone mode works only with harness code and optional diff files for general vulnerability discovery.
By leveraging LLM reasoning on minimal code context,
it transforms pre-trained knowledge into effective seeds that broaden fuzzing campaigns.

\PP{Architecture and Operation}
The standalone mode replaces the multi-agent coordination of the full mode
with a single Generator Agent that analyzes harness code directly using LLM-driven semantic reasoning.
This lightweight design prioritizes rapid deployment
and broad applicability over deep static analysis.
As a result, this makes it particularly effective for exploratory fuzzing scenarios
where quick seed generation is more valuable than targeted exploitation.

Similar as the Generator Agent in \autoref{ss:generator},
the standalone mode produces Python functions, \ccw{generate},
but applies semantic reasoning from harness code analysis
rather than comprehensive static analysis results.
The system generates 20 sample payloads
and employs coverage-guided iterative improvement,
merging coverage information from all variations
to refine generation strategies.
For the competition, we chose 4 iterations.

Standalone mode includes automatic sanitizer selection capabilities,
choosing appropriate sanitizers based on harness analysis
and generating targeted payloads for vulnerability detection.
However, we disabled this capability for the competition
to avoid restricting LLM path exploration,
as the full \sysmlla pipeline already provides targeted vulnerability analysis.
All execution occurs within isolated Docker containers
with configurable timeouts and resource limits
to ensure system stability during payload generation and testing.
This approach integrates seamlessly with the \sysuniafl infrastructure (\autoref{ss:uniafl}),
contributing generated seeds to the unified fuzzing pipeline
alongside traditional input generators.
The system utilizes Claude Sonnet 4 (\ccw{claude-sonnet-4-20250514})
with temperature \ccw{0.4} for consistent code generation,
using the same model configuration as the full \sysmlla pipeline
based on our performance-cost evaluation
(see \autoref{tab:domain-knowledge-evaluation}).

\clearpage

\subsection{Shared Utils \& Libraries}
\label{ss:shared-utils}

As described in \autoref{ss:multilang-overview}, \sys-Multilang shares utility components across multiple nodes targeting the same CP.
A dedicated node hosts the \functracer and \retriever, responsible for collecting function call traces and retrieving source code, respectively.
These components are reusable across different fuzzing harnesses within the same CP.
By decoupling them from the nodes executing fuzzing tasks, \sys-Multilang effectively eliminates redundancy and reduces unnecessary overhead.
Furthermore, this subsection introduces \libfdp, a library designed for encoding and decoding inputs based on the FuzzedDataProvider (FDP) interface, which is used by both libFuzzer and Jazzer.

\subsubsection{\functracer}
To obtain a more accurate function call graph in \sysmlla, MCGA~(\autoref{ss:mcga}) leverages the dynamic function tracing capabilities implemented in \functracer.
\functracer instruments both the target fuzzing harnesses and the given CP to produce function call traces during input execution.
Similar as \textsc{Input Executor} in \autoref{ss:uniafl}, \functracer provides a language-agnostic interface, although its implementations vary depending on the programming language of the CP.

\PS{Function Tracer for C}
Instead of relying on compiler-based instrumentation, we employ dynamic instrumentation using \ccw{DynamoRIO}.
This approach eliminates concerns related to compilation failures.
Furthermore, this enables reliable extraction of function traces by executing the given inputs under the provided CP and associated fuzzing harness.

\PS{Function Tracer for Java}
Thanks to the flexibility of the JVM and the design of Jazzer, we are able to obtain function traces in Java-based CPs by modifying the coverage logging module of Jazzer.
In particular, we modified Jazzer to record function call traces instead of populating the coverage map, enabling seamless integration with our dynamic tracing infrastructure.

\subsubsection{\retriever}
\label{ss:retriever}

In multiple components within \sys-Multilang such as \sysmlla and
\textsc{Testlang}-based module, the Code Retriever tools form the foundation for
extracting and
structuring program information at scale. 
These tools provide complementary capabilities for retrieving semantic and
structural views of large codebases. They
establish a unified retrieval layer that supplies higher-level agents with
precise code facts, enabling effective vulnerability analysis, 
and reasoning without requiring each agent to implement them again.

\PP{Joern}
Joern is a widely used open-source framework for building Code Property Graphs
(CPGs), a unified representation that merges abstract syntax trees, control-flow
graphs, and data-flow graphs into a single graph structure.
It natively supports C, C++, and Java, making it applicable across the diverse
set of languages targeted by \sysmlla.
In \sysmlla, Joern was employed to extract rich program facts such as
function-level control-flow structures.
These graph-based insights were integrated into the agent workflow to enhance
vulnerability reasoning, for example by assisting CGPA in providing more
accurate function call information.
This integration allowed \sysmlla to leverage Joern’s multi-language analysis
strengths while combining them with LLM-based reasoning for complex
codebase understanding.

\PP{Language Server Protocol (LSP)}
LSP provides a standardized interface for language-aware tooling, enabling
features such as symbol resolution, definition
lookup, reference search, and type information across different programming
languages. 
In \sys-Multilang, we adopted multilspy, a Python client for LSP, and
extended its implementation to fully support \ccw{clangd}, thereby ensuring
robust coverage for C and C++ in addition to Java.
This adaptation allowed components in \sys-Multilang to uniformly query semantic
information across all three target languages, 
eliminating the need for custom parsers. 
Especially, in \sysmlla, 
LSP was primarily used to retrieve
precise function boundaries, call sites, and cross-references, which were
essential for building accurate reference-to-definition mappings and for supplying
downstream agents such as CGPA and MCGA with context-rich code facts.
By integrating LSP into the retrieval layer, \sysmlla achieved language-agnostic
yet semantically precise navigation of large codebases, which proved critical
for scalable and accurate vulnerability analysis.

\PP{Code Indexer}
Code Indexer is a component that parses codebases written in multiple programming languages.
It supports C, C++, and Java, using language-specific Tree-sitter parsers to asynchronously process source code files and extract functions, as well as other elements such as structures and unions.
The extracted data is stored in a Redis backend, with separate namespaces for each project to support efficient multi-project management.
For functions, the stored data includes the function signature, function body, file path, and line number.
Code Indexer uses both the simple name and the full signature as keys, allowing users to query function bodies or structural information based on either.
Each full signature serves as a unique identifier for a function, although multiple functions can share the same simple name.
For example, a \ccw{sum} function in class \ccw{A} and a \ccw{sum} function in class \ccw{B} share the same simple name but have different full signatures.
In such cases, Code Indexer returns all matching functions, and it is the responsibility of the consuming component to distinguish between them using additional context, such as the file path.

\begin{figure}[t]
  \centering
  \begin{promptbox}{FDP encoder semantic error example}
  \begin{promptcontent}\input{code/libfdp-bad-usage-example.py}\end{promptcontent}
  \end{promptbox}
  \caption{FDP encoder example with semantic error which LLM may produce}
  \label{fig:libfdp-bad-example}
\end{figure}

\subsubsection{\libfdp}
\label{ss:libfdp}

\libfdp is a codec provider and encoder library for well-known FuzzedDataProvider (FDP) implementations, supporting both libFuzzer and Jazzer.
It enables deterministic encoding of fuzzing inputs for cross-language and cross-platform fuzzing workflows, and provides both Rust and Python bindings for integration.

The challenge that \libfdp addresses is to fill the gap between the complexity of byte processing in the behind scenes of FDP APIs and the weakness of mathematical processing in LLMs.
Though some implementations between the two FDPs differs, one common point is that they both have a complex backend which requires LLMs to have a deep understanding of byte level processings to mimic the behavior, and this is not definitely a strong point of LLMs.
Catching errors from LLMs is also one of the interests of \libfdp.
\libfdp can provide semantic errors related to FDP usages which can be used to guide LLMs.
This allows to generate inputs that are more likely to be accepted by the target harness, thus supporting the effectiveness of fuzzers using FDP as inputs.
For example, given that LLM provided encoding sequence like in \autoref{fig:libfdp-bad-example}, there is a semantic error on line 5 because it requests to produce a floating point value after consuming all remaining bytes on line 4.
When the code in \autoref{fig:libfdp-bad-example} gets executed, \libfdp may raise the error that can be utilized for providing feedback to LLMs to correct the sequence.

\PP{Internal library structure}%
\begin{squishitemize}
    \item \ccw{libfdp}: The core Rust library implementing encoder logic for LLVM and Jazzer FDPs.
    \item \ccw{pyfdp}: Python FFI bindings, exposing encoder APIs for Python-based fuzzing or harness scripting.
    \item \ccw{fdp-reference}: Reference C++ implementations and headers for testing implementation correctness and compatibility with LLVM and Jazzer FDPs.
\end{squishitemize}

\begin{figure}[t]
  \centering
  \begin{promptbox}{FDP fuzzer example}
  \begin{promptcontent}\input{code/libfdp-consumer-example.c}\end{promptcontent}
  \end{promptbox}
  \begin{promptbox}{FDP encoder example}
  \begin{promptcontent}\input{code/libfdp-usage-example.py}\end{promptcontent}
  \end{promptbox}
  \caption{FDP fuzzer and its corresponding encoder example with given values}
  \label{fig:libfdp-example}
\end{figure}

\PP{Supported FDP Encoders}%
\begin{squishitemize}
    \item \ccw{LlvmFdpEncoder}: For FuzzedDataProvider in libfuzzer (targets written in C/C++)
    \item \ccw{JazzerFdpEncoder}: For FuzzedDataProvider in Jazzer (targets written in Java)
\end{squishitemize}

Each encoder exposes a set of producer functions that correspond one-to-one with the consumer functions in the target harness.
%For instance, `\ccw{produce_jint_in_range}' encodes an integer for Jazzer's `\ccw{consumeInt(min, max)}'.
\autoref{fig:libfdp-example} shows the example of a fuzzing harness and the corresponding encode with \libfdp.
To produce the bytes consumed by the functions in the harness, the FDP encoder calls the corresponding encoding functions (\eg \ccw{produce_unsigned_short_in_range} for \ccw{ConsumeIntegralInRange<uint}).
As as result, the encoder will produce \ccw{1}, \ccw{128}, \ccw{true}, \ccw{1}, \ccw{"abcd"}, respectively for variable \ccw{a}, \ccw{b}, \ccw{c}, \ccw{d}, \ccw{e} in the fuzzing code.

\PP{Reference Implementations}%
\libfdp includes reference C++ code for both LLVM and Jazzer FDPs, ensuring compatibility and correctness. The encoder logic is carefully designed to match the semantics of the original consumer APIs, but some caveats apply (see below).

\PP{Caveats and Limitations}%
\libfdp was used by Testlang-based Generation/Mutation module and \sysmlla, which generates python scripts for mutating and generating inputs.
Concolic execution also utilizes \ccw{fdp-reference} for modeling FDP related functions.
However, \libfdp has some caveats and limitations.
\begin{squishitemize}
    \item \textit{Encoding is not always invertible}: Due to information loss in some consumer APIs (e.g., string and char consumption), the encoded blob may not exactly match the original target input.
    \item \textit{Floating point handling}: Floating point encoding may yield different results depending on platform or build options.
    \item \textit{Floating point quantization}: Encoding arbitrary floating point numbers is not possible due to the nature of FDP implementations. \libfdp can only encode exact target values when the value is producible by the corresponding consumer API.
    \item \textit{Unchecked APIs}: For cases where strict checks would reject legal sequences (e.g., due to indeterminism in FDP), \libfdp provides `\ccw{_unchecked}' variants of encoder functions.
\end{squishitemize}

\clearpage
\section{\syspatching}
\label{s:crs-patching}

\subsection{Overview}

\begin{figure*}[t]
  \centering
  \resizebox{\textwidth}{!}{
\begin{tikzpicture}[
  module/.style={anchor=north west, rectangle, draw, minimum width=4.5cm, minimum height = 0.5cm, very thick, font=\small, fill=gray!10},
  bmodule/.style={anchor=north west, rectangle, draw, minimum width=9cm, minimum height = 0.5cm, very thick, font=\small, fill=gray!10},
  moduletxt/.style={font = \small},
  smodule/.style={anchor=north west, rectangle, draw, minimum width=3.75cm, minimum height = 0.5cm, very thick, font=\footnotesize, fill=white},
  smoduletxt/.style={font = \footnotesize},
  ssmodule/.style={rectangle, draw, rounded corners = 1mm, minimum width=3.25cm, , minimum height=0.5cm, thick, font=\scriptsize, align=center, fill=gray!10},
  ssmoduleLLM/.style={ssmodule, fill=GreenYellow},
  arrow/.style={-{Latex[length=2mm, width=2mm]}, ultra thick},
  arrowBi/.style={{Latex[length=2mm, width=2mm]}-{Latex[length=2mm, width=2mm]}, ultra thick},
  arrowBig/.style={-{Latex[length=3mm, width=3mm]}, line width=3pt},
  arrowBigBi/.style={{Latex[length=3mm, width=3mm]}-{Latex[length=3mm, width=3mm]}, line width=3pt},
  arrowTxtR/.style={right, font=\scriptsize, pos=0.5, align=left},
  arrowTxtB/.style={right, font=\footnotesize, pos=0.5, align=left},
  db/.style={cylinder, draw, minimum width = 4.5cm, minimum height = 5cm, anchor=north west, very thick, shape border rotate=90, shape aspect=.90, font=\small, fill=gray!10}
]

\newcommand*{\yInitGap}{0.8}
\newcommand*{\yGap}{1.15}
\newcommand*{\ySmallGap}{0.7}
\newcommand*{\cx}{-11}
\newcommand*{\ax}{-7.5}
\newcommand*{\bx}{0}

\newcommand*{\txtX}{0.05}

\newcommand*{\intY}{1.1}
\newcommand*{\ay}{0}

% CRS-patch main
\newcommand*{\yMain}{\ay}
\node[module, minimum height = 4.25cm] at (\ax, \yMain) (main) {};
\node[moduletxt, below] at (main.north) {\textsc{\textbf{\sys-Patching main}}};
\newdimen\xMain
\pgfextractx{\xMain}{\pgfpointanchor{main}{north}};

\node[smodule, align=center, anchor=north] at (\xMain, \yMain-\yInitGap)(dispatcher) {
  \textbf{Dispatcher}
};

\node[smodule, align=center, anchor=north] at (\xMain, \yMain-\yInitGap-\yGap)(patchmgr) {
  \textbf{Patch Manager}\\(PoV Deduplication)
};

\node[smodule, align=center, anchor=north] at (\xMain, \yMain-\yInitGap-\yGap*2-0.35)(submitter) {
  \textbf{Submitter}
};

\draw[arrowBi] (dispatcher.south) -> (patchmgr.north);
\draw[arrowBi] (patchmgr.south) -> (submitter.north);

% CRS-patch sub
\newcommand*{\ySub}{\ay}
\node[module, minimum height = 6.25cm] at (\bx, \ySub) (sub) {};
\node[moduletxt, below] at (sub.north) {\textsc{\textbf{\sys-Patching sub}}};
\newdimen\xSub
\pgfextractx{\xSub}{\pgfpointanchor{sub}{north}};

\node[smodule, align=center, anchor=north] at (\xSub, \ySub-\yInitGap)(scheduler) {
  \textbf{Scheduler}
};

\newcommand*{\yCrete}{\ySub-\yInitGap-\yGap}
\node[smodule, align=center, anchor=north, minimum height = 4cm] at (\xSub, \yCrete) (crete) {};
\node[smoduletxt, below] at (crete.north) {
  \textsc{\textbf{\syscrete}}~(\autoref{ss:crete})
};

\draw[arrowBi] (scheduler.south) -> (crete.north);

\newdimen\xCrete
\pgfextractx{\xCrete}{\pgfpointanchor{crete}{north}};

\node[ssmodule] at (\xCrete, \yCrete-\yInitGap) {
  \textbf{\sysmartian}~(\autoref{ss:martian})
};
\node[ssmodule] at (\xCrete, \yCrete-\yInitGap-\ySmallGap) {
  \textbf{\sysmr}~(\autoref{ss:multiretrieval})
};
\node[ssmodule] at (\xCrete, \yCrete-\yInitGap-\ySmallGap*2) {
  \textbf{\sysprism}~(\autoref{ss:prism})
};
\node[ssmodule] at (\xCrete, \yCrete-\yInitGap-\ySmallGap*3) {
  \textbf{\sysclaudelike}~(\autoref{ss:claudelike})
};
\node[ssmodule] at (\xCrete, \yCrete-\yInitGap-\ySmallGap*4) {
  \textbf{\syseraser}~(\autoref{ss:custom-model})
};

% Detection
\newdimen\yDispatcher
\pgfextracty{\yDispatcher}{\pgfpointanchor{dispatcher}{west}};

\node[smodule, minimum width=2.6cm, minimum height=1.5cm, anchor=west] at (\cx, \yDispatcher) (detection) {};
\node[smoduletxt] at (detection.center) {
  \textsc{\textbf{PatchRequest}}
};

\draw[arrowBig](detection.east) -> (detection.east -| main.west);

\newdimen\xArrow
\newdimen\yArrow
\pgfextractx{\xArrow}{\pgfpointanchor{crete}{west}};
\pgfextracty{\yArrow}{\pgfpointanchor{patchmgr}{east}};
\draw[arrowBig](dispatcher.east) -> (scheduler.west) node[arrowTxtB, above] {\textbf{Patch Request}};
\draw[arrowBig](\xArrow, \yArrow) -> (patchmgr.east) node[arrowTxtB, above] {\textbf{Generated Patch}};

% VAPI
\newcommand*{\yVapi}{\ay-5.35}
\node[db, minimum height = 1cm] at (\ax+1.68, \yVapi) (vapi) {};
\newdimen\xVapi
\pgfextractx{\xVapi}{\pgfpointanchor{vapi}{north}};

\node[moduletxt, below] at (\xVapi, \yVapi-0.4) {\textsc{\textbf{\syscpmgr}}};

\draw[arrowBig](main.south) -> (main.south |- vapi.north) node[arrowTxtB, right] {\textbf{Verified Patch}};

\newcommand*{\gap}{0.5}
\draw[ultra thick, rounded corners=5pt]
     (\ax-\gap,\ay+\gap) -- (\bx+4.5+\gap,\ay+\gap) -- (\bx+4.5+\gap, \ay-6.15-\gap)
     -- (\bx-1,\ay-6.15-\gap) -- (\bx-1, \ay-4.1-\gap) -- (\ax-\gap, \ay-4.1-\gap)
     -- cycle;
    
\newcommand*{\xPatch}{-1.5}
\node[smodule, align=center, anchor=center, font=\normalsize] at (\xPatch, \ay+\gap) {
  \textsc{\textbf{\sys-Patching}~(\autoref{s:crs-patching})}
};

\end{tikzpicture}
}
    \caption{The overall architecture of \syspatching.}
    \vspace*{-4px}
  \label{fig:crs-patch-arch}
\end{figure*}

\syspatching is a subsystem within the \sys framework designed to automatically generate and deliver security patches. It leverages proof-of-vulnerability (PoV) and repository-level information to construct patches, and is implemented as a web server that exposes a set of APIs for handling patch requests and submitting validated patches to \syscpmgr.

\subsubsection{Workflow} The overall architecture of \syspatching is shown in
\autoref{fig:crs-patch-arch}. When other subsystems detect a vulnerability, this
information is delivered to the main node, which we referred to as
\cc{PatchRequest}. \cc{PatchRequest} includes the project name, sanitizer type,
harness name, and crashing input, which are essential for reproducing the
vulnerability.  Then, the main node first deduplicates the request as \syscpmgr
failed to duplicate the PoV properly.  Thus, the main node uses previous patches
for further deduplication.  If the request is unique, it is forwarded to sub
nodes, and the scheduler distributes the \cc{PatchRequest} to multiple patching
agents, each of which generates candidate patches. These patches are returned to
the main node, where they are validated, deduplicated, and finally submitted to
\syscpmgr.

\subsubsection{Key Ideas}
In this section, we highlight the key ideas behind the design of \syspatching.

\PP{Ensemble Agents.}
One of the key ideas of \syspatching is the use of ensemble agents, which combine multiple patching methods through ensembling. Ensembling~\cite{opitz1999popular} is a well-established approach in machine learning that integrates diverse models to improve overall performance, and it is particularly effective in automatic patch generation. This effectiveness arises from the fact that generated patches can be validated through post-generation checks, such as compilation success, PoV mitigation, and functional testing. As these checks are incomplete and cannot guarantee correctness --- since even validated patches may still be incorrect --- we refer to such patches as \emph{plausible patches}~\cite{qi2015analysis}. Thus, in AIxCC, manual verification is further applied to address this limitation. Nonetheless, this partial verifiability of patches makes ensembling highly beneficial, as leveraging multiple agents increases the likelihood of producing a valid patch.

Ensembling provides two primary benefits: performance improvement and reliability. First, in terms of performance, ensembling enhances the system by integrating the strengths of each agent. Our observations indicate that LLMs do not operate in a principled or predictable manner. Our internal experiments further revealed that no single agent consistently outperformed all others across workloads; rather, the best-performing agent varied depending on the task. To address this, \syspatching employs an ensemble strategy, running diverse agents in parallel so that success is achieved if any one of them generates a valid patch.

Second, in terms of reliability, ensembling enables the system to remain robust across diverse workloads, which is crtical in the AIxCC setting. In AIxCC, our system should work autonomously across a wide range of projects and languages, without human intervention. Unfortunately, it is extremely difficult to construct such a reliable system as individual components often fail under specific conditions. For instance, we found that certain projects may not be compiled with a certain compiler (e.g., \cc{clang}) or with certain flags (e.g., \cc{-g}). Moreover, there can be projects that dynamically generate code at build time or even merge multiple code into a single file (e.g., \cc{sqlite3}), which breaks a certain static analysis. Thus, a system that depends on a single method is vulnerable to such failures. Ensembling can mitigate this risk by distributing reliance across multiple implementations, analogous to N-versioning~\cite{chen1978n}. Consequently, even if one implementation fails, the overall system can continue to operate reliably.

\PP{Universal Framework for Agent Development.}
To implement ensembling effectively, it is essential to provide a framework that enables efficient development and integration of diverse patching agents. For this purpose, we introduce \syscrete, a universal framework that unifies core functionalities required by agents. Although individual agents are designed differently, they share common operations such as building and PoV testing. In addition, several functionalities --- such as file handling, code parsing, and patch generation --- can be shared among agents. While individual developers could independently implement these capabilities, it may duplicate effort and reduce overall efficiency. By offering \syscrete as a shared framework, we ensure that common functionalities are reusable and accessible across agents. This allows developers to focus on designing core algorithms and specialized strategies, thereby accelerating the development process and improving the overall quality of the system.

\PP{Two-level Policy Enforcement}
Since agents are developed independently, some may behave abnormally, potentially undermining the reliability of the overall system. To mitigate this risk, patches must adhere to a set of mandatory policies. Specifically, a patch must first be plausible (i.e., a patch should be compilable and prevents PoV). Moreover, it must not modify harness files, and the changes must be restricted to source code files such as C or Java. Furthermore, patches should aim to be correct rather than merely plausible, since all submissions are manually validated after the competition. In practice, however, agents often generate plausible but incorrect patches, for instance, by removing functionality entirely or suppressing errors through a large \cc{try-catch} block. While some of these policies can be enforced through rule-based checks, others are more difficult to validate in such a manner.

To address this challenge, \syspatching employs a \emph{two-level policy enforcement mechanism}. The first level is agent-level. We explicitly communicated the required policies to all agent developers, who incorporated them into their prompts. This strategy strengthens patch generation by providing feedback and helps enforce policies that are not easily rule-based. However, it remains prone to developer error and to the inherent unpredictability of LLM-generated outputs. Thus, we implement a second level of verification at the system level. Once a patch is generated, the main node of \syspatching applies rule-based checks to validate the patch. Particularly, it revalidates plausibility of a patch by recompilation and PoV testing. Additionally, it ensures that it modifies only source code files and not harness files. Through this two-level verification process, the system compensates for agent-level errors and ensures robustness across the entire patching workflow.

% \PP{Detection \& Patch request.} \cc{Detection} and \cc{PatchRequest} are objects that carry all the information necessary to reproduce and triage the bug. They include project name, project mode --- full mode or diff mode ---, sanitizer type, harness name, crashing inputs. If available, they also include the SARIF report, which would be validated by the CP Manager before submission. These objects are used by each agent and node as the main input for patch generation.

% \begin{squishitemize}
%   \item 
%   \item Sanitizer type used to detect the issue
%   \item Fuzzing harness name
%   \item Crashing inputs that trigger the vulnerability
%   \item Associated SARIF report (if available)
% \end{squishitemize}

% \noindent This is submitted by the CP Manager in the form of \TODO{...} + more detail

\subsection{Node Architecture}
The design of \syspatching requires an execution environment that supports parallelized builds and patch validation. A primary obstacle in adopting the OSS-Fuzz framework~\cite{oss-fuzz} within the AIxCC competition setting lies in its build process. The original OSS-Fuzz build scripts are not designed for parallel execution, as build artifacts are stored on mounted volumes, which complicates process isolation without modifying the build scripts.
Unfortunately, it is risky to modify these scripts, as we cannot guarantee that the modified scripts will work correctly across all projects.

To overcome this limitation, we deployed multiple Azure nodes, each operating within its own isolated environment and equipped with dedicated hardware resources, such as network bandwidth. This setup enables fully parallelized execution of multiple agents, allowing OSS-Fuzz builds and reproduction tasks to run concurrently. Particularly, we provisioned one node for the \syspatching main node and four nodes for the sub nodes, each equipped with 32 vCPUs and 128 GB RAM. Furthermore, we assigned agents to nodes according to their underlying LLMs to avoid triggering excessive LLM rate limits. For example, we deployed \sysmr with claude-3.7-sonnet and \sysclaudelike at the same node, as both agents utilize Claude LLMs. Meanwhile, we deployed \sysprism and \sysaider at a separate node, as both agents rely on OpenAI LLMs. This strategic allocation of agents to nodes helps mitigate the risk of rate limiting and ensures smooth operation across the system.

\subsubsection{Main node} 

The \syspatching main node coordinates the entire patching workflow. 
Given each incoming request, the main node processes it through three stages: de-duplication, dispatch, and submission.

\PP{Deduplication}
\label{ss:deduplication}
When a new request arrives, the main node first performs de-duplication. This step is necessary because the de-duplication conducted by \syscpmgr is inherently incomplete. The \syscpmgr relies on organizer-provided methods and internal heuristics to identify duplicate vulnerabilities. However, as vulnerability de-duplication is a fundamentally hard problem~\cite{cui:rept,cui2016retracer,jiang2021igor}, it is possible that some duplicates are not detected via these methods.

To mitigate this limitation, we employ a de-duplication strategy using patches. Specifically, \syspatching checks whether previously generated patches can block the newly submitted PoV. If an earlier patch successfully mitigates the new PoV, this implies that both PoVs originate from the same underlying vulnerability. The new PoV is therefore classified as a duplicate, and no new patch is generated. Through this mechanism, \syspatching efficiently eliminates redundant work while ensuring that no unique vulnerabilities are overlooked.

\PP{Dispatch} If a PoV is determined to be unique, the main node dispatches it to the sub nodes for patch generation. The dispatch mechanism is implemented in a multi-threaded manner, allowing the main node to continue receiving new PoVs while awaiting responses from sub nodes.

Among the patches generated by sub nodes, \syspatching selects the first valid patch returned. While more sophisticated selection policies are possible, 
we chose this simple approach for two reasons. First, as aforementioned, our system should be robust and reliable. If we wait for multiple nodes to respond, we should handle risks that some nodes may not respond indefinitely due to failures or network issues. If we failed to handle such cases, the entire system may become unresponsive. Thus, we prioritize responsiveness and simplicity in our design.
Second, in the AIxCC competition, the scoring is influenced by a time-based multiplier, making early patch submission advantageous.
Therefore, we just select the earliest patch and submit it upstream, instead of using more complex policies (e.g., LLM-as-a-judge~\cite{gu2024survey} or self-consistency~\cite{wang2022self}).

\PP{Submission} Once a valid patch is identified, the main node proceeds with the submission stage. Before sending the patch upstream to \syscpmgr, it performs two types of checks: de-duplication and policy enforcement.

First, the main node performs an additional round of de-duplication. Since \syspatching operates in parallel across multiple PoVs, it is possible that the current PoV is already resolved by a patch generated for another PoV. To handle such race conditions, the main node re-checks whether the patch is redundant.

Second, the main node enforces policies through a set of pre-defined rules to ensure the integrity of the submission. Specifically, it verifies that the patch compiles successfully, blocks the PoV, and modifies only valid source code files. Even though we instructed agents to adhere to these policies, it is possible that some agents may not fully comply due to developer error or the inherent unpredictability of LLMs. To this end, we leverage an ML-based file type detection tool, which is provided by the organizer, to confirm that the modified file is indeed source code (e.g., C or Java) rather than auxiliary artifacts. Additionally, we applied heuristic checks to ensure that harness files are not altered. Then, we submit the patch to \syscpmgr only if it passes all these checks.

\PP{Sub node} 
Each sub node holds a set of agents and executes them iteratively to generate patches for incoming requests. As described earlier, we configured the agent set of each sub node for diversity and to avoid LLM rate limits. Upon receiving a patch request, the scheduler within a sub node iterates through its set of agents, invoking each agent to attempt patch generation. If a patch is successfully generated, it is returned to the main node, where it validates, enforces policies, and finally submits the patch.

\subsection{\syscrete: A Unified Framework for Patch Generation Framework}
\label{ss:crete}

In this section, we describe \syscrete, a framework for developing automated patching agents, and we briefly introduce the agents built on top of it.
After this section, we describe each agent in detail in \autoref{ss:martian}--\autoref{ss:opensourceagents}.

\subsubsection{Overview}

\syscrete is a framework designed to provide a standardized environment and essential 
utilities for developing automated patching agents. It serves as a foundational layer 
that abstracts repository management, build configuration, and patch validation, 
thereby enabling heterogeneous agents to operate on a common platform. The framework initializes target 
repositories for specific vulnerabilities, delivers them to designated agents, and 
subsequently validates the resulting patches through an integrated evaluation process.  

The primary objective of \syscrete is to facilitate the development of diverse patching 
agents while preventing environmental conflicts and reducing redundant implementation. 
By offering a stable and unified interface, the framework minimizes the risk of errors 
and improves development efficiency, allowing developers to focus on agent-specific 
logic. This design has enabled the concurrent operation of multiple, distinct agents 
within our system.  

\subsection{Core Components}

\syscrete comprises several core components that collectively support the patching workflow. These components include:

\begin{itemize}
  \item \textbf{Environment.} Provides managed interfaces for interacting with challenge projects, 
  provisioning distinct build environments, and leveraging caching mechanisms to avoid redundant 
  compilations. This accelerates both testing and development cycles.  

  \item \textbf{Evaluator.} Integrates with OSS-Fuzz to rigorously assess generated patches. 
  It verifies compilation, reproduction, and functional correctness to ensure that patches 
  mitigate vulnerabilities without introducing regressions.  

  \item \textbf{Fault Localizer.} Analyzes crash logs and SARIF reports to identify likely 
  fault locations, offering a baseline that agents can refine for precise fault localization 
  and patch synthesis.  

  \item \textbf{Code Retriever.} Supplies an API for code navigation, enabling agents to query 
  program elements such as functions or variables. It leverages tools like \cc{tree-sitter}~\cite{tree-sitter} and 
  \cc{ctags}~\cite{ctags} to support efficient source-code analysis.  

  \item \textbf{Common Analyzers.} Provides reusable analyzers for crash logs, call traces, 
  and commits, allowing agents to extract actionable insights.  
\end{itemize}

\noindent In the following, we briefly describe several interesting features that \syscrete provides.

\PP{Environment Pool}
To optimize the build process, \syscrete employs an \emph{environment pool}. Unlike fuzzing, which generally requires a single build, patch generation involves repeated builds for testing and validation. The environment pool addresses this by maintaining pre-configured environments consisting of reusable artifacts such as compiled binaries and libraries, thereby avoiding redundant builds. \syscrete supports multiple types of environments for fault tolerance. For instance, some environments use \cc{ccache} to reduce build time, while others exclude \cc{ccache} to handle cases where it is unreliable. When an agent requests an environment, \syscrete allocates one from the pool according to the agent’s preferences and current availability. The same environments are also used for static analyses, such as call-trace or debug-symbol extraction.

\PP{Cache Everywhere}
To improve efficiency, \syscrete adopts a \emph{cache-everywhere} strategy. It caches all results that are computationally expensive yet deterministic, such as crash reproduction, patch validation, and selected static analyses. This mechanism eliminates redundant computations, allowing agents to reuse previously obtained results transparently. For example, a developer can invoke the patch validation API multiple times with the same patch without incurring additional overhead. By automating caching, \syscrete reduces performance concerns and allows developers to focus on the core logic of their agents.

\PP{Special Handling for Crash Logs}
\syscrete also includes specialized handling for certain types of crash logs. For instance, stack overflow crashes in Java often produce repetitive stack frames, which hinder effective analysis. To mitigate this, \syscrete preprocesses stack traces to remove redundant frames. Similarly, timeout bugs in Java typically do not generate stack traces, complicating the initial analysis. To address this, \syscrete leverages \cc{jstack}~\cite{jstack} to capture stack traces when a timeout occurs. In particular, if the crash is a timeout for Java, \syscrete reproduces the crash while periodically capturing stack traces 30 times using \cc{jstack}.
Then, the collected stack traces are provided to agents for further analysis. As timeout bugs are often caused by infinite loops, the stack traces can help agents identify the loop location. By doing so, \syscrete can supply valuable diagnostic information independent of agent design and eliminates the need for redundant implementations.

\subsubsection{Agents built on \syscrete}
On top of \syscrete, we implemented multiple automated patching agents, each with distinct 
specializations and architectural designs (see \autoref{tab:patch-agents-comparison}). 
In the AIxCC competition setting, the primary goal is to produce at least one correct patch among many attempts. To maximize this 
likelihood, \syspatching exploits the diversity of agents via emsembling, leveraging differences in 
their underlying LLMs, reasoning strategies, and tool integrations. This diversity enables 
agents to explore different portions of the codebase and generate complementary patches, 
increasing the probability of finding a correct fix.  

In addition to our in-house agents, we incorporated open-source agents such as 
\sysaider~\cite{aider} and \syssweagent~\cite{yang2024swe}. Leveraging these external 
tools allows us to build on established implementations that have been validated by the 
developer community. Their inclusion also enhances patch diversity, as they are trained or 
tuned with different datasets, prompting strategies, and reasoning heuristics. Detailed 
descriptions of \sysaider and \syssweagent are provided in \autoref{ss:opensourceagents}.
In the following sections, we describe the agents that \syspatching employs, including those built on \syscrete as well as open-source agents.

\begin{table*}[!t]
\centering
\footnotesize
 \resizebox{\textwidth}{!}{
    \begin{tabular}{l l p{8cm} l}
\toprule
\textbf{Agent} & \textbf{Architecture} & \textbf{Motivation} & \textbf{Used models} \\
\midrule
\sysmartian (\autoref{ss:martian}) & Workflow & Introducing architectural diversity for robustness and to enable flexible integration of external tools via a ReAct-style design & o4-mini and claude-4-sonnet \\
\midrule
\sysmr (\autoref{ss:multiretrieval}) & Agent & Autonomously exploring codebases and generate vulnerability patches by iteratively building context through interaction & claude-3.7-sonnet or o4-mini \\
\midrule
\sysprism (\autoref{ss:prism}) & Multi-agent & Addressing context length limitations and prompt complexity in MULTIRETRIEVAL through a team-based multi-agent system. & o4-mini \\
\midrule
\sysvincent (\autoref{ss:vincent}) & Workflow & Incorporates project-specific properties from software verification to guide patching while maintaining generality & gemini-2.5-pro \\
\midrule
\sysclaudelike (\autoref{ss:claudelike}) & Agent & Inspired by Claude Code, featuring advanced file editor tools and sub-agent delegation to manage context efficiently and streamline patching tasks & claude-3.7-sonnet \\
\midrule
\syseraser (\autoref{ss:custom-model}) & Workflow & Aiming to be used for a custom model & Custom model \\
\bottomrule
\end{tabular}
}
\caption{Patching agents in \syspatching (except for open-source agents). Notably, an architecture in this table is just a high-level classification of the agents, and does not imply that the agents are implemented. For that, please refer to the individual agent descriptions.}
\label{tab:patch-agents-comparison}
\end{table*}

\subsection{Agent: \sysmartian}
\label{ss:martian}

\sysmartian agent is an AI agent that fixes security bugs with a fixed workflow that mimic the approach of human developers.
The agent decomposes the patching process into two stages: Fault Localization and Patch Generation.
Each stage is handled by a ReAct-based agent, equipped with specialized tools.
The core design philosophy behind \sysmartian agent is having a straightforward workflow with easily extendable external tools.
With this design, it can not only ensure effective vulnerability analysis and patch generation, but also provide the flexiblity to adapt new tools.

\subsubsection{Motivation}

The motivation behind the development of the \sysmartian agent was twofold.
First, we aimed to explore a different architectural approach compared to existing agents, such as the \sysmr agent.
By introducing architectural diversity, we intended to make our \syspatching framework more robust and adaptable across a wider range of problems.
Second, we wanted to simplify the integration and testing of external tools.
By leveraging a ReAct-style node with tool calls, the \sysmartian agent enables easy addition and removal of tools, providing the flexibility needed for rapid experimentation and adaptation.

\subsubsection{System Architecture}

\begin{figure}[t]
  \centering
  \includegraphics[width=.9\textwidth]{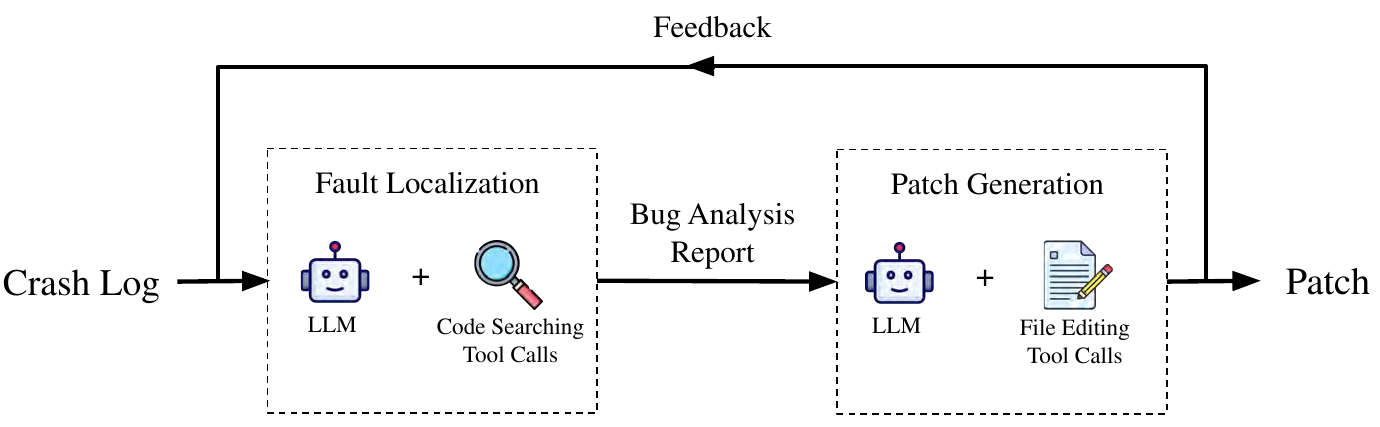}
    \caption{\sysmartian agent architecture}
  \label{fig:martian-architecture}
\end{figure}

The overall workflow of \sysmartian agent is shown in \autoref{fig:martian-architecture}.
First, it generates a crash log from the given security bug.
The crash log should include a sanitizer report or at least a stacktrace that clearly describe the bug.
The fault localization module then analyzes the crash log by searching the codebase and find the root cause of the bug.
As output, it identifies the buggy function and generate a plan to fix it.
These outputs are passed to the patch generation module, which finally generates a patch for the identified buggy function.
The generated patch is then evaluated, and if it fails validation, feedback is sent back and retries the process.

\PP{Fault Localization}
\sysmartian's fault localization module is inspired by \syscoderovers \cite{patch:CodeRover-S}.
Given that \syscoderovers generated 52.6\% plausible patches on the ARVO dataset, we selected it as a baseline.
Since the code search APIs of \syscoderovers are closed-source, we developed our own language-agnostic APIs for code search.
The fault localization module includes the following three APIs as tools:

\begin{squishitemize}
    \item \cc{\textbf{search_symbol(symbol_name, file_or_directory_path)}}: Given a symbol name, it finds a symbol definition in a given path and return the source code of the symbol definition.
    \item \cc{\textbf{search_string(string, file_or_directory_path)}}: Given a string, it returns the file name and line number(s) where the string appears in the codebase.
    \item \cc{\textbf{view_file(file_path, offset, limit)}}: Given a file name, offset, and limit, it returns the file content from the offset to the limit.
\end{squishitemize}

The code search APIs provide high-level, abstracted interface for robust and efficient code navigation.
The \cc{search_symbol} tool works similarly to "Go to Definition" in modern code editors, while \cc{search_string} works similarly to "Ctrl+Shift+F" for full-text search.
These tools serve as powerful primitives for the code exploration -- even for the human developers.
The \cc{search_symbol} tool operates on abstract symbols without requiring the LLM to specify their types (e.g., function, variable, or class).
This design keeps the interface simple, offloading the complexity of type handling to the implementation side of the tool.
Moreover, if \cc{search_symbol} fails, the LLM can fall back to the \cc{search_string} tool to continue the analysis.
These tool combinations enable flexible and resilent code search.

\PP{Patch Generation}
Generating an appliable patch is a non-trival task, as it must adhere to a strict format to be correctly applied.
To address this challenge, we separate the patch generation process from fault localization and have an independent ReAct agent with its own dedicated context.
This separation allows the patch generation module to focus more effectively on its task by adapting the context.

The patch generation module employs a search/replace strategy at the function level, implemented via tool calls.
This approach is inspired by Claude Code \cite{ClaudeCode}; however, unlike Claude Code, which operates at the file level, \sysmartian operates at the function-level.
By narrowing the scope of the patch generation, this method offers advantages in terms of context length.
Additionally, as other agents operate at the file level, \sysmartian agent focuses on a narrower, more precise scope, contributing to the overall diversity of our patch system. The patch generation module includes the following three APIs:

\begin{squishitemize}
    \item \cc{\textbf{view_function(function_name)}}: Given a function name, it returns the source code of the function.
    \item \cc{\textbf{edit_function(function_name, old_string, new_string)}}: Given a function name and search/replace code snippets, it edits the source code of the function.
    \item \cc{\textbf{add_import_module(module_name, file_path)}}: Add import statement in the given file. This is only used for JAVA.
\end{squishitemize}

\noindent With these APIs, the patch generator is able to fix the function and finally we use \cc{git diff} command to get the patch diff.
We need \cc{add_import_module} tool for JAVA because empirically there were few compilation error because of missing module imports.

\subsection{Agent: \sysmr}
\label{ss:multiretrieval}

The \sysmr represents an automated vulnerability patching system that combines multiple code retrieval strategies with iterative patch generation and validation.
This system leverages state-machine based workflows, diverse retrieval backends, and verifiable evaluations to generate patches for software vulnerabilities.
By integrating Abstract Syntax Tree (AST) augmented retrieval, grep-based retrieval, and direct file retrieval as tools, the agent achieves robust patch generation across C and Java codebases.

\subsubsection{Motivation}

The motivation behind \sysmr is to create an agent that can effectively explore the codebase and generate patches for software vulnerabilities by iteratively interacting with the codebase and evaluation system.
This approach allows the agent to build its own context incrementally, refining its understanding of the codebase and the vulnerability at hand.
This was inspired by the agentic capabilities of LLMs, where we only give right tools and goals to the agent, and it can explore the codebase and generate patches autonomously.

\subsubsection{System Architecture}

\begin{figure}[t]
    \centering
    \begin{minipage}[t]{0.46\textwidth}
        \centering
        \includegraphics[width=0.8\textwidth]{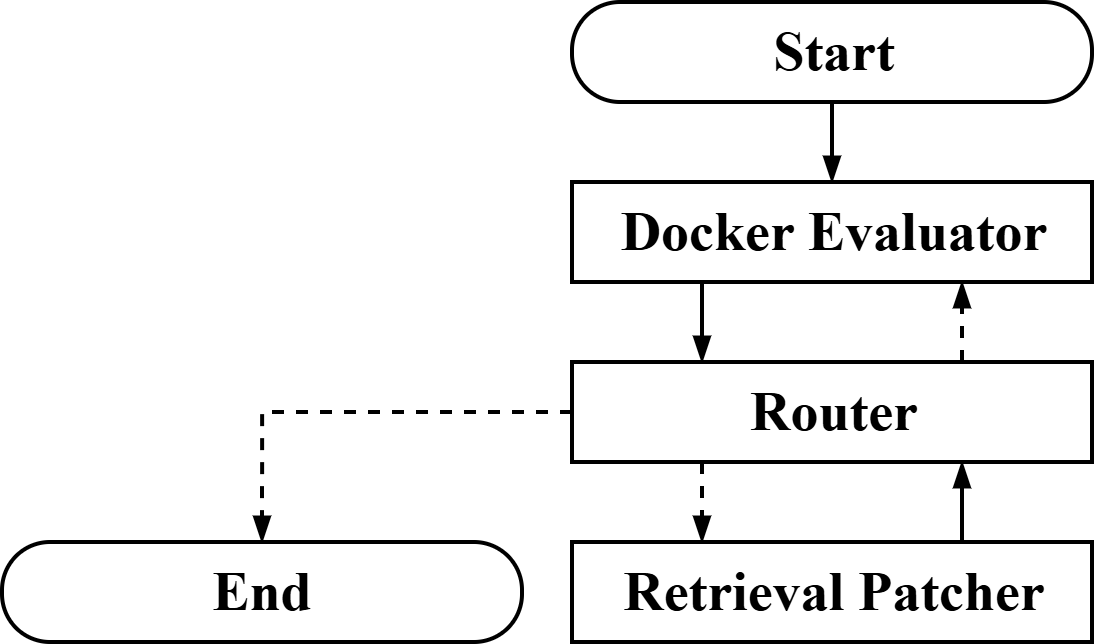}
    \end{minipage}
    \hfill
    \begin{minipage}[t]{0.46\textwidth}
        \centering
        \includegraphics[width=0.8\textwidth]{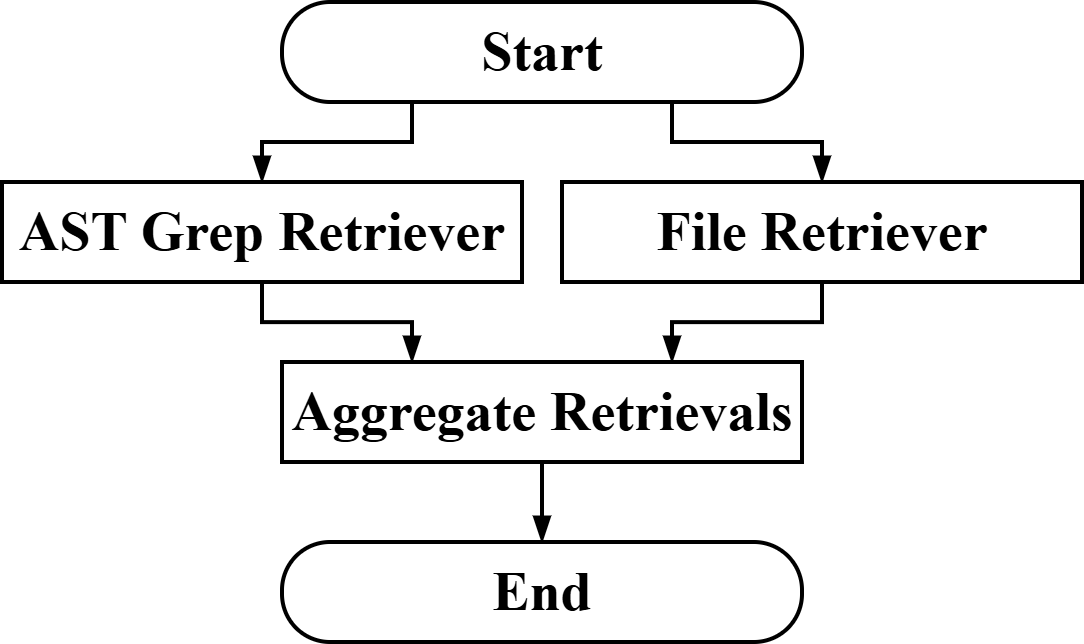}
    \end{minipage}
    \caption{The architecture of the \sysmr system, illustrating the interactions between the specialized nodes. The left figure shows the overall architecture, while the right figure focuses on the \cc{CodeRetriever} that manages multi-source code retrieval operations which be used by the \cc{SystemGuidedPatcher}.}
    \label{fig:multiret-arch-and-coderet}
\end{figure}

The \sysmr follows an iterative process where the agent queries the codebase to retrieve relevant context, generates patches based on the retrieved information to get a feedback from the evaluation system, and refines the patches iteratively until a valid patch is produced or the maximum number of evaluations is reached.
Although the agent operates within a single context with multiple turns of interaction, it uses three specialized nodes to handle different aspects of the patching process which are the \cc{SystemGuidedPatcher}, \cc{DockerEvaluator}, and \cc{CodeRetriever} illustrated in \autoref{fig:multiret-arch-and-coderet}.

\begin{algorithm}
\footnotesize
\DontPrintSemicolon
\SetKwSty{algokeywordsty}
\SetFuncSty{algofuncsty}
\SetArgSty{algoargsty}

\caption{MultiRetrieval Agent Workflow for Iterative Patch Generation}
\label{alg:multiret-agent-workflow}

Initialize PatchState with repository path\;
Set initial action to \cc{EVALUATE}\;
Initialize \cc{n\_evals} $\leftarrow$ 0, \cc{max\_n\_evals} $\leftarrow$ 10\;
\While{\cc{action} $\neq$ \cc{DONE} and \cc{n\_evals} $<$ \cc{max\_n\_evals}}{
    \If{\cc{action} == \cc{EVALUATE}}{
        Increment \cc{n\_evals}\;
        \If{\cc{status} == \cc{INITIALIZED}}{
            Run proof-of-vulnerability test to get initial crash log\;
        }
        \Else{
            Run \cc{DockerEvaluator} to assess current patch\;
            Update patch status based on evaluation results\;
        }
        \If{\cc{status} == \cc{PLAUSIBLE}}{
            Run internal tests\;
            \If{tests pass}{
                \cc{action} $\leftarrow$ \cc{DONE}\;
            }
            \Else{
                \cc{action} $\leftarrow$ \cc{ANALYZE\_ISSUE}\;
            }
        }
        \Else{
            \cc{action} $\leftarrow$ \cc{ANALYZE\_ISSUE}\;
        }
    }
    \ElseIf{\cc{action} == \cc{ANALYZE\_ISSUE}}{
        \cc{SystemGuidedPatcher} analyzes the issue\;
        Generate exploration plan\;
        \cc{action} $\leftarrow$ \cc{RETRIEVE}\;
    }
    \ElseIf{\cc{action} == \cc{RETRIEVE}}{
        \cc{SystemGuidedPatcher} analyzes current retrieval context and decides next step\;
        \If{needs more context}{
            Execute retrieval queries\;
            Aggregate and format results\;
        }
        \Else{
            Generate patches\;
            \cc{action} $\leftarrow$ \cc{EVALUATE}\;
        }
    }
}
\Return{final diff}\;
\end{algorithm}

\PP{SystemGuidedPatcher}
The \cc{SystemGuidedPatcher} is the main node that analyzes the current state and decides the next action based on the current patch status and retrieved context.
It uses a state machine to manage the patching process, which includes actions such as \cc{EVALUATE}, \cc{ANALYZE\_ISSUE}, and \cc{RETRIEVE} based on the decision made by the LLM.
The system prompt for the \cc{SystemGuidedPatcher} is designed to contain all the necessary information to guide the LLM to interact with the codebase and the evaluation system effectively.

\PP{DockerEvaluator}
The \cc{DockerEvaluator} is responsible for validating the generated patches in isolated containerized environments.
It runs the patch against the codebase and evaluates its effectiveness by checking whether the patch resolves the vulnerability and passes the internal test suite if available.
The evaluation results are categorized into multiple states such as \cc{PLAUSIBLE}, \cc{UNCOMPILABLE}, and \cc{VULNERABLE} to provide feedback to the \cc{SystemGuidedPatcher} for further actions.

\PP{CodeRetriever}
The \cc{CodeRetriever} is a specialized node that manages multi-source code retrieval operations.
It leverages various retrieval techniques, including AST-based retrieval, text-based retrieval, and direct file retrieval, to gather relevant code snippets and context for the patching process.
By integrating these diverse retrieval methods, the agent can explore the codebase more effectively and retrieve the most relevant information for patch generation.

\PP{Multi-turn State Machine.}
The \sysmr follows a multi-turn state machine workflow as illustrated in \autoref{alg:multiret-agent-workflow}.
The agent starts with an initial crash log and decides the next action based on the current status.
The system automatically returns the requested information whether it is the code snippets or the evaluation results as if the agent is interacting with a human developer.

\subsubsection{Code Retrieval Strategy}

The \cc{CodeRetriever} implements a code retrieval strategy that combines three complementary approaches to work with a variety of codebases:

\begin{squishitemize}
    \item \textbf{AST-based Retrieval:} Uses grep to locate possibly related code and Abstract Syntax Tree analysis augments the retrieval if available. 
    \item \textbf{Text-based Retrieval:} Employs regex-capable text search for fast pattern matching and is used as a fallback when AST-based retreival is failed.
    \item \textbf{Direct File Retreival:} Retrieves complete files when specific context is needed.
\end{squishitemize}

\noindent The retrieval system processes LLM-generated queries by executing them in parallel across available retrievers.
Results are aggregated with a predefined priority based filtering to ensure the most relevant snippets are returned to the agent.
For example, AST-based retrieval is prioritized over text-based retreival where if returns the whole function definition instead of few surrounding lines when queried with a function name.
This augmented retrieval strategy allows the agent to efficiently gather context from large codebases, improving the quality of generated patches. 

\subsection{Agent: \sysprism}
\label{ss:prism}

\sysprism is a multi-agent system that employs a hierarchical team-based architecture with three specialized agent teams—Analysis, Patch, and Evaluation—collaborating to analyze vulnerabilities, generate patches, and validate their effectiveness.

\subsubsection{Motivation}

During the development of \sysmr, we observed that in some cases, the agent could not generate a valid patch within the maximum context length.
Also, the overall system prompt had multiple responsibilities, which made it difficult to manage its effectiveness.
To address these issues, we designed \sysprism as a multi-agent system that separates the responsibilities of each agent into specialized teams.
This team-based architecture allows for more focused and efficient interactions, where the workflow naturally compacts the context by sending only the relevant information and reduces the cognitive load on individual agents.

\subsubsection{System Architecture}

\begin{figure}[t]
  \centering
  \includegraphics[width=0.85\textwidth]{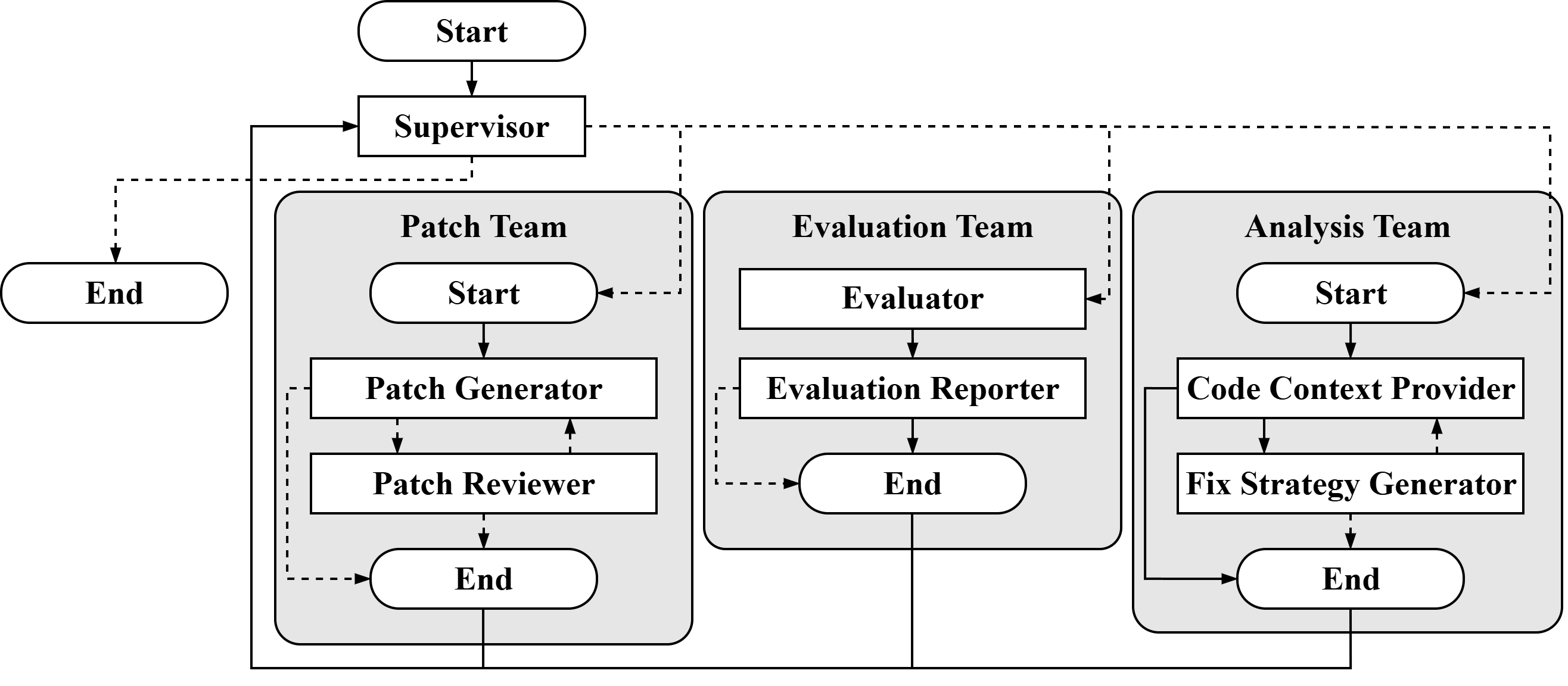}
    \caption{The architecture of the \sysprism system, illustrating the interactions between the specialized agent teams.}
  \label{fig:prism-architecture}
\end{figure}

Three specialized teams within \sysprism are coordinated by a supervisor which executes a iterative deterministic workflow of analysis, patch generation, and evaluation processes as illustrated in \autoref{fig:prism-architecture}. The system follows a deterministic workflow pattern which iterates through the teams until a valid patch is generated or the maximum number of iterations is reached. It iterates through evaluation, analysis, and patch generation steps, with each team handing off cumulative results to the next team in the cycle.

\subsubsection{Team Implementations}

\PP{Evaluation Team}
The Evaluation Team generates an evaluation report to provide a general description of the crash log and the result of the patch evaluation for next iteration if needed. It consists of two agents: \cc{Evaluator} and \cc{EvaluationReporter}.
The \cc{Evaluator} executes patches against the codebase, running proof-of-vulnerability tests and internal test suites to determine patch effectiveness. It handles various failure modes, from compilation errors to internal test failures, and processes logs with appropriate filtering and formatting to extract relevant information from verbose output, handling language-specific error formats and filtering noise to increase the token efficiency and overall accuracy by reducing distractors. The \cc{EvaluationReporter} generates comprehensive reports for failed patches, analyzing failure patterns and providing actionable feedback. These reports guide subsequent analysis and patch generation cycles, enabling the system to learn from unsuccessful attempts and progressively refine its approach.

\PP{Analysis Team.}
The Analysis Team is responsible for building the relevant code context while finding the root cause of the current issue and suggesting a comprehensive fix strategy structured as an analysis report. It consists of two agents: \cc{CodeContextProvider} and \cc{FixStrategyGenerator}.
The \cc{CodeContextProvider} explores the codebase to gather relevant code snippets and build a comprehensive understanding of the current context. This agent employs both top-down and bottom-up exploration strategies, using AST augmented grep-based searches and file retrieval to construct an analysis notebook incrementally where it will serve as the code context to the \cc{FixStrategyGenerator}. The \cc{FixStrategyGenerator} takes the collected code snippets and analysis to generate a detailed fixing strategy that can guide the Patch Team. The agent can iterate multiple times to refine its strategies by retrieving additional context.

\PP{Patch Team}
The Patch Team transforms analysis and evaluation reports into a concrete patch formatted as an unified diff. It consists of two agents: \cc{PatchGenerator} and \cc{PatchReviewer}.
The \cc{PatchGenerator} creates patches based on the fix strategy and evaluation results described in the reports, maintaining awareness of previous patch attempts to avoid repetitions or ineffective changes. The \cc{PatchReviewer} validates the generated patches for correctness and adherence to predefined coding standards, ensuring that they are syntactically correct and semantically meaningful. This review process helps catch potential issues before evaluation, improving the efficiency of the overall system by reducing the number of obviously flawed patches that reach the evaluation stage.

\subsubsection{State Management}

\sysprism employs a hierarchical state management system that maintains context across team boundaries while allowing for team-specific extensions. The maintained context mainly includes code snippets, analysis report, evaluation report, and patch diff. These are shared across the teams and condensed, replaced, or updated throughout the iterative process giving the system a coherent understanding of the current state and maintaining continuity. The team-specific state extensions allow each team to focus on its specific tasks with additional context. This hierarchical state management enables efficient communication and collaboration between teams, ensuring that only the relevant information is passed along. For example, the Analysis Team contructs an code analysis notebook that contains multiple code snippets and analysis for each code snippet, which is only used to generate a fix strategy. The Patch Team, on the other hand, only needs the final fix strategy and the code snippets where it does not need to access the entire analysis notebook which increases the operating efficiency of the system. This state management system combined with multiple layers of error handling and feedback loops allows \sysprism to robustly navigate through the codebase, iteratively refining its understanding and improving the quality of generated patches.

\subsection{Agent: \sysvincent}
\label{ss:vincent}

\sysvincent is an agent that includes a property analysis step in its pipeline, aiming to generate more context-aware patches.

\subsubsection{Motivation} % \IY{Maybe motivation?}
In our experiments of agents in \syspatching, we identified a challenging case that LLMs fail to generate a valid patch. 
In \cc{nginx}, there is a function named \cc{ngx_http_process_prefer} that handles \cc{Prefer} headers in HTTP requests.
As shown in \autoref{fig:nginx-buggy-function}, the function checks whether the \cc{Prefer} header has already been found (line 8). If the pointer is not \cc{NULL}, it frees the existing pointer and returns \cc{NGX_OK}.
However, a problem arises when the HTTP request contains multiple \cc{Prefer} headers, as illustrated in \autoref{fig:nginx-double-free-poc}.
The malformed HTTP request contains three \cc{Prefer} headers, which is not an allowed situation for \cc{nginx}.
Due to the first \cc{Prefer} header, memory is allocated to \cc{r->headers_in.prefer}.
When the program tries to handle the second header, it enters the \cc{if}-block (line 8 in \autoref{fig:nginx-buggy-function}) since \cc{r->headers_in.prefer} already has a value.
However, the problematic pointer is not properly nullified, which fails to prevent future access to the dangling pointer.
Finally, when the third header is processed, the \cc{if}-block is entered again, triggering the double-free of \cc{r->headers_in.prefer}.

\begin{figure}[t]
  \centering
  \begin{promptbox}{}
  \begin{promptcontent}\input{code/ngx_http_process_prefer.c}\end{promptcontent}
  \end{promptbox}
    \caption{The part of the \cc{ngx_http_process_prefer} function. This function contains the double-free bug (line 10) due to improper handling of \cc{Prefer} headers in the invalid HTTP request.}
  \label{fig:nginx-buggy-function}
\end{figure}

\begin{figure}[t]
  \centering
\colorbox{lightgray!30}{%
  \parbox{0.95\linewidth}{\ttfamily\small
    GET / HTTP/1.1\\
    Host: localhost\\
    Prefer: FirstPrefer\\
    Prefer: SecPrefer\\
    Prefer: ThirdPrefer\\
    Accept: */*
  }
} 
    \caption{The PoC for \cc{nginx} double-free vulnerability. This HTTP request contains three prefer headers, leading to entering the \cc{if}-block in \autoref{fig:nginx-buggy-function} (line 8) multiple times.}
  \label{fig:nginx-double-free-poc}
\end{figure}

To fix this bug, the patch needs to (i) nullify the problematic \mbox{\cc{r->headers_in.prefer}} pointer, (ii) finalize the current request with \cc{NGX_HTTP_BAD_REQUEST}, and (iii) return \cc{NGX_ERROR} instead of \cc{NGX_OK}. The intended patch is presented in \autoref{fig:nginx-double-free-patch}.
However, our agents kept failing to generate a working patch, leading to incomplete patches like \autoref{fig:nginx-double-free-patch-wrong}.
The main reason for failure was that the generated patches only nullified the problematic pointer, missing the proper cleanup process for an invalid HTTP request (\textit{i.e.,} steps (ii) and (iii)).
In other words, LLMs only focused on fixing the superficial symptom presented in the ASAN report.

\begin{figure}[t]
  \centering
  \begin{promptbox}{}
  \begin{promptcontent}\input{code/nginx-double-free-patch.diff}\end{promptcontent}
  \end{promptbox}
    \caption{The intended patch for \cc{nginx}'s double-free bug. The patch requires the function to (i) free the problematic pointer, (ii) finalize the request with \cc{NGX_HTTP_BAD_REQUEST}, and (iii) return \cc{NGX_ERROR} to prevent further processing of the given request.}
  \label{fig:nginx-double-free-patch}
\end{figure}

\begin{figure}[t]
  \centering
  \begin{promptbox}{}
  \begin{promptcontent}\input{code/nginx-double-free-patch-wrong.diff}\end{promptcontent}
  \end{promptbox}
    \caption{The typical patch generated by LLMs. It nullifies the problematic pointer, but fails to consider nginx-specific cleanup process for the invalid HTTP request.}
  \label{fig:nginx-double-free-patch-wrong}
\end{figure}

To deal with such cases, we built an LLM agent named \sysvincent.
The core idea is to enable LLMs to consider project-specific aspects without losing generality.
To achieve this, \sysvincent borrows the concept of \emph{properties}~\cite{4556678}, which refers to a condition or behavior that a program must satisfy, from the field of software verification and engineering.
% \IY{divide into subsection}

\begin{figure}[t]
  \centering
  \includegraphics[width=0.9\textwidth]{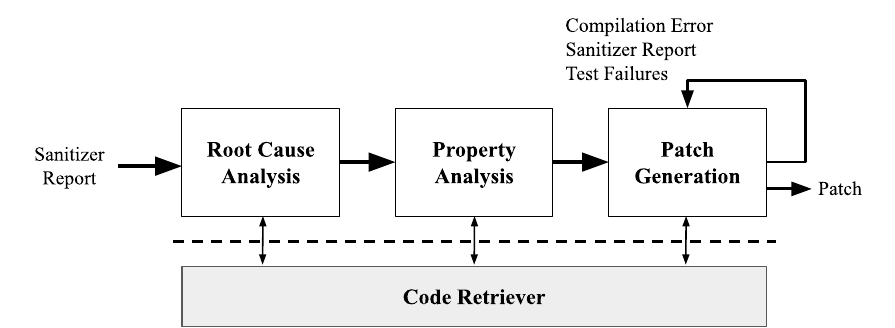}
    \caption{The overall workflow of \sysvincent agent. Given a sanitizer report and PoC, it conducts root cause analysis and property analysis. Based on the analysis result, it generate security patch in an iterative approach.}
  \label{fig:vincent-workflow}
\end{figure}

\subsubsection{Agent Workflow}
At a high-level, \sysvincent consists of three analysis steps, as presented in \autoref{fig:vincent-workflow}.
During the workflow, each analysis step can retrieve code information from the code retriever whenever the LLM wants.
For the first step, \sysvincent conducts root cause analysis given a sanitizer report and PoC.
Then, \sysvincent extracts the program properties related to the given bug.
Finally, based on the root cause and property analysis report, \sysvincent generates a patch. 
As with other agents in \syspatching, \sysvincent repeats patch generation until it produces a plausible patch.

% \IY{add more detail}

\subsubsection{Code Retriever}
During the analysis step, \sysvincent allows the LLM to request any code information whenever needed. 
Instead of returning source files as a whole, \sysvincent combines \cc{ctags} utility~\cite{ctags} and \cc{tree-sitter}~\cite{tree-sitter} library to serve the code snippets essential to LLM's analysis.
Specifically, given a target symbol name such as a function or variable, \sysvincent first locates where the requested code snippet exists using the \cc{ctags}.
Then, it extracts the actual code information, including the surrounding code context (\textit{e.g.,} the full implementation of a function, struct, enum, etc) using \cc{tree-sitter}.

Beyond this basic strategy, we observed occaisional failures of \cc{tree-sitter}'s parser when dealing with complex codebases.
For example, if the target source code contains complex compile directives or macros, \cc{tree-sitter} often fails to parse the intended snippets properly.
To handle such potential errors from external libraries, 
\sysvincent allows LLMs to request additional code lines when the initial query result is not sufficient.
This two-step strategy effectively mitigates both parsing failures caused by the \cc{tree-sitter} and excessive token usage caused by naive line-based file exploration.

To further assist the LLM in understanding the codebase, \sysvincent supports both symbol-reference search and code-embedding-based~\cite{neelakantan2022text} search features, providing LLMs with richier context for reasoning about the codebase.
LLMs can utilize this information to infer program behaviors, function usages, and essential code patterns needed to know for patching.
For example, to correctly understand the behavior of a certain function \cc{foo}, it may be necessary to inspect locations that call \cc{foo}, or functions with similar structures to \cc{foo}.
For the symbol reference search, \sysvincent adopts a string-based strategy, which retrieves code snippets where the requested symbol name appears.
For the code-embedding-based search, \sysvincent converts functions in the codebase and stores the results in the database.
When a specific function symbol is requested, \sysvincent provides the code snippets similar to the given function symbol.
In the current implementation, \sysvincent leverages OpenAI's \cc{text-embedding-3-large} model to embed each code snippet.
To calculate the similarity, \sysvincent adopts cosine similarity, which is a widely-used metric for this purpose.
However, naively embedding the entire codebase may consume excessive computing resources and significantly increase LLM costs.
To mitigate the issue, \sysvincent limits embedding calcualations to the source files that contain symbols previously requested for code retrieval. 

\subsubsection{Root Cause Analysis}

When \sysvincent accepts the sanitizer report and PoC, it performs an intial root cause analysis.
Essentially,
\sysvincent lets the LLM explore the codebase on its own as much as needed.
This strategy allows \sysvincent to fully utilize the LLM's reasoning capabilities, enabling it to infer the fault location---even when that location is not explicitly presented in the sanitizer report's callstack.
In addition to the sanitizer report, \sysvincent provides the PoC bytes to the LLM using the \cc{xxd} utility to further assist LLM's reasoning process.
To prevent excessive token usage, the PoC bytes are provided only if their size is less than 25,000 bytes, a value determined heuristically.

\subsubsection{Property Analysis}

As mentioned before, \sysvincent extracts program's properties from the given codebase.
Note that \sysvincent utilizes properties represented in natural language, unlike their typical usage in the field of software verification.
For example, the following is a partial list of properties generated by the LLM regarding the \cc{nginx} double-free bug. 

\begin{squishenumerate}
  \item \textbf{Memory Safety}: For any header field in \cc{ngx_http_headers_in_t}, the pointer must either be NULL or point to a valid ngx_table_elt_t structure.
  \item \textbf{Header Processing Consistency}: All single-value HTTP headers (like Host, From, Content-Length) must maintain exactly one instance throughout the request processing lifecycle.
  \item \textbf{Error Response Consistency}: When encountering invalid headers, the system must either: (1) Return \cc{NGX_ERROR} and call \cc{ngx_http_finalize_request} with \cc{NGX_HTTP_BAD_REQUEST} (2) Log the issue and return \cc{NGX_OK} But never both or neither.
\end{squishenumerate}

\noindent To be specific, the ``Error Response Consistency'' is one that LLMs failed to consider in patches like the one shown in \autoref{fig:nginx-double-free-patch-wrong}.
In the patch step prompt, \sysvincent instructs LLMs to consider such properties to generate more appropriate for the given program context.

Based on this concept of properties, \sysvincent performs this analysis step as follows.
Similar to the previous root cause analysis, \sysvincent allows the LLM to explore the codebase on its own.
When the LLM decides that the collected information is sufficient, it outputs the list of properties that must be considered regarding the future patch generation.
Regarding prompt engineering, 
it was found that LLMs could infer properties without any specialized tools or a detailed prompt that explains what a property is.
This is likely because the LLM already has knowledge of program properties, allowing for the use of relatively simple prompts.

\subsubsection{Patch Generation}
In this step, \sysvincent requests the LLM to generate a patch considering the previous analysis results.
Similar to other agents in \syspatching, 
\sysvincent adopts a feedback-based strategy for patch generation.
In particular, whenever the LLM generates a patch, \sysvincent evaluates the patch by using the default OSS-Fuzz-based evaluator of \syspatching. 
If the evaluation fails (\textit{e.g.,} compilation error, another crash, or internal test failures), \sysvincent provides the failure information to the LLM and retries the patch generation.
As in the previous analysis steps, the LLM is allowed to request additional code information during the patch generation.

To translate LLM's patches into actual diffs, \sysvincent employs a line-replace strategy.
To elaborate, \sysvincent instructs the LLM to submit a patch as a tuple of three items: (i) the target source filename, (ii) the lines to replace, and (iii) code content to replace the specified lines.

\subsection{Agent: \sysclaudelike}
\label{ss:claudelike}

\sysclaudelike is an agent that performs patching by using the ReAct agent \sysclaudelike coder, inspired by Claude Code~\cite{ClaudeCode}.

\subsubsection{Motivation}
During the preparation phase for AIxCC, Anthropic released Claude Code, one of the most 
powerful LLM-based tools for programming assistance. Preliminary experiments demonstrated 
that Claude Code achieved competitive performance in understanding project structures and 
performing code modifications, making it a promising reference for patching tasks. 

Claude Code’s strength lies in two design choices: (i) well-engineered file editor tools 
that allow the agent to efficiently analyze and modify project files, and (ii) the use of 
sub-agents to offload tasks requiring frequent tool invocations, thereby reducing 
irrelevant context retained by the main agent. This design enables the main agent to focus 
on essential reasoning for patch generation.  

Motivated by these findings, we designed \sysclaudelike to incorporate the strengths of 
Claude Code into the patching domain. Specifically, \sysclaudelike integrates a diverse 
set of file editor tools and supports the spawning of sub-agents to handle specialized 
tasks. This design not only improves efficiency in patch generation but also enhances 
scalability by mitigating the context overhead that often limits LLM agents.

\subsubsection{Agent Workflow}

The \sysclaudelike agent operates through an iterative workflow inspired by Claude Code.  
First, (i) the agent constructs a prompt by incorporating insights extracted from the PoC and SARIF reports, and forwards this prompt to the coder module.  
Next, (ii) the coder generates a candidate patch in the form of a \cc{diff} using the ReAct model.  
Once the patch is produced, (iii) the agent evaluates its validity by rebuilding the project, executing the functional tests, and running the PoV to confirm whether the vulnerability is mitigated.  
Finally, (iv) if the patch fails any of these checks, feedback is provided back to the coder, which then attempts to synthesize a new patch.  
This iterative process continues until a plausible and valid patch is produced or a predefined termination condition is reached.

\subsubsection{Coder}

\begin{figure}[t]
  \centering
  \includegraphics[width=\textwidth]{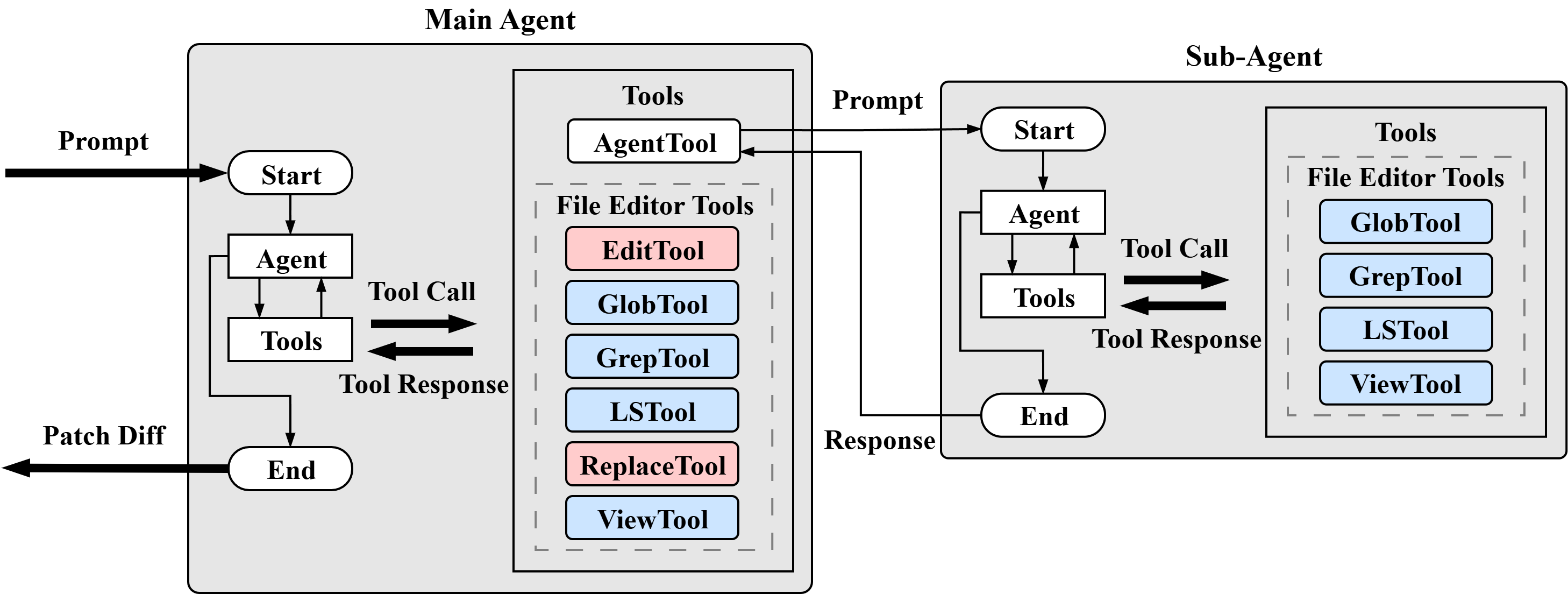}
    \caption{The overall structure of the \sysclaudelike coder. The read-only file editor tools that can be invoked by the sub-agent are highlighted with blue boxes, while the file editor tools that modify files are highlighted with red boxes.}
  \label{fig:claudelike-coder-overview}
\end{figure}

The workflow of \sysclaudelike is largely similar to that of other simple agents employing a feedback loop. The key difference, however, lies in the internal mechanism of the \sysclaudelike coder. As shown in \autoref{fig:claudelike-coder-overview}, the overall structure of the coder is organized around two ReAct models: a main agent and a sub-agent. Each of these agents is able to interact with the file system through a common set of tools, enabling them to analyze and modify project files as needed. In addition, the \cc{AgentTool} provides the capability for the main agent to spawn a sub-agent dynamically, delegating specific tasks to it when necessary. This design separates responsibilities between the two agents and improves the efficiency of patch generation.

\PP{File Editor Tools}
The file editor tools provide coders with flexible capabilities to navigate project files, inspect their contents, and apply necessary modifications, thereby supporting efficient patch generation. The following describes each tool in detail:

\begin{squishitemize}
  \item \cc{\textbf{EditTool(file_path, old_string, new_string)}}: Replaces the \cc{old_string} from the file in \cc{file_path} into \cc{new_string}. The \cc{old_string} must exist in the file and be uniquely matched in the file.
  \item \cc{\textbf{GlobTool(pattern, path)}}: Returns the list of files in the \cc{path} directory whose filenames match the glob pattern.
  \item \cc{\textbf{GrepTool(pattern, path, include)}}: Returns the list of files in the \cc{path} directory whose content match the given regex expression \cc{pattern}. The result only includes files whose filenames match the glob pattern \cc{include}.
  \item \cc{\textbf{LSTool(path)}}: Returns the list of files in the \cc{path} directory.
  \item \cc{\textbf{ReplaceTool(file_path, content)}}: Replaces the content of file \cc{file_path} into given \cc{content}. If the file \cc{file_path} does not exist, \cc{ReplaceTool} creates it and sets the content of the file as \cc{content}.
  \item \cc{\textbf{ViewTool(file_path, offset, limit)}}: Returns the maximum \cc{limit} lines of the file \cc{file_path} starts from line number \cc{offset}.
\end{squishitemize}

\PP{AgentTool}
The primary feature of the \sysclaudelike coder is the \cc{AgentTool}, which decomposes complex tasks by dispatching a sub-agent. Invoked as \cc{AgentTool(prompt)}, it spawns a ReAct-based sub-agent to perform the task described in the \cc{prompt} and return the result. This mechanism is particularly useful for operations such as searching for keywords in files or identifying the role of specific source code, which frequently arise during patching but require multiple tool calls. By offloading these operations to a sub-agent, the main agent can avoid retaining unnecessary context and focus on essential reasoning for patch generation.

The sub-agent operates with the same interface as the main agent but with a restricted tool set to prevent unintended file modifications. Specifically, it may call only \cc{GlobTool}, \cc{GrepTool}, \cc{LSTool}, and \cc{ViewTool}, ensuring that its role is limited to navigation and inspection. Furthermore, the sub-agent cannot invoke \cc{AgentTool}, preventing recursive spawning of agents and eliminating the risk of infinite loops. These restrictions preserve the integrity of the codebase while maintaining controlled and predictable task delegation.

% The main feature of \sysclaudelike coder is the \cc{AgentTool}, which breaks down complicated jobs by dispatching a new sub-agent. \cc{AgentTool(prompt)} spawns the new ReAct agent that performs the job given by \cc{prompt}. The spawned agent conducts the job and returns the result.
% 
% Some tasks, such as searching for specific keywords in files or identifying the role of particular source code, frequently occur during the patch operation but require multiple tool calls to perform. It leads to an increase in the context for the agent to keep. By using \cc{AgentTool}, the main agent can delegate these tasks to the sub-agent and discard the unnecessary context.
% 
% The sub-agent can call the same tools as the main agent; however, it has a limited set of tools to prevent unexpected file edits. The sub-agent can only call \cc{GlobTool}, \cc{GrepTool}, \cc{LSTool}, and \cc{ViewTool}. This limitation ensures that the sub-agent can only perform its tasks without modifying files, preventing unexpected changes to the codebase that the main agent does not intend.
% 
% Additionally, the sub-agent cannot call \cc{AgentTool}, which prevents the sub-agent from spawning another sub-agent. This restriction ensures that the agent does not create an infinite loop of agent dispatching.
% 
\subsubsection{Differences from Claude Code}

The \sysclaudelike coder is inspired by the architecture and tool design of Claude Code but incorporates several key differences. First, we removed tools that are not well suited for patching, such as those designed for Jupyter notebooks (e.g., \cc{ReadNotebook}, \cc{NotebookEditCell}) or tools requiring an online environment (e.g., \cc{WebFetchTool}). This reduction decreases both the context size and engineering overhead. Furthermore, we redesigned the system prompts and tool prompts to better align the agent’s behavior with the requirements of automated patch generation.

% The \sysclaudelike coder is inspired by Claude Code, but it has several differences from Claude Code.
% 
% First, we removed tools that are difficult to use for patching—such as tools for Jupyter notebooks (e.g., \cc{ReadNotebook}, \cc{NotebookEditCell}) or those requiring an online environment (e.g., \cc{WebFetchTool})—to reduce context size and engineering costs. In addition, we designed the agent to be more suitable for patching by employing our own system prompts and tool prompts.
% 
\subsection{Open-Source Agents}
\label{ss:opensourceagents}
In addition to our internally developed agents, \syspatching integrates selected
open-source agents to further enhance the diversity and robustness of patch
generation. These agents bring complementary strengths developed and refined by
the broader community, enabling our system to explore problem-solving approaches
that differ from our in-house designs. In particular, we employ
\sysaider~\cite{aider} and \syssweagent~\cite{yang2024swe}, each of which
specializes in distinct patching strategies and workflows. 

However, integrating these open-source agents also introduces practical
considerations: their embedded prompts are typically required to have long
context windows, which can lead to excessive context usage. To mitigate these
issues, we adopted more affordable models like \textit{Gemini 2.5 Pro} and
\textit{O4 Mini}.

\subsubsection{Aider}
\sysaider~\cite{aider} is an open-source agent designed for precise, instruction-guided code modifications. 
It is particularly effective at handling straightforward bugs where the fix can be expressed as a small, localized edit. 
While it may lack the advanced reasoning capabilities needed for complex, multi-step problems, \sysaider excels in applying clear, direct patches with a high success rate when the task is simple. 
In \syspatching, we use \sysaider as a reliable ``lifeguard'' agent; if a problem is easy but happens to be overlooked or overcomplicated by our more sophisticated agents, \sysaider can step in and resolve it quickly. 
This role ensures that trivial fixes are not missed, allowing the system to capture low-hanging opportunities without unnecessary reasoning overhead.

\subsubsection{SWE-Agent}
\syssweagent~\cite{yang2024swe} is an open-source agent built for comprehensive, iterative bug repair. 
It employs a multi-step reasoning process that includes analyzing code context, executing relevant commands, and refining its proposed patches based on observed outcomes. 
While \syspatching already incorporates agents with a variety of strategies, including multi-step reasoning, we include \syssweagent because it is one of the most widely used open-source multi-step repair agents. 
Its workflow has been exercised and refined by a large community of users, giving us confidence that its reasoning patterns and iterative refinement loop are effective in practice. 
By incorporating \syssweagent, we add a well-validated, community-tested problem solver to our agent pool, further diversifying the approaches explored and increasing the likelihood of finding a correct patch.

\subsection{Evaluation}
This section presents the performance evaluation of \syspatching, our automated patching system. We evaluate both individual agents and the end-to-end system. To assess the practicality and efficiency of each agent, we use three metrics: patch success rate, Time-to-Patch (TTP), and LLM cost. For fair comparison, we run all benchmarks on a uniform hardware platform with Intel(R) Xeon(R) Gold 6346 CPU (16 cores @ 3.10GHz), 256 GB of DDR4 RAM, installed with Ubuntu 22.04.

% This section presents a performance evaluation of agents, our proposed automated
% patching system. The experiments are designed to compare the performance of the
% different agents within \syscrete, focusing on two critical metrics for
% assessing the practicality and efficiency of such solutions: Time-to-Patch (TTP)
% and LLM Cost. To ensure fair and reproducible comparisons, all benchmarks were
% conducted on a uniform hardware platform equipped with an Intel(R) Xeon(R) Gold
% 6346 CPU (16 cores @ 3.10GHz), 256 GB of DDR4 RAM.

\subsubsection{Evaluation Metrics}
To quantitatively assess the performance of each agent, we employ three key metrics: patch success rate, Time-to-Patch (TTP), and LLM cost. These metrics provide a comprehensive view of each agent's effectiveness, efficiency, and resource utilization in the context of automated patch generation.
% We evaluated the performance of each agent based on the following key metrics.

\PP{Patch Success Rate}
In our evaluation, a patch is considered plausible if it resolves the identified vulnerability without introducing new issues, as verified by the provided proof-of-concept (PoC) exploit and ensuring the internal test passes successfully. However, if a patch fails to compile, does not fix the vulnerability, or exceeds the time limit, it is deemed unsuccessful. We planned to report the internal test failure cases separately since they are unable to be verified in the competition environment, but in our evaluation, we only encountered such cases during the final competition postmortem analysis. The patch success rate is calculated as the ratio of successfully generated plausible patches to the total number of cases attempted by each agent. This metric is crucial for understanding the effectiveness of each agent in generating plausible and reliable patches.

\PP{Time-to-Patch (TTP)}
% % This metric measures the end-to-end wall-clock time from the initial problem
% statement to the generation and validation of a correct patch. Since TTP is
% dominated by the time spent on LLM requests and project builds, a higher TTP
% generally implies that an agent performed more build attempts or made more
% frequent and complex LLM requests during its operation. It is crucial within the
% context of the AIxCC competition due to its strict time and resource
% limitations. A lower TTP directly enhances the ability to execute more patch
% attempts, thereby maximizing the probability of success.
%
Time-to-Patch (TTP) is the elapsed time required for an agent to generate a patch. Most of this duration arises from LLM interactions and project builds, so a higher TTP often indicates more build attempts or more frequent and complex LLM calls. In the AIxCC competition~\cite{aixcc-website}, shorter TTP is beneficial because it increases the score through time multipliers~\cite{aixcc-scoring}. Beyond competitive settings, reducing TTP improves system reliability and maintainability by minimizing downtime, enhancing service availability, and lowering operational overhead in large-scale deployments.

\PP{LLM Cost} 
% LLM Cost metric quantifies the total number of tokens processed by
% the language model to arrive at a final, correct patch. It is particularly
% important given the AIxCC competition's strict budget on LLM usage. Minimizing
% token consumption is essential for operating within these constraints, making a
% lower token count a direct reflection of a system's cost-efficiency and its
% viability under the competition rules.
LLM cost measures the total expenditure associated with language model usage to generate a patch. Generally, the cost depends on the model; in our system, each agent is tied to a fixed model, as listed in \autoref{tab:patch-agents-comparison}, and thus directly incurs the corresponding cost. In the AIxCC competition, agents must operate under a strict LLM budget, so reducing this cost is desirable because it allows more effective use of the limited budget.

\subsubsection{Microbenchmark}
To evaluate the performance of each agent, we conducted a microbenchmark using the Round 3 dataset from the AIxCC competition.

\PP{Dataset - Exhibition Round 3}
We used the Round 3 dataset from the AIxCC competition~\cite{aixcc-website} for our microbenchmark.
The Round 3 dataset contains 34 vulnerabilities drawn from nine challenge projects, comprising 16 in C and 18 in Java. This dataset was officially released by the organizers as part of an unscored exhibition round. Since our evaluation used the final version of the agent, which had been further developed after Round 3 was released, the results can suffer from data contamination due to overfitting. Nevertheless, because the dataset was selected independently by the organizers and includes diverse vulnerabilities in both C and Java, we believe that it remains a useful benchmark for examining relative performance of agents.
% The Round 3 dataset contains 34 vulnerabilities drawn from nine challenge

\begin{figure*}[]
\centering
\subfigure[CPVs with bug type, changed files, hunks, and added and removed lines per diff]{\
  \footnotesize
  \begin{tabular}{l l r r r r r}
  \toprule
  \textbf{CPV Name}                & \textbf{Bug Type}           & \textbf{Files} & \textbf{Hunks} & \textbf{+} & \textbf{-} & \textbf{Total} \\
  \midrule
  \textbf{c-curl-1}                & Stack Buffer Overflow & 1              & 1              & 1          & 1          & 2              \\
  \textbf{c-curl-2}                & Null Pointer Dereference    & 1              & 1              & 1          & 1          & 2              \\
  \textbf{c-freerdp-1}             & Integer Overflow            & 2              & 2              & 4          & 1          & 5              \\
  \textbf{c-freerdp-2}             & Backdoor                    & 1              & 1              & 0          & 10         & 10             \\
  \textbf{c-freerdp-3}             & Heap Buffer Overflow  & 1              & 1              & 4          & 4          & 8              \\
  \textbf{c-libpng-1}              & Stack Buffer Overflow & 1              & 1              & 1          & 1          & 2              \\
  \textbf{c-libxml2-1}             & Heap Buffer Overflow  & 1              & 1              & 1          & 1          & 2              \\
  \textbf{c-libxml2-2}             & Double Free                 & 1              & 1              & 0          & 1          & 1              \\
  \textbf{c-libexif-1}             & Heap Buffer Overflow  & 1              & 1              & 3          & 9          & 12             \\
  \textbf{c-libexif-2}             & Heap Buffer Overflow  & 1              & 1              & 5          & 0          & 5              \\
  \textbf{c-libexif-3}             & Heap Buffer Overflow  & 1              & 3              & 9          & 4          & 13             \\
  \textbf{c-sqlite3-1}             & Stack Buffer Overflow & 1              & 1              & 1          & 1          & 2              \\
  \textbf{c-sqlite3-2}             & Heap Buffer Overflow  & 2              & 3              & 3          & 0          & 3              \\
  \textbf{c-sqlite3-3}             & Heap Buffer Overflow  & 1              & 1              & 1          & 1          & 2              \\
  \textbf{c-sqlite3-4}             & Null Pointer Dereference    & 1              & 1              & 1          & 1          & 2              \\
  \textbf{c-sqlite3-5}             & Stack Buffer Overflow & 1              & 1              & 1          & 1          & 2              \\
  \midrule
  \textbf{java-commons-compress-0} & Server Side Request Forgery & 1              & 2              & 1          & 13         & 14             \\
  \textbf{java-commons-compress-1} & Backdoor                    & 1              & 1              & 1          & 3          & 4              \\
  \textbf{java-commons-compress-2} & Backdoor                    & 1              & 3              & 24         & 3          & 27             \\
  \textbf{java-commons-compress-3} & Zip Slip                    & 1              & 1              & 1          & 1          & 2              \\
  \textbf{java-commons-compress-4} & Denial of Service           & 1              & 1              & 1          & 1          & 2              \\
  \textbf{java-commons-compress-5} & Out of Memory               & 1              & 1              & 4          & 2          & 6              \\
  \textbf{java-tika-1}             & XXE Injection               & 1              & 3              & 2          & 19         & 21             \\
  \textbf{java-tika-2}             & Remote Code Execution       & 1              & 2              & 5          & 3          & 8              \\
  \textbf{java-tika-3}             & Tar Slip                    & 1              & 2              & 5          & 1          & 6              \\
  \textbf{java-tika-4}             & Server Side Request Forgery & 1              & 1              & 0          & 21         & 21             \\
  \textbf{java-tika-5}             & OS Command Injection        & 1              & 2              & 1          & 42         & 43             \\
  \textbf{java-tika-6}             & OS Command Injection        & 1              & 2              & 12         & 0          & 12             \\
  \textbf{java-tika-7}             & Server Side Request Forgery & 1              & 1              & 0          & 11         & 11             \\
  \textbf{java-tika-8}             & Server Side Request Forgery & 1              & 1              & 1          & 13         & 14             \\
  \textbf{java-tika-9}             & File Path Traversal         & 1              & 2              & 24         & 28         & 52             \\
  \textbf{java-tika-10}            & Zip Slip                    & 1              & 2              & 6          & 5          & 11             \\
  \textbf{java-zookeeper-1}        & Backdoor                    & 1              & 4              & 1          & 51         & 52             \\
  \textbf{java-zookeeper-2}        & Denial of Service           & 1              & 1              & 5          & 6          & 11             \\
  \bottomrule
\end{tabular}

}
\subfigure[Cumulative distribution of changed lines]{
  \centering
  \includegraphics[height=0.3\linewidth]{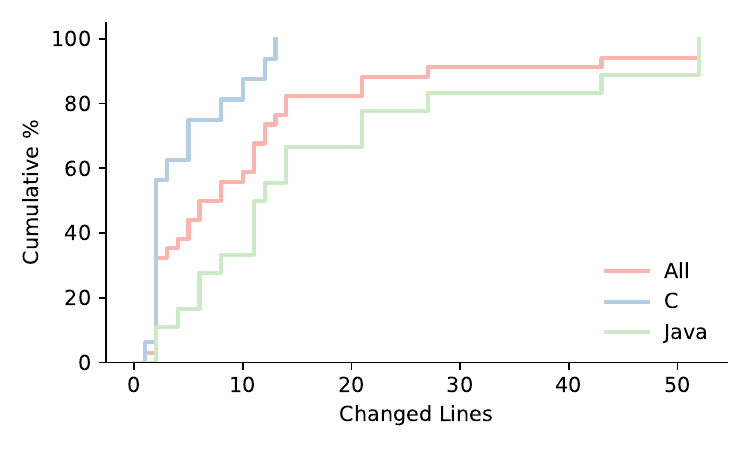}
}
\subfigure[Single vs multi-hunk patches]{
  \centering
  \includegraphics[height=0.4\linewidth]{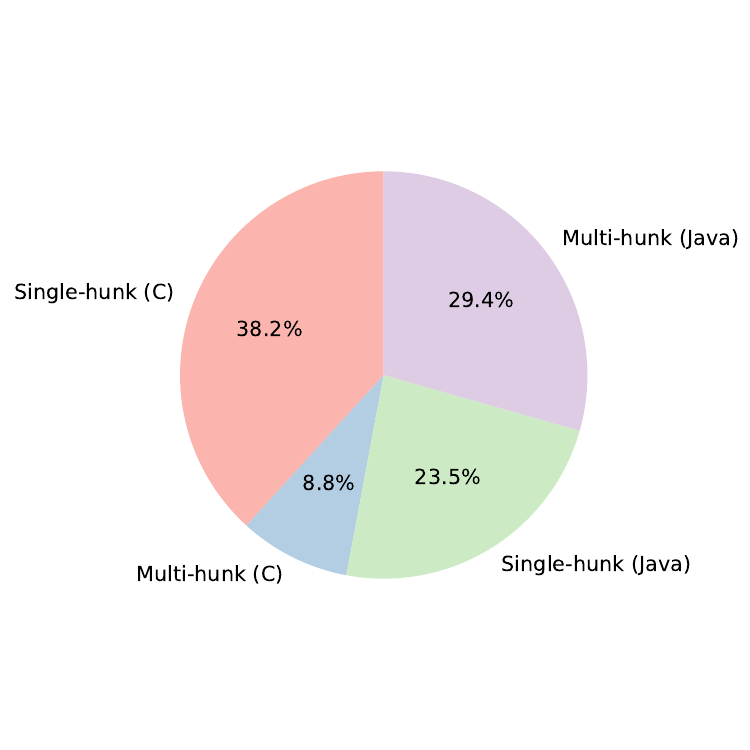}
}

\caption{Statistics of the Round 3 benchmark, showing the distributions of
changed files, hunks, and added and removed lines per diff.  Notably, this
distribution is from the ground truth patches provided by the organizers.}
\label{fig:bench-3-stats}
\vspace*{10pt}
\end{figure*}

\autoref{fig:bench-3-stats} shows the statistics of the Round 3 dataset.
As illustrated, the dataset includes a wide range of bug types with varying patch sizes.
Specifically, it includes memory safety issues (e.g., buffer overflows, null pointer dereferences) and also logic errors (e.g., server side request forgery, XXE injection, Zip Slip). Moreover, the dataset covers not only single-hunk patches but also multi-hunk and even multi-file patches, 
which are often challenging for existing APR techniques.

\PP{Results}
\autoref{tab:patch-performance-3} shows the summary of the patch success rate of
each agent on the Round 3 dataset, while \autoref{fig:patch-performance-3-graph}
visualizes the LLM cost and TTP of each agent.  We also include the raw data in
\autoref{tab:patch-performance-comparison-llm-cost} and
\autoref{tab:patch-performance-comparison-ttp}.
In summary, \sysprism achieved the highest patch success rate of 100\% (34/34), followed by
\sysmr and \sysprism (33/34, 97.1\%). Notably, all of our newly designed agents
outperformed the open-source agents (\sysaider and \syssweagent).
Nevertheless, as mentioned earlier, we further refined the agents after
Round 3 was released, so the results may be influenced by overfitting to this
dataset.  Therefore, the results should be interpreted as relative performance
trends rather than absolute measures of capability.

In terms of TTP and LLM cost, most agents result in similar performance, except
for \sysprism, which incurred significantly higher TTP and LLM cost.  This is
because \sysprism uses expensive, reasoning models (i.e., o4-mini), and its
multi-agent design is more complex than other agents, involving more LLM
requests and build attempts. Moreover, \sysaider used an extremely low LLM cost;
this is because \sysaider is not designed for automatic patch generation but
for interactive use. As a result, it does not perform iterative patch generation
like other agents, leading to fewer LLM requests and lower LLM cost.

% \autoref{fig:patch-performance-3-pie}
% shows the visualization of each agent's performance on the Round 3 dataset.
% Moreover, 
% Each agent was evaluated on all 34
% cases, and we recorded both the time to generate a patch and the associated LLM
% cost. We observed that the agents generally achieved high
% performance, which can be attributed to the fact that they were tuned for the R3
% benchmark. A summary of results is provided in
% \autoref{tab:patch-performance-3}, while detailed breakdowns of
% Time-to-Patch (TTP) and LLM cost per agent and project are presented in
% Tables~\ref{tab:patch-performance-comparison-ttp} and
% \ref{tab:patch-performance-comparison-llm-cost}.

\subsubsection{End-to-End Performance}

\PP{Dataset - Exhibition Round 3.5} 
To evaluate the end-to-end performance of \syspatching, we used Exibition Round
3.5, an internal benchmark we constructed.  This dataset consists of 22 CPVs
from four challenge projects, all implemented in C. Unlike those in Round 3, it
was created after the final version of the agents had been developed, so no
tuning or modifications were applied. As a result, the evaluation results are
more meaningful than those from Round 3.  As mentioned earlier, we used this
benchmark to evaluate the end-to-end performance of \syspatching, rather than
individual agents.  As \syspatching does not run subsequent agents if one agent
successfully produces a patch, some agents were not executed for certain cases.
For this reason, we do not report TTP or LLM cost here; instead, we present the
patch success rate of the executed agents and the overall end-to-end
performance.

\PP{Results}
\autoref{tab:patch-performance-extra} presents the overall results. An interesting observation is that \emph{there is no agent that consistently works well across all cases}. In terms of patch success rate alone, \sysmr performs the best, but even \sysmr failed to fix two patches. Interestingly, one of these patches was successfully fixed by \sysmartian and \sysclaudelike, despite their patch success ratios being only in the 20\% range.
Moreover, the other was fixed by \sysaider. Considering that \sysaider is relatively simple without an interactive process, this is a particularly interesting result. It suggests that, in certain cases, complex architectures and tools may actually become a burden for LLMs.

\begin{table}
  \centering
  \footnotesize
  \begin{tabular}{clccc}
\toprule
\textbf{Rank} & \textbf{Team} & \textbf{Vulnerabilities Found} & \textbf{Correct Patches} & \textbf{Patch Rate (\%)} \\
\midrule

1 & Team Atlanta & 43 & 31 & 72.1 \\
2 & Trail of Bits & 28 & 19 & 67.9 \\
3 & Theori & 34 & 20 & 58.8 \\
4 & All You Need IS A Fuzzing Brain & 28 & 14 & 50.0 \\
5 & Shellphish & 28 & 11 & 39.3 \\
6 & 42-b3yond-6ug & 41 & 3 & 7.3 \\
7 & Lacrosse & 1 & 1 & 100.0 \\
\bottomrule
\end{tabular}

  \caption{The summary of patch success rate of each team on the AIxCC final competition.}
  \label{tab:patch-final-summary}
\end{table}

\subsubsection{Postmortem Analysis on AIxCC Final Competition}
In this subsection, we present a postmortem analysis of the AIxCC final competition. \autoref{tab:patch-final-summary} shows a summary of final results. Our system, \sys, not only discovered the largest number of vulnerabilities (43 CPVs in total) but also \emph{produced the most correct patches (31 in total)}. In particular, \syspatching achieved a patch success rate of 72.1\%. \emph{This is the highest rate among competing systems except for Lacrosse, which only patched a single vulnerability.} These results highlight that \syspatching attains leading  performance among the competition teams.

% In more detail, \syspatching could find 31 correct patches via the following process.
% \syspatching received \textbf{117 PoVs} from \sys's bug finding system,
% After the initial de-duplication process, we attempted to patch \textbf{47 PoVs},
% Two PoVs were filtered out further by de-duplication after patching, leaving \textbf{45 CPVs},
% Among the 45 PoVs, we failed to generate any plausible patches for \textbf{2 PoVs} from the \cc{hertzbeat} challenge project,
% For remaining \textbf{43 CPVs}, \syspatching submitted \textbf{48 patches},
% where 42 patches passed the internal test while 6 patches failed.
% In particular, \syspatching successfully submitted sound patches for \cc{mongoose-3}, \cc{pdfbox-5}, and \cc{tika-1} 2, 3, and 3 times respectively. Fortunately, we submitted sound patches for \cc{mongoose-3} and \cc{tika-1}, but failed to submit a sound patch for \cc{pdfbox-5} in all 3 attempts.
% Finally, we submitted sound patches for \textbf{42 CPVs}. After manual inspection by the AIxCC organizers, we received scores for \textbf{31 of them}.
% However, we do not have access to the detailed manual inspection results, so we cannot further examine which specific patches received scores.

In more detail, \syspatching produced 31 correct patches through the following steps. In total, \syspatching received 117 PoVs from \sys's bug-finding system and, after the initial de-duplication, attempted to patch 47 PoVs. Post-patch de-duplication removed two additional duplicates, leaving 45 CPVs. Among these, we could not patch 2 PoVs from the \cc{hertzbeat} project. For the remaining 43 CPVs, \syspatching submitted 48 patches; 42 passed internal tests and 6 failed. In particular, \syspatching attempted \cc{mongoose-3}, \cc{pdfbox-5}, and \cc{tika-1} 2, 3, and 3 times, respectively. It produced plausible patches in the final attempts for \cc{mongoose-3} and \cc{tika-1}, but failed in all three attempts for \cc{pdfbox-5}. Overall, we submitted plausible patches for 42 CPVs, and the AIxCC organizers awarded scores to 31 of them after manual inspection. Since the organizers did not disclose the detailed inspection results, we cannot determine which specific patches received scores.

% We also analyzed the end-to-end performance of \syspatching during the AIxCC Final Competition from the competition logs. The results appear in \autoref{tab:patch-performance-final}.
% We collected 42 CPVs from the final competition, where 25 CPVs are in C and 17 CPVs are in Java.
% Unfortunately, due to unexpected log truncation, some entries are missing. For example, \syspatching submitted successful patches for CPVs \texttt{mongoose-1, 3, 4} and \texttt{xz-1}, but we could not attribute the fixes to specific agents.
% Nevertheless, we believe that the results are still meaningful and reflect a large-scale evaluation.

\PP{Per-agent Analysis}
Similar to the previous experiment, we analyzed the performance of each agent during the AIxCC final competition. To this end, we examined both LLM logs and system logs after the competition. Unfortunately, due to unexpected log truncation, several entries are missing.
For instance, \syspatching successfully submitted patches for CPVs, which are \cc{mongoose-1, 3, 4} and \cc{xz-1}, but we could not attribute these fixes to specific agents.
Despite these missing records, we believe that the available data are still meaningful and reflect a large-scale evaluation.

As shown in \autoref{tab:patch-performance-final}, the results of the final competition are consistent with the previous experiments.
% As we previously mentioned, it is worth noting that we are unable to analyze the results of 5 out of 47 POVs due to the unexpected log truncation: \cc{c-curl-1}, \cc{java-jsoup-1}, \cc{c-libexif-1}, \cc{c-libexif-2}, and \cc{c-libxml2-1}. We submitted sound patches for 4 of these POVs (excluding \cc{c-libxml2-1}), but we are unable to identify which agent generated the fixes.
\sysmr, \sysprism, and \sysvincent again achieved strong performance; \sysmr recorded the highest patch submission rate of 80.56\% (29/36), followed by \sysprism at 74.36\% (29/39) and \sysvincent at 72.97\% (27/37). \sysmartian showed a relatively low rate of 46.51\% (20/43), but it uniquely fixed \cc{shadowsocks-1} and also repaired \cc{pdfbox-5}, where both \sysmr and \sysvincent failed in internal functional tests.
Across the final competition, \syspatching produced plausible patches for 42 CPVs, even though we could not identify the responsible agents for 5 patches
due to log truncation.
This is far more than the results of the best single agent, \sysmr, which produced 29 patches. This demonstrates that no single agent could solve all problems, making our ensembling crucial for achieving high overall success.

\clearpage

\clearpage

\begin{table*}[]
  \footnotesize
    \begin{adjustbox}{width=\linewidth}
    \scriptsize
    \newcommand{\RowHead}[1]{\textbf{#1}}
    \newcommand{\ColHead}[1]{\textbf{#1}}
    \newcommand{\Best}[1]{\textbf{#1}}
    \resizebox{\linewidth}{!}{%
      \begin{tabularx}{\linewidth}{@{}ll*{7}{>{\centering\arraybackslash}m{0.97cm}}@{}}
        \toprule
        \RowHead{CPV Name}                &
        \RowHead{Bug Type}               &
        \RowHead{\sysmartian}             &
        \RowHead{\makecell{\textsc{Multi}                                                                                                     \\\textsc{Retrieval}}} &
        \RowHead{\sysprism}               &
        \RowHead{\sysvincent}             &
        \RowHead{\makecell{\textsc{Claude}                                                                                                    \\\textsc{Like}}} &
        \RowHead{\sysaider}               &
        \RowHead{\makecell{\textsc{SWE}                                                                                                       \\\textsc{Agent}}} \\
        \midrule
        \ColHead{c-curl-1}                & Stack Buffer Overflow & \Sound  & \Sound  & \Sound  & \Sound  & \Sound  & \Sound  & \Sound  \\
        \ColHead{c-curl-2}                & Null Pointer Dereference    & \Sound  & \Sound  & \Sound  & \Sound  & \Sound  & \Fail   & \Sound  \\
        \ColHead{c-freerdp-1}             & Integer Overflow            & \Sound  & \Sound  & \Sound  & \Sound  & \Sound  & \Sound  & \Fail   \\
        \ColHead{c-freerdp-2}             & Backdoor                    & \Sound  & \Sound  & \Sound  & \Sound  & \Sound  & \Fail   & \Fail   \\
        \ColHead{c-freerdp-3}             & Heap Buffer Overflow  & \Sound  & \Sound  & \Sound  & \Sound  & \Sound  & \Fail   & \Fail   \\
        \ColHead{c-libexif-1}             & Heap Buffer Overflow  & \Sound  & \Sound  & \Sound  & \Sound  & \Sound  & \Sound  & \Sound  \\
        \ColHead{c-libexif-2}             & Heap Buffer Overflow  & \Sound  & \Sound  & \Sound  & \Sound  & \Sound  & \Sound  & \Fail   \\
        \ColHead{c-libexif-3}             & Heap Buffer Overflow  & \Sound  & \Sound  & \Sound  & \Sound  & \Sound  & \Sound  & \Sound  \\
        \ColHead{c-libpng-1}              & Stack Buffer Overflow & \Sound  & \Sound  & \Sound  & \Sound  & \Fail   & \Sound  & \Fail   \\
        \ColHead{c-libxml2-1}             & Heap Buffer Overflow  & \Sound  & \Sound  & \Sound  & \Sound  & \Sound  & \Fail   & \Sound  \\
        \ColHead{c-libxml2-2}             & Double Free                 & \Sound  & \Sound  & \Sound  & \Sound  & \Sound  & \Sound  & \Sound  \\
        \ColHead{c-sqlite3-1}             & Stack Buffer Overflow & \Sound  & \Sound  & \Sound  & \Sound  & \Sound  & \Fail   & \Sound  \\
        \ColHead{c-sqlite3-2}             & Heap Buffer Overflow  & \Sound  & \Sound  & \Fail   & \Sound  & \Fail   & \Fail   & \Fail   \\
        \ColHead{c-sqlite3-3}             & Heap Buffer Overflow  & \Sound  & \Sound  & \Sound  & \Sound  & \Sound  & \Fail   & \Sound  \\
        \ColHead{c-sqlite3-4}             & Null Pointer Dereference    & \Sound  & \Sound  & \Sound  & \Sound  & \Fail   & \Fail   & \Fail   \\
        \ColHead{c-sqlite3-5}             & Stack Buffer Overflow & \Sound  & \Sound  & \Sound  & \Sound  & \Fail   & \Fail   & \Fail   \\
        \ColHead{java-commons-compress-0} & Server Side Request Forgery & \Sound  & \Sound  & \Sound  & \Sound  & \Sound  & \Sound  & \Sound  \\
        \ColHead{java-commons-compress-1} & Backdoor                    & \Sound  & \Sound  & \Sound  & \Sound  & \Sound  & \Sound  & \Sound  \\
        \ColHead{java-commons-compress-2} & Backdoor                    & \Fail   & \Fail   & \Sound  & \Sound  & \Fail   & \Fail   & \Fail   \\
        \ColHead{java-commons-compress-3} & Zip Slip                    & \Sound  & \Sound  & \Sound  & \Sound  & \Sound  & \Sound  & \Sound  \\
        \ColHead{java-commons-compress-4} & Denial of Service           & \Sound  & \Sound  & \Sound  & \Sound  & \Sound  & \Sound  & \Fail   \\
        \ColHead{java-commons-compress-5} & Out of Memory               & \Fail   & \Sound  & \Sound  & \Sound  & \Fail   & \Fail   & \Fail   \\
        \ColHead{java-tika-1}             & XXE Injection               & \Sound  & \Sound  & \Sound  & \Sound  & \Sound  & \Sound  & \Fail   \\
        \ColHead{java-tika-2}             & Remote Code Execution       & \Sound  & \Sound  & \Sound  & \Sound  & \Sound  & \Sound  & \Fail   \\
        \ColHead{java-tika-3}             & Tar Slip                    & \Sound  & \Sound  & \Sound  & \Sound  & \Sound  & \Sound  & \Sound  \\
        \ColHead{java-tika-4}             & Server Side Request Forgery & \Sound  & \Sound  & \Sound  & \Sound  & \Sound  & \Fail   & \Sound  \\
        \ColHead{java-tika-5}             & OS Command Injection        & \Sound  & \Sound  & \Sound  & \Sound  & \Sound  & \Fail   & \Fail   \\
        \ColHead{java-tika-6}             & OS Command Injection        & \Fail   & \Sound  & \Sound  & \Sound  & \Fail   & \Fail   & \Fail   \\
        \ColHead{java-tika-7}             & Server Side Request Forgery & \Sound  & \Sound  & \Sound  & \Sound  & \Sound  & \Sound  & \Fail   \\
        \ColHead{java-tika-8}             & Server Side Request Forgery & \Sound  & \Sound  & \Sound  & \Sound  & \Sound  & \Fail   & \Sound  \\
        \ColHead{java-tika-9}             & File Path Traversal         & \Fail   & \Sound  & \Sound  & \Sound  & \Sound  & \Sound  & \Sound  \\
        \ColHead{java-tika-10}             & Zip Slip                    & \Sound  & \Sound  & \Sound  & \Sound  & \Sound  & \Sound  & \Sound  \\
        \ColHead{java-zookeeper-1}        & Backdoor                    & \Sound  & \Sound  & \Sound  & \Sound  & \Sound  & \Sound  & \Sound  \\
        \ColHead{java-zookeeper-2}        & Denial of Service           & \Sound  & \Sound  & \Sound  & \Sound  & \Sound  & \Fail   & \Fail   \\ \midrule
        % \midrule
        \ColHead{Patch Success Rate}     &                             & 30 / 34 (88.24\%) & 33 / 34 (97.06\%) & 33 / 34 (97.06\%) & 34 / 34 (100.00\%) & 27 / 34 (79.41\%) & 18 / 34 (52.94\%) & 17 / 34 (50.00\%) \\
        \bottomrule
      \end{tabularx}
    }
  \end{adjustbox}

  \caption{Patch success results for 34 CPVs from the Round~3 dataset. A checkmark (\Sound) indicates a plausible patch, while a cross (\Fail) denotes failure due to various reasons including patches that are still vulnerable, uncompilable, or exceed the time limit. The bottom row summarizes the total number of plausible patches out of the total attempts for each agent.}
  \label{tab:patch-performance-3}
\end{table*}

\begin{figure*}[]
  \footnotesize
  % Local helpers for multi-line tick labels (no extra packages)
\newcommand{\lblMulti}{\shortstack{\textsc{Multi}\\\textsc{Retrieval}}}
\newcommand{\lblClaude}{\shortstack{\textsc{Claude}\\\textsc{Like}}}
\newcommand{\lblSWE}{\shortstack{\textsc{SWE}\\\textsc{Agent}}}

% ---------- (a) LLM Cost (top) ----------
\begin{minipage}{\linewidth}
\centering
\begin{tikzpicture}
\begin{axis}[
  width=\linewidth, height=7cm,
  ymajorgrids,
  ylabel={LLM Cost (\$)},
  xtick={1,2,3,4,5,6,7},
  xticklabels={\sysmartian,\lblMulti,\sysprism,\sysvincent,\lblClaude,\sysaider,\lblSWE},
  xticklabel style={align=center},
  enlarge x limits=0.08,
  ymin=0, ymax=2.85,
  legend to name=patchLegend,
  legend columns=2,
  legend style={draw=none, fill=none, font=\small, /tikz/column sep=8pt}
]

\addplot+[only marks, mark=star*, mark size=2.4pt,
          draw=blue, mark options={fill=blue}, forget plot]
table[row sep=\\, x=x, y=tm]{
x  tm    \\
1  0.213 \\
2  0.281 \\
3  0.962 \\
4  0.315 \\
5  0.645 \\
6  0.050 \\
7  0.205 \\
};

% P10--P90 range as error bars only — BLUE, and do not contribute to legend
\addplot+[only marks, mark=none, draw=none,
          forget plot,                 % <-- move this BEFORE error bars/.cd
          error bars/.cd, y dir=both, y explicit,
          error mark=|, error bar style={line width=0.9pt, draw=blue}]
table[row sep=\\, x=x, y=tm, y error plus=yplus, y error minus=yminus]{
x  tm      yplus    yminus \\
1  0.213   1.157    0.160  \\
2  0.281   0.383    0.142  \\
3  0.962   1.742    0.617  \\
4  0.315   0.209    0.126  \\
5  0.645   1.183    0.512  \\
6  0.050   0.044    0.027  \\
7  0.205   0.290    0.168  \\
};
% ---- custom legend icons (explicit, always blue) ----
% mean: blue star marker
\addlegendimage{only marks, mark=*, mark size=2.4pt, draw=blue, fill=blue}
\addlegendentry{Trimmed mean}

% range: blue error-bar icon (vertical line with caps)
\addlegendimage{
  legend image code/.code={
    \draw[blue,line width=0.9pt] (0.0,0.0)--(0.44,0);
    % \draw[blue,line width=0.9pt] (0.05,0.22)--(0.21,0.22);
    % \draw[blue,line width=0.9pt] (0.05,-0.22)--(0.21,-0.22);
  }
}
\addlegendentry{P10--P90 range}

\end{axis}
\end{tikzpicture}

\small (a) LLM Cost per patch -- trimmed mean with P10--P90 range
\end{minipage}

\vspace{2em}

% ---------- (b) Time to Patch (bottom) ----------
\begin{minipage}{\linewidth}
\centering
\begin{tikzpicture}
\begin{axis}[
  width=\linewidth, height=7cm,
  ymajorgrids,
  ylabel={Time to Patch (s)},
  xtick={1,2,3,4,5,6,7},
  xticklabels={\sysmartian,\lblMulti,\sysprism,\sysvincent,\lblClaude,\sysaider,\lblSWE},
  xticklabel style={align=center},
  enlarge x limits=0.08,
  ymin=0, ymax=2450
]

% mean points — BLUE, no legend
\addplot+[only marks, mark=star*, mark size=2.4pt,
          draw=blue, mark options={fill=blue}, forget plot]
table[row sep=\\, x=x, y=tm]{
x  tm     \\
1  617.3  \\
2  544.9  \\
3  925.4  \\
4  582.9  \\
5  487.0  \\
6  528.0  \\
7  590.4  \\
};

% P10--P90 error bars — BLUE, no legend
\addplot+[only marks, mark=none, draw=none,
          forget plot,                 % <-- BEFORE .cd
          error bars/.cd, y dir=both, y explicit,
          error mark=|, error bar style={line width=0.9pt, draw=blue}]
table[row sep=\\, x=x, y=tm, y error plus=yplus, y error minus=yminus]{
x  tm     yplus   yminus \\
1  617.3  639.0   382.8  \\
2  544.9  646.5   274.5  \\
3  925.4  1384.4  552.4  \\
4  582.9  639.4   277.3  \\
5  487.0  312.8   274.4  \\
6  528.0  599.9   305.8  \\
7  590.4  490.0   251.6  \\
};
\end{axis}
\end{tikzpicture}
\vspace{2pt}
\small (b) TTP per patch -- trimmed mean with P10--P90 range
\end{minipage}

% shared legend printed once here
\begin{center}
\pgfplotslegendfromname{patchLegend}
\end{center}
  \caption{Per-agent summary across 34 CPVs (Round~3). Dots show \textbf{the
  per-method mean of the central 80\%} (i.e., with the lowest and highest 10\%
  removed) for LLM cost and time to patch (TTP). Vertical bars show \textbf{the
  central 80\% range (P10--P90)} of observed values, indicating distributional
  spread. Missing entries (--) were excluded. Lower is better.}
  \label{fig:patch-performance-3-graph}
\end{figure*}

\begin{table*}[]
  \footnotesize
    \begin{adjustbox}{width=\linewidth}
    \newcommand{\RowHead}[1]{\textbf{#1}}
    \newcommand{\ColHead}[1]{\textbf{#1}}
    \newcommand{\Best}[1]{\textbf{#1}}
    \begin{tabularx}{\linewidth}{@{}l*{7}{>{\centering\arraybackslash}m{1.4cm}}@{}}
      \toprule
      \RowHead{CPV Name}                       &
      \RowHead{\sysmartian}                    &
      \RowHead{\makecell{\textsc{Multi}                                                                                             \\\textsc{Retrieval}}} &
      \RowHead{\sysprism}                      &
      \RowHead{\sysvincent}                    &
      \RowHead{\makecell{\textsc{Claude}                                                                                            \\\textsc{Like}}} &
      \RowHead{\sysaider}                      &
      \RowHead{\makecell{\textsc{SWE}                                                                                               \\\textsc{Agent}}} \\
      \midrule
      \ColHead{c-curl-1}                       & 818        & 300        & 456  & 335        & 641        & \Best{276} & 374        \\
      \ColHead{c-curl-2}                       & 287        & \Best{272} & 359  & 336        & 329        & -          & 301        \\
      \ColHead{c-freerdp-1}                    & 582        & 423        & 426  & 390        & 1026       & \Best{224} & -          \\
      \ColHead{c-freerdp-2}                    & \Best{329} & 618        & 534  & 330        & 459        & -          & -          \\
      \ColHead{c-freerdp-3}                    & 348        & 285        & 417  & 309        & \Best{274} & -          & -          \\
      \ColHead{c-libexif-1}                    & 273        & 272        & 319  & 331        & 212        & \Best{183} & 768        \\
      \ColHead{c-libexif-2}                    & 217        & 270        & 364  & 307        & \Best{167} & 268        & -          \\
      \ColHead{c-libexif-3}                    & 238        & 307        & 409  & 305        & 453        & \Best{218} & 438        \\
      \ColHead{c-libpng-1}                     & 1002       & 358        & 581  & 467        & -          & \Best{303} & -          \\
      \ColHead{c-libxml2-1}                    & \Best{228} & 328        & 511  & 283        & 424        & -          & 307        \\
      \ColHead{c-libxml2-2}                    & \Best{230} & 369        & 581  & 262        & 234        & 281        & 394        \\
      \ColHead{c-sqlite3-1}                    & 1304       & 610        & 1016 & \Best{472} & 655        & -          & 495        \\
      \ColHead{c-sqlite3-2}                    & 599        & 997        & -    & \Best{566} & -          & -          & -          \\
      \ColHead{c-sqlite3-3}                    & 1251       & 650        & 768  & \Best{538} & 603        & -          & 1084       \\
      \ColHead{c-sqlite3-4}                    & 724        & 640        & 798  & \Best{587} & -          & -          & -          \\
      \ColHead{c-sqlite3-5}                    & 524        & 688        & 898  & \Best{511} & -          & -          & -          \\
      \ColHead{java-commons-compress-0} & \Best{370} & 564        & 852  & 428        & 398        & 396        & 404        \\
      \ColHead{java-commons-compress-1} & 427        & 522        & 706  & 429        & 579        & 541        & \Best{411} \\
      \ColHead{java-commons-compress-2} & -          & -          & 2503 & \Best{725} & -          & -          & -          \\
      \ColHead{java-commons-compress-3} & 518        & 732        & 672  & \Best{473} & 722        & 548        & 542        \\
      \ColHead{java-commons-compress-4} & \Best{518} & 623        & 948  & 475        & 538        & 676        & -          \\
      \ColHead{java-commons-compress-5} & -          & 1604       & 1749 & \Best{847} & -          & -          & -          \\
      \ColHead{java-tika-1}                    & \Best{235} & 1157       & 313  & 826        & 303        & 522        & -          \\
      \ColHead{java-tika-2}                    & 1916       & 379        & 644  & 1273       & \Best{185} & 1144       & -          \\
      \ColHead{java-tika-3}                    & \Best{545} & 1845       & 1337 & 911        & 717        & 779        & 901        \\
      \ColHead{java-tika-4}                    & 867        & \Best{199} & 2450 & 813        & 213        & -          & 638        \\
      \ColHead{java-tika-5}                    & 1080       & \Best{166} & 1317 & 1070       & 813        & -          & -          \\
      \ColHead{java-tika-6}                    & -          & \Best{573} & 2852 & 1826       & -          & -          & -          \\
      \ColHead{java-tika-7}                    & 718        & \Best{236} & 693  & 1160       & 791        & 1160       & -          \\
      \ColHead{java-tika-8}                    & 1350       & 1200       & 1704 & 1249       & \Best{457} & -          & 1078       \\
      \ColHead{java-tika-9}                    & -          & \Best{600} & 1303 & 1262       & 1557       & 1121       & 662        \\
      \ColHead{java-tika-10}                    & 1122       & 1232       & 1717 & \Best{224} & 603        & 593        & 360        \\
      \ColHead{java-zookeeper-1}               & 563        & \Best{402} & 2691 & 504        & 421        & 558        & 1755       \\
      \ColHead{java-zookeeper-2}               & 877        & 572        & 1134 & 628        & \Best{362} & -          & -          \\ \bottomrule
    \end{tabularx}
  \end{adjustbox}

  \caption{Measured TTP (in seconds) for 34 CPVs from the Round~3 dataset,
comparing the performance of each integrated agent. Rows correspond to
individual CPVs and columns to the respective agents. Dash (-) indicates that
the agent did not produce a patch.}
  \label{tab:patch-performance-comparison-ttp}
\end{table*}

\begin{table*}[]
  \footnotesize
    \begin{adjustbox}{width=\linewidth}
      \newcommand{\RowHead}[1]{\textbf{#1}}
      \newcommand{\ColHead}[1]{\textbf{#1}}
      \newcommand{\Best}[1]{\textbf{#1}}
      \begin{tabularx}{\linewidth}{@{}l*{7}{>{\centering\arraybackslash}m{1.4cm}}@{}}
          \toprule
          \RowHead{CPV Name}                       &
          \RowHead{\sysmartian}                    &
          \RowHead{\makecell{\textsc{Multi}                                                                                                          \\\textsc{Retrieval}}} &
          \RowHead{\sysprism}                      &
          \RowHead{\sysvincent}                    &
          \RowHead{\makecell{\textsc{Claude}                                                                                                         \\\textsc{Like}}} &
          \RowHead{\sysaider}                      &
          \RowHead{\makecell{\textsc{SWE}                                                                                                            \\\textsc{Agent}}} \\
          \midrule
          \ColHead{c-curl-1}                       & 0.382        & 0.138        & 0.836 & 0.257        & 0.875        & \Best{0.032} & 0.110        \\
          \ColHead{c-curl-2}                       & 0.086        & 0.101        & 0.341 & 0.208        & 0.264        & -            & \Best{0.027} \\
          \ColHead{c-freerdp-1}                    & 0.119        & 0.398        & 0.463 & 0.485        & 4.450        & \Best{0.093} & -            \\
          \ColHead{c-freerdp-2}                    & \Best{0.106} & 0.309        & 1.180 & 0.295        & 0.984        & -            & -            \\
          \ColHead{c-freerdp-3}                    & 0.272        & \Best{0.153} & 0.385 & 0.279        & 0.206        & -            & -            \\
          \ColHead{c-libexif-1}                    & 0.077        & 0.319        & 0.286 & 0.366        & 0.268        & \Best{0.037} & 0.602        \\
          \ColHead{c-libexif-2}                    & \Best{0.055} & 0.350        & 0.400 & 0.357        & 0.131        & 0.089        & -            \\
          \ColHead{c-libexif-3}                    & 0.062        & 0.641        & 0.522 & 0.483        & 1.440        & \Best{0.053} & 0.276        \\
          \ColHead{c-libpng-1}                     & 6.130        & 0.174        & 0.664 & 0.456        & -            & \Best{0.095} & -            \\
          \ColHead{c-libxml2-1}                    & \Best{0.054} & 0.204        & 0.461 & 0.211        & 0.320        & -            & 0.097        \\
          \ColHead{c-libxml2-2}                    & \Best{0.051} & 0.389        & 0.365 & 0.161        & 0.177        & 0.289        & 0.146        \\
          \ColHead{c-sqlite3-1}                    & 1.460        & \Best{0.093} & 1.210 & 0.187        & 1.750        & -            & 0.098        \\
          \ColHead{c-sqlite3-2}                    & 0.629        & 0.626        & -     & \Best{0.366} & -            & -            & -            \\
          \ColHead{c-sqlite3-3}                    & 4.040        & \Best{0.252} & 1.310 & 0.294        & 1.900        & -            & 0.424        \\
          \ColHead{c-sqlite3-4}                    & 0.171        & \Best{0.159} & 0.715 & 0.353        & -            & -            & -            \\
          \ColHead{c-sqlite3-5}                    & 0.328        & 0.670        & 1.900 & \Best{0.244} & -            & -            & -            \\
          \ColHead{java-commons-compress-0} & 0.053        & 0.184        & 0.834 & 0.194        & 0.106        & 0.038        & \Best{0.029} \\
          \ColHead{java-commons-compress-1} & 0.052        & 0.162        & 0.491 & 0.183        & 0.205        & \Best{0.026} & 0.044        \\
          \ColHead{java-commons-compress-2} & -            & -            & 2.830 & \Best{0.499} & -            & -            & -            \\
          \ColHead{java-commons-compress-3} & 0.391        & 0.186        & 0.325 & 0.593        & 2.040        & \Best{0.023} & 0.174        \\
          \ColHead{java-commons-compress-4} & 0.113        & 0.143        & 0.434 & 0.201        & 0.740        & \Best{0.068} & -            \\
          \ColHead{java-commons-compress-5} & -            & 1.140        & 2.200 & \Best{0.536} & -            & -            & -            \\
          \ColHead{java-tika-1}                    & 0.109        & 0.191        & 1.090 & 0.308        & 1.760        & \Best{0.023} & -            \\
          \ColHead{java-tika-2}                    & 1.360        & 0.883        & 1.730 & 0.745        & 0.154        & \Best{0.040} & -            \\
          \ColHead{java-tika-3}                    & 0.063        & 0.161        & 0.991 & 0.203        & 0.081        & \Best{0.017} & 0.043        \\
          \ColHead{java-tika-4}                    & \Best{0.109} & 0.204        & 3.040 & 0.202        & 0.303        & -            & 0.363        \\
          \ColHead{java-tika-5}                    & 0.123        & \Best{0.084} & 0.704 & 0.377        & 0.660        & -            & -            \\
          \ColHead{java-tika-6}                    & -            & \Best{0.413} & 2.940 & 2.340        & -            & -            & -            \\
          \ColHead{java-tika-7}                    & 0.071        & 0.934        & 0.441 & 0.212        & 0.135        & \Best{0.048} & -            \\
          \ColHead{java-tika-8}                    & \Best{0.085} & 0.273        & 0.988 & 0.186        & 1.780        & -            & 0.305        \\
          \ColHead{java-tika-9}                    & -            & 0.278        & 0.871 & 0.345        & 0.286        & \Best{0.040} & 0.203        \\
          \ColHead{java-tika-10}                    & 0.059        & 0.192        & 0.332 & 0.200        & 0.191        & \Best{0.050} & 0.162        \\
          \ColHead{java-zookeeper-1}               & 0.047        & 0.162        & 5.790 & 0.319        & 0.167        & \Best{0.044} & 1.480        \\
          \ColHead{java-zookeeper-2}               & 0.227        & 0.262        & 1.630 & 0.396        & \Best{0.140} & -            & -            \\ \bottomrule
      \end{tabularx}
  \end{adjustbox}

  \caption{Measured LLM cost (USD) for 34 CPVs from the Round~3 dataset,
comparing the performance of each integrated agent. Rows correspond to
individual CPVs and columns to the respective agents. Dash (-) indicates that
the agent did not produce a patch.}
  \label{tab:patch-performance-comparison-llm-cost}
\end{table*}

\begin{table*}[]
  \footnotesize
    \begin{adjustbox}{width=\linewidth}
    \scriptsize
    \newcommand{\RowHead}[1]{\textbf{#1}}
    \newcommand{\ColHead}[1]{\textbf{#1}}
    \newcommand{\Best}[1]{\textbf{#1}}
    \resizebox{\linewidth}{!}{%
      \begin{tabularx}{\linewidth}{@{}ll*{7}{>{\centering\arraybackslash}m{1.05cm}}@{}}
        \toprule
        \RowHead{CPV Name}           &
        \RowHead{Bug Type}          &
        % RowHead{Sanitizer Type} &
        \RowHead{\sysmartian}        &
        \RowHead{\makecell{\textsc{Multi}                                                                                                                                                              \\\textsc{Retrieval}}} &
        \RowHead{\sysprism}          &
        \RowHead{\sysvincent}        &
        \RowHead{\makecell{\textsc{Claude}                                                                                                                                                             \\\textsc{Like}}} &
        \RowHead{\sysaider}          &
        \RowHead{\makecell{\textsc{SWE}                                                                                                                                                                \\\textsc{Agent}}} \\
        \midrule
        \ColHead{c-binutils-1}       & Use After Free              & \Fail            & \Sound            & \Sound            & \Fail             & \Fail           & -               & \Fail          \\
        \ColHead{c-binutils-2}       & Use After Free              & \Fail            & \Sound            & \Fail             & \Fail             & \Fail           & \Fail           & -              \\
        \ColHead{c-binutils-3}       & Heap Buffer Overflow  & \Fail            & \Sound            & \Sound            & \Sound            & -               & -               & -              \\
        \ColHead{c-binutils-4}       & Null Pointer Dereference    & \Sound           & \Sound            & \Sound            & \Sound            & -               & -               & -              \\
        \ColHead{c-ffmpeg-1}         & Heap Buffer Overflow  & \Fail            & \Sound            & \Sound            & \Sound            & -               & -               & -              \\
        \ColHead{c-ffmpeg-2}         & Heap Buffer Overflow  & \Fail            & \Sound            & \Sound            & \Sound            & -               & -               & -              \\
        \ColHead{c-ffmpeg-3}         & Heap Buffer Overflow  & \Sound           & \Sound            & \Sound            & \Sound            & -               & -               & -              \\
        \ColHead{c-ffmpeg-4}         & Stack Buffer Overflow & \Fail            & \Sound            & -                 & \Sound            & -               & -               & -              \\
        \ColHead{c-ffmpeg-5}         & Heap Buffer Overflow  & \Sound           & \Sound            & \Sound            & \Sound            & -               & -               & -              \\
        \ColHead{c-ffmpeg-6}         & Heap Buffer Overflow  & \Fail            & \Sound            & \Sound            & \Sound            & -               & -               & -              \\
        \ColHead{c-ffmpeg-7}         & Timeout                     & \Fail            & \Sound            & \Sound            & \Sound            & -               & -               & -              \\
        \ColHead{c-ffmpeg-8}         & Heap Buffer Overflow  & \Sound           & \Sound            & \Sound            & \Sound            & -               & -               & -              \\
        \ColHead{c-ffmpeg-9}         & Heap Buffer Overflow  & \Sound           & \Sound            & \Sound            & \Sound            & -               & -               & -              \\
        \ColHead{c-ffmpeg-10}        & Stack Buffer Overflow & \Fail            & \Sound            & \Fail             & \Sound            & -               & \Sound          & -              \\
        \ColHead{c-ffmpeg-11}        & Stack Buffer Overflow & \Fail            & \Sound            & \Sound            & \Sound            & -               & -               & -              \\
        \ColHead{c-pcre2-1}          & Heap Buffer Overflow  & \Fail            & \Sound            & \Sound            & \Sound            & -               & -               & -              \\
        \ColHead{c-pcre2-2}          & Heap Buffer Overflow  & \Fail            & \Fail             & \Fail             & \Fail             & -               & \Sound          & \Fail          \\
        \ColHead{c-pcre2-3}          & Heap Buffer Overflow  & \Fail            & \Sound            & \Sound            & \Fail             & \Fail           & -               & \Fail          \\
        \ColHead{c-pcre2-4}          & Heap Buffer Overflow  & \Fail            & \Sound            & \Sound            & \Sound            & -               & -               & -              \\
        \ColHead{c-sleuthkit-1}      & Null Pointer Dereference    & \Fail            & \Sound            & \Fail             & \Sound            & -               & \Fail           & -              \\
        \ColHead{c-sleuthkit-2}      & Timeout                     & \Fail            & \Sound            & \Fail             & \Sound            & -               & \Sound          & -              \\
        \ColHead{c-sleuthkit-3}      & Timeout                     & \Sound           & \Fail             & \Fail             & \Fail             & \Sound          & \Fail           & \Fail          \\
        \midrule
        \ColHead{Patch Success Rate} &                             & 6 / 22 (27.27\%) & 20 / 22 (90.91\%) & 15 / 21 (71.43\%) & 17 / 22 (77.27\%) & 1 / 4 (25.00\%) & 3 / 6 (50.00\%) & 0 / 4 (0.00\%) \\
        \bottomrule
      \end{tabularx}
    }
  \end{adjustbox}
  
% \begin{table}[h]
%   \caption{Average TTP per patch (seconds) across challenge projects}
%   \label{tab:patch-performance-comparison-ttp}
%   \begin{tabular}{@{}l *{7}{C{1.4cm}}@{}}
%     \toprule
%     \multicolumn{1}{l}{Project Name}         &
%     \multicolumn{1}{M{1.4cm}}{\sysmartian} &
%     \multicolumn{1}{M{1.4cm}}{\makecell{\textsc{Multi}\\\textsc{Retrieval}}} &
%     \multicolumn{1}{M{1.4cm}}{\sysprism}   &
%     \multicolumn{1}{M{1.4cm}}{\sysvincent} &
%     \multicolumn{1}{M{1.4cm}}{\makecell{\textsc{Claude}\\\textsc{Like}}} &
%     \multicolumn{1}{M{1.4cm}}{\sysaider}   &
%     \multicolumn{1}{M{1.4cm}}{\makecell{\textsc{SWE}\\\textsc{Agent}}} \\
%     \midrule
%     curl (C)                & 552.5  & 286.0  & 407.5  & 335.5  & 485.0  & 257.0 & 337.5  \\
%     freerdp (C)             & 419.7  & 442.0  & 459.0  & 343.0  & 586.3  & 147.0 & 2437.7 \\
%     libexif (C)             & 242.7  & 283.0  & 364.0  & 314.3  & 277.3  & 223.0 & 510.7  \\
%     libpng (C)              & 1002.0 & 358.0  & 581.0  & 467.0  & 221.0  & 303.0 & 309.0  \\
%     libxml2 (C)             & 229.0  & 348.5  & 546.0  & 272.5  & 329.0  & 211.5 & 350.5  \\
%     sqlite3 (C)             & 880.4  & 717.0  & 1031.0 & 534.8  & 1194.4 & 62.8  & 2493.2 \\
%     apache commons compress (Java)                & 2699.5 & 1094.0 & 1238.3 & 562.8  & 650.2  & 493.5 & 1232.5 \\
%     tika (Java)             & 1084.8 & 758.7  & 1433.0 & 1061.4 & 656.3  & 721.8 & 1236.7 \\
%     zookeeper (Java)        & 720.0  & 487.0  & 1912.5 & 566.0  & 391.5  & 580.5 & 1003.5 \\ \bottomrule
%   \end{tabular}
% \end{table}

  \caption{Patch success results for 22 CPVs from the Round~3.5 dataset. A checkmark (\Sound) indicates a plausible patch, while a cross (\Fail) denotes failure due to various reasons including patches that are still vulnerable, uncompilable, or exceed the time limit. Dash (-) in the tables indicates that the agent was not executed for that CPV, as a plausible patch had already been found and submitted. The bottom row summarizes the total number of plausible patches out of the total attempts for each agent.}
  \label{tab:patch-performance-extra}
\end{table*}

\begin{table*}[]
  \footnotesize
    \begin{adjustbox}{width=\linewidth}
    \scriptsize
    \newcommand{\RowHead}[1]{\textbf{#1}}
    \newcommand{\ColHead}[1]{\textbf{#1}}
    \newcommand{\Best}[1]{\textbf{#1}}
    \resizebox{\linewidth}{!}{%
      \begin{tabularx}{\linewidth}{@{}ll*{7}{>{\centering\arraybackslash}m{1.05cm}}@{}}
        \toprule
        \RowHead{CPV Name}            & \RowHead{Bug Type}           & \RowHead{\sysmartian}          & \RowHead{\makecell{\textsc{Multi}\\\textsc{Retrieval}}}          & \RowHead{\sysprism}          & \RowHead{\sysvincent}          & \RowHead{\makecell{\textsc{Claude}\\\textsc{Like}}}          & \RowHead{\sysaider}          & \RowHead{\makecell{\textsc{SWE}\\\textsc{Agent}}}          \\
        \midrule
        \ColHead{java-compress-1}       & File Path Traversal              & \Sound & \Sound & \Sound & \Sound & - & - & - \\
        \ColHead{java-compress-2}       & Null Pointer Dereference              & \Sound & \Sound & \Sound & \Sound & - & - & - \\
        \ColHead{java-compress-3}       & File Path Traversal              & \Fail & \Sound & \Sound & \Sound & - & - & - \\
        \ColHead{java-compress-4}       & File Path Traversal              & \Sound & \Sound & \Sound & \Sound & - & - & - \\
        % \ColHead{curl-1}     & Heap Buffer Overflow              & - & - & - & - & - & - & - \\
        \ColHead{c-curl-2}       & Heap Buffer Overflow              & \Fail & \Sound & \Fail & \Fail & - & \Fail & - \\
        \ColHead{c-curl-3}       & Format String Bug              & \Sound & \Sound & \Sound & \Sound & - & - & - \\
        \ColHead{c-curl-4}       & Heap Buffer Overflow              & \Sound & \Fail & \Sound & \Sound & \Fail & - & - \\
        \ColHead{c-curl-5}       & Null Pointer Dereference              & \Fail & \Sound & \Sound & \Sound & - & - & - \\
        \ColHead{java-dicoogle-1}       & Number Format Exception              & \Sound & \Sound & \Sound & \Sound & - & - & - \\
        \ColHead{c-freerdp-1}       & Heap Buffer Overflow              & \Fail & \Sound & \Sound & \Sound & - & - & - \\
        \ColHead{java-healthcare-1}       & Out Of Memory              & \Fail & \Sound & \Fail & \Fail & - & \Fail & - \\
        \ColHead{java-healthcare-2}       & Index Out Of Bounds              & \Fail & \Sound & \Fail & \Sound & - & - & \Fail \\
        \ColHead{java-hertzbeat-1}       & Out Of Memory              & \Fail & \Fail & \Fail & \Fail & - & \Fail & - \\
        \ColHead{java-hertzbeat-2}       & Index Out Of Bounds              & \Fail & \Fail & \Fail & \Fail & \Fail & \Fail & \Fail \\
        % \ColHead{jsoup-1}       & Index Out Of Bounds              & - & - & - & - & - & - & - \\
        % \ColHead{libexif-1}       & f01ed9b8-7d81-4122-b6c6-f5116c83bebc ()              & \Fail & - & - & - & - & - & - \\
        % \ColHead{libexif-2}       & Heap Buffer Overflow             & - & - & - & - & - & - & - \\
        % \ColHead{libxml2-1}       & 3d69259c-69f4-4342-aeda-5c76c6af9fa1 ()              & \Fail & - & - & - & - & - & - \\
        \ColHead{java-log4j2-1}       & Remote JNDI Lookup              & \Fail & \Sound & - & - & - & - & - \\
        \ColHead{c-mongoose-1}       & Memory Leak              & \Fail & \Fail & \Fail & \Fail & - & \Fail & - \\
        \ColHead{c-mongoose-2}       & Stack Buffer Overflow              & \Sound & \Sound & \Sound & \Sound & - & - & - \\
        \ColHead{c-mongoose-3}       & Heap Buffer Overflow             & \Fail & - & \Fail & \Fail & - & \Fail & - \\
        \ColHead{c-mongoose-4}       & Stack Buffer Overflow              & \Fail & - & \Fail & - & - & - & - \\
        \ColHead{java-pdfbox-1}       & Stack Overflow              & \Sound & \Sound & \Sound & \Sound & - & - & - \\
        \ColHead{java-pdfbox-2}       & Stack Overflow              & \Fail & \Sound & \Sound & \Fail & - & - & - \\
        \ColHead{java-pdfbox-3}       & Timeout              & \Fail & \Sound & \Sound & \Sound & - & - & - \\
        \ColHead{java-pdfbox-4}       & No Class Def Found Error               & \Sound & \Sound & \Sound & \Sound & - & - & - \\
        \ColHead{java-pdfbox-5}       & OS Command Injection              & \Sound & \Warning & \Sound & \Warning & - & - & - \\
        \ColHead{java-pdfbox-6}       & Invocation Target Exception              & \Sound & \Sound & \Sound & \Sound & - & - & - \\
        \ColHead{c-shadowsocks-1}       & Heap Buffer Overflow             & \Sound & - & - & - & - & - & - \\
        \ColHead{java-tika-1}       & Timeout              & \Fail & \Sound & \Sound & \Sound & - & - & - \\
        \ColHead{c-wireshark-1}       & Use After Free              & \Sound & \Sound & \Sound & \Fail & - & \Sound & - \\
        \ColHead{c-wireshark-2}       & Stack Buffer Overflow              & \Sound & \Sound & \Sound & \Sound & - & - & - \\
        \ColHead{c-wireshark-3}       & Null Pointer Dereference              & \Sound & \Sound & \Sound & \Sound & - & - & - \\
        \ColHead{c-wireshark-4}       & Format String Bug              & \Fail & \Fail & \Fail & \Sound & \Fail & - & \Fail \\
        \ColHead{c-wireshark-5}       & Heap Buffer Overflow             & \Fail & \Fail & \Fail & \Sound & - & - & \Fail \\
        \ColHead{c-wireshark-6}       & Null Pointer Dereference              & \Fail & \Sound & \Sound & \Sound & - & - & - \\
        \ColHead{c-wireshark-7}       & Stack Buffer Overflow              & \Fail & \Sound & \Sound & \Sound & - & - & - \\
        \ColHead{c-wireshark-8}       & Use After Free              & \Sound & \Sound & \Sound & \Sound & \Fail & - & - \\
        \ColHead{c-wireshark-9}       & Stack Buffer Overflow              & \Fail & \Sound & \Sound & \Fail & - & \Fail & - \\
        \ColHead{c-wireshark-10}       & Heap Buffer Overflow             & \Sound & \Sound & \Sound & \Sound & - & - & - \\
        \ColHead{c-wireshark-11}       & Stack Buffer Overflow              & \Sound & \Sound & \Sound & \Sound & - & - & - \\
        \ColHead{c-wireshark-12}       & Heap Buffer Overflow             & \Sound & \Sound & \Sound & \Sound & - & - & - \\
        \ColHead{c-wireshark-13}       & Global Buffer Overflow              & \Sound & \Sound & \Sound & \Sound & - & - & - \\
        \ColHead{c-wireshark-14}       & Global Buffer Overflow              & - & - & \Sound & - & - & - & - \\
        \ColHead{c-xz-1}       & Use After Free              & \Fail & - & - & - & - & - & - \\
        \midrule
        \ColHead{Patch Success Rate} &                             & 20 / 41 (48.78\%) & 29 / 36 (80.56\%) & 29 / 39 (74.36\%) & 27 / 37 (72.97\%) & 0 / 4 (0.00\%) & 1 / 8 (12.50\%) & 0 / 4 (0.00\%) \\
        \bottomrule
      \end{tabularx}
    }
  \end{adjustbox}

  \caption{Patch success results for 42 CPVs from the AIxCC Final Competition. A checkmark (\Sound) indicates a plausible patch, while a cross (\Fail) denotes failure due to various reasons including patches that are still vulnerable, uncompilable, or exceed the time limit. A warning (\Warning) indicates a patch that was submitted but scored as incorrect due to the hidden functionality tests. Dash (-) in the tables indicates that the agent was not executed for that CPV, as a plausible patch had already been found and submitted. The bottom row summarizes the total number of plausible patches out of the total attempts for each agent.}
  \label{tab:patch-performance-final}
\end{table*}

\clearpage
\section{Custom LLMs in \sys-Patching}
\label{ss:custom-model}

Large language models (LLMs) have recently demonstrated superior performance on code-related tasks \cite{yang2024swe,wang2025openhands}, \eg achieving success rates of 60-76\% on SWE-Bench Verified \cite{jimenez2024swebench}. Despite these advancements, \emph{effectively} and \emph{efficiently} providing LLMs with relevant context for complex coding tasks, such as \emph{security patch generation} for AIxCC, remains challenging.

Practical experience demonstrates that effective \emph{context engineering}—the strategic selection and presentation of task-relevant information—is crucial for agent performance \cite{wei2022chain}. Even the most advanced agents struggle when deprived of appropriate documentation, API references, or source code context, much like skilled developers facing unfamiliar codebases \cite{mei2025surveycontextengineeringlarge}.  

We \emph{reconfirm that context engineering is also crucial for security patch generation} during the postmortem analysis of the AIxCC Semifinal.
In particular, as a controlled experiment, we examine the one of AIxCC Semifinal challenges, \ccw{challenge-004-nginx-cp/cpv15} \footnote{\url{https://github.com/aixcc-public/challenge-004-nginx-cp/blob/main/.internal_only/cpv15}}.
In this experiment, we observed that our baseline patching agent Aider \cite{aider} generates a patch without retrieving the \ccw{typedef} definitions for two key structures from the Nginx codebase (\ie \autoref{fig:nginx-typedef-example}) consistently produced an uncompilable patched codebase.
However, enforcing to retrieve the key \ccw{typedef} definitions yielded the successful compilation of patched codebase.
We empirical validated this by conducting 20 random experiments where the patching performance is in \autoref{fig:custom-model-motivation}.
As can be seen, the clear separation between “with \ccw{typedef}” and “without \ccw{typedef}” runs demonstrates the decisive role of accurate context retrieval in enabling valid patch generation.
% focusing retrieving undefined terms -- connection to multi-turn retrieval
Based on this critical performance leap, our direction of patching agents in \textsc{Crete} is more focused on retrieving missing definitions, \ie \emph{context engineering for missing context in patching}, leading to our first performant agent \ccw{MultiRetrieval}.

%As a controlled demonstration—independent from the performance of our deployed system—we examine the AIxCC Semifinals challenge \ccw{challenge-004-nginx-cp/cpv15} \footnote{\url{https://github.com/aixcc-public/challenge-004-nginx-cp/blob/main/.internal_only/cpv15}}. In this experiment, generating a patch without retrieving the \ccw{typedef} definitions for two key structures from the Nginx codebase consistently produced uncompilable output, whereas inclusion of these typedefs yielded the successful compilation of patched codebase. Using the Aider \cite{aider}, the same query was repeated 20 times on a copy of the repository, with outcomes summarized in \autoref{fig:custom-model-motivation}. The clear separation between “with typedef” and “without typedef” runs demonstrates the decisive role of accurate context retrieval in enabling valid patch generation.

% engineering to learning
Building on this achievement, we further adances our direction from \emph{code context engineering}—a largely manual, heuristic-driven process—towards \emph{code context learning}, in which an agent automatically learns to identify and retrieve missing and relevant code artifacts for a patching task at hand. This shift enables a learnable approach that adapts dynamically to varying bug contexts and codebases.

\begin{figure}[t]
  \centering
  \begin{promptbox}{}
  \begin{promptcontent}\input{code/nginx_typedef_example.c}\end{promptcontent}
  \end{promptbox}
  \caption{Necessary \ccw{Typedef} definitions of \ccw{challenge-} \ccw{004-nginx-cp/cpv15} for successful  patching.}
  \label{fig:nginx-typedef-example}
\end{figure}

% \begin{figure}[t]
%   \centering
%   \includegraphics[width=\columnwidth]{fig/custom-model-motivation.png}
%   \caption{Experiment on \ccw{challenge-004-nginx-cp} \ccw{/cpv15} showing the effect of retrieving typedef definitions. Each bar represents one of 20 repeated Aider queries; patches generated without typedefs consistently failed to compile, while those with typedefs compiled successfully.}
%   \label{fig:custom-model-motivation2}
% \end{figure}

\begin{table}[t]
  \newcommand{\BuildSuccess}{\ding{109}}
  \newcommand{\BuildFail}{\ding{56}}
  \centering
  \resizebox{\linewidth}{!}{%
    \footnotesize
    \begin{tabular}{@{}lccccccccccccccccccccc@{}}
      \toprule
      \multirow{2}{*}{\textbf{}} & \multicolumn{20}{c}{\textbf{Build Attempts}} & \multirow{2}{*}{\textbf{Success Rate}}                                                                                                                                                                                                                                                                                                              \\
      \cmidrule(lr){2-21}
                                 & \textbf{1}                                   & \textbf{2}                             & \textbf{3}    & \textbf{4}    & \textbf{5}    & \textbf{6}    & \textbf{7} & \textbf{8}    & \textbf{9}    & \textbf{10}   & \textbf{11}   & \textbf{12}   & \textbf{13}   & \textbf{14}   & \textbf{15}   & \textbf{16}   & \textbf{17}   & \textbf{18}   & \textbf{19}   & \textbf{20}   &               \\
      \midrule
      \textbf{w/o \ccw{typedef}} & \BuildFail                                   & \BuildSuccess                          & \BuildFail    & \BuildFail    & \BuildFail    & \BuildFail    & \BuildFail & \BuildFail    & \BuildFail    & \BuildFail    & \BuildFail    & \BuildFail    & \BuildFail    & \BuildSuccess & \BuildSuccess & \BuildFail    & \BuildFail    & \BuildSuccess & \BuildSuccess & \BuildFail    & \textbf{5/20} \\
      \textbf{w/ \ccw{typedef}}  & \BuildSuccess                                & \BuildFail                             & \BuildSuccess & \BuildSuccess & \BuildSuccess & \BuildSuccess & \BuildFail & \BuildSuccess & \BuildSuccess & \BuildSuccess & \BuildSuccess & \BuildSuccess & \BuildSuccess & \BuildSuccess & \BuildSuccess & \BuildSuccess & \BuildSuccess & \BuildSuccess & \BuildSuccess & \BuildSuccess & \textbf{18/20} \\
      \bottomrule
    \end{tabular}
  }
  \caption{Ablation study on \ccw{challenge-004-nginx-cp} \ccw{/cpv15} using
    \sysaider, evaluating the impact of type definitions on build success rate.
    Each cell shows whether a generated patch compiled successfully (\BuildSuccess) or
    failed (\BuildFail) across 20 repeated queries. Including type definitions led to a
    markedly different success rate compared to excluding them.}
  \label{fig:custom-model-motivation}
\end{table}

\subsection{Problem: Code Context Learning}

We consider LLM-based security patch generation, \ie given a code base and a vulnerability-reporting or crash log, generate a patch to fix the vulnerability while maintaining functionalities. Inspired by our observation in \autoref{fig:custom-model-motivation}, we focus on selecting missing but right contexts for patching with the following requirements.
\begin{squishitemize}
    \item \textbf{Context window limitations.} Even with today's long-context models, supplying an entire large-scale codebase is infeasible due to attention complexity and window size limits. To avoid this issue, a patching agent must operate with a window that is orders of magnitude smaller than the full repository.
    \item \textbf{Prohibitive API costs.} Processing very large inputs with agent is financially expensive. Repeatedly sending thousands lines to a agent for patching would be unsustainable.
\end{squishitemize}

To this end, we formulate our code context learning in reinforcement learning (RL).
% basic notations
In particular,
let
$\Delta(\mathcal{S})$ be a distribution over a set $\mathcal{S}$,
$T$ be the number of challenge project vulnerabilities (CPVs),
$H$ be the number of turns,
and
$\mathcal{X}$ be a set of token sequences, which is a set of states in the standard RL.
Here, 
$x_{t, h} \in \mathcal{X}$ represents an input at the $t$-th CPV and the $h$-th turn.
For example,
$x_{t, 1}$ represents an instruction for LLMs, meta data of the $t$-th CPV, and the summary of a crash log, and
$x_{t, h}$ for $h > 1$ includes the summary of $x_{t, 1}$ and retrieved code contexts after $h$ turns.
Finally, 
$\mathcal{A} \coloneqq \mathcal{X}$ be a set of actions for code context retrieval, consist of tokens sequences to represent the list of symbol names to be retrieved, \eg function and struct definitions.

%  policy
We consider a policy for code context retrieval $\pi: \mathcal{X} \to \Delta(\mathcal{A})$, which chooses to retrieve necessary code contexts.
In particular,
This policy chooses a set of actions, requesting to retrieves the definitions of symbols (\eg retrieve the definition of a \ccw{ngx_http_userid_ctx_t} struct). 
Note that different to a conventional policy in RL, our context policy can select multiple actions, where the policy is learned to select actions of retrieving undefined symbols useful for correct patch generation.

% learning goal
In code context learning, we find a policy $\pi$ for code context retrieval that maximizes the average return over $T$ episodes of CPVs. 
\begin{equation*}
  \max_{\pi} \frac{1}{T} \sum_{t=1}^T \sum_{h=1}^H R(x_{t, h}, a_{t, h}),
\end{equation*}
where $R: \mathcal{X} \times \mathcal{A} \to [0, 1]$ is a reward function, which will be defined in the following section,
and
$a_{t, h}$ is an action chosen by $\pi$ given $x_{t, h}$.

%% \begin{squishitemize}
%% \item $\mathcal{X}$: a set of token sequences.
%% \item $\mathcal{A} \subseteq \mathcal{X}$: a set of code context retrieval actions
%%   \begin{squishitemize}
%%   \item We specifically consider a set of symbol names to be retrieved, \eg function and struct definitions.
%%   \end{squishitemize}
%% \item $T$: the number of challenge project vulnerabilities (CPVs)
%% \item $H$: the number of turns
%% \item $x_{t, h} \in \mathcal{X}$: an input at the $t$-th CPV and the $h$-th turn.
%%   \begin{squishitemize}
%%   \item $x_{t, 1}$: an initial input, including a {\color{red}system prompt} and crash log\SP{beta, what else?}.
%%   \item $x_{t, >1}$: an input, including a {\color{red}system prompt}, crash log, and selected context\SP{beta, what else?}.
%%   \item the context includes function and struct definitions.
%%   \end{squishitemize}
%% \item $\pi: \mathcal{X} \to 2^\mathcal{A}$: a code context policy, which selects code contexts.
%%   \begin{squishitemize}
%%   \item an action retrieves the definition of symbols (\eg retrieve the definition of a \ccw{read} function)\SP{beta, use the previous (will-be-added) example}
%%   \item our context policy can select multiple actions.
%%   \item only select undefined symbols necessary for patch generation.
%%   \end{squishitemize}
%% \item objective: find a code policy $\pi$ that maximizes the average expected returns over $T$ episodes of CPVs. 
%%   \begin{equation*}
%%     \max_{\pi} \frac{1}{T} \sum_{t=1}^T \sum_{h=1}^H R(x_{t, h}, \pi(x_{t, h}))
%%   \end{equation*}
%% \end{squishitemize}

\begin{remark*}
We can significantly reduce code context size via code context learning.
In particular, feeding the entire code context for patching may provide successful patch generation
but prohibitive due to the limited context size and API costs for LLMs.
We overcome these challenges by learning to retrieve necessary code contexts for vulnerability patching via RL. 
\end{remark*}

Note that in contrast to \emph{in-context learning}, which relies on passively conditioning on provided input sequences, our \emph{code context learning} framework actively acquires and refines relevant source code artifacts through an explicit retrieval policy. This enables adaptive, targeted construction of context tailored to each vulnerability.

\subsection{Solution: Learning to Retrieve Code Context}

Our approach pursues three primary objectives:

\begin{squishitemize}
    \item \textbf{Modeling code context retrieval agent.}  We develop a specialized retrieval module that identifies concise, highly relevant code snippets for downstream patch generation. Unlike existing \emph{generic} methods, this retrieval agent is trainable to be \emph{specialized} for security patching on specific bug contexts.
    \item \textbf{Modeling a multi-turn interactive retrieval.}
      %Multi-turn interaction enables the retrieval process to be iterative and adaptive.
      Human developers rarely locate a bug's root cause in one step; instead, they form a hypothesis, examine some code, run tests, gather clues, and then refine their search. We want our agent to emulate this process with multiple retrieval turns. At each step, it can utilize information gathered in previous steps to decide what to read or fetch next. By explicitly modeling retrieval as a multi-turn dialogue with the codebase, our system aims to replicate the way a person would gradually narrow down the fault, leading to have concise code context.
    \item \textbf{Learning the retrieval policy for patching success.}
      We train our multi-turn code retrieval agent via reinforcement learning to tailor our agent for efficiently and effectively adapt to security patch generation. To this end, we consider repair success our reward function. 
      %We train a lightweight retrieval model to optimize directly for successful bug-fixing outcomes. This enables smaller models to deliver strong performance through task-specific fine-tuning, reducing the necessity for larger, computationally expensive models.
\end{squishitemize}

\subsubsection{Learning Setup}

\begin{figure}[t]
  \centering
  \includegraphics[width=\columnwidth]{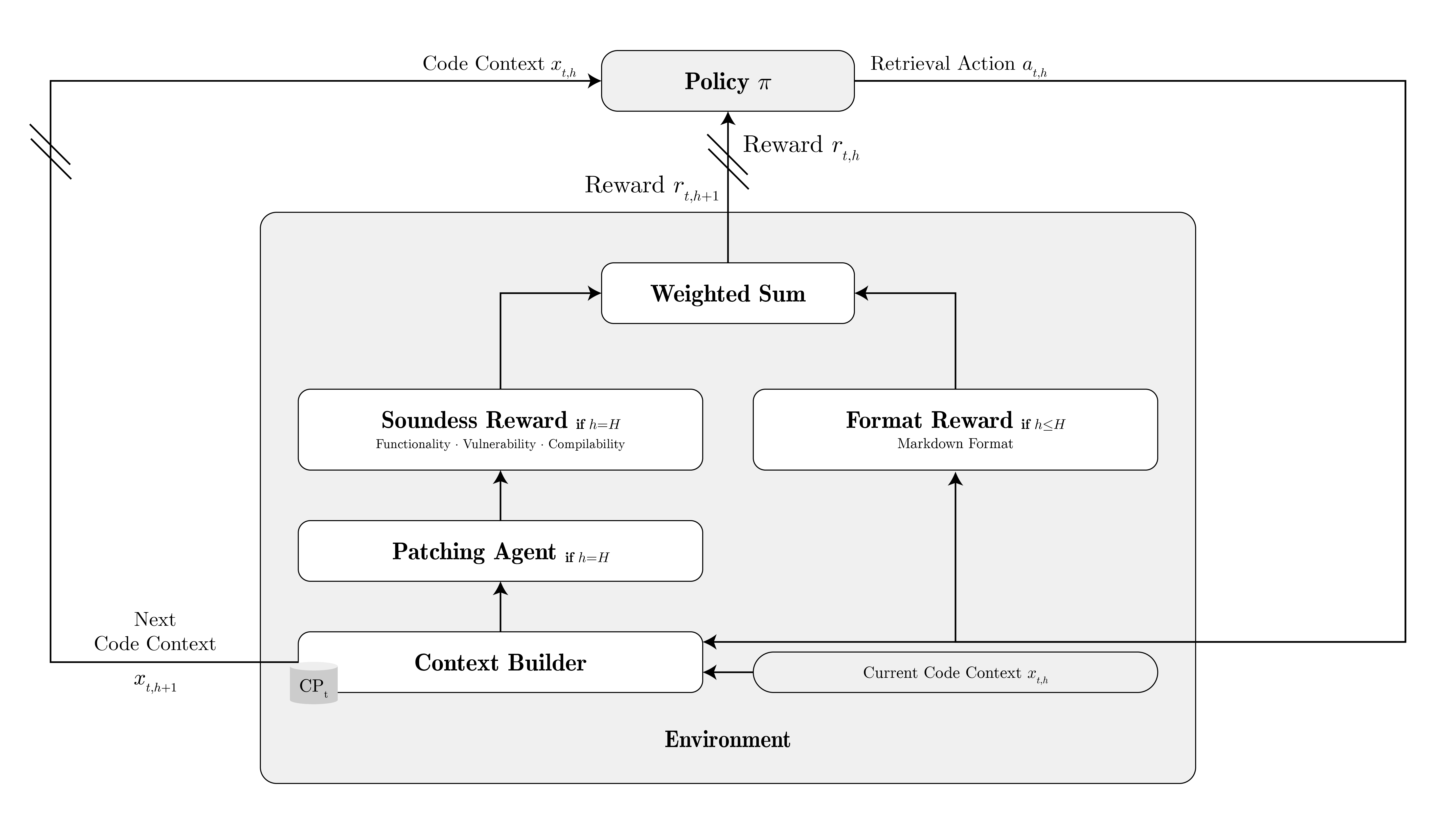}
  \caption{\textbf{RL setup for code patching via multi-turn retrieval.} The policy $\pi$ iteratively conditions on contextual states $x_{t,h}$ to produce retrieval actions $a_{t,h}$. The environment integrates retrieved artifacts with a challenge project $\text{CP}_t$ to construct subsequent states $x_{t,h+1}$. The reward $r_{t,h+1}$ combines format validation (structural compliance) of retrieval actions  and soundness assessment (patch quality) of a patched codebase, with aggregated rewards guiding policy optimization across multi-turn interactions for code retrieval.}
  \label{fig:custom-model-overview}
\end{figure}

Our approach is multi-turn retrieval-based code patching via RL, as illustrated its setup in \autoref{fig:custom-model-overview}. The learning process begins with an initial code context $x_{t,1}$ derived from a crash log along with a system prompt. At each turn $h \in \{1, \ldots, H\}$, our retrieval agent, represented by a policy $\pi$, conditions on the current contextual state $x_{t,h}$ to produce a set of retrieval actions $a_{t,h}$. The action is defined as the retrieval actions of code symbols (\eg functions, structures, or types) from the target codebase where the format validity of retrieval actions is measured by a format reward $r_{t, h+1}$. The environment then integrates these retrieved artifacts with the challenge project $\text{CP}_t$, constructing the subsequent state $x_{t,h+1}$ (\ie aggregated code context). After $H$ turns, the final aggregated context $x_{t,H}$ is passed to an external general-purpose language model (\eg GPT-4, Claude, or Gemini) for patch synthesis where the soundness of the patched codebase is measured by the soundness reward $r_{t, H+1}$. Importantly, this architecture deliberately decouples context acquisition from patch generation, allowing us to leverage state-of-the-art LLMs while maintaining precise control over the information flow.

\subsubsection{Code Context Learning via Multi-turn GRPO}
\PP{Overview}
At a high level, our training framework adapts Group Relative Policy Optimization (GRPO) \cite{DBLP:journals/corr/abs-2402-03300} to the multi-turn retrieval setting. 
Unlike standard single-turn GRPO, where each prompt produces a single completion, our agent generates a trajectory of retrieval actions over multiple turns, each conditioned on the evolving context. 
To ensure effective learning, we define a composite reward that provides both intermediate feedback on the structural validity of retrievals and a final signal tied to patch soundness. 
This design allows the agent to refine its retrieval policy step by step, gradually constructing the minimal yet sufficient context required for secure patch generation. 
The following sections detail the reward modeling, training objective, and online learning procedure.

\PP{Reward Modeling}
To guide the agent toward effective retrieval strategies, we employ a composite reward signal. For a given state-action pair $(x_{t,h}, a_{t,h})$, the reward is defined as:
\begin{align*}
R(x_{t,h}, a_{t,h}) = \lambda_{\mathrm{fmt}} \, R_{\mathrm{fmt}}(x_{t,h}, a_{t,h}) + \lambda_{\mathrm{snd}} \, R_{\mathrm{snd}}(x_{t,h}, a_{t,h})
\end{align*}
where $\lambda_{\mathrm{fmt}}$ and $\lambda_{\mathrm{snd}}$ are non-negative weighting coefficients. The reward consists of two components:
\begin{squishitemize}
    \item A \emph{format reward function} ($R_{\mathrm{fmt}}$), evaluated at every turn ($h \leq H$), which encourages structurally correct retrieval actions that adhere to Markdown formatting and contain properly identified code symbols.
    \item A \emph{soundness reward function} ($R_{\mathrm{snd}}$), evaluated at the final turn ($h=H$), which assesses the quality of the generated patch based on compilation success, functionality preservation, and vulnerability remediation. This directly incentivizes the agent to retrieve context that is most useful for generating a correct patch.
\end{squishitemize}

\begin{figure*}[t]
    \centering
    \includegraphics[width=\textwidth]{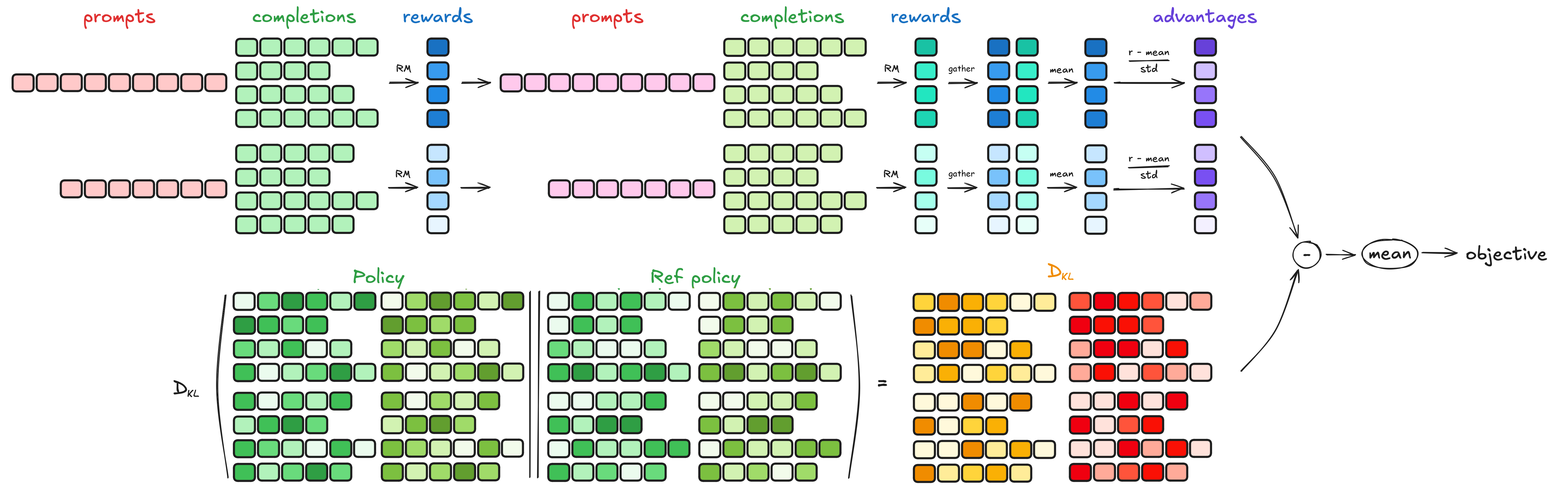}
    \caption{\textbf{Multi-turn GRPO training flow (adapted and re-colored from the original Hugging Face diagram).} For each batch of prompts (pink), the policy generates multi-turn completions (green); a reward model assigns per-turn rewards (shades of blue). Rewards are centered and scaled to yield advantages (purple) that drive the policy update. In parallel, token-wise $D_{\text{KL}}$ between the current and reference policies is computed (orange$\to$red). The final loss balances reward maximization and regularization.}
    \label{fig:custom-model-multi-turn-grpo}
\end{figure*}

\PP{Training Objective}
We then find a policy $\pi$ that maximizes the average return using Group Relative Policy Optimization (GRPO) \cite{DBLP:journals/corr/abs-2402-03300}  adapted for our multi-turn setting. As summarized in \autoref{fig:custom-model-multi-turn-grpo}, the training process involves generating multi-turn completions from the policy, using our reward model to assign per-turn rewards, and computing advantages to guide the policy update. 

%Let $G$ be the number of outputs in a batch, $o_i$ the $i$-th output sequence, $q$ the prompt, and $\hat{A}_{i,t}$ the token-level advantage. The single-turn GRPO loss is shown in Equation (\ref{eq:grpo-single}), and we extend this objective to our multi-turn scenario by aggregating the loss across all turns $h=1, \ldots, H_i$ for each instance $i$. With turn-conditional prompts $q_h$ and per-turn outputs $o_{i,h}$, the multi-turn GRPO objective is given in Equation (\ref{eq:grpo-multiturn}), where $N = \sum_{i=1}^{G} \sum_{h=1}^{H_i} |o_{i,h}|$ is the total number of tokens in the batch. This aggregated reward signal propagates through the entire multi-turn trajectory to optimize the policy.
Let $G$ be the number of groups and $\hat{A}_{t,g}$ is the advantage where: $$
\hat{A}_{t,g} = \frac{r_{t,g} - \mathrm{mean}(\{r_{t,1}, r_{t,2}, \cdots, r_{t,G}\})}
           {\mathrm{std}(\{r_{t,1}, r_{t,2}, \cdots, r_{t,G}\})}.
$$ when $r_{t,g}$ is the reward of the $g$-th group at the $t$-th CPV.
% single to multi
The conventional single-turn GRPO objective is shown as following:
%% \begin{equation}
%%   \label{eq:grpo-single}
%%   \mathcal{L}_{\mathrm{GRPO}}(\theta) = -\frac{1}{\sum_{i=1}^{G} |o_i|} \sum_{i=1}^{G} \sum_{t=1}^{|o_i|} \left( \frac{\pi_{\theta}(o_{i,t} \mid q, o_{i,<t})}{[\pi_{\theta}(o_{i,t} \mid q, o_{i,<t})]_{\mathrm{no\;grad}}} \, \hat{A}_{i,t} - \beta \, D_{\mathrm{KL}}\!\left[ \pi_{\theta}(\cdot \mid q, o_{i,<t}) \,\|\, \pi_{\mathrm{ref}}(\cdot \mid q, o_{i,<t}) \right] \right)
%% \end{equation}
\begin{equation*}
  %\label{eq:grpo-single}
  \mathcal{L}_{\mathrm{GRPO}}(\pi) = \frac{1}{T} \sum_{t=1}^T \frac{1}{G} \sum_{g=1}^{G} \left( \frac{\pi(a_{t,1} \mid x_{t, 0})}{\pi_{\text{old}}(a_{t,1} \mid x_{t, 0})} \, \hat{A}_{t,g} - \beta \, D_{\mathrm{KL}}\!\left[ \pi(\cdot \mid x_{t, 1}) \,\|\, \pi_{\mathrm{ref}}(\cdot \mid x_{t, 0}) \right] \right),
\end{equation*}
where $\pi(a \mid x)$ is the probability of an action $a$ computed from from an action distribution $\pi(x)$,
$D_{\mathrm{KL}}$ is the Kullback-Leibler (KL) divergence,
$\beta$ is a regularization parameter,
$\pi_{\text{old}}$ is a ``frozen'' $\pi$ (\ie $\pi$ without allowing update),
and
$\pi_{\mathrm{ref}}$ is an initial policy. 
Note that the KL divergence term is included to regularize the policy, preventing it from deviating too far from a frozen reference policy $\pi_{\mathrm{ref}}$.

% multi
We extend the original GRPO objective to our multi-turn scenario by aggregating the loss across all turns $h=1, \ldots, H$, as follows:
\begin{equation*}
  \mathcal{L}_{\mathrm{GRPO}}^{\mathrm{multi}}(\pi) = \frac{1}{T} \sum_{t=1}^T \sum_{h=1}^H \frac{1}{G} \sum_{g=1}^{G} \left( \frac{\pi(a_{t,h} \mid x_{t, h})}{\pi_{\text{old}}(a_{t,h} \mid x_{t, h})} \, \hat{A}_{t,g} - \beta \, D_{\mathrm{KL}}\!\left[ \pi(\cdot \mid x_{t, h}) \,\|\, \pi_{\mathrm{ref}}(\cdot \mid x_{t, h}) \right] \right).
  %\label{eq:grpo-multiturn}
\end{equation*}
This aggregated reward signal propagates through the entire multi-turn trajectory to optimize the policy.
Note that the clipping term from the original GRPO formulation \cite{DBLP:journals/corr/abs-2402-03300} is omitted here for notational simplicity.

%With turn-conditional prompts $q_h$ and per-turn outputs $o_{i,h}$, the multi-turn GRPO objective is given in Equation (\ref{eq:grpo-multiturn}), where $N = \sum_{i=1}^{G} \sum_{h=1}^{H_i} |o_{i,h}|$ is the total number of tokens in the batch. This aggregated reward signal propagates through the entire multi-turn trajectory to optimize the policy.
%% \begin{align}
%% \label{eq:grpo-multiturn}
%% \mathcal{L}^{\mathrm{multi}}_{\mathrm{GRPO}}(\theta) = -\frac{1}{N} \sum_{i=1}^{G} \sum_{h=1}^{H_i} \sum_{t=1}^{|o_{i,h}|} \Bigg( &\frac{\pi_{\theta}(o_{i,h,t} \mid q_h, o_{i,h,<t})}{[\pi_{\theta}(o_{i,h,t} \mid q_h, o_{i,h,<t})]_{\mathrm{no\;grad}}} \, \hat{A}_{i,h,t} \nonumber \\
%% &- \beta \, D_{\mathrm{KL}}\!\left[ \pi_{\theta}(\cdot \mid q_h, o_{i,h,<t}) \,\|\, \pi_{\mathrm{ref}}(\cdot \mid q_h, o_{i,h,<t}) \right] \Bigg)
%% \end{align}

\PP{Online Learning}
As optimizing $\mathcal{L}_{\mathrm{GRPO}}(\pi)$ over all $T$ CPVs is memory-intensive, we update the policy in an online manner.
In particular, each CPV is sequentially provided and then
the learning agent concentrates on the single CPV repeatedly until a success plateau, then advances to the next CPV. This mirrors developer workflows, yielding sharp intra-case improvements while occasionally resetting performance on new cases.

%We adopt a relaxed online learning schedule (\autoref{fig:custom-model-online}): the agent concentrates on a single case repeatedly until a success plateau, then advances to the next case. This mirrors developer workflows, yielding sharp intra-case improvements while occasionally resetting performance on new cases.

\subsection{Use Case: \ccw{babynginx/cpv-0}}

We now present a representative case study of the learned policy with $H=2$ on \ccw{babynginx/cpv-0}, which is part of our internal benchmark rather than an official AIxCC challenge. This CPV is motivated by CVE-2022-0995 (\ie an out-of-bounds access in the Linux kernel) and adapts its root cause to a simplified Nginx setting. Specifically, the bug arises in the parsing of a custom request header \ccw{X-Feature}, where a numeric value is converted using \ccw{ngx\_atoi} and then used to set a bitmap index. The range check incorrectly uses \ccw{sizeof(bitmap) * BITS\_PER\_LONG}, which permits out-of-bounds access. The intended fix location corresponds to the first frame of the crash stack trace, making it tractable for automated patching. However, without retrieval of the relevant type definitions, an agent is unlikely to produce a correct patch. Moreover, the function contains a subtle trap: unlike the standard \ccw{atoi}, \ccw{ngx\_atoi} maps both negative inputs and strings like ``$+1$'' to \ccw{$-1$}, which often misleads naïve patching strategies. This case thus provides a controlled yet realistic testbed to evaluate whether multi-turn retrieval can recover the missing type definitions and avoid misleading local fixes.

\PP{Turn 1.}
The process begins with an AddressSanitizer report, which clearly indicates the failure site:

\begin{promptbox}{Code Context $x_{t, 1}$}
  \begin{promptcontent}
    \begin{Verbatim}[commandchars=\\\{\},codes={\catcode`\$=3\catcode`\^=7\catcode`\_=8\relax}]
==32==ERROR: AddressSanitizer: heap\PYZhy{}buffer\PYZhy{}overflow on address 0x50200000b588 
at pc 0x555909e7d03d bp 0x7ffc299a95c0 sp 0x7ffc299a95b8
READ of size 8 at 0x50200000b588 thread T0
SCARINESS: 33 (8\PYZhy{}byte\PYZhy{}read\PYZhy{}heap\PYZhy{}buffer\PYZhy{}overflow\PYZhy{}far\PYZhy{}from\PYZhy{}bounds)
    \PYZsh{}0 0x555909e7d03c in ngx\PYZus{}http\PYZus{}process\PYZus{}custom\PYZus{}features /src/nginx/src/http/ngx\PYZus{}http\PYZus{}request.c:2006:5
    \PYZsh{}1 0x555909e912af in ngx\PYZus{}http\PYZus{}process\PYZus{}request\PYZus{}headers /src/nginx/src/http/ngx\PYZus{}http\PYZus{}request.c:1507:23
    \PYZsh{}2 0x555909e8e7d0 in ngx\PYZus{}http\PYZus{}process\PYZus{}request\PYZus{}line /src/nginx/src/http/ngx\PYZus{}http\PYZus{}request.c:1202:13
    (additional stack frames elided)
\end{Verbatim}

  \end{promptcontent}
\end{promptbox}

\begin{promptbox}{Retrieved Action $a_{t, 1}$ with Reasoning}
  \begin{promptcontent}
    \begin{Verbatim}[commandchars=\\\{\},codes={\catcode`\$=3\catcode`\^=7\catcode`\_=8\relax}]
\PY{g+gu}{\PYZsh{}\PYZsh{} Reasoning}

The potential vulnerability is an out\PYZhy{}of\PYZhy{}bounds read error in the \PY{l+s+sb}{`ngx\PYZus{}http\PYZus{}process\PYZus{}custom\PYZus{}features`} function.
(ommitted)

\PY{g+gu}{\PYZsh{}\PYZsh{} Relevant Symbols}

\PY{k}{1.} \PY{l+s+sb}{`ngx\PYZus{}http\PYZus{}process\PYZus{}custom\PYZus{}features`}
(ommitted)
\end{Verbatim}

  \end{promptcontent}
\end{promptbox}

The log establishes the precise function requiring modification: \ccw{ngx\_http\_process\_custom\_features}. Without this initial localization, subsequent retrievals would lack direction, resulting in irrelevant or incomplete context.

\PP{Turn 2.} At this stage, partial context has been assembled, but crucial information about valid feature indices is still missing.

\begin{promptbox}{Augmented Code Context for $x_{t, 2}$}
  \begin{promptcontent}
    \begin{Verbatim}[commandchars=\\\{\},codes={\catcode`\$=3\catcode`\^=7\catcode`\_=8\relax}]
\PY{c+c1}{// nginx/src/http/ngx\PYZus{}http\PYZus{}request.c}

\PY{k}{static}\PY{+w}{ }\PY{n}{ngx\PYZus{}int\PYZus{}t}
\PY{n+nf}{ngx\PYZus{}http\PYZus{}process\PYZus{}custom\PYZus{}features}\PY{p}{(}\PY{n}{ngx\PYZus{}http\PYZus{}request\PYZus{}t}\PY{+w}{ }\PY{o}{*}\PY{n}{r}\PY{p}{,}\PY{+w}{ }\PY{n}{ngx\PYZus{}table\PYZus{}elt\PYZus{}t}\PY{+w}{ }\PY{o}{*}\PY{n}{h}\PY{p}{,}
\PY{+w}{    }\PY{n}{ngx\PYZus{}uint\PYZus{}t}\PY{+w}{ }\PY{n}{offset}\PY{p}{)}
\PY{p}{\PYZob{}}
\PY{+w}{    }\PY{c+c1}{// (ommitted)}
\PY{p}{\PYZcb{}}

\PY{c+c1}{// (additional code elided)}
\end{Verbatim}

  \end{promptcontent}
\end{promptbox}

\begin{promptbox}{Retrieved Action $a_{t, 2}$ with Reasoning}
  \begin{promptcontent}
    \begin{Verbatim}[commandchars=\\\{\},codes={\catcode`\$=3\catcode`\^=7\catcode`\_=8\relax}]
\PY{g+gu}{\PYZsh{}\PYZsh{} Reasoning}

The following symbols are relevant to the provided C code snippet: (ommitted)

\PY{g+gu}{\PYZsh{}\PYZsh{} Relevant Symbols}

\PY{k}{1.} ngx\PYZus{}http\PYZus{}features\PYZus{}t
(omitted)
\end{Verbatim}

  \end{promptcontent}
\end{promptbox}

\PP{Before Turn 3.}
The context builder at the phase of Turn 2 retrieves the header file, uncovering the missing structure.
\begin{promptbox}{Augmented Code Context for $x_{t, 3}$}
  \begin{promptcontent}
    \begin{Verbatim}[commandchars=\\\{\},codes={\catcode`\$=3\catcode`\^=7\catcode`\_=8\relax}]
\PY{c+c1}{// nginx/src/http/ngx\PYZus{}http\PYZus{}request.h}

\PY{k}{typedef}\PY{+w}{ }\PY{k}{struct}\PY{+w}{ }\PY{p}{\PYZob{}}
\PY{+w}{    }\PY{n}{DECLARE\PYZus{}BITMAP}\PY{p}{(}\PY{n}{bitmap}\PY{p}{,}\PY{+w}{ }\PY{n}{NGX\PYZus{}CUSTOM\PYZus{}FEATURE\PYZus{}\PYZus{}NR}\PY{p}{)}\PY{p}{;}
\PY{p}{\PYZcb{}}\PY{+w}{ }\PY{n}{ngx\PYZus{}http\PYZus{}features\PYZus{}t}\PY{p}{;}

\PY{c+c1}{// (additional code elided)}
\end{Verbatim}

  \end{promptcontent}
\end{promptbox}
Importantly, this discovery is decisive: the macro \ccw{NGX\_CUSTOM\_FEATURE\_\_NR} specifies the upper bound of valid feature indices. Our empirical evaluation showed that without retrieving this symbol, LLM-generated patches consistently failed, either uncompilable or logically unsound. Its retrieval demonstrates the necessity of multi-turn interaction—single-pass retrieval would likely terminate prematurely.

As $H=2$, this augmented code context for $x_{t, 3}$ is fed into the patching agent to generate a patched codebase as follows.

\begin{promptbox}{Generated Patch with Code Context $x_{t, 3}$}
  \begin{promptcontent}
    \begin{Verbatim}[commandchars=\\\{\},codes={\catcode`\$=3\catcode`\^=7\catcode`\_=8\relax}]
\PY{g+gd}{\PYZhy{}\PYZhy{}\PYZhy{} a/nginx/src/http/ngx\PYZus{}http\PYZus{}request.c}
\PY{g+gi}{+++ b/nginx/src/http/ngx\PYZus{}http\PYZus{}request.c}
\PY{g+gu}{@@ \PYZhy{}1995,7 +1995,7 @@}
\PY{+w}{ }
\PY{+w}{ }    features = r\PYZhy{}\PYZgt{}headers\PYZus{}in.features;
\PY{+w}{ }
\PY{g+gd}{\PYZhy{}    if (custom\PYZus{}flag\PYZus{}value \PYZgt{}= sizeof(features\PYZhy{}\PYZgt{}bitmap) * BITS\PYZus{}PER\PYZus{}LONG) \PYZob{}}
\PY{g+gi}{+    if (custom\PYZus{}flag\PYZus{}value \PYZlt{} 0 || custom\PYZus{}flag\PYZus{}value \PYZgt{}= NGX\PYZus{}CUSTOM\PYZus{}FEATURE\PYZus{}\PYZus{}NR) \PYZob{}}
\PY{+w}{ }        ngx\PYZus{}log\PYZus{}error(NGX\PYZus{}LOG\PYZus{}INFO, r\PYZhy{}\PYZgt{}connection\PYZhy{}\PYZgt{}log, 0,
\PY{+w}{ }                      \PYZdq{}Unknown custom feature value\PYZdq{});
\PY{+w}{ }        ngx\PYZus{}http\PYZus{}finalize\PYZus{}request(r, NGX\PYZus{}HTTP\PYZus{}BAD\PYZus{}REQUEST);
\end{Verbatim}

  \end{promptcontent}
\end{promptbox}

The key improvement is replacing a brittle size-based bound with a semantically correct check against \ccw{NGX\_CUSTOM\_FEATURE\_\_NR}. This ensures both memory safety and logical correctness.

The \ccw{babynginx/cpv-0} case exemplifies the role of multi-turn retrieval:
\begin{squishitemize}
    \item \textbf{Fault localization to \ccw{ngx\_http\_process\_custom\_features}.} The initial retrieval step successfully identifies the precise function containing the vulnerability, providing essential direction for subsequent context gathering efforts.
    \item \textbf{Expansion of surrounding function and type context.} The agent systematically retrieves additional code artifacts surrounding the identified fault location, including related function definitions and structural type information necessary for comprehensive understanding.
    \item \textbf{Identification of the critical macro \ccw{NGX\_CUSTOM\_FEATURE\_\_NR}.} Through multi-turn interaction, the retrieval process discovers the essential boundary definition that specifies valid feature indices, which proves crucial for generating semantically correct patches.
    \item \textbf{Patch generation leveraging the retrieved bound.} The final patching step utilizes the complete assembled context to generate a fix that replaces brittle size-based bounds with semantically appropriate checks against the retrieved macro definition.
\end{squishitemize}

This pipeline underscores that retrieval of \emph{all} relevant symbols—not merely local function bodies—is essential for valid patching. The case validates our claim that multi-turn context learning is indispensable for automated vulnerability remediation.

\subsection{Evaluation}

\subsubsection{Overview}

To complement our evaluations with general-purpose LLMs, we developed a lightweight yet task-specialized retrieval agent tailored to AIxCC-style patching scenarios. The motivation was twofold: (i) to validate whether a parameter-efficient adaptation of an open-source base model can internalize retrieval heuristics from a limited number of CPV instances, and (ii) to examine training dynamics when reinforcement signals are derived from multi-turn context construction rather than direct patch supervision. This custom agent was fine-tuned on a curated set of CPVs under controlled compute resources, enabling us to probe the trade-offs between scalability, efficiency, and retrieval quality in a reproducible manner.

\subsubsection{Experimental Setup}
We fine-tuned a custom model based on Meta's Llama-3.2-3B-Instruct. The model was trained on a proprietary dataset comprising 7 CPV instances. Training was conducted on a high-performance computing environment with 8× NVIDIA A100 (80GB) GPUs, an AMD EPYC 7513 32-Core processor, and 1.96TB RAM.

Our hyperparameter configuration was optimized for code retrieval tasks:
\begin{squishitemize}
    \item \textbf{Context Window}: 8,192 tokens maximum prompt length with 10,240 tokens total sequence length to accommodate lengthy code contexts
    \item \textbf{Parameter-Efficient Fine-Tuning}: LoRA adaptation with the rank of 32 and alpha of 32, striking a balance between expressivity and computational efficiency
    \item \textbf{Optimization}: AdamW optimizer with a learning rate of $5 \times 10^{-6}$ and constant scheduling across 3 training epochs
    \item \textbf{GRPO Group Size}: 12 parallel generations per step, enabling diverse exploration trajectories
    \item \textbf{$H$}: 4 retrieval steps per episode, aligned with the typical depth required for vulnerability localization
    \item \textbf{$T$}: 7 CPVs for training.
    \item \textbf{Downstream Evaluation}: GPT-4.1 as the patch generation model to assess retrieval quality
\end{squishitemize}

This configuration allowed us to efficiently fine-tune the retrieval model while maintaining reasonable computational requirements.

\subsubsection{Training Dynamics}
\begin{figure*}[t]
    \centering
    \includegraphics[width=\textwidth]{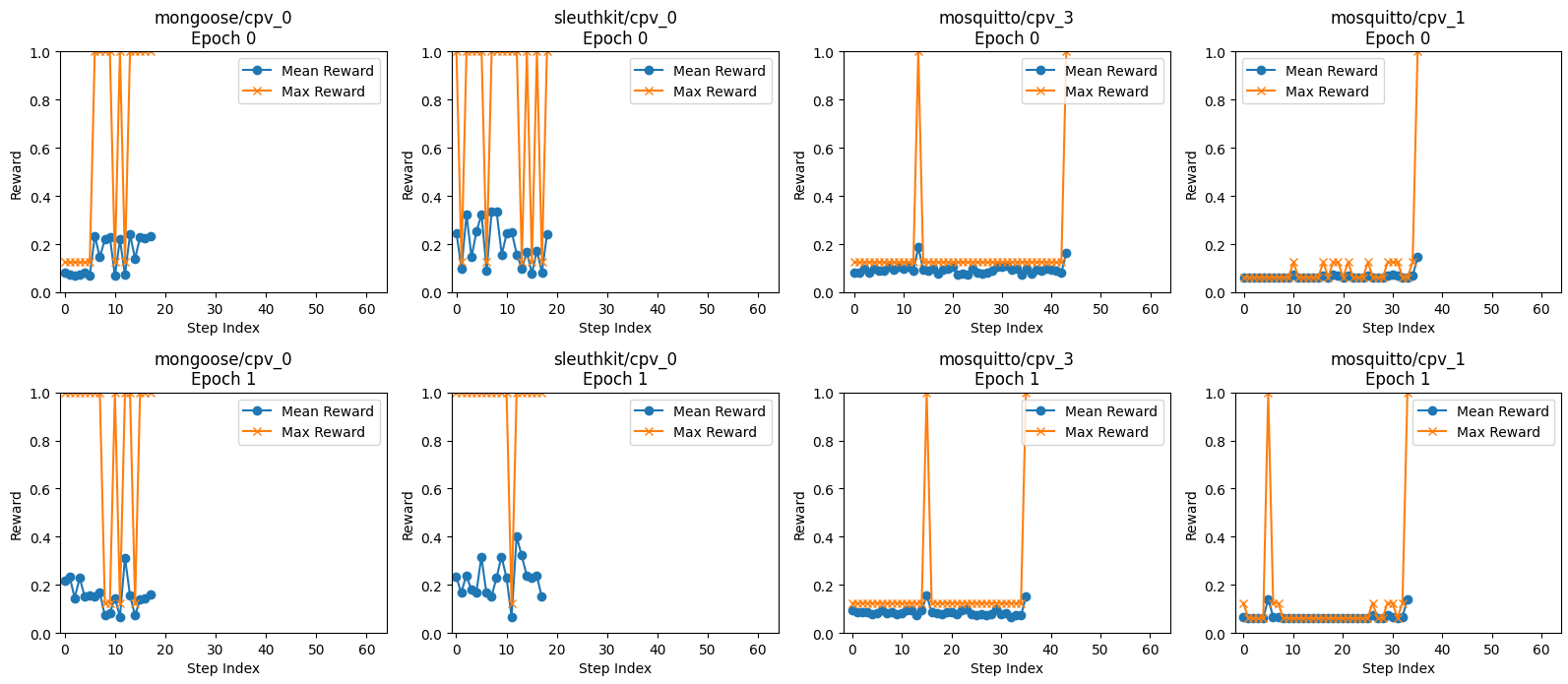}
    \caption{\textbf{Training dynamics.} Each subplot shows the reward trajectory for a specific project and CPV pair over two epochs. The blue line represents the mean reward across all agents in a group, while the orange line shows the maximum reward achieved by any agent in that group at each step.}
    \label{fig:custom-model-trend}
\end{figure*}

\autoref{fig:custom-model-trend} shows per-project reward trajectories over two epochs. Each subplot corresponds to a (project, CPV) pair; blue traces denote the mean reward across agents in a group, while orange traces denote the max reward attained by any agent in that group at a given step. We observe the following points.

\begin{squishitemize}
    \item \textbf{Early sparse successes.} In many panels the orange curve spikes to 1.0 well before the blue curve rises, indicating that at least one rollout quickly discovers a high-soundness trajectory while the cohort average remains low. This confirms the utility of group-based optimization: even when the policy is immature, diversity in rollouts can surface a correct retrieval path.
    \item \textbf{Staircase-like improvement.} The mean reward typically increases in discrete jumps rather than smoothly. This reflects the sequential nature of retrieval: once the policy learns to fetch a critical symbol or file, downstream steps become substantially easier, lifting the average.
    \item \textbf{Project-dependent difficulty.} Some projects (e.g., \ccw{mosquitto/cpv\_2} and \ccw{swftools/cpv\_0}) exhibit long flat regions with zero reward, suggesting that either (i) the necessary context is harder to localize, or (ii) the final patch is more brittle. Others (\eg \ccw{mongoose/cpv\_0}) show frequent reward spikes, implying richer intermediate signals or easier-to-exploit structure.
    \item \textbf{Persistence matters.} The recurrence of max-reward spikes across epochs indicates that repeated attempts on the same instance in our online learning do not merely overfit one trajectory; instead, they stabilize the policy so that success can be rediscovered consistently.
\end{squishitemize}

\subsubsection{Patch Outcomes}

\begin{figure}[t]
    \centering
    \includegraphics[width=\columnwidth]{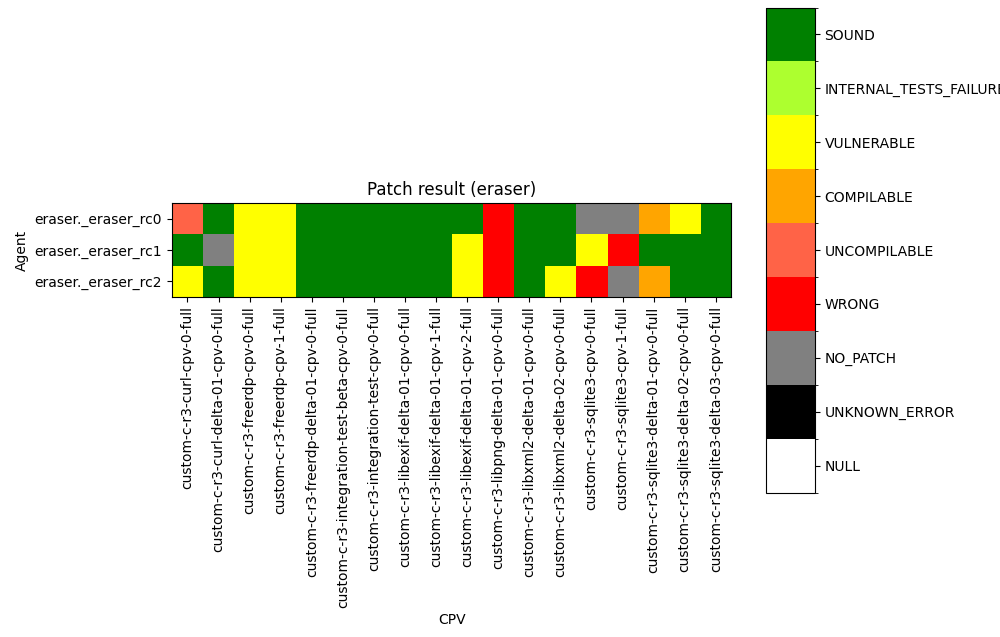}
    \caption{\textbf{Patch outcomes across CPVs and agent checkpoints.} Each cell represents the final patch outcome for a given CPV (columns) and agent checkpoint. Color encodes categorical evaluation results.}
    \label{fig:custom-model-evaluation}
\end{figure}

\autoref{fig:custom-model-evaluation} presents a heatmap of final patch outcomes across CPVs (columns) and agent checkpoints (rows: rc0, rc1, rc2). Color encodes categorical evaluation results (e.g., SOUND, VULNERABLE, WRONG, NO_PATCH). The following includes our findings. 

\begin{squishitemize}
    \item \textbf{High density of SOUND patches.} Green dominates across many CPVs, indicating that the retrieved contexts were sufficient for the downstream LLM to generate correct, test-passing fixes.
    \item \textbf{Failure mode taxonomy.}
    \begin{squishitemize}
        \item \textbf{UNCOMPILABLE / WRONG (reds).} Typically arise when the retriever omits a dependency or surfaces stale code, causing syntactic or semantic mismatches.
        \item \textbf{VULNERABLE (yellow-orange).} Patches compile but fail security or vulnerability checks, suggesting the retriever surfaced code relevant to functionality but not to the root cause of the vulnerability.
        \item \textbf{NO\_PATCH / UNKNOWN\_ERROR (gray/black).} Reflect cases where the patching model either abstained or the pipeline failed externally (tooling/timeouts), highlighting engineering, not policy, limitations.
    \end{squishitemize}
    \item \textbf{Checkpoint progression.} Later checkpoints (rc1, rc2) show fewer severe failures (e.g., UNCOMPILABLE, WRONG) and more SOUND or at least COMPILABLE outcomes, consistent with policy refinement over training.
    \item \textbf{Instance heterogeneity.} Certain CPVs remain problematic across all checkpoints (persistent reds/yellows), pointing to classes of bugs where our current retrieval actions (or reward shaping) are insufficient—prime targets for future ablations (e.g., deeper call graph exploration, semantic diffing).
\end{squishitemize}

\subsection{Discussion}

We discuss the contributions, key findings, and limitations of custom LLMs in \sys-Patching.

\PP{Contributions}
This work introduces a specialized context retrieval agent addressing the challenges for LLM-driven automated security patch generation. Our key contributions include the following:

\begin{squishitemize}
\item \textbf{Efficacy in providing missing context for patching.} We share a key observation for successful patching, \ie providing missing code context (\eg definitions on undefined symbols within a given context) is crucial for sound patch generation by LLMs. 
\item \textbf{Multi-turn retrieval agent for patching.} We propose and learn a novel multi-turn retrieval agent that iteratively retrieves concise, targeted code context, enabling effective utilization of powerful yet context-limited commercial LLMs for patch generation.
%\item \textbf{Task-driven reinforcement learning.} Optimizes the code retrieval policy specifically for successful patch generation through format and soundness rewards, surpassing static heuristic methods.
\item \textbf{Demonstration of effectiveness on real-world benchmarks.} We demonstrate the efficacy of learned multi-turn retrieval agent through successful participation in benchmarks such as the DARPA AIxCC competition.
\end{squishitemize}

\PP{Key Findings}
Our findings emphasize code context retrieval as integral to successful security patch generation, advocating a broader perspective on prompt engineering wherein structured context provisioning significantly enhances model outcomes.

The following summarizes our key findings in developing custom LLMs for patching. 
\begin{squishitemize}
\item \textbf{With proper coding context, commercial LLMs are able to generate secure patches.} When provided with comprehensive and relevant code context through our multi-turn retrieval approach, state-of-the-art commercial language models demonstrate strong capability in producing functionally correct and security-compliant patches for complex vulnerabilities.
\item \textbf{A pretrained model has fair context selection performance.} Base models such as Meta's Llama-3.2-3B-Instruct exhibit reasonable baseline performance in identifying and retrieving relevant code artifacts, providing a solid foundation for further specialization through reinforcement learning techniques.
\item \textbf{RL fine-tuning improves context selection performance.} Reinforcement learning optimization using our multi-turn GRPO framework significantly enhances the model's ability to select optimal code contexts, leading to measurable improvements in downstream patch generation quality and success rates.
\item \textbf{Forgetting in learning hinders RL advancement.} The online learning approach introduces catastrophic forgetting challenges, where knowledge acquired from earlier challenge project vulnerabilities degrades when adapting to new instances, limiting the overall effectiveness of the training process.
\end{squishitemize}

\PP{Limitations}
Our approach has several practical limitations.

\begin{squishitemize}
    \item \textbf{Language scope.} The current agent learning code is specialized for C codebases and has not been extended to other programming languages such as Java.
    \item \textbf{Restricted retrieval scope.} The retrieval process focuses only on function definitions, method definitions, and type definitions, potentially overlooking other useful artifacts such as build scripts or configuration files.
    \item \textbf{Tooling overhead.} Our use of LSP/parser-based tools required manual, per-rule adapter development, which limits scalability; broader, language-agnostic tools would improve generality.
    \item \textbf{Forgetting risk.} The online learning approach can lead to catastrophic forgetting, \ie a retrieval policy learned on earlier CPVs is degraded when adapting to new or more CPVs.
    \item \textbf{Competition constraints.} Late changes in competition rules limited available custom model learning time and reduced the breadth of the model validation.
\end{squishitemize}

\clearpage
\section{\sys-SARIF}
\label{s:crs-sarif}

\begin{figure*}[t]
    \centering
    \includegraphics[width=\textwidth]{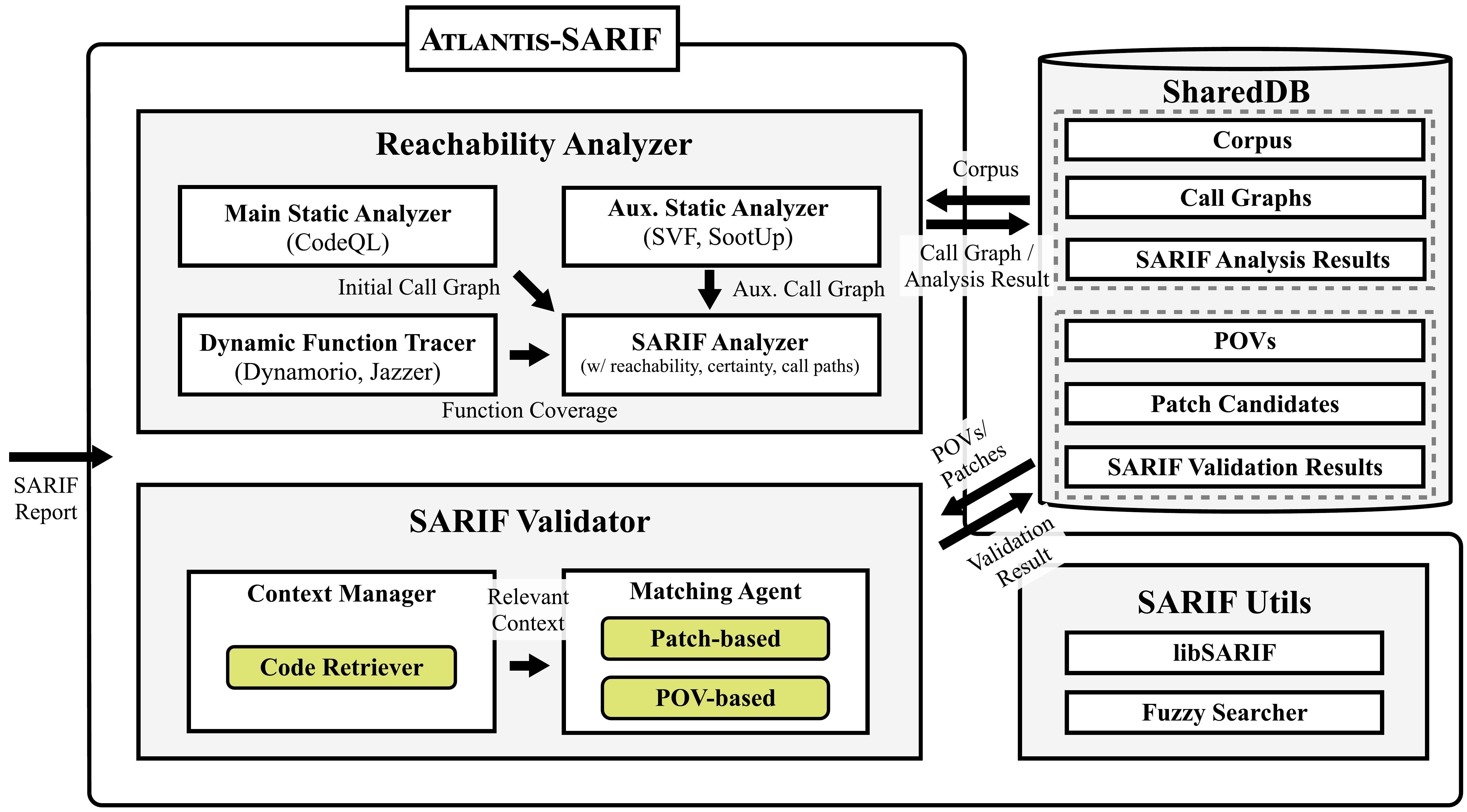}
      \caption{SARIF Validation Architecture. LLM-based modules are highlighted with green boxes.
      The system processes incoming SARIF reports through reachability analysis and LLM-based validation.      
      Results are stored in the SharedDB and accessed through our custom API interface.}
    \label{fig:sarif}
\end{figure*}

SARIF (Static Analysis Results Interchange Format) is a standardized,
JSON-based format for reporting static analysis findings.
It enables consistent and automated integration across tools.
In \sys, 
SARIF reports are provided by competition systems as broadcasts,
describing potential vulnerabilities,
and accurate assessment is critical for scoring.
However, 
the reported code locations 
and vulnerability descriptions are not always accurate.
This limitation necessitates a Reachability Analysis module (\autoref{ss:sarif-reachability}),
which verifies whether a vulnerability is reachable from harness's entry points.
Moreover, SARIF descriptions are often ambiguous or incomplete.
To address this, a SARIF Matching module aligns external reports
with internally discovered bugs using heuristics and LLM-based analysis (\autoref{ss:sarif-validity}).
These modules together enhance the reliability,
scoring accuracy, and practical utility of SARIF
in automated vulnerability analysis.

\autoref{fig:sarif} presents an overview of the SARIF module.
Broadcasted SARIF reports first pass through the \emph{Reachability Analyzer},
which merges a primary static call graph (via \cc{CodeQL}),
an auxiliary static graph (via \cc{SVF} or \cc{SootUp}),
and dynamic function-trace coverage (using \cc{DynamoRIO} or \cc{Jazzer})
into a unified call graph.
This combined graph is stored in the shared database,
alongside SARIF analysis results.
%
% Generated call graphs and analysis results are subsequently leveraged by other Bug Finding CRSes for targeted bug discovery.
%
% The database also maintains PoV inputs,
% patch candidates,
% and validation outcomes.
%
The \emph{SARIF Validator} then analyzes and processes incoming SARIF reports.
It includes a \emph{Context Manager} with a file retriever
that extracts relevant code snippets,
and a \emph{Matching Agent} that applies both patch-based
and PoV-based logic to assess report correctness.
The \emph{Matching Agent} uses an LLM to make final validation decisions.
The \emph{SARIF Validator} begins by checking whether the target SARIF report corresponds to any stored artifacts (PoVs or patches).
In the absence of a match, it waits for new artifacts to be added and repeats the matching procedure accordingly.
Once the report aligns with at least one artifact, it is classified as ``Correct''.
Validation results are written back to the database
and made accessible through our custom designed API interface.
The \emph{SARIF Utils} component contains helper libraries.
This includes \cc{libSarif} for programmatic call graph queries
and a fuzzy searcher to handle edge cases,
such as amalgamated codebases.

\PP{SARIF Scoring Rules}
When a SARIF broadcast is received,
a CRS may submit a SARIF Assessment.
The assessment is 
``correct'' if the issue exists at the specified file, line, and CWE/bug type, 
or ``incorrect'' otherwise.
Each assessment includes the SARIF ID, a binary verdict, and a justification.
Correct assessments earn 
$1 \times \tau_{\text{assessment}}$ points,
where $\tau_{\text{assessment}}$ decays 
from 1 to 0.5 over time.
Incorrect or outdated assessments score zero and reduce accuracy.
Only the last assessment per broadcast is scored;
earlier ones count as inaccuracies. \looseness=-1

% YJ: Remove unnecessary information
% \sys‐generated SARIF reports are not directly scored,
% but must follow the SARIF standard, 
% including a \cc{rules} section and rule IDs.
%
CRS also should bundle SARIF broadcasts with PoVs and patches.
A bundle scores only if it correctly links the broadcast UUID 
to a PoV or patch for the same vulnerability.
This yields up to 3 extra points, while incorrect links are penalized.

% YJ: Add our core strategy
\PP{Our Approach: Conservative Validation}
SARIF broadcasts frequently contain false positives,
and wrong assessments are penalized
while also reducing accuracy.
To maximize scoring reliability,
\sys-SARIF adopts a \emph{conservative} policy:
\sys determines a SARIF report as \emph{Correct}
only when there is concrete evidence from either
(1) a PoV that reaches the SARIF location,
or (2) a patch whose changes are logically consistent with the SARIF report.
This choice prioritizes precision over recall,
protecting the accuracy multiplier 
and avoiding penalties
from spurious submissions.

Because validation hinges on 
concrete evidence from PoVs or patches,
the primary objective is 
to discover a PoV at the code location of the SARIF report,
leveraging the targeted bug-finding features of our CRSes.
In this process, 
to prevent unnecessary resource consumption,
an initial reachability analysis filters out 
locations that are not exploitable.
When a relevant PoV or patch is identified,
the LLM-based validator integrates 
crash logs, patch diffs, and code context
to validate the SARIF report.
Importantly, 
this validator never operates 
without such explicit evidence, 
thereby preserving the conservative strategy
that disallows \emph{Correct} matches without a PoV or patch.

\subsection{Reachability Analysis Module}
\label{ss:sarif-reachability}

Reachability analysis determines whether a specific code location
can be exercised from one or more entry points (\eg harnesses's main function).
By constructing and traversing a call graph,
we identify all invocation paths that may lead to a reported vulnerability site.
This analysis is essential for the \sys-SARIF
as static analysis tools often report issues in dead code
or in paths unreachable from any harness.
Filtering out such false positives saves time
and prevents wasted effort during dynamic validation and patch generation.

At the core of our reachability module
is a unified call graph construction pipeline.
It aggregates results from multiple analyzers:
\cc{CodeQL} for primary static analysis,
\cc{SVF} for Andersen-style points-to analysis in C,
and \cc{SootUp} for Class Hierarchy Analysis (CHA) and Rapid Type Analysis (RTA) in Java.
% YJ: Add more detailed dynamic call graph construction. But I think it's too long.
% Purely static analysis proved insufficient, 
% so we refined the graph 
% by executing fuzzing corpora and collecting function traces. 
% %
% Our function tracer was built on 
% \cc{DynamoRIO} for C 
% and extended \cc{Jazzer} for Java. 
% %
% The call graph was continuously updated 
% with these runtime traces, 
% allowing us to capture indirect calls 
% that static analysis alone often misses.
%
The module supports both static-only graphs and dynamic updates 
that augment this graph with coverage information obtained from function-trace logs 
collected during fuzzing by other CRS components.
From this combined graph, we generate for each SARIF report
a \emph{reachable harness list} and a \emph{partial call graph}
connecting harness entry points to the vulnerability location.

To ensure robustness on large and complex codebases,
we employ a tiered fallback strategy:
(1) whole-program points-to analysis,
(2) restricted extraction of only harness-reachable nodes,
and (3) direct-edge-only extraction under memory or time constraints.
The fallback strategy guaranteed 
consistent call graph generation 
for multiple OSS-Fuzz projects 
and thereby enhanced the overall stability of \sys-SARIF.

Also, we assign each result one of three reachability confidence levels:
\emph{Certain}, where at least one path contains only \emph{strong} edges;
\emph{Possible}, where all paths include one or more \emph{weak} edges;
and \emph{Unlikely}, where no path is found.
Here, \emph{weak} and \emph{strong} edges are determined 
by the reliability of the method 
used to infer the edge.
For example, 
edges from the results of function tracer are treated as \emph{strong}, 
since they reflect actual execution.
In static call graphs, 
direct calls are treated as \emph{strong} edges, 
but edges inferred from pointer analysis are treated as \emph{weak} edges.
This \emph{weak}/\emph{strong} edge labeling and three-level confidence schem reflects our trust in the analysis
and guides downstream filtering and prioritization.
Other CRS components can apply their own policies based on these confidence levels.
%  for example, skipping \emph{Unlikely} findings or requiring manual review for \emph{Possible} paths.

To maintain precision, we address several edge cases.
For amalgamated codebases (\eg \cc{SQLite3}),
where source lines shift after build,
we apply fuzzy similarity matching to locate the correct function.
For projects with identical harness paths,
such as multiple \cc{curl} variants,
we disambiguate by inspecting each harness's linking targets.
The module emits reachability annotations with every SARIF result.
These annotations include harness names,
confidence levels,
and the filtered subgraph.
Downstream components
including LLM-based matchers and patch generators
consume this information directly.
The module also integrates with the \syscpmgr interface.
This allows the CRS to retrieve call graph fragments on demand,
supporting seamless end-to-end submissions
and continuous integration pipelines.

\PP{{libSARIF}: Interface for Other CRSs}
\cc{libSarif} is a standalone library that exposes
\sys-SARIF's call graph and reachability functionality as a reusable API.
It parses the merged static-and-dynamic graph format produced by our module
and supports several core queries.
The library provides functions to determine
whether a given source-line location is reachable,
to enumerate all call paths from harness entry points to that location,
and to compute the shortest such path.
By abstracting call graph logic,
libSarif enables other CRS components,
such as directed fuzzer of \sys-C,
to query and traverse exactly the subgraph they need,
without re-implementing underlying analysis logic.

\subsection{SARIF Validity Check}
\label{ss:sarif-validity}

In the \sys-SARIF workflow,
each incoming SARIF report must be matched to 
an internally discovered artifacts
(\ie either a PoV or a patch)
and then validated as correct or incorrect.
This approach is consistent with our conservative policy 
aimed at reducing incorrect assessments.
Reports often contain only file paths, line ranges,
natural-language messages, and CWE identifiers.
These fields may be incomplete or misleading.
Simplistic matching based on filename and line-number proximity
can produce false positives,
especially when multiple vulnerabilities affect the same code region.
Syntactic matching alone fails to account for semantic context,
such as crash stack traces or patch logic.
Validation must therefore integrate structural, dynamic,
and logical evidence to ensure that only true positives are accepted.
This preserves scoring accuracy and prevents wasted effort in exploitation or remediation.

Our initial approach treated matching as a statistical correlation problem.
For SARIF--PoV pairs,
we computed the fraction of SARIF locations exercised by the PoV's execution trace.
For SARIF--patch pairs,
we measured the overlap between patched lines and SARIF locations.
A high overlap ratio suggested a match;
a low ratio indicated a mismatch.
Although fast and model-free,
this approach struggled with common utility lines
which may appear across unrelated vulnerabilities
and with small patches or deep call graph paths,
which reduce apparent overlap.
Since the depth of crash stack traces and the count of patched lines differ widely across projects, 
we must tune decision thresholds in a project-specific manner. 
Edge cases, such as multi-harness scenarios,
also remained problematic.

To overcome these limitations,
we deployed an LLM-based agent as the final matcher and validator.
We supply the full SARIF report, crash logs (for PoVs), and patch diffs when available.
Relevant source context is also provided; 
we use a LLM-based file retriever to pull in code snippets, referenced files, and dependent functions.
We instruct the model to:
(1) identify root causes and trigger conditions,
(2) assess logical correlation,
and (3) decide among \emph{Matched}, \emph{Not Matched}, or \emph{Uncertain}.
We only confirm a SARIF report as ``Correct'' 
when the model's decision is \emph{Matched}, 
and submit our assessment accordingly.
By leveraging the reasoning ability of the LLM
instead of raw line counts,
we achieve higher precision.

We standardize prompts and limit input to relevant context snippets,
then validate performance on benchmark datasets
to calibrate model confidence.
%
% We verified that the approach operates robustly across codebases written in diverse languages and frameworks.
%
The LLM's decision is supplemented with quantitative metadata,
including reachability confidence from the call graph module.
We also log the model's rationale for auditability.
To control cost,
we selectively route cases to the LLM
particularly those with borderline overlap
or high-impact vulnerabilities
and throttle invocations when necessary.
Integration into \sys's \syscpmgr interface
allows downstream CRS components
to programmatically request match or validation results. \looseness=-1

\subsection{Evaluation}
\label{ss:sarif-eval}

This section evaluates two core components of the \sys-SARIF workflow:
the Reachability Analysis Module,
which determines whether reported code locations are actually reachable,
and the SARIF Validity Check,
which verifies whether each SARIF report corresponds to a true vulnerability.
Both components are evaluated on custom benchmark datasets,
and we analyze their contribution to matching accuracy
and overall scoring precision. \looseness=-1

\begin{table*}[!t]
  \centering
  \footnotesize
  \begin{tabular}{l  ccc  ccc}
    \toprule
     & \multicolumn{3}{c}{\textbf{C benchmark}} & \multicolumn{3}{c}{\textbf{Java benchmark}} \\
    \cmidrule(lr){2-4} \cmidrule(lr){5-7}
           & \textbf{CodeQL (basic)} & \textbf{SVF (Ander)} & \textbf{CodeQL + SVF} 
           & \textbf{CodeQL (basic)} & \textbf{SootUp (CHA)} & \textbf{CodeQL + SootUp} \\
    \midrule
    \textbf{Total}  & 28/31 & 23/31 & 28/31 & 20/57 & 44/57 & 45/57 \\
    \textbf{Acc}    & 90\%  & 74\%  & 90\%  & 35\%  & 77\%  & 79\%  \\
    \bottomrule
\end{tabular}

  \caption{Reachability analysis success rates for C and Java benchmarks under 
  different call graph configurations.}
  \label{tab:sarif-reachability}
\end{table*}

\PP{Reachability Analysis Module}
\autoref{tab:sarif-reachability} compares reachability analysis success rates
across our C and Java benchmarks under different call graph configurations.
We converted the PoVs of our benchmarks into SARIF reports 
and examined whether the reachability analysis classified them as reachable.
In the C suite,
\cc{CodeQL} alone (basic forward analysis) correctly resolves 28 of 31 true positives (90\%).
\cc{SVF}, using Andersen-style points-to analysis,
reaches only 23 of 31 (74\%) when used in isolation.
Merging \cc{CodeQL} and \cc{SVF} graphs yields the same 28/31 (90\%) as \cc{CodeQL} alone,
suggesting that \cc{SVF}'s additional points-to information does not improve coverage
beyond \cc{CodeQL}'s basic analysis in this benchmark.

On the Java side,
basic \cc{CodeQL} analysis succeeds on only 20 of 57 cases (35\%),
due to the complexity of virtual dispatch and class loading.
Integrating \cc{SootUp}'s Class Hierarchy Analysis (CHA) raises the true positive rate to 44/57 (77\%).
Combining \cc{CodeQL} with \cc{SootUp} edges provides a modest further gain to 45/57 (79\%).
These results show that auxiliary analyses, 
especially for Java, 
are essential to handle dynamic dispatch.
An ensemble of static tools yields 
the most robust reachability coverage 
in large, real-world codebases. \looseness=-1
Note that this evaluation did not employ 
call graph refinement based on dynamic function traces. 
Therefore, performance may improve in the actual competition environment.

\begin{table*}[!t]
  \centering
  \footnotesize
  \begin{tabular}{l cc  cc  cc}
      \toprule
      & \multicolumn{2}{c}{\textbf{Crash‑only}} 
      & \multicolumn{2}{c}{\textbf{Patch‑only}} 
      & \multicolumn{2}{c}{\textbf{Crash+Patch}} \\
      \cmidrule(lr){2-3} \cmidrule(lr){4-5} \cmidrule(lr){6-7}
      \textbf{Decision} & \textbf{Matched} & \textbf{Not Matched} 
      & \textbf{Matched} & \textbf{Not Matched} 
      & \textbf{Matched} & \textbf{Not Matched} \\
      \midrule
      \textbf{Matched}        &  50 &   5 &  51 &   2 &  53 &   4 \\
      \textbf{Uncertain}      &   3 &  20 &   7 &   6 &   0 &   0 \\
      \textbf{Not Matched}    &   5 & 209 &   0 & 226 &   5 & 230 \\
      \bottomrule
\end{tabular}
        
  \caption{Confusion matrices for SARIF matching under three scenarios: 
  \textbf{Crash‑only}, \textbf{Patch‑only}, and \textbf{Crash+Patch}.}
  \label{tab:sarif-matching}
\end{table*}

\PP{SARIF Validity Check}
Our matching accuracy across three scenarios 
(\emph{Crash-only}, \emph{Patch-only}, and \emph{Crash with Patch}) 
is summarized in \autoref{tab:sarif-matching}.
We manually constructed a benchmark of SARIF reports, 
carefully labeling each as correctly or incorrectly matched 
to establish ground truth for evaluation.
In the \emph{Crash-only} case,
we correctly matched 50 reports and incorrectly matched 5. 
We deferred 23 reports
including 3 true positives labeled as ``Uncertain''
and 20 cases correctly rejected as ``Uncertain''
while missing 5 true positives (false negatives).
This yields robust but imperfect coverage.
The \emph{Patch-only} scenario shows similar trends:
51 true positives,
2 false positives,
7 uncertain cases,
and no false negatives among 226 negatives.
Combining crash and patch evidence further improves precision.
We correctly matched 53 reports,
with only 4 false positives and no uncertain decisions.
However, 5 true positives remained unmatched.
Overall, the LLM-based matcher achieves high true-positive rates
and low false-positive rates,
especially when both crash and patch inputs are present.

% \subsection{Future Work}
% \sys-SARIF's validation relies heavily on PoVs 
% obtained through dynamic analysis. 
% %
% This dependence poses a limitation in scenarios when dynamic execution is infeasible.
% %
% For example, in scenarios where the SARIF location is extremely difficult to explore, 
% or where a harness for dynamic analysis is unavailable.

% To overcome this limitation, 
% future work could explore validation methods \emph{without} PoVs. 
% %
% We internally prototyped an LLM-agentic validation module that directly assesses SARIF reports. 
% %
% Although we did not deploy this module during the competition to maintain our conservative validation policy, 
% internal evaluations with recent LLMs demonstrated promising potential. 

\clearpage
\clearpage
\section{Benchmark}
\label{s:benchmark}

\sys integrates several modules for automatic bug findings and automatic patch generation as described in \autoref{ss:overview}.
To systematically evaluate their improvement, robustness, capability, and limitation, we developed 56 C/C++ benchmarks with 130 vulnerabilities and 40 Java benchmarks with 152 vulnerabilities.
In addition to target repositories, the benchmarks also provide descriptions, POVs, and correct patches about vulnerabilities:
\url{https://github.com/Team-Atlanta/aixcc-afc-benchmark}.

\subsection{Component-Specific Benchmark Design}
The benchmark is designed to evaluate both the overall robustness of \sys and the capability of each module.
To assess robustness, we introduced examples targeting edge cases such as excessive standard output, file descriptor leaks, and unprintable byte sequences, all of which are intended to stress or disrupt the system.
To evaluate capability, we constructed examples that may appear trivial for one component but prove infeasible for another, with some components even failing entirely on specific classes of inputs.
Furthermore, we carefully designed multi-level benchmarks by analyzing coverage gaps observed when running \sys on real-world projects and deliberately inserting vulnerabilities into previously unreachable code regions.

\PP{For Concolic Executor}
The concolic executor symbolically executes program inputs to solve branch conditions, thereby enabling exploration of branches that are often difficult for traditional fuzzers.
To assess its effectiveness, we developed benchmarks involving complex arithmetic operations, which are typically challenging for fuzzing alone.
However, the concolic executor also suffers from well-known limitations, such as path explosion and limited robustness.
To stress test its capabilities, we included benchmarks with special instructions (e.g., AVX) and scenarios where vulnerabilities are located beyond deep loop nests.

\PP{For Directed Fuzzer}
The directed fuzzer is particularly effective when the suspected location of a vulnerability is known and the search space of program is large.
To evaluate its capability, we constructed benchmarks having diverse call depths and control-flow structures, with vulnerabilities intentionally placed deep within call chains.
However, when the codebase is very large, locating the precise target location can itself become difficult, which undermines reliable evaluation.
To address this, we provide the target location in delta mode by supplying diffs, enabling evaluation even when the vulnerable location is hard to identify.

\PP{For LLM-powered components}
\sys has many LLM-powered components for not only automatic bug findings but also automatic patch generation.
While LLMs have shown strong empirical performance, several limitations must be addressed to ensure fair and robust evaluation.
First, evaluation must account for potential data leakage from training.
If a vulnerability or challenge was present in the LLM’s training corpus, its performance may not reflect true generalization.
To mitigate this, our benchmark suite includes newly discovered and synthetic vulnerabilities unlikely to appear in training data.
Second, LLMs are constrained by fixed context windows.
Designing benchmarks that require effective context reduction or summarization is essential to assess how well LLM-based components handle long-range dependencies.
Finally, we include challenges that test robustness of LLMs against misleading or incomplete natural language descriptions.
In real-world scenarios, comments and documentation may not accurately describe program behavior, and LLMs must avoid overreliance on such information to make sound predictions or patches.

\PP{For Patching Agents}
The difficulty of bug detection does not necessarily correlate with the difficulty of patch generation.
For example, stack buffer overflows may be challenging to trigger due to input constraints, yet the corresponding patch such as bounds checking, is often straightforward.
More compelling patching scenarios arise when the crash log provides limited diagnostic value, requiring extensive static or dynamic analysis to locate the root cause and generate a correct patch without breaking existing functionality. 
To capture these cases, we included benchmarks where the bug may be easily discovered by a simple fuzzer, but repairing it demands non-trivial reasoning and code understanding.

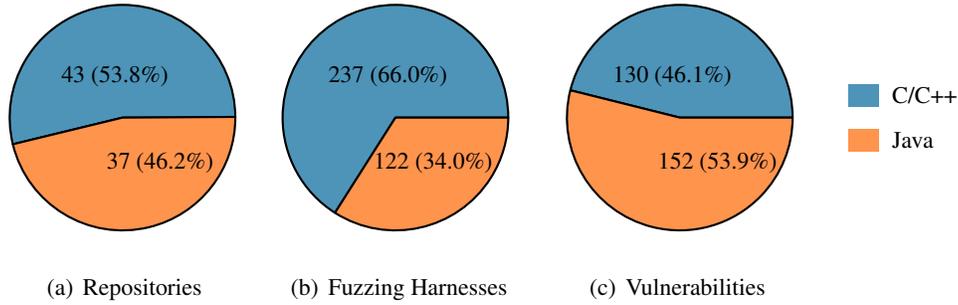
\begin{figure*}[!t]
\centering

\subfigure[Repositories]{%
\raisebox{0pt}[60pt][60pt]{%
\begin{tikzpicture}[scale=0.5, baseline=(pie center)]
\coordinate (pie center) at (0,0);
\pie[color={teal!80!blue!70,orange!80!red!70}, hide number, font=\footnotesize]{
  53.8/,
  46.3/
}
\node at (-0.2,1.1) [font=\footnotesize] {43 (53.8\%)};
\node at (1.0,-1.2) [font=\footnotesize] {37 (46.2\%)};
\end{tikzpicture}%
}%
\label{f:repo-dist}
}%
\hspace{0.5cm}%
\subfigure[Fuzzing Harnesses]{%
\raisebox{0pt}[60pt][60pt]{%
\begin{tikzpicture}[scale=0.5, baseline=(pie center)]
\coordinate (pie center) at (0,0);
\pie[color={teal!80!blue!70,orange!80!red!70}, hide number, font=\footnotesize]{
  66.0/,
  34.0/
}
\node at (-0.2,1.1) [font=\footnotesize] {237 (66.0\%)};
\node at (1.0,-1.2) [font=\footnotesize] {122 (34.0\%)};
\end{tikzpicture}%
}%
\label{f:harness-dist}
}%
\hspace{0.5cm}%
\subfigure[Vulnerabilities]{%
\raisebox{0pt}[60pt][60pt]{%
\begin{tikzpicture}[scale=0.5, baseline=(pie center)]
\coordinate (pie center) at (0,0);
\pie[color={teal!80!blue!70,orange!80!red!70}, hide number, font=\footnotesize]{
  46.1/,
  53.9/
}
\node at (-0.2,1.1) [font=\footnotesize] {130 (46.1\%)};
\node at (1.0,-1.2) [font=\footnotesize] {152 (53.9\%)};
\end{tikzpicture}%
}%
\label{f:vuln-dist}
}%
\hspace{0.5cm}%
\begin{tikzpicture}[scale=0.5, baseline=(legend center)]
\coordinate (legend center) at (0,0);
\node at (0,0.6) [anchor=center, font=\footnotesize] {\textcolor{teal!80!blue!70}{\rule{0.4cm}{0.3cm}}};
\node at (0.5,0.6) [anchor=west, font=\footnotesize] {C/C++};
\node at (0,-0.6) [anchor=center, font=\footnotesize] {\textcolor{orange!80!red!70}{\rule{0.4cm}{0.3cm}}};
\node at (0.5,-0.6) [anchor=west, font=\footnotesize] {Java};
\end{tikzpicture}%
\caption{Distribution of benchmark components across C/C++ and Java}
\label{f:benchmark-dist}
\end{figure*}

\subsection{Benchmark Statistics}
Each benchmark comprises a repository, and a repository may include multiple fuzzing harnesses that exercise the codebase and its injected vulnerabilities.
Notably, most repositories in our benchmarks are based on well-known open-source projects rather than small synthetic examples.
\autoref{f:benchmark-dist} summarizes the benchmark statistics across C/C++ and Java. 
Out of the 80 repositories, 43 (53.8\%) are implemented in C/C++ and 37 (46.2\%) in Java.
Fuzzing harnesses are more prevalent in the C/C++ set, with 237 harnesses (66.0\%) versus 122 (34.0\%) for Java.
For vulnerabilities, we included 130 cases in C/C++ (46.1\%) and 152 in Java (53.9\%), for a total of 282 vulnerabilities.
Overall, the benchmarks capture a balanced representation across the two languages while ensuring coverage in all three categories.

\PP{C/C++ Benchmarks}
Following the OSS-Fuzz standard, our C/C++ benchmarks include vulnerabilities detected by ASan, MSan, and UBSan.
In total, we constructed 56 C/C++ benchmarks comprising 130 vulnerabilities, 2.3 vulnerabilities per benchmark on average.
Of these, 122 were detected by ASan, 4 by UBSan, and 4 by MSan.
\autoref{tab:cvulntypes} lists the vulnerability types included in our suite, with brief descriptions and occurrence counts;
the most frequent types correspond to those commonly observed in real-world software, reflecting our combination of real-world vulnerability backports and synthetic examples.
In addition, our benchmarks are large enough to evaluate the capability of \sys.
Each of them contains 1,715 source files on average while
the mean lines of code per benchmark is 454K.

\PP{Java Benchmarks}
OSS-Fuzz uses Jazzer for fuzzing Java projects and employs Jazzer’s sanitizers to detect Java-specific vulnerabilities such as command injection and file-path traversal.
It also leverages ASan to capture memory corruption bugs in native code invoked via the Java Native Interface.
Based on this setup, we constructed 40 Java benchmarks comprising 152 vulnerabilities, 3.8 vulnerabilities per benchmark.
\autoref{tab:javavulntypes} summarizes the vulnerability types included in the suite.
In addition, we made our benchmarks large enough to test capability of \sys.
Each of them contains 2,473 source files and 295K lines of code on average.

\subsection{Constructing Realistic and Diverse Benchmark}
Our benchmark combines backported vulnerabilities, newly added projects, and synthetic cases to balance realism and diversity, avoiding overfitting to narrow dataset. 
We backport CVEs and OSS-Fuzz vulnerabilities into their original projects, grounding the benchmark in real bug distributions.
Even if patches appear in an LLM’s pretraining data, solving tasks remains challenging since patching systems must correctly adapt and reapply fixes in new contexts.
To broaden coverage, we extend beyond OSS-Fuzz by adding projects from new domains and injecting synthetic vulnerabilities in challenging locations to reach. 
By maintaining 60–70\% backported vulnerabilities with synthetic ones, the benchmark ensures fairness against pretraining leakage while still providing fresh challenges for automated vulnerability discovery and repair.

\begin{table*}[!t]
\centering
\resizebox{\textwidth}{!}{
    \renewcommand{\arraystretch}{1.16}
    \begin{tabular}{llr}
        \toprule
        \textbf{Vulnerability Type} & \textbf{Description} & \textbf{Count (\%)} \\
        \midrule
        ABRT & Program aborted, typically due to an assertion failure or fatal runtime error. & 1 (0.8\%) \\
        \rowcolor{gray!5}DoubleFree & Attempting to deallocate the same memory region more than once. & 4 (3.1\%) \\
        DynamicStackBufferOverflow & Overflow on a dynamically allocated stack buffer, such as one created with \cc{alloca}. & 2 (1.5\%) \\
        \rowcolor{gray!5}FPE & Floating-point exception due to invalid arithmetic operation, such as division by zero. & 3 (2.3\%) \\
        GlobalBufferOverflow & Out-of-bounds access to a global or static buffer. & 5 (3.8\%) \\
        \rowcolor{gray!5}HeapBufferOverflow & Writing or reading beyond the bounds of a heap-allocated buffer. & 46 (35.1\%) \\
        HeapUseAfterFree & Accessing heap memory after it has been deallocated. & 9 (6.9\%) \\
        \rowcolor{gray!5}ILL & Illegal instruction encountered, possibly due to corrupt or invalid code execution. & 2 (1.5\%) \\
        IntraObjectOverflow & Overflow within fields of a single object. & 1 (0.8\%) \\
        \rowcolor{gray!5}MemcpyParamOverlap & Source and destination buffers in \cc{memcpy} overlap. & 1 (0.8\%) \\
        MemoryLeak & Allocated memory is not freed. & 1 (0.8\%) \\
        \rowcolor{gray!5}NegativeSizeParam & A function received a negative value for a size parameter. & 1 (0.8\%) \\
        OutOfMemory & Excessive memory usage beyond the allowed threshold. & 1 (0.8\%) \\
        \rowcolor{gray!5}SEGV & Segmentation fault due to invalid memory access. & 31 (23.7\%) \\
        SignedIntegerNegation & Negating the most negative 64-bit signed integer. & 1 (0.8\%) \\
        \rowcolor{gray!5}SignedIntegerOverflow & Signed integer exceeds its maximum or minimum representable value. & 2 (1.5\%) \\
        StackBufferOverflow & Buffer overflow in stack memory. & 11 (8.4\%) \\
        \rowcolor{gray!5}StackBufferUnderflow & Accessing memory before the start of a buffer on the stack. & 1 (0.8\%) \\
        StackUseAfterReturn & Accessing stack memory from a function that has already returned. & 1 (0.8\%) \\
        \rowcolor{gray!5}StackUseAfterScope & Accessing stack memory outside its declared lifetime. & 1 (0.8\%) \\
        Timeout & Execution exceeded time limits. & 3 (2.3\%) \\
        \rowcolor{gray!5}UnknownCrash & A crash with an undetermined cause. & 1 (0.8\%) \\
        UseOfUninitializedValue & Use of a variable before it has been initialized. & 1 (0.8\%) \\
        \bottomrule
    \end{tabular}
}
\caption{C/C++ vulnerability types with counts and descriptions}
\label{tab:cvulntypes}
\end{table*}

\begin{table*}[!ht]
\centering
\resizebox{\textwidth}{!}{
    \renewcommand{\arraystretch}{1.16}
    \begin{tabular}{llr}
        \toprule
        \textbf{Vulnerability Type} & \textbf{Description} & \textbf{Count (\%)} \\
        \midrule
        ArbitraryLibraryLoad & Loading attacker-specified native libraries via \cc{System.load()} or similar APIs. & 2 (1.3\%) \\
        \rowcolor{gray!5}ArrayIndexOutOfBounds & Accessing an array index outside its valid range. & 1 (0.7\%) \\
        ArraySizeLimitExceeded & Attempting to allocate an array larger than the JVM allows. & 1 (0.7\%) \\
        \rowcolor{gray!5}CommandInjection & Unsanitized input is used in command execution. & 26 (17.1\%) \\
        HeapBufferOverflow & Writing or reading beyond the bounds of a heap-allocated buffer. & 1 (0.7\%) \\
        \rowcolor{gray!5}IllegalState & Inconsistency in decompression or data processing logic. & 1 (0.7\%) \\
        IndexOutOfBounds & Attempt to access a list or string with an invalid index. & 3 (2.0\%) \\
        \rowcolor{gray!5}InvalidFree & Simulated in Java via incorrect use of custom memory management or JNI. & 1 (0.7\%) \\
        LDAPInjection & Injection into LDAP queries. & 2 (1.3\%) \\
        \rowcolor{gray!5}OutOfMemory & Excessive memory usage beyond the allowed threshold. & 9 (5.9\%) \\
        PathTraversal & Manipulating file paths to access unintended files. & 9 (5.9\%) \\
        \rowcolor{gray!5}RegexInjection & Malicious input causes catastrophic backtracking in regex evaluation. & 5 (3.3\%) \\
        RemoteCodeExecution & Input leads to arbitrary code execution via unsafe reflection or dynamic evaluation. & 52 (34.2\%) \\
        \rowcolor{gray!5}RemoteJNDILookup & Remote lookup via JNDI for arbitrary object deserialization. & 2 (1.3\%) \\
        ScriptEngineInjection & Unsafe evaluation of user-controlled input in scripting engines. & 2 (1.3\%) \\
        \rowcolor{gray!5}ServerSideRequestForgery & The application makes unauthorized HTTP requests due to attacker-controlled input. & 17 (11.2\%) \\
        SQLInjection & Manipulated input in SQL queries. & 2 (1.3\%) \\
        \rowcolor{gray!5}StackOverflow & Deep or infinite recursion exceeding call stack limits. & 8 (5.3\%) \\
        Timeout & Execution exceeded time limits. & 2 (1.3\%) \\
        \rowcolor{gray!5}XPathInjection & Injection of input into XPath queries. & 6 (3.9\%) \\
        \bottomrule
    \end{tabular}
}
\caption{Java vulnerability types with counts and descriptions}
\label{tab:javavulntypes}
\end{table*}

\section{0-day Bugs}
\label{s:0day}

\subsection{\textsc{SQLite3}: Off-by-one Read}

We discovered
a 0-day vulnerability in SQLite3's FTS5 (Full-Text Search 5) module,
specifically within the trigram tokenizer implementation.
The trigram tokenizer breaks
text into overlapping three-character sequences
for substring matching in full-text searches.

The vulnerability occurs
when users create a virtual table with incomplete tokenizer configuration.
While FTS5 expects
tokenizer options as key-value pairs (e.g., \cc{case\_sensitive 1}),
it fails to validate
argument counts when a key lacks its corresponding value
before accessing array elements.

This results in
an off-by-one read beyond the argument array bounds
(\cc{azArg[i+1]} when \cc{i+1 >= nArg}).
While this could potentially lead
to information disclosure in other contexts,
SQLite3's defensive programming practices mitigate
the impact.
The library uses
\cc{sqlite3\_malloc} (which zero-initializes memory)
for the argument array allocation,
ensuring that out-of-bounds reads return
NULL rather than arbitrary memory content.
Consequently, the vulnerability manifests
as a NULL pointer dereference when the code attempts
to examine the non-existent value.

\PP{Root Cause Analysis}%
The vulnerability stems from
insufficient bounds checking in the \cc{fts5TriCreate} function.
The code iterates through
tokenizer arguments in pairs (key-value),
incrementing the loop counter by 2:

\begin{promptbox}{}
\begin{promptcontent}\begin{Verbatim}[commandchars=\\\{\},codes={\catcode`\$=3\catcode`\^=7\catcode`\_=8\relax}]
\PY{+w}{ }\PY{k}{static}\PY{+w}{ }\PY{k+kt}{int}\PY{+w}{ }\PY{n+nf}{fts5TriCreate}\PY{p}{(}
\PY{+w}{   }\PY{k+kt}{void}\PY{+w}{ }\PY{o}{*}\PY{n}{pUnused}\PY{p}{,}
\PY{+w}{   }\PY{k}{const}\PY{+w}{ }\PY{k+kt}{char}\PY{+w}{ }\PY{o}{*}\PY{o}{*}\PY{n}{azArg}\PY{p}{,}
\PY{+w}{   }\PY{k+kt}{int}\PY{+w}{ }\PY{n}{nArg}\PY{p}{,}
\PY{+w}{   }\PY{n}{Fts5Tokenizer}\PY{+w}{ }\PY{o}{*}\PY{o}{*}\PY{n}{ppOut}
\PY{+w}{ }\PY{p}{)}\PY{p}{\PYZob{}}
\PY{+w}{   }\PY{k+kt}{int}\PY{+w}{ }\PY{n}{rc}\PY{+w}{ }\PY{o}{=}\PY{+w}{ }\PY{n}{SQLITE\PYZus{}OK}\PY{p}{;}
\PY{+w}{   }\PY{n}{TrigramTokenizer}\PY{+w}{ }\PY{o}{*}\PY{n}{pNew}\PY{+w}{ }\PY{o}{=}\PY{+w}{ }\PY{n}{sqlite3\PYZus{}malloc}\PY{p}{(}\PY{k}{sizeof}\PY{p}{(}\PY{o}{*}\PY{n}{pNew}\PY{p}{)}\PY{p}{)}\PY{p}{;}
\PY{+w}{   }\PY{n}{UNUSED\PYZus{}PARAM}\PY{p}{(}\PY{n}{pUnused}\PY{p}{)}\PY{p}{;}
\PY{+w}{   }\PY{k}{if}\PY{p}{(}\PY{+w}{ }\PY{n}{pNew}\PY{o}{=}\PY{o}{=}\PY{l+m+mi}{0}\PY{+w}{ }\PY{p}{)}\PY{p}{\PYZob{}}
\PY{+w}{     }\PY{n}{rc}\PY{+w}{ }\PY{o}{=}\PY{+w}{ }\PY{n}{SQLITE\PYZus{}NOMEM}\PY{p}{;}
\PY{+w}{   }\PY{p}{\PYZcb{}}\PY{k}{else}\PY{p}{\PYZob{}}
\PY{+w}{     }\PY{k+kt}{int}\PY{+w}{ }\PY{n}{i}\PY{p}{;}
\PY{+w}{     }\PY{n}{pNew}\PY{o}{\PYZhy{}}\PY{o}{\PYZgt{}}\PY{n}{bFold}\PY{+w}{ }\PY{o}{=}\PY{+w}{ }\PY{l+m+mi}{1}\PY{p}{;}
\PY{+w}{     }\PY{n}{pNew}\PY{o}{\PYZhy{}}\PY{o}{\PYZgt{}}\PY{n}{iFoldParam}\PY{+w}{ }\PY{o}{=}\PY{+w}{ }\PY{l+m+mi}{0}\PY{p}{;}
\PY{+w}{     }\PY{k}{for}\PY{p}{(}\PY{n}{i}\PY{o}{=}\PY{l+m+mi}{0}\PY{p}{;}\PY{+w}{ }\PY{n}{rc}\PY{o}{=}\PY{o}{=}\PY{n}{SQLITE\PYZus{}OK}\PY{+w}{ }\PY{o}{\PYZam{}}\PY{o}{\PYZam{}}\PY{+w}{ }\PY{n}{i}\PY{o}{\PYZlt{}}\PY{n}{nArg}\PY{p}{;}\PY{+w}{ }\PY{n}{i}\PY{o}{+}\PY{o}{=}\PY{l+m+mi}{2}\PY{p}{)}\PY{p}{\PYZob{}}
\PY{+w}{      }\PY{c+c1}{// NOTE. off\PYZhy{}by\PYZhy{}one}
$\star$\PY{+w}{     }\PY{k}{const}\PY{+w}{ }\PY{k+kt}{char}\PY{+w}{ }\PY{o}{*}\PY{n}{zArg}\PY{+w}{ }\PY{o}{=}\PY{+w}{ }\PY{n}{azArg}\PY{p}{[}\PY{n}{i}\PY{o}{+}\PY{l+m+mi}{1}\PY{p}{]}\PY{p}{;}\PY{+w}{ }
\PY{+w}{      }\PY{k}{if}\PY{p}{(}\PY{+w}{ }\PY{l+m+mi}{0}\PY{o}{=}\PY{o}{=}\PY{n}{sqlite3\PYZus{}stricmp}\PY{p}{(}\PY{n}{azArg}\PY{p}{[}\PY{n}{i}\PY{p}{]}\PY{p}{,}\PY{+w}{ }\PY{l+s}{\PYZdq{}}\PY{l+s}{case\PYZus{}sensitive}\PY{l+s}{\PYZdq{}}\PY{p}{)}\PY{+w}{ }\PY{p}{)}\PY{p}{\PYZob{}}
\PY{+w}{        }\PY{c+c1}{// NOTE. null dereference }
$\star$\PY{+w}{       }\PY{k}{if}\PY{p}{(}\PY{+w}{ }\PY{p}{(}\PY{n}{zArg}\PY{p}{[}\PY{l+m+mi}{0}\PY{p}{]}\PY{o}{!}\PY{o}{=}\PY{l+s+sc}{\PYZsq{}}\PY{l+s+sc}{0}\PY{l+s+sc}{\PYZsq{}}\PY{+w}{ }\PY{o}{\PYZam{}}\PY{o}{\PYZam{}}\PY{+w}{ }\PY{n}{zArg}\PY{p}{[}\PY{l+m+mi}{0}\PY{p}{]}\PY{o}{!}\PY{o}{=}\PY{l+s+sc}{\PYZsq{}}\PY{l+s+sc}{1}\PY{l+s+sc}{\PYZsq{}}\PY{p}{)}\PY{+w}{ }\PY{o}{|}\PY{o}{|}\PY{+w}{ }\PY{n}{zArg}\PY{p}{[}\PY{l+m+mi}{1}\PY{p}{]}\PY{+w}{ }\PY{p}{)}\PY{p}{\PYZob{}}\PY{+w}{ }
\PY{+w}{          }\PY{n}{rc}\PY{+w}{ }\PY{o}{=}\PY{+w}{ }\PY{n}{SQLITE\PYZus{}ERROR}\PY{p}{;}
\PY{+w}{        }\PY{p}{\PYZcb{}}\PY{k}{else}\PY{p}{\PYZob{}}
\PY{+w}{          }\PY{n}{pNew}\PY{o}{\PYZhy{}}\PY{o}{\PYZgt{}}\PY{n}{bFold}\PY{+w}{ }\PY{o}{=}\PY{+w}{ }\PY{p}{(}\PY{n}{zArg}\PY{p}{[}\PY{l+m+mi}{0}\PY{p}{]}\PY{o}{=}\PY{o}{=}\PY{l+s+sc}{\PYZsq{}}\PY{l+s+sc}{0}\PY{l+s+sc}{\PYZsq{}}\PY{p}{)}\PY{p}{;}
\PY{+w}{        }\PY{p}{\PYZcb{}}
\PY{+w}{      }\PY{p}{\PYZcb{}}
\PY{+w}{  }\PY{p}{.}\PY{p}{.}\PY{p}{.}
\PY{p}{\PYZcb{}}
\end{Verbatim}
\end{promptcontent}
\end{promptbox}

At line 18, the code accesses
\cc{azArg[i+1]} without verifying that \cc{i+1 < nArg}.
When users provide
an odd number of arguments (e.g., only a key without a value),
this results in reading
beyond the argument array bounds.
The subsequent dereference at line 21 (\cc{zArg[0]}) triggers
a NULL pointer dereference,
causing the program to crash.

\PP{PoC}%
The following SQL statements demonstrate
the vulnerability:

\begin{promptbox}{}
\begin{promptcontent}\begin{Verbatim}[commandchars=\\\{\},codes={\catcode`\$=3\catcode`\^=7\catcode`\_=8\relax}]
\PY{k}{CREATE}\PY{+w}{ }\PY{n}{VIRTUAL}\PY{+w}{ }\PY{n+nc}{TABLE}\PY{+w}{ }\PY{n}{t}
\PY{+w}{  }\PY{k}{USING}\PY{+w}{ }\PY{n}{fts5}\PY{p}{(}\PY{n}{s}\PY{p}{,}\PY{+w}{ }\PY{n}{tokenize}\PY{o}{=}\PY{l+s+s1}{\PYZsq{}trigram case\PYZus{}sensitive\PYZsq{}}\PY{p}{)}\PY{p}{;}
\PY{k}{CREATE}\PY{+w}{ }\PY{n}{VIRTUAL}\PY{+w}{ }\PY{n+nc}{TABLE}\PY{+w}{ }\PY{n}{t}
\PY{+w}{  }\PY{k}{USING}\PY{+w}{ }\PY{n}{fts5}\PY{p}{(}\PY{n}{s}\PY{p}{,}\PY{+w}{ }\PY{n}{tokenize}\PY{o}{=}\PY{l+s+s1}{\PYZsq{}trigram remove\PYZus{}diacritics\PYZsq{}}\PY{p}{)}\PY{p}{;}
\end{Verbatim}
\end{promptcontent}
\end{promptbox}

Both statements omit
the required value for their respective options
(\cc{case\_sensitive} and \cc{remove\_diacritics}),
triggering the off-by-one read and subsequent crash.
While this vulnerability might appear
straightforward in hindsight,
it represents
a genuine security issue in one of the world's most widely deployed database engines.
Our team was the only one to identify
this 0-day vulnerability during the competition,
demonstrating the effectiveness of our automated bug-finding approach---particularly
notable given that we had no prior knowledge
that SQLite3 was the target project.

\PP{Patch Analysis}%
\sys generated a patch addressing
the vulnerability with proper bounds checking:

\begin{promptbox}{}
\begin{promptcontent}\begin{Verbatim}[commandchars=\\\{\},codes={\catcode`\$=3\catcode`\^=7\catcode`\_=8\relax}]
\PY{g+gh}{diff \PYZhy{}\PYZhy{}git a/ext/fts5/fts5\PYZus{}tokenize.c b/ext/fts5/fts5\PYZus{}tokenize.c}
\PY{g+gh}{index f12056170..552f14be9 100644}
\PY{g+gd}{\PYZhy{}\PYZhy{}\PYZhy{} a/ext/fts5/fts5\PYZus{}tokenize.c}
\PY{g+gi}{+++ b/ext/fts5/fts5\PYZus{}tokenize.c}
\PY{g+gu}{@@ \PYZhy{}1299,8 +1299,10 @@ static int fts5TriCreate(}
\PY{+w}{ }    pNew\PYZhy{}\PYZgt{}bFold = 1;
\PY{+w}{ }    pNew\PYZhy{}\PYZgt{}iFoldParam = 0;
\PY{+w}{ }    for(i=0; rc==SQLITE\PYZus{}OK \PYZam{}\PYZam{} i\PYZlt{}nArg; i+=2)\PYZob{}
\PY{g+gd}{\PYZhy{}      const char *zArg = azArg[i+1];}
\PY{g+gd}{\PYZhy{}      if( 0==sqlite3\PYZus{}stricmp(azArg[i], \PYZdq{}case\PYZus{}sensitive\PYZdq{}) )\PYZob{}}
\PY{g+gi}{+      const char *zArg = (i+1 \PYZlt{} nArg) ? azArg[i+1] : NULL;}
\PY{g+gi}{+      if (zArg == NULL) \PYZob{}}
\PY{g+gi}{+        rc = SQLITE\PYZus{}ERROR;}
\PY{g+gi}{+      \PYZcb{} else if( 0==sqlite3\PYZus{}stricmp(azArg[i], \PYZdq{}case\PYZus{}sensitive\PYZdq{}) )\PYZob{}}
\PY{+w}{ }        if( (zArg[0]!=\PYZsq{}0\PYZsq{} \PYZam{}\PYZam{} zArg[0]!=\PYZsq{}1\PYZsq{}) || zArg[1] )\PYZob{}
\PY{+w}{ }          rc = SQLITE\PYZus{}ERROR;
\PY{+w}{ }        \PYZcb{}else\PYZob{}
\end{Verbatim}
\end{promptcontent}
\end{promptbox}

The patch implements
a two-layer defense:
\begin{squishitemize}
\item \textbf{Bounds checking}: Before accessing \cc{azArg[i+1]},
      the patch verifies
      that \cc{i+1 < nArg} using a ternary operator,
      preventing the off-by-one read.
\item \textbf{NULL validation}: If the bounds check fails
      or if the argument is missing,
      \cc{zArg} is set to NULL,
      and an explicit check returns
      \cc{SQLITE\_ERROR} before any dereference occurs.
\end{squishitemize}

This fix ensures
that incomplete tokenizer configurations are properly rejected
with an error rather than causing a crash.
The patch was generated
in approximately 15 minutes,
including the entire build, patch generation,
iterative refinement, and correctness validation process.

\subsection{\textsc{SQLite3}: Use-after-free}

We identified
another 0-day vulnerability in SQLite3's LSM1 (Log-Structured Merge-tree) extension,
specifically a UAF bug
affecting virtual table cursor management.
This bug manifests
when executing queries with multiple equality conditions
(e.g., \cc{WHERE key IN ('key\_0','key\_1')})
against an LSM1 virtual table.

The vulnerability arises from
improper cache invalidation in the LSM1 cursor implementation.
When processing unique key lookups,
the cursor caches decoded column data
but fails to reset
this cache between iterations,
leading to use-after-free conditions
when the underlying memory is reallocated.

\PP{Root Cause Analysis}%
The bug occurs due to
an optimization in the \cc{lsm1Next} function
that only clears cached data when the result is non-unique.
\cc{bUnique} indicates true if no more than one row of output.

\begin{promptbox}{}
\begin{promptcontent}\begin{Verbatim}[commandchars=\\\{\},codes={\catcode`\$=3\catcode`\^=7\catcode`\_=8\relax}]
\PY{c+c1}{// ext/lsm1/lsm\PYZus{}vtab.c}
\PY{k}{static}\PY{+w}{ }\PY{k+kt}{int}\PY{+w}{ }\PY{n+nf}{lsm1Next}\PY{p}{(}\PY{n}{sqlite3\PYZus{}vtab\PYZus{}cursor}\PY{+w}{ }\PY{o}{*}\PY{n}{cur}\PY{p}{)}\PY{p}{\PYZob{}}
\PY{+w}{  }\PY{n}{lsm1\PYZus{}cursor}\PY{+w}{ }\PY{o}{*}\PY{n}{pCur}\PY{+w}{ }\PY{o}{=}\PY{+w}{ }\PY{p}{(}\PY{n}{lsm1\PYZus{}cursor}\PY{o}{*}\PY{p}{)}\PY{n}{cur}\PY{p}{;}
\PY{+w}{  }\PY{k+kt}{int}\PY{+w}{ }\PY{n}{rc}\PY{+w}{ }\PY{o}{=}\PY{+w}{ }\PY{n}{LSM\PYZus{}OK}\PY{p}{;}
\PY{+w}{  }\PY{k}{if}\PY{p}{(}\PY{+w}{ }\PY{n}{pCur}\PY{o}{\PYZhy{}}\PY{o}{\PYZgt{}}\PY{n}{bUnique}\PY{+w}{ }\PY{p}{)}\PY{p}{\PYZob{}}
\PY{+w}{    }\PY{n}{pCur}\PY{o}{\PYZhy{}}\PY{o}{\PYZgt{}}\PY{n}{atEof}\PY{+w}{ }\PY{o}{=}\PY{+w}{ }\PY{l+m+mi}{1}\PY{p}{;}
\PY{+w}{  }\PY{p}{\PYZcb{}}\PY{k}{else}\PY{p}{\PYZob{}}
\PY{+w}{    }\PY{c+cm}{/* ...code continues... */}
$\star$\PY{+w}{   }\PY{n}{pCur}\PY{o}{\PYZhy{}}\PY{o}{\PYZgt{}}\PY{n}{zData}\PY{+w}{ }\PY{o}{=}\PY{+w}{ }\PY{l+m+mi}{0}\PY{p}{;}\PY{+w}{ }\PY{c+c1}{// BUG: Only when NOT bUnique}
\PY{+w}{  }\PY{p}{\PYZcb{}}
\PY{+w}{  }\PY{k}{return}\PY{+w}{ }\PY{n}{rc}\PY{o}{=}\PY{o}{=}\PY{n}{LSM\PYZus{}OK}\PY{+w}{ }\PY{o}{?}\PY{+w}{ }\PY{n}{SQLITE\PYZus{}OK}\PY{+w}{ }\PY{o}{:}\PY{+w}{ }\PY{n}{SQLITE\PYZus{}ERROR}\PY{p}{;}
\PY{p}{\PYZcb{}}
\end{Verbatim}
\end{promptcontent}
\end{promptbox}

The \cc{pCur->zData} pointer caches
decoded value data from the current row.
When \cc{bUnique} is set,
the cache is not cleared in \cc{lsm1Next}.
However, subsequent queries reuse
the same virtual table cursor,
and if the new value requires more memory,
\cc{realloc} frees the old buffer
while \cc{pCur->zData} still points to it.

\PP{PoC}%
The following SQL demonstrates
the vulnerability:

\begin{promptbox}{}
\begin{promptcontent}\begin{Verbatim}[commandchars=\\\{\},codes={\catcode`\$=3\catcode`\^=7\catcode`\_=8\relax}]
\PY{p}{.}\PY{k}{load}\PY{+w}{ }\PY{l+s+ss}{\PYZdq{}./lsm.so\PYZdq{}}

\PY{k}{CREATE}\PY{+w}{ }\PY{n}{VIRTUAL}\PY{+w}{ }\PY{n+nc}{TABLE}\PY{+w}{ }\PY{n}{test\PYZus{}1337}\PY{+w}{ }\PY{err}{\PYZbs{}}
\PY{+w}{  }\PY{k}{USING}\PY{+w}{ }\PY{n}{lsm1}\PY{+w}{ }\PY{p}{(}\PY{l+s+s1}{\PYZsq{}test\PYZus{}1337.lsm\PYZsq{}}\PY{p}{,}\PY{+w}{ }\PY{k}{key}\PY{p}{,}\PY{+w}{ }\PY{n+nc}{TEXT}\PY{p}{,}\PY{+w}{ }\PY{k}{value}\PY{+w}{ }\PY{n+nc}{TEXT}\PY{p}{)}\PY{p}{;}

\PY{k}{INSERT}\PY{+w}{ }\PY{k}{INTO}\PY{+w}{ }\PY{n}{test\PYZus{}1337}\PY{+w}{ }\PY{k}{VALUES}\PY{+w}{ }\PY{p}{(}\PY{l+s+s1}{\PYZsq{}key\PYZus{}0\PYZsq{}}\PY{p}{,}\PY{+w}{ }\PY{l+s+s1}{\PYZsq{}value\PYZus{}0\PYZsq{}}\PY{p}{)}\PY{p}{;}
\PY{k}{INSERT}\PY{+w}{ }\PY{k}{INTO}\PY{+w}{ }\PY{n}{test\PYZus{}1337}\PY{+w}{ }\PY{k}{VALUES}\PY{+w}{ }\PY{p}{(}\PY{l+s+s1}{\PYZsq{}key\PYZus{}1\PYZsq{}}\PY{p}{,}\PY{+w}{ }\PY{l+s+s1}{\PYZsq{}value\PYZus{}1A\PYZsq{}}\PY{p}{)}\PY{p}{;}

\PY{k}{SELECT}\PY{+w}{ }\PY{k}{value}\PY{+w}{ }\PY{k}{FROM}\PY{+w}{ }\PY{n}{test\PYZus{}1337}\PY{+w}{ }\PY{k}{WHERE}\PY{+w}{ }\PY{k}{key}\PY{+w}{ }\PY{o+ow}{IN}\PY{+w}{ }\PY{p}{(}\PY{l+s+s1}{\PYZsq{}key\PYZus{}0\PYZsq{}}\PY{p}{,}\PY{l+s+s1}{\PYZsq{}key\PYZus{}1\PYZsq{}}\PY{p}{)}\PY{p}{;}
\end{Verbatim}
\end{promptcontent}
\end{promptbox}

The first \cc{SELECT} for \cc{key\_0} sets
\cc{bUnique=1} and caches the value data.
The second lookup for \cc{key\_1} triggers
a reallocation due to the longer value,
but \cc{lsm1DecodeValues} uses
the stale \cc{pCur->zData} pointer,
resulting in a use-after-free read.

\PP{Security Impact}%
This vulnerability enables
information disclosure through controlled heap manipulation.
By carefully sizing the value data,
we showed that attackers can leak sensitive information:

\begin{squishitemize}
\item \textbf{Heap addresses}: Through tcache metadata (16-byte allocations)
\item \textbf{Libc addresses}: Via unsorted bin pointers (1024-byte allocations)
\end{squishitemize}

The leak is particularly powerful
on Ubuntu 20.04 (glibc 2.31),
where \cc{tcache\_perthread\_struct} pointers reveal
heap base addresses.
On newer systems (glibc 2.34+),
the \cc{tcache\_key} provides
less useful information but still indicates heap activity.

\PP{Patch Analysis}%
The fix ensures
\cc{pCur->zData} is always reset when advancing the cursor:

\begin{promptbox}{}
\begin{promptcontent}\begin{Verbatim}[commandchars=\\\{\},codes={\catcode`\$=3\catcode`\^=7\catcode`\_=8\relax}]
\PY{g+gh}{diff \PYZhy{}\PYZhy{}git a/ext/lsm1/lsm\PYZus{}vtab.c b/ext/lsm1/lsm\PYZus{}vtab.c}
\PY{g+gh}{index 8c21923e1a..cfc80de883 100644}
\PY{g+gd}{\PYZhy{}\PYZhy{}\PYZhy{} a/ext/lsm1/lsm\PYZus{}vtab.c}
\PY{g+gi}{+++ b/ext/lsm1/lsm\PYZus{}vtab.c}
\PY{g+gu}{@@ \PYZhy{}389,7 +389,7 @@ static int lsm1Next(sqlite3\PYZus{}vtab\PYZus{}cursor *cur)\PYZob{}}
\PY{+w}{ }        if( c\PYZgt{}0 ) pCur\PYZhy{}\PYZgt{}atEof = 1;
\PY{+w}{ }      \PYZcb{}
\PY{+w}{ }    \PYZcb{}
\PY{g+gd}{\PYZhy{}    pCur\PYZhy{}\PYZgt{}zData = 0;}
\PY{+w}{ }  \PYZcb{}
\PY{g+gi}{+  pCur\PYZhy{}\PYZgt{}zData = 0;}
\PY{+w}{ }  return rc==LSM\PYZus{}OK ? SQLITE\PYZus{}OK : SQLITE\PYZus{}ERROR;
\PY{+w}{ }\PYZcb{}
\end{Verbatim}
\end{promptcontent}
\end{promptbox}

This simple one-line change moves
the cache reset outside the conditional block,
ensuring stale pointers are cleared
regardless of the \cc{bUnique} flag.
The patch maintains
backward compatibility while eliminating
the use-after-free condition.

\subsection{\textsc{Apache Commons Compress}: Out-of-bounds Write}

We discovered
a 0-day vulnerability in Apache Commons Compress,
specifically in the LZW decompression system
used for processing ZIP archives with unshrinking compression.
The vulnerability leads to
an infinite loop during decompression,
ultimately causing a negative array index write attempt.

\PP{Root Cause Analysis}%
The bug exists in
the \cc{expandCodeToOutputStack} method of \cc{LZWInputStream}.
This method traverses
a prefix lookup table to decompress LZW codes,
decrementing an index while writing to an output buffer:

\begin{promptbox}{}
\begin{promptcontent}\begin{Verbatim}[commandchars=\\\{\},codes={\catcode`\$=3\catcode`\^=7\catcode`\_=8\relax}]
\PY{k+kd}{protected}\PY{+w}{ }\PY{k+kt}{int}\PY{+w}{ }\PY{n+nf}{expandCodeToOutputStack}\PY{p}{(}\PY{k+kd}{final}\PY{+w}{ }\PY{k+kt}{int}\PY{+w}{ }\PY{n}{code}\PY{p}{,}\PY{+w}{ }\PY{k+kd}{final}\PY{+w}{ }\PY{k+kt}{boolean}\PY{+w}{ }\PY{n}{addedUnfinishedEntry}\PY{p}{)}\PY{+w}{ }
\PY{+w}{    }\PY{k+kd}{throws}\PY{+w}{ }\PY{n}{IOException}\PY{+w}{ }\PY{p}{\PYZob{}}
\PY{+w}{    }\PY{k}{for}\PY{+w}{ }\PY{p}{(}\PY{k+kt}{int}\PY{+w}{ }\PY{n}{entry}\PY{+w}{ }\PY{o}{=}\PY{+w}{ }\PY{n}{code}\PY{p}{;}\PY{+w}{ }\PY{n}{entry}\PY{+w}{ }\PY{o}{\PYZgt{}}\PY{o}{=}\PY{+w}{ }\PY{l+m+mi}{0}\PY{p}{;}\PY{+w}{ }\PY{n}{entry}\PY{+w}{ }\PY{o}{=}\PY{+w}{ }\PY{n}{prefixes}\PY{o}{[}\PY{n}{entry}\PY{o}{]}\PY{p}{)}\PY{+w}{ }\PY{p}{\PYZob{}}
\PY{+w}{        }\PY{n}{outputStack}\PY{o}{[}\PY{o}{\PYZhy{}}\PY{o}{\PYZhy{}}\PY{n}{outputStackLocation}\PY{o}{]}\PY{+w}{ }\PY{o}{=}\PY{+w}{ }\PY{n}{characters}\PY{o}{[}\PY{n}{entry}\PY{o}{]}\PY{p}{;}
\PY{+w}{    }\PY{p}{\PYZcb{}}
\PY{+w}{    }\PY{k}{if}\PY{+w}{ }\PY{p}{(}\PY{n}{previousCode}\PY{+w}{ }\PY{o}{!}\PY{o}{=}\PY{+w}{ }\PY{o}{\PYZhy{}}\PY{l+m+mi}{1}\PY{+w}{ }\PY{o}{\PYZam{}}\PY{o}{\PYZam{}}\PY{+w}{ }\PY{o}{!}\PY{n}{addedUnfinishedEntry}\PY{p}{)}\PY{+w}{ }\PY{p}{\PYZob{}}
\PY{+w}{        }\PY{n}{addEntry}\PY{p}{(}\PY{n}{previousCode}\PY{p}{,}\PY{+w}{ }\PY{n}{outputStack}\PY{o}{[}\PY{n}{outputStackLocation}\PY{o}{]}\PY{p}{)}\PY{p}{;}
\PY{+w}{    }\PY{p}{\PYZcb{}}
\PY{+w}{    }\PY{n}{previousCode}\PY{+w}{ }\PY{o}{=}\PY{+w}{ }\PY{n}{code}\PY{p}{;}
\PY{+w}{    }\PY{n}{previousCodeFirstChar}\PY{+w}{ }\PY{o}{=}\PY{+w}{ }\PY{n}{outputStack}\PY{o}{[}\PY{n}{outputStackLocation}\PY{o}{]}\PY{p}{;}
\PY{+w}{    }\PY{k}{return}\PY{+w}{ }\PY{n}{outputStackLocation}\PY{p}{;}
\PY{p}{\PYZcb{}}
\end{Verbatim}
\end{promptcontent}
\end{promptbox}

The vulnerability occurs when
the \cc{addEntry} method creates
a cycle in the prefix lookup table.
During normal operation,
\cc{addEntry} builds the compression dictionary
by linking previous codes:

\begin{promptbox}{}
\begin{promptcontent}\begin{Verbatim}[commandchars=\\\{\},codes={\catcode`\$=3\catcode`\^=7\catcode`\_=8\relax}]
\PY{k+kd}{protected}\PY{+w}{ }\PY{k+kt}{int}\PY{+w}{ }\PY{n+nf}{addEntry}\PY{p}{(}\PY{k+kd}{final}\PY{+w}{ }\PY{k+kt}{int}\PY{+w}{ }\PY{n}{previousCode}\PY{p}{,}\PY{+w}{ }\PY{k+kd}{final}\PY{+w}{ }\PY{k+kt}{byte}\PY{+w}{ }\PY{n}{character}\PY{p}{,}\PY{+w}{ }
\PY{+w}{                      }\PY{k+kd}{final}\PY{+w}{ }\PY{k+kt}{int}\PY{+w}{ }\PY{n}{maxTableSize}\PY{p}{)}\PY{+w}{ }\PY{p}{\PYZob{}}
\PY{+w}{    }\PY{k}{if}\PY{+w}{ }\PY{p}{(}\PY{n}{tableSize}\PY{+w}{ }\PY{o}{\PYZlt{}}\PY{+w}{ }\PY{n}{maxTableSize}\PY{p}{)}\PY{+w}{ }\PY{p}{\PYZob{}}
\PY{+w}{        }\PY{n}{prefixes}\PY{o}{[}\PY{n}{tableSize}\PY{o}{]}\PY{+w}{ }\PY{o}{=}\PY{+w}{ }\PY{n}{previousCode}\PY{p}{;}
\PY{+w}{        }\PY{n}{characters}\PY{o}{[}\PY{n}{tableSize}\PY{o}{]}\PY{+w}{ }\PY{o}{=}\PY{+w}{ }\PY{n}{character}\PY{p}{;}
\PY{+w}{        }\PY{k}{return}\PY{+w}{ }\PY{n}{tableSize}\PY{o}{+}\PY{o}{+}\PY{p}{;}
\PY{+w}{    }\PY{p}{\PYZcb{}}
\PY{+w}{    }\PY{k}{return}\PY{+w}{ }\PY{o}{\PYZhy{}}\PY{l+m+mi}{1}\PY{p}{;}
\PY{p}{\PYZcb{}}
\end{Verbatim}
\end{promptcontent}
\end{promptbox}

Through crafted input,
we can manipulate the decompression state
to create a 2-cycle between entries 257 and 295
in the prefix table.
When \cc{expandCodeToOutputStack} traverses
this cyclic structure,
it enters an infinite loop
where \cc{outputStackLocation} decrements indefinitely,
eventually attempting to write
at index -1.

Our analysis revealed
the exact sequence leading to the crash.
The decompression initially proceeds normally,
building the dictionary with entries 0--256.
At a critical point,
the following operations occur:

\begin{squishitemize}
\item Entry 257 is added with \cc{previousCode=295}
\item Entry 295 is added with \cc{previousCode=257}
\item When code 295 is expanded, the traversal alternates:
      295 → 257 → 295 → 257 → ...
\end{squishitemize}

The output stack location decrements
from 8192 down through 0,
then attempts to access index -1,
triggering an \cc{ArrayIndexOutOfBoundsException}.

\PP{PoC}%
We crafted a malicious ZIP archive
that triggers the vulnerability.
The 255-byte payload contains
carefully constructed LZW-compressed data
that creates the cyclic dependency:

{\tiny
\begin{verbatim}
0000: 504b 0304 2e00 0000 0c00 84b6 ba46 72b6  PK...........Fr.
0010: 0063 7700 fe00 6b00 0000 0300 1c00 6262  .cw...k.......bb
0020: 6255 5409 0003 e7ce 6455 f3ce 6455 7578  bUT.....dU..dUux
0030: 0b00 5704 8800 5c13 f904 0100 0042 5a68  ..W...\......BZh
0040: 3931 4159 2653 5962 e44f 5100 000d d180  91AY&SYb.OQ.....
\end{verbatim}
}

\PP{Security Impact}%
While the immediate effect is
a denial of service through an unhandled exception,
the vulnerability has broader implications:

\begin{squishitemize}
\item \textbf{Memory corruption potential}: The negative index write
      could corrupt adjacent memory structures
      in languages without bounds checking.
\item \textbf{Attack surface}: Any application using
      Apache Commons Compress to handle
      untrusted ZIP files is vulnerable.
\item \textbf{Bypass potential}: The infinite loop occurs
      before size validation checks,
      potentially bypassing security controls.
\end{squishitemize}

This vulnerability demonstrates
how subtle logic errors in compression algorithms
can lead to severe security issues,
particularly when handling
complex data structures like cyclic graphs
in a linear traversal context.

\subsection{\textsc{SQLite3}: SIGBUS in LSM1 Extension}

We discovered
a SIGBUS vulnerability in SQLite3's LSM1 extension
that manifests when multiple virtual tables
reference the same database file through different paths.
The vulnerability stems from
improper handling of shared database connections
when the same file is accessed
via both absolute and relative paths.

\PP{Root Cause Analysis}%
The bug occurs due to
flawed path comparison logic in the LSM1 database connection manager.
When SQLite3 opens
an LSM1 virtual table,
it attempts to reuse existing database connections
by comparing canonical paths.
However, the path resolution mechanism
fails to properly normalize paths,
treating \cc{/tmp/tmp.lsm} and \cc{../../../../../../tmp/tmp.lsm}
as distinct databases despite referencing
the same underlying file.

This leads to
a race condition during database closure.
When the first connection closes,
it truncates the shared file to zero bytes.
The second connection,
unaware of this modification,
continues accessing the now-invalid memory mapping,
resulting in a SIGBUS when attempting
to read from the truncated file's former address space.

\PP{PoC}%
The following SQL demonstrates
the vulnerability:

\begin{promptbox}{}
\begin{promptcontent}\begin{Verbatim}[commandchars=\\\{\},codes={\catcode`\$=3\catcode`\^=7\catcode`\_=8\relax}]
\PY{p}{.}\PY{k}{load}\PY{+w}{ }\PY{p}{.}\PY{o}{/}\PY{n}{lsm}

\PY{k}{CREATE}\PY{+w}{ }\PY{n}{VIRTUAL}\PY{+w}{ }\PY{n+nc}{TABLE}\PY{+w}{ }\PY{n}{lsm\PYZus{}table\PYZus{}1}\PY{+w}{ }\PY{k}{USING}\PY{+w}{ }\PY{n}{lsm1}\PY{+w}{ }
\PY{+w}{    }\PY{p}{(}\PY{l+s+s1}{\PYZsq{}/tmp/tmp.lsm\PYZsq{}}\PY{p}{,}\PY{+w}{ }\PY{n}{id}\PY{p}{,}\PY{+w}{ }\PY{n+nc}{TEXT}\PY{p}{,}\PY{+w}{ }\PY{k}{data}\PY{p}{)}\PY{p}{;}
\PY{k}{CREATE}\PY{+w}{ }\PY{n}{VIRTUAL}\PY{+w}{ }\PY{n+nc}{TABLE}\PY{+w}{ }\PY{n}{lsm\PYZus{}table\PYZus{}2}\PY{+w}{ }\PY{k}{USING}\PY{+w}{ }\PY{n}{lsm1}\PY{+w}{ }
\PY{+w}{    }\PY{p}{(}\PY{l+s+s1}{\PYZsq{}../../../../../../tmp/tmp.lsm\PYZsq{}}\PY{p}{,}\PY{+w}{ }\PY{n}{id}\PY{p}{,}\PY{+w}{ }\PY{n+nc}{TEXT}\PY{p}{,}\PY{+w}{ }\PY{k}{data}\PY{p}{)}\PY{p}{;}
\end{Verbatim}
\end{promptcontent}
\end{promptbox}

Executing these statements triggers
a SIGBUS during database shutdown
as SQLite3 attempts to access
memory that was unmapped when the first table's connection
truncated the shared file.

The vulnerability manifests through
the following sequence of events:

\begin{squishitemize}
\item Both virtual tables map the same file
      but LSM1 treats them as separate databases
\item During shutdown, the first connection acquires
      an exclusive lock (\cc{LSM\_LOCK\_DMS2})
\item The \cc{doDbDisconnect} function truncates
      the database file via \cc{dbTruncateFile}
\item The second connection still holds
      a memory mapping to the now-truncated file
\item When \cc{lsmCheckpointId} attempts to read
      from \cc{0x7ffff6f32000}, it triggers SIGBUS
\end{squishitemize}

\PP{Security Impact}%
While this vulnerability primarily causes
denial of service through application crashes,
it reveals deeper architectural issues:

\begin{squishitemize}
\item \textbf{Path confusion}: Applications using
      relative paths may inadvertently create
      conflicting database connections
\item \textbf{Resource exhaustion}: Attackers could create
      multiple virtual tables with varying paths
      to the same file, causing repeated crashes
\item \textbf{Data integrity}: The unexpected truncation
      could lead to data loss if other processes
      are accessing the same database file
\end{squishitemize}

The vulnerability highlights
the importance of proper path canonicalization
and resource tracking in database systems,
particularly when dealing with
memory-mapped files and shared resources.

\section{Conclusion}
\label{s:conclusion}

Our team started this competition as AI skeptics
but has now become a strong advocate
for using LLMs in hardcore, traditional security tasks.
Thanks to AIxCC,
we had the opportunity to gain first-hand experience
with LLMs---in fact,
evolving together with state-of-the-art LLM services
over the past two years,
and realized that the majority of security tasks
once believed impossible
are now feasible with LLMs.
As the speed of improvement
in foundation models and surrounding tools
is unprecedentedly fast,
it is difficult to imagine
how this will transform our security industries and research
in the coming years.
Our team will continue innovating in this area of research
and seeking ways to conduct follow-up research and development
together with various stakeholders---%
OpenAI has already started collaborating with our team,
and we hope to contribute to the mission of DARPA
and government organizations
in the future.

The implementation and benchmarks
will be made publicly available:
\url{https://github.com/Team-Atlanta/}.
We plan to release the details of our CRS
via a series of blog postings at \url{https://team-atlanta.github.io/}
as well as through publication
in academic conferences.

\section{Acknowledgment}
\label{s:ack}

As an open track team,
it would not have been possible to dedicate our time
for this journey
without the support of our organizations.
We sincerely thank
Georgia Tech, Samsung Research, KAIST, and POSTECH.

\vspace{30pt}
\begin{figure*}[h]
  \centering
  \includegraphics[width=0.8\linewidth]{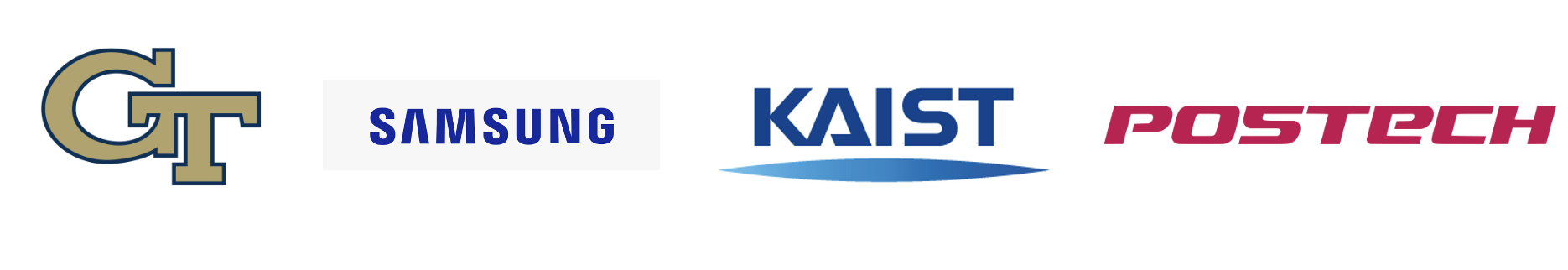}
\end{figure*}
\clearpage

\bibliographystyle{abbrvnat}
\footnotesize
\setlength{\bibsep}{3pt}
\bibliography{p,sslab,conf}

\printindex

\begin{appendix}
\label{s:appendix}

\section{Bug-Finding Module Performance by Harness}
\label{s:appendix-multilang-targets}

This appendix provides detailed bug-finding performance data
for fuzzing harnesses in the AIxCC final competition based on available logs.
The results are organized by programming language:
\autoref{t:c-harness-performance} covers 46 C harnesses
(35 individual harnesses plus 11 grouped with zero passed PoVs),
\autoref{t:java-harness-performance} covers 25 Java harnesses  
(13 individual harnesses plus 12 grouped with zero passed PoVs),
plus 1 unknown harness (1 PoV, 1 passed, 1 patch).
For each harness, we report five key metrics:

\PP{Total PoVs}
The total number of proof-of-vulnerabilities generated by all bug-finding modules
for each harness, regardless of their final verification status.

\PP{Passed PoVs}
PoVs that successfully triggered vulnerabilities and passed all verification checks:
(1) crashed the harness with appropriate return codes,
(2) for delta-mode harnesses, crashed only the HEAD version but not the BASE version,
and (3) passed deduplication analysis using stack trace signatures.
These contribute directly to the team's competition score.

\PP{Duplicate PoVs}
PoVs that successfully triggered real vulnerabilities
but were identified as duplicates of previously submitted PoVs
through stack trace analysis and crash signature matching.
While these represent legitimate vulnerability discoveries,
they do not contribute to scoring due to redundancy.

\PP{Passed Patches}
Patches that successfully remediated identified vulnerabilities.
These must pass verification showing that
(1) the patched version no longer crashes with the original PoV
and (2) the patch does not break existing functionality.
Successful patches contribute additional points to the competition score.

\PP{Successful Bug-Finding Modules}
The specific \sys components that successfully generated passed or duplicate PoVs for each harness.
This shows which approaches successfully generated passed or duplicate PoVs
for different harness types.

These granular results show
how different bug-finding modules performed across harness types.

\begin{table*}[htbp]
\centering
\scriptsize
\begin{threeparttable}
\begin{tabular}{lrrrr>{\raggedright\arraybackslash}p{5cm}}
\toprule
Harness & \multicolumn{3}{c}{PoVs} & Patches & Successful Bug-Finding \\
\cmidrule(lr){2-4}
        & Total & Passed & Dup. & Passed & Modules \\
\midrule
TestFuzzCryptoCertificateDataSetPEM & 7 & 1 & 0 & 1 & Multilang.testlang\_input\_gen \\
curl\_fuzzer & 8 & 1 & 0 & 1 & Multilang.given\_fuzzer \\
curl\_fuzzer\_dict & 16 & 13 & 0 & 1 & Multilang.concolic\_input\_gen, Multilang.given\_fuzzer \\
curl\_fuzzer\_ftp & 3 & 1 & 0 & 1 & Multilang.mlla \\
curl\_fuzzer\_http & 8 & 1 & 0 & 0 & C \\
curl\_fuzzer\_rtsp & 9 & 5 & 0 & 1 & C \\
curl\_fuzzer\_tftp & 23 & 2 & 0 & 0 & Multilang.given\_fuzzer \\
exif\_from\_data\_fuzzer & 104 & 18 & 2 & 1 & Multilang.given\_corpus, Multilang.concolic\_input\_gen, Multilang.given\_fuzzer, Multilang.shared\_corpus, Multilang.testlang\_input\_gen, C \\
exif\_loader\_fuzzer & 38 & 7 & 0 & 0 & Multilang.given\_fuzzer \\
fuzz & 84 & 6 & 66 & 4 & Multilang.given\_corpus, Multilang.given\_fuzzer, Multilang.mlla.gen, C \\
fuzz-catalog & 11 & 3 & 0 & 0 & Multilang.given\_fuzzer \\
fuzz-link-parser & 5 & 3 & 0 & 0 & Multilang.given\_corpus \\
fuzz-netdev-parser & 5 & 2 & 0 & 0 & Multilang.given\_corpus, Multilang.testlang\_input\_gen \\
fuzz-network-parser & 5 & 2 & 0 & 0 & Multilang.given\_corpus, Multilang.mlla \\
fuzz-udev-rule-parse-value & 3 & 2 & 0 & 0 & Multilang.given\_fuzzer \\
fuzz-unit-file & 1 & 1 & 0 & 0 & Multilang.given\_corpus \\
fuzz\_encode\_stream & 4 & 1 & 0 & 1 & Multilang.given\_fuzzer \\
handler\_aim & 1 & 1 & 0 & 1 & Multilang.given\_fuzzer \\
handler\_ber & 2 & 1 & 0 & 1 & Multilang.given\_fuzzer \\
handler\_gvcp & 1 & 1 & 0 & 1 & Multilang.given\_corpus \\
handler\_icmp\_extension & 1 & 1 & 0 & 1 & Multilang.given\_fuzzer \\
handler\_irc & 1 & 1 & 0 & 1 & Multilang.given\_fuzzer \\
handler\_json & 2 & 1 & 0 & 1 & Multilang.given\_fuzzer \\
handler\_netbios & 2 & 2 & 0 & 1 & Multilang.given\_fuzzer \\
handler\_telnet & 9 & 5 & 0 & 3 & Multilang.given\_fuzzer \\
handler\_wlan\_centrino & 4 & 1 & 0 & 0 & Multilang.given\_fuzzer \\
handler\_wlan\_noqos & 3 & 1 & 0 & 1 & Multilang.mlla.mut \\
handler\_wlan\_withfcs & 4 & 1 & 0 & 1 & Multilang.given\_fuzzer \\
handler\_zbee\_zdp & 1 & 1 & 0 & 1 & Multilang.given\_fuzzer \\
json\_fuzz & 1 & 1 & 0 & 1 & Multilang.given\_fuzzer \\
lint & 1 & 1 & 0 & 0 & Multilang.given\_fuzzer \\
reader & 1 & 1 & 0 & 0 & Multilang.given\_fuzzer \\
schema & 8 & 3 & 0 & 0 & Multilang.given\_fuzzer, C \\
xml & 1 & 1 & 0 & 0 & Multilang.given\_fuzzer \\
xpath & 2 & 1 & 0 & 0 & Multilang.given\_fuzzer \\
\textit{Other C harnesses (11)} & 23 & 0 & 1 & 0 & - \\
\midrule
\textbf{Total C harnesses (46)} & \textbf{402} & \textbf{94} & \textbf{69} & \textbf{25} & \textbf{Multiple finders} \\
\bottomrule
\end{tabular}
\begin{tablenotes}
\item Multilang: \sys-Multilang, C: \sys-C.
\end{tablenotes}
\end{threeparttable}
\caption{Performance breakdown for C harnesses in the AIxCC final competition.}
\label{t:c-harness-performance}
\end{table*}

\begin{table*}[htbp]
\centering
\scriptsize
\begin{threeparttable}
\begin{tabular}{lrrrr>{\raggedright\arraybackslash}p{5cm}}
\toprule
Harness & \multicolumn{3}{c}{PoVs} & Patches & Successful Bug-Finding Modules \\
\cmidrule(lr){2-4}
        & Total & Passed & Dup. & Passed &  \\
\midrule
CompressorGzipFuzzer & 84 & 2 & 25 & 1 & Java \\
DomXfaParserFuzzer & 2 & 1 & 0 & 1 & Java \\
DomXmpParserFuzzer & 3 & 1 & 0 & 1 & Java \\
ExcelImExportServiceFuzzer & 4 & 2 & 0 & 0 & Multilang.given\_fuzzer, Java \\
ExpanderFuzzer & 208 & 6 & 41 & 3 & Multilang.given\_fuzzer, Multilang.shared\_corpus, Multilang.testlang\_input\_gen, Java \\
HtmlFuzzer & 19 & 1 & 0 & 1 & Multilang.given\_fuzzer \\
JsonFuzzer & 10 & 3 & 0 & 2 & Multilang.given\_fuzzer, Multilang.testlang\_input\_gen, Java \\
PDFExtractTextFuzzer & 4 & 1 & 0 & 1 & Java \\
PDFOCRFuzzer & 3 & 2 & 0 & 1 & Java \\
PDFStreamParserFuzzer & 3 & 1 & 0 & 1 & Java \\
PrivateDictionaryFuzzer & 6 & 1 & 0 & 1 & Multilang.mlla.gen \\
SimpleLoggerFuzzer & 2 & 1 & 0 & 1 & Multilang.given\_fuzzer \\
TextAndCSVParserFuzzer & 2 & 1 & 0 & 1 & Java \\
\textit{Other Java harnesses (12)} & 250 & 0 & 16 & 0 & - \\
\midrule
\textbf{Total Java harnesses (25)} & \textbf{600} & \textbf{23} & \textbf{82} & \textbf{15} & \textbf{Multiple finders} \\
\bottomrule
\end{tabular}
\begin{tablenotes}
\item Multilang: \sys-Multilang, Java: \sys-Java
\end{tablenotes}
\end{threeparttable}
\caption{Performance breakdown for Java harnesses in the AIxCC final competition.}
\label{t:java-harness-performance}
\end{table*}

The harness-level data reveals several key insights.
Among C harnesses,
\cc{exif\_from\_data\_fuzzer} (18 passed PoVs) and \cc{curl\_fuzzer\_dict} (13 passed PoVs) 
emerged as the most productive targets,
both benefiting from multiple active bug-finding modules working in combination.
Harnesses with multiple successful modules, such as \cc{exif\_from\_data\_fuzzer},
engaged up to 6 different approaches simultaneously.

Java harnesses showed lower pass rates relative to total PoVs generated.
\cc{ExpanderFuzzer} generated 208 PoVs with 6 passed verification,
while \cc{CompressorGzipFuzzer} generated 84 PoVs with 2 passed.

Based on the available log data, 
\sys generated 1,003 PoVs total,
with 118 passing final verification and 151 identified as duplicates.
Of the harnesses captured in these logs, 23 (32\%) produced no verified vulnerabilities.

\end{appendix}
\end{document}